\documentclass[english,aip,letter,superscriptaddress,twocolumn,floatfix,jcp,nourl,nofootinbib]{revtex4}
\usepackage[utf8]{inputenc}
\usepackage[T1]{fontenc}
\usepackage{upgreek}
\usepackage{graphicx}
\usepackage{ae,aecompl}
\usepackage{color}
\usepackage{pstricks,pst-node}
\usepackage{psfrag}
\usepackage{verbatim}
\usepackage{bm}
\usepackage{pdflscape}
\usepackage{rotating}
\usepackage{stmaryrd}
\usepackage{amsfonts}
\usepackage{amsmath}
\usepackage{amssymb}
\usepackage{bbold}
\usepackage{endnotes}
\usepackage{ifthen}
\usepackage{bm}
\usepackage{setspace}
\usepackage{endnotes}
\usepackage{fancyhdr}
\usepackage{fancybox}
\usepackage{makeidx}
\usepackage{framed}
\usepackage{listings}

\usepackage{alphalph}
\definecolor{shadecolor}{rgb}{0,.7,0}
\definecolor{shadecolor}{gray}{0.7}
\definecolor{shadecolor}{gray}{0.95}
\definecolor{ogreen}{rgb}{0,0.8,0}
\definecolor{magenta}{rgb}{1,0,1}
\definecolor{brown}{rgb}{0.7,0.4,0.2}
\definecolor{shadecolor}{gray}{0.9}

\newcommand{\GK}[2]{\langle#1 || #2\rangle}

\newcommand{\llangle}{\left\langle}
\newcommand{\rrangle}{\right\rangle}
\newcommand{\esc}{\!\cdot\!}

\RequirePackage{dsfont}

\usepackage{xr-hyper}
\usepackage{hyperref}

\graphicspath{{./Figures/}}
\begin{document}

\title{The role of thermal fluctuations in the motion of a free body}

\author{Pep Espa\~nol}
\affiliation{ Dept.   F\'{\i}sica Fundamental, Universidad Nacional
  de Educaci\'on a Distancia, Madrid, Spain}
\author{Mark Thachuk}
\affiliation{ Department of Chemistry, University of British Columbia, Vancouver, Canada}
\author{J.A. de la Torre}
\affiliation{ Dept.   F\'{\i}sica Fundamental, Universidad Nacional
  de Educaci\'on a Distancia, Madrid, Spain}

\date{\today}
\begin{abstract}
 The motion of a rigid body is described in Classical Mechanics with the venerable Euler's equations which are based on the assumption that the relative distances among the constituent particles are fixed in time. Real bodies, however, cannot satisfy this property, as a consequence of thermal fluctuations. We generalize Euler's equations for a free body in order to describe dissipative and thermal fluctuation effects in a thermodynamically consistent way. The origin of these effects is internal, i.e. not due to an external thermal bath. The stochastic differential equations governing the orientation and central moments of the body are derived from first principles through the theory of coarse-graining. Within this theory, Euler's equations emerge as the reversible part of the dynamics. For the irreversible part, we identify two distinct dissipative mechanisms; one associated with diffusion of the orientation, whose origin lies in the difference between the spin velocity and the angular velocity, and one associated with the damping of dilations, i.e. inelasticity. We show that a deformable body with zero angular momentum will explore uniformly, through thermal fluctuations, all possible orientations. When the body spins, the equations describe the evolution towards the alignment of the body's major principal axis with the angular momentum vector. In this alignment process, the body increases its temperature. We demonstrate that the origin of the alignment process is not inelasticity but rather orientational diffusion. The theory also predicts the equilibrium shape of a spinning body.
\end{abstract}
\maketitle
\section{Introduction}

Understanding  how a  body  moves in  space is  a  central problem  in
Classical Mechanics, with strong  implications in astrophysics ranging
from  the  dynamics  of   interstellar  dust  \cite{Lazarian2007},  to
asteroid                                                      dynamics
\cite{Efroimsky2000,Efroimsky2001,Warner2009,Breiter2012,Frouard2018}
and  the  motion  of   the  Earth  \cite{Souchay2003,Chen2010},  among
others. At  the scale of  the laboratory, the field  of levitodynamics
aims  to   control  levitated   nano-  and  micro-objects   in  vacuum
\cite{Gonzalez-Ballestero2021}.    These  systems   provide  exquisite
sensors      for      torques,     forces,      and      accelerations
\cite{Prat-Camps2017,Ricci2019}, as  well as help  unravel fundamental
questions  in quantum  physics \cite{Millen2020},  and the  stochastic
thermodynamics of small systems  \cite{Gieseler2014}.  Given the small
size  of   the  bodies   involved,  understanding  the   interplay  of
dissipation  and  thermal fluctuations  is  a  crucial issue,  treated
phenomenologically so far \cite{VanDerLaan2020}.

\begin{figure}[h]
    \includegraphics[width=\columnwidth]{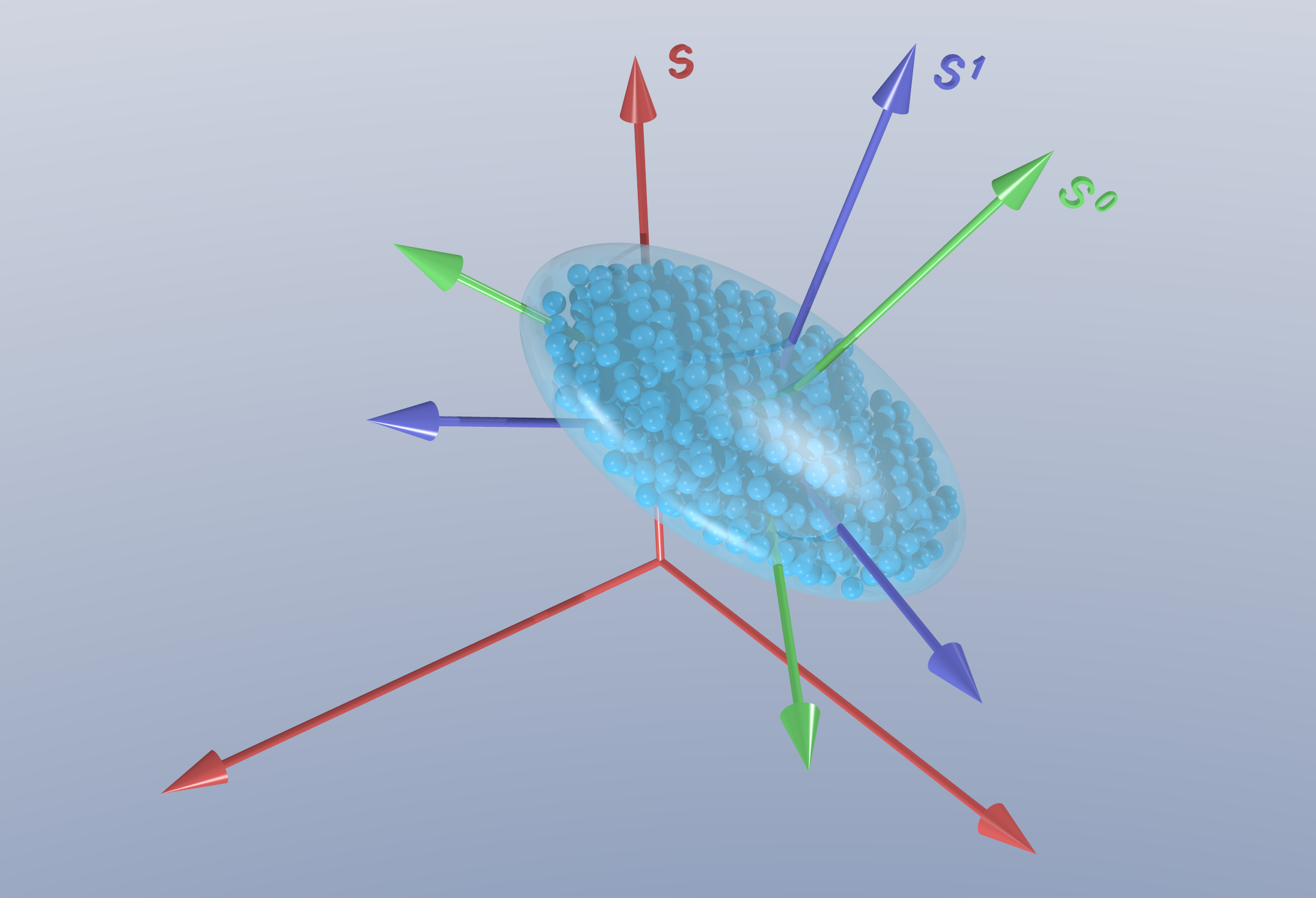}
    \caption{A body made of moving  bonded atoms (little blue spheres)
      is  coarse-grained with  the gyration  tensor, represented  here
      with a  translucent ellipsoid. Also  shown are the  inertial lab
      frame ${\cal  S}$ (red),  the principal axis  non-inertial frame
      ${\cal S}^0$ (green), and the zero angular momentum non-inertial
      frame ${\cal S}^1$ (blue).}
\label{Fig1}
\end{figure}

The  starting point  for  describing the  motion of  a  rigid body  in
Classical  Mechanics  are  Euler's  equations.  A  rigid  body  is  an
idealization  for a  system of  classical particles  that assumes  the
relative    distances    among    them   are    constant    in    time
\cite{Goldstein1983,Arnold1989}.   Under  this assumption,  there  are
reference  systems in  which all  the  particles are  at rest.   Euler
obtained the  equations from  the relationship between  two orthogonal
reference systems in  motion, the observation that  the inertia tensor
diagonalizes  in   the  principal  axis  reference   system,  and  the
conservation   laws,   in   particular  that   of   angular   momentum
\cite{Descamps2008,Gautschi2008}.

In   reality  the   particles   in  a   solid   object  (sketched   in
Fig.~\ref{Fig1}) at finite temperature are moving and oscillating very
rapidly around their equilibrium positions  and have no fixed relative
distances.  The dynamics  of the particles is  governed classically by
Hamilton's equations.   To our  knowledge, the conceptual  gap between
Hamilton's equations for the particles of a body and Euler's equations
for a  rigid body  has not yet  been closed.  We  aim at  deriving the
equations for a  rigid body directly from Hamilton's  equations of the
evolving  interacting  particles  that constitute  the  body,  without
recourse to the rigid body idealization.   This is not only a pleasant
intellectual achievement,  but also  sheds light  on the  behaviour of
realistic   systems  where   the  rigid   body  idealization   is  not
accurate.  As an  example,  we identify  a  dissipative mechanism  not
considered so far, that of orientational diffusion, which turns out to
be the one actually responsible  for the alignment of rotating bodies.
Also, as thermal fluctuations are  important in molecular systems, and
on many occasions molecules or portions of molecules \cite{Goujon2020}
are  treated  as rigid  entities,  a  clarification of  the  interplay
between  rigidity  and  thermal  fluctuations is  important  in  these
systems.  In fact, the  modelling of a body made of  atoms in terms of
constraints  (like constant  distances  between atoms)  is subtle  and
problematic \cite{Fixman1974,Fixman1978,vanKampen1981,Espanol2011b}.

A number  of works  have treated orientation  and shape  variables for
deformable         bodies        in         Classical        Mechanics
\cite{Guichardet1984,Shapere1989,Littlejohn1997}.    They   coordinate
transform the $n$-body problem into two distinct sets that capture the
overall orientation and  the shape.  In this approach,  the freedom in
choosing  the  body  frame  leads  to  gauge  invariance,  and  to  an
interesting  mathematical  structure \cite{Shapere1989}.   The  theory
allows one to understand how a closed  path in shape space leads to an
overall  rotation,  explaining how  a  falling  cat can  maneuver  its
internal degrees of freedom to land  on its feet, or how a torque-free
trampolinist can perform turns in the air.  The fact a deformable body
may  rotate in  the  absence  of angular  momentum  has been  observed
recently in  very small molecular systems  \cite{SaportaKatz2019}, and
has   also   provided   examples   of   a   classical   time   crystal
\cite{Peng2021}.  However,  these microscopic  theories do  not obtain
closed equations  for the  ``macroscopic shape''  as captured  by, for
example, the principal moments of the body.

A  different approach  is given  from the  point of  view of  rational
continuum mechanics  in the  so called  theory of  pseudo-rigid bodies
\cite{Cohen1988},\cite{OReilly2003}.   In  this   theory,  a  body  is
characterized as a zero dimensional directed continuum consistent with
non-linear  elasticity theory.   In that  respect such  a theory  is a
coarse version of a continuum  theory. Being formulated at an abstract
level,  the  statistical  mechanics  underpinning  of  the  theory  of
pseudo-rigid bodies  is lacking. Dissipation  is also lacking  in this
approach.

There are  recent attempts to introduce  stochasticity and dissipation
in rigid body dynamics, from the  point of view of geometric mechanics
\cite{Arnaudon2017,Arnaudon2018}.   In  this  mathematically  oriented
approach, a  stochastic dynamics  for the  angular momentum  vector is
proposed.  Physically, this means the solid body is interacting with a
bath of particles, otherwise, the angular momentum would be conserved.
In fact, the  literature on so-called Brownian rotors,  that is, rigid
bodies  immersed  in a  fluid  and  experiencing thermal  fluctuations
arising   from  collisions   with   fluid   molecules,  is   extensive
\cite{Galkin2008,Walter2010,Shrestha2015,Martinetz2018},  dating  back
to the relaxation model of Debye.  The thermal fluctuations considered
in the  present paper are,  however, of  a very different  nature from
those of Brownian rotors. They do not arise from an external bath, but
from  eliminated degrees  of  freedom  of the  body  in  favor of  the
coarse-grained (CG)  variables. The  dissipation is  ``internal'', the
body is isolated, and angular  momentum and total energy are conserved
in our work.

The way we  proceed in order to  derive the equations of  motion for a
quasi-rigid  body is  based on  non-equilibrium statistical  mechanics
\cite{Green1952,Zwanzig1961,Grabert1982}.  This is  a theory of coarse
graining  that  provides  the  tools  for the  reduction  of  all  the
information about rapidly varying atomic  variables to a few variables
governing  the behaviour  of the  system as  a whole.   Typically, the
selected coarse-grained (CG)  variables are assumed to be  slow on the
atomic time scale, thus allowing  for a description of their evolution
as an overall  slow motion with superimposed  small rapid fluctuations
modelled as white  noise.  The theory of  coarse-graining provides the
Fokker-Planck Equation (FPE) for  the evolution of the non-equilibrium
probability  distribution  of the  CG  variables,  and its  associated
stochastic differential equation (SDE).  The drift and diffusion terms
of the FPE and SDE are given in terms of conditional expectations over
the  microscopic  degrees  of   freedom  {and  are,  therefore,  fully
  expressed in microscopic terms}.  Thermal fluctuations are naturally
described  in  this  approach  and,  concomitant  with  them,  the  CG
description fully describes dissipation which is in agreement with the
Second Law.  In the limit of small fluctuations, valid for macroscopic
bodies, the approach leads to deterministic equations.

In this work, we coarse-grain a free body by using the eigenvalues and
eigenvectors  of the  gyration tensor  as the  coarse variables,  from
which the stochastic differential  equations governing the dynamics of
the orientation and central moments of a free body are derived.  These
equations generalize Euler's equations for a free solid body by taking
into account deformation, dissipation,  and thermal fluctuations.  The
original Euler's equations turn out to  be just the reversible part of
the whole dynamics.  Therefore, to our knowledge, this constitutes the
first derivation  of Euler's equations from  Hamilton's equations.  We
make  explicit two  crucial  but questionable  assumptions in  Euler's
treatment.  First,  we distinguish between the  spin velocity, related
to  the angular  momentum, and  the angular  velocity, related  to the
rotation  of  the principal  axis  frame.   These two  quantities  are
assumed to coincide in a rigid body,  but they are different in a real
body.  This difference is at the heart of orientational dissipation in
the system.  Second,  we describe an additional  source of dissipation
due to the unavoidable deformation (non-constant principal moments) of
a real body that can be identified with viscoelasticity.

The SDE obtained in this work predicts the rotating body will follow a
Brownian  motion  that directs  the  orientation  towards the  maximum
entropy state, in  accordance with the Second Law. In  this state, the
body's  principal  axis  with  the  largest  moment  aligns  with  the
conserved  angular  momentum vector,  an  effect  known as  precession
relaxation   \cite{Frouard2018}   and   also   as   nutation   damping
\cite{Sharma2005}.    This  effect   is  predicted   from  equilibrium
thermodynamics (see \S 26 of  Ref.  \cite{Landau1980}) but the present
theory describes  the \textit{evolution}  towards the  maximum entropy
equilibrium state.  The need to include dissipation in the dynamics of
rotating bodies arises very explicitly  in the description of rotating
astronomical   objects   like   interstellar  dust,   asteroids,   and
satellites.  The vast majority of  asteroids, for example, are in pure
rotation \cite{Lamy1972,Warner2009,Breiter2012}.   Previous approaches
describing  the alignment  process  and  estimating the  corresponding
relaxation  times are  based  on the  idea  that inelastic  relaxation
arises from  alternating elastic stresses generated  inside a wobbling
body by  the transversal  and centripetal  acceleration of  its parts.
The alternate  stresses deform the  body, and inelastic  effects cause
energy                dissipation               \cite{Prendergast1958,
  Efroimsky2002,Sharma2005,Sharma2017,Kwiecinski2020}.  At the coarser
level  of description  selected  in  the present  work,  based on  the
gyration tensor, the alignment of a free body towards pure rotation is
captured through orientational diffusion, while dilational friction --
that would represent viscoelasticity at this level of description-- is
seen to play no role in the mechanism of precession relaxation.

The  paper  is structured  as  follows.   In Sec.~\ref{Sec:Review}  we
review the Classical Mechanics for the  motion of a free body from the
point of  view of  Hamiltonian dynamics.   This allows  us to  set the
notation  and  pinpoint subtle  issues  appearing  in the  rigid  body
idealization, as  elaborated in  Sec.~\ref{Sec:Rig.Sol} where  we also
announce   the   generalization   of   Euler's   equations   including
dissipation.   In  Sec.~\ref{sec:theo}  we  summarize  the  theory  of
coarse-graining  used in  the present  work.  Section~  \ref{Sec:free}
discusses the CG  variables used at the present  level of description.
Sections~\ref{Sec:Entropy}-\ref{Sec:Noise}   present   the   different
building blocks entering  into the SDE governing the motion  of a free
solid body.  We are able to formulate all the terms in such a way that
they are either  analytically known, can be computed  explicitly by MD
simulations, or  eventually fitted from observations  of real systems.
The final  form of the SDE  is given in Sec.~\ref{Sec:FinSDE},  with a
summary of the approximations in Sec.~\ref{Sec:Summary}.  The physical
interpretation   of    the   different    terms   is    presented   in
Sec.~\ref{Sec:Det}, and  discussion and  conclusions are  presented in
Sec.~\ref{Sec:Conclusions}.    The   calculations  involved   in   the
evaluation  of   the  different  building  blocks,   along  with  some
instrumental  mathematical  results,  are  presented  as  Supplemental
Material  \cite{SM}.

\section{Review  of Hamiltonian  dynamics for a body  made of  bonded particles}
\label{Sec:Review}
In order to set the notation,  in this section we review the Classical
Mechanics for a system of  particles interacting with an interparticle
potential but otherwise free from external forces.

Let $z=\{r,p\}=\{{\bf  r}_i,{\bf p}_i,  i=1,\cdots,N\}$ be the  set of
microscopic degrees of  freedom for a body with  $N$ particles, where
${\bf  r}_i$ is  the  position of  particle $i$  and  ${\bf p}_i$  its
momentum.   We will  refer to  $z$ as  the microstate  and $r$  as the
configuration. The  set of all  $z$ constitutes the phase  space.  The
degrees  of  freedom $z$  are  defined  with  respect to  an  inertial
reference frame  ${\cal S}$ in  which Hamilton's equations  are valid.
We assume the Hamiltonian of the system to be of the usual form
\begin{align}
  \label{eq:1}
  \hat{H}(z)&=\sum_i^N\frac{{\bf p}_i^2}{2m_i}+\hat{\Phi}(r)
\end{align}
with          the           potential          of          interaction
$\hat{\Phi}(r)=\hat{\Phi}({\bf r}_1,\cdots,{\bf r}_N)$.  In this work,
circumflexed symbols denote functions in phase space.  The Hamiltonian
is a dynamic  invariant, as are the linear and  angular momenta of the
system. When  the total linear  momentum is  zero, the center  of mass
position is also a dynamic invariant.

We introduce  the  first  geometrical  \textit{moments}  of  the
distribution  of   particles, which are   the  total  mass
${\rm M}$, the  center of mass position $\hat{\bf R}$,  and the tensor
of inertia $\hat{\bf I}$ of the body.  These phase functions depend on
the position of the particles but not on their momenta. They are given
by
\begin{align}
{\rm M}&\equiv\sum^{N}_{i}m_{i}
\nonumber\\
  \hat{\bf R} &\equiv 
\frac{1}{M}\sum^{N}_{i}m_{i}{\bf r}_{i}
\nonumber\\
\hat {\bf I} &\equiv  \sum^{N}_{i}m_{i}
[{\bf r}_{i}-\hat{\bf R}]_\times^T
\esc[{\bf r}_{i}-\hat{\bf R}]_\times
\label{Geomoments}
\end{align}
where  the  superscript  $T$  denotes  the  matrix  transpose.  The  cross product matrix  $[{\bf A}]_\times$ is defined  as the
antisymmetric     matrix      constructed     from      an arbitrary     vector
${\bf A}=({\bf A}^x,{\bf A}^y,{\bf A}^z)$ as
\begin{align}
[{\bf A}]_\times&=\left(\begin{array}{ccc}
 0  &-{\bf A}^z& {\bf A}^y\\
 {\bf A}^z& 0  &-{\bf A}^x\\
-{\bf A}^y& {\bf A}^x& 0
\end{array}\right)
\label{times}
\end{align}
or                  in                 component                  form
$[{\bf
  A}]_\times^{\alpha\beta}=\sum_\gamma\epsilon^{\alpha\gamma\beta}{\bf
  A}^\gamma$, where $\epsilon^{\alpha\gamma\beta}$  is the Levi-Civita
symbol.  Expression  (\ref{Geomoments})  for  the  inertia  tensor  is
identical to the usual definition given by
\begin{align}
  \hat{\bf I} &= \sum_{i}m_{i}\left[\left({\bf r}_{i}-\hat{\bf R}\right)^2\mathbb{1}
-\left({\bf r}_{i}-\hat{\bf R}\right)\left({\bf r}_{i}-\hat{\bf R}\right)^T\right]
\label{s-Iusual}
\end{align}
where $\mathbb{1}$ is the identity matrix.
However,  form  (\ref{Geomoments})  is  more  convenient  in  explicit
calculations.  It  also shows  manifestly that  the inertia  tensor is
symmetric and  positive definite.

The linear momentum $\hat{\bf P}$ and spin $\hat{\bf S}$ of the system are
\begin{align}
  \hat{\bf P} &\equiv \sum_{i}{\bf p}_{i}
\nonumber\\
  \hat{\bf S} &\equiv \sum_{i}\left({\bf r}_{i}-\hat{\bf R}\right)\times{\bf p}_{i}
\label{PS}
\end{align}
The  spin is  the angular  momentum of  the body  with respect  to the
center of  mass of the  body.  It is  convenient to define  the linear
velocity $\hat{\bf  V}$ and spin  velocity $\hat{\boldsymbol{\Omega}}$
of the body as
\begin{align}
  \hat{\bf V} &\equiv {\rm M}^{-1} \hat{\bf P}
\nonumber\\
  \hat{\boldsymbol{\Omega}} &\equiv \hat{\bf I}^{-1}\esc\hat{\bf S} 
\label{vel}
\end{align}
We refer to $ \hat{\boldsymbol{\Omega}}$  as the spin velocity and not
as the  angular velocity because the  latter name is reserved  for the
angular  velocity $\hat{\boldsymbol{\omega}}$  of  the principal  axis
frame with respect to the lab frame, as defined below. These two phase
functions are different in general.

Throughout  this  work and  in  the  Supplementary Material,  vectors,
matrices or tensors are rendered in bold font ${\bf A}$, with diagonal
matrices  additionally   rendered  with  voided   fonts  $\mathbb{A}$.
Indices representing  Cartesian components of these  quantities appear
in  Greek  font either  as  subscripts  or superscripts,  while  those
representing  particle  number appear  as  subscripts  in Roman  font.
Unless  stated otherwise  or  shown explicitly,  repeated indices  are
assumed  to  be  summed  over,  that is  the  Einstein  convention  is
employed,  unless  the  indices  are  underlined,  in  which  case  no
summation is implied.

\subsection{The orientation and principal moments}
Because the inertia tensor is  a symmetric positive definite matrix it
can be diagonalized. The principal axis system ${\cal S}^0$ is defined
as the reference system with  origin at the center  of mass in
which the inertia  tensor diagonalizes.  Let the basis  vectors of the
inertial laboratory  reference system ${\cal S}$  and the non-inertial
principal   axis  reference   system  ${\cal   S}^0$  be   denoted  by
$\hat{\bf e}_\alpha$,  and $\hat{\bf e}^0_\alpha$,  respectively, with
$\alpha=1,2,3$.   The  components of the rotation   matrix  of
${\cal  S}^0$ with  respect to  ${\cal S}$  are defined  as 
\begin{align}
  \label{eq:142}
    \hat{\boldsymbol{\cal R}}_{\alpha\beta}&={\bf  e}^T_\beta\esc\hat{\bf  e}^0_\alpha
\end{align}
In ${\cal S}^0$ the inertia tensor takes the form
 \begin{align}
\hat{\boldsymbol{\cal R}}\esc\hat{\bf I} \esc {\hat{\boldsymbol{\cal R}}}^{T}&= \hat{\mathbb{I}}
\label{diagonalization}
\end{align}
where $\hat{\mathbb{I}}$ is  a diagonal matrix whose  elements are the
principal moments $\hat{I}_1,\hat{I}_2,\hat{I}_3$. Note  that $\hat{\bf  e}^0_\alpha$ are  the normalized
eigenvectors  of   the  inertia   tensor. 

The  rotation  matrix  can  be  expressed in terms of the exponential matrix
\begin{align}
  \label{eq:206}
\hat{\boldsymbol{\cal R}}=e^{-[\hat{\boldsymbol{\Lambda}}]_\times}
\end{align}
Properties   of   the  rotation   matrix   are   summarized  in   Sec.
J  of  the  Supplemental  Material.   The
parameters   $\hat{\boldsymbol{\Lambda}}$    in   (\ref{eq:206})   are
sometimes  referred to  as  the Euler  vector,  or attitude  parameter
\cite{Olguin2019},  and the  resulting  representation as  exponential
coordinates   \cite{Gallego2015}.    We    will   usually   refer   to
$\hat{\boldsymbol{\Lambda}}$  as  the orientation.   The  conventional
minus  sign  in  the  definition  of  $\hat{\boldsymbol{\Lambda}}$  in
(\ref{eq:206})  leads   to  more  natural  expressions   relating  the
orientation  and  the  angular  velocity later  on.   The  orientation
$\hat{\boldsymbol{\Lambda}}$  allows one  to introduce  the angle/axis
representation  of   rotations  \cite{Olguin2019}.   By   writing  the
orientation as $\hat{\boldsymbol{\Lambda}}=\hat{\Lambda} \hat{\bf n}$,
where the  modulus $\hat{\Lambda}=|\hat{\boldsymbol{\Lambda}}|$  is an
angle and $\hat{\bf n}$ is a unit vector, Rodrigues' formula expresses
the rotation matrix in the angle/axis representation as
\begin{align}
  \label{eq:143}
\hat{\boldsymbol{\cal R}}&=\mathbb{1}-\sin{\hat{\Lambda}} [\hat{\bf n}]_\times+\left(1-\cos{\hat{\Lambda}}\right) [\hat{\bf n}]_\times[\hat{\bf n}]_\times
\end{align}
 The
orientation $\hat{\boldsymbol{\Lambda}}$ is invariant under the action
of the particular rotation 
\begin{align}
  \label{eq:432}
\hat{\boldsymbol{\cal R}}\esc  \hat{\boldsymbol{\Lambda}}=\hat{\boldsymbol{\Lambda}}
\end{align}
as is easily  seen from  Eq. (\ref{eq:143}).   This allows  one to
interpret  $\hat{\boldsymbol{\Lambda}}$ as  defining  a rotation  axis
around which a  rotation of the lab frame  leads to the principal
axis  of the  body. We  choose the  convention that a  rotation
matrix  is   given  by   the  anticlockwise   rotation  of   an  angle
$0\le\Lambda\le\pi$ around the unit vector ${\bf n}$.

Due to the trigonometric  functions, the relationship between rotation
matrices  and attitude  parameters  is not  one-to-one.  According  to
(\ref{eq:143}),            the             two            orientations
$\boldsymbol{\Lambda}'=(\Lambda+\pi){\bf            n}$            and
$\boldsymbol{\Lambda}''=(\Lambda-\pi){\bf n}$  give the  same rotation
matrix.  This means that for every point $\boldsymbol{\Lambda}$ within
a sphere  of radius $\pi$  we have  a unique rotation,  whereas points
outside  this   sphere  correspond  to  rotations   that  are  already
represented by  a point  within the  sphere.  Note  also that  the two
antipodal points $\pm\pi{\bf n}$ on  the surface of  the sphere
 also  give the  same rotation.   With antipodal
points identified,  any point $\boldsymbol{\Lambda}$ in  the sphere of
radius $\pi$ gives a unique  rotation, and any rotation is represented
by one point in this sphere.

From  (\ref{diagonalization}), the  inertia  tensor  can be  expressed
uniquely in terms  of the orientation $\hat{\boldsymbol{\Lambda}}$
and the principal moment matrix $ \hat{\mathbb{I}}$ as 
\begin{align}
  \hat{\bf I}&=e^{[\hat{\boldsymbol{\Lambda}}]_\times}\esc
\hat{\mathbb{I}}\esc e^{-[\hat{\boldsymbol{\Lambda}}]_\times}
\label{ID}
\end{align}
This expression relates the six  independent elements of the symmetric
inertia    tensor   $\hat{\bf    I}$,   with    the   three    numbers
$\hat{\boldsymbol{\Lambda}}$ giving the orientation  of the body and
the three principal moments.  Both $\hat{\boldsymbol{\Lambda}}$ and
$\hat{\mathbb{I}}$ are constructed  from the inertia tensor,
and   they  are   phase   functions  depending   on  the   microscopic
configuration through the positions of the particles of the body, that
is,
\begin{align}
  \hat{\boldsymbol{\Lambda}}
  &=\hat{\boldsymbol{\Lambda}}(\{{\bf r}_{i}\})
\nonumber\\
  \hat{\mathbb{I}}&=\hat{\mathbb{I}}(\{{\bf r}_{i}\})
\end{align}

Observe  that the  inertia  tensor  in (\ref{ID})  is  denoted with  a
circumflexed symbol  because it  is a phase  function. We  may express
(\ref{ID}) in the form
\begin{align}
  \label{eq:52}
  \hat{\bf I}&={\bf I}(\hat{\boldsymbol{\Lambda}},\hat{\mathbb{I}})
\end{align}
where the matrix valued function (with no circumflex symbol) is
\begin{align}
  \label{eq:403}
  {\bf      I}(\boldsymbol{\Lambda},\mathbb{I})
  =e^{[\boldsymbol{\Lambda}]_\times}\esc\mathbb{I}\esc
e^{-[\boldsymbol{\Lambda}]_\times}  
\end{align}

\subsection{The angular velocity}
The coordinates ${\bf  r}_{0i}$ of  a particle in   ${\cal  S}^0$ and the
coordinates ${\bf r}_i$ of the same particle in  ${\cal S}$ are related
by
\begin{align}
  \label{eq:74}
  {\bf r}_{0i}&=\hat{\boldsymbol{\cal R}}\esc({\bf r}_i-\hat{\bf R})
\end{align}
The relationship between the particle  velocity ${\bf v}_i$ in the lab
frame ${\cal S}$ and the velocity  ${\bf v}_{0i}$ in the principal axis
frame ${\cal S}^0$ follows from (\ref{eq:74}) through differentiation
\begin{align}
  \label{eq:41}
  {\bf v}_{0i}&\equiv\frac{d{\bf r}_{0i}}{dt}=\hat{\boldsymbol{\cal R}}\esc\left( {\bf v}_i-\hat{\bf V}-[\hat{\boldsymbol{\omega}}]_\times\esc({\bf r}_i-\hat{\bf R})\right)
\end{align}
where the  antisymmetric angular velocity  matrix is defined  from the
rotation  matrix  as
\begin{align}
  [\hat{\boldsymbol{\omega}}]_\times
  &\equiv-{\hat{\boldsymbol{\cal R}}}^{T}\esc\frac{d{\hat{\boldsymbol{\cal R}}}}{dt}
    =\frac{d{\hat{\boldsymbol{\cal R}}}^{T}}{dt}\esc{\hat{\boldsymbol{\cal R}}}
\label{omega1}
\end{align}
where the axial vector $\hat{\boldsymbol{\omega}}$ is the angular velocity of ${\cal S}^0$
as viewed from ${\cal S}$. For future reference, we introduce the following matrix
  \begin{align}
      [\hat{\boldsymbol{\omega}}_0]_\times
  &\equiv{\hat{\boldsymbol{\cal R}}}\esc\frac{d{\hat{\boldsymbol{\cal R}}}^{T}}{dt}
    =-\frac{d{\hat{\boldsymbol{\cal R}}}}{dt}\esc{\hat{\boldsymbol{\cal R}}}^T
\label{omega0}
  \end{align}
  that satisfies
  \begin{align}
    \label{eq:348}
      [\hat{\boldsymbol{\omega}}_0]_\times&=\hat{\boldsymbol{\cal R}}\esc      [\hat{\boldsymbol{\omega}}]_\times\esc \hat{\boldsymbol{\cal R}}^T
  \end{align}
  which shows the angular velocity vectors are related through
  \begin{align}
    \label{eq:4}
       \hat{\boldsymbol{\omega}}_0&=\hat{\boldsymbol{\cal R}}\esc      \hat{\boldsymbol{\omega}}
= e^{-[{\hat{\boldsymbol{\Lambda}}}]_\times}\esc\hat{\boldsymbol{\omega}}
  \end{align}
  As this  is the  way a  vector transforms under  a rotation,  we may
  interpret  $ \hat{\boldsymbol{\omega}}_0$  as  the angular  velocity
  referred to the  principal axis frame, that is  the angular velocity
  of $\cal S$ as viewed from ${\cal S}^0$.

The angular velocity vector is  closely related to the time derivative
of  the orientation  $\frac{d\hat{\boldsymbol{\Lambda}}}{dt}$.  As  we
show  in  Sec. K.3 of  the  Supplemental  Material, the explicit connection is given by
\begin{align}
  \label{eq:29}
  \hat{\boldsymbol{\omega}}
  &={\bf B}^{-1}(\hat{\boldsymbol{\Lambda}})  \esc \frac{d{\hat{\boldsymbol{\Lambda}}}}{dt}
\\
  \label{eq:29b}  \frac{d{\hat{\boldsymbol{\Lambda}}}}{dt}
  &={\bf B}(\hat{\boldsymbol{\Lambda}})  \esc   \hat{\boldsymbol{\omega}}
    \end{align}
where the matrices are
\begin{align}
  \label{eq:53}
  {\bf B}(\boldsymbol{\Lambda})
  &=\mathbb{1}+{p} [{\bf n}]_\times+{q}[{\bf n}]_\times[{\bf n}]_\times
    \nonumber\\
  {\bf B}^{-1}({\boldsymbol{\Lambda}})
  & =\mathbb{1}+{a} [{\bf n}]_\times+{b}[{\bf n}]_\times[{\bf n}]_\times
\end{align}
where  $a,b,p,q$ are the following functions of the modulus of  the orientation ${\Lambda}=|{\boldsymbol{\Lambda}}|$ 
\begin{align}
      \label{eq:306}
{a}&=\left(\frac{1-\cos{{\Lambda}}}{{\Lambda}}\right),
&
{b}=\left(1-\frac{\sin{\Lambda}}{{\Lambda}}\right)
     \nonumber\\
{p}&=-\frac{{\Lambda}}{2},
   &
{q}=1-\frac{{\Lambda}}{2}\frac{\sin{\Lambda}}{(1-\cos{\Lambda})}
\end{align}
The matrix ${\bf B}(\boldsymbol{\Lambda})$  is referred to as Attitude
Kinematic Operator in Ref. \cite{Olguin2019}.
In Sec. L of the  Supplemental Material, we
show that  the list  of three numbers  $\hat{\boldsymbol{\Lambda}}$ is
not a vector as it does not transform as such.

\section{Euler's equations for a rigid body}
\label{Sec:Rig.Sol}
When the particles  of the body remain strongly bonded, the body
may be considered as quasi-rigid. In  this case, we may derive Euler's
equation for  a rigid body in  a way that unveils  the two assumptions
 implicitly taken when  deriving them.

 Let ${\cal  S}^1$ be a  non-inertial reference system with  origin at
 the center of mass  of the body in which the  angular momentum of the
 body vanishes.   As we show below,  ${\cal S}^1$ can always  be found
 and it  is, in  principle, different from  the principal  axis system
 ${\cal  S}^0$,  as shown  schematically  in  Fig. \ref{Fig1}.   Using
 equations analogous  to (\ref{eq:74})  and (\ref{eq:41}),  along with
 (\ref{Geomoments}), the angular momentum  $\hat{\bf S}_1$ of the body
 with  respect to  ${\cal  S}^1$  is given  in  terms  of the  angular
 momentum $\hat{\bf S}$ in ${\cal S}$ as
\begin{align}
  \label{eq:111}
 \hat{\bf S}_1\equiv\sum_im_i{\bf r}_{1i}\times{\bf v}_{1i}=
\hat{\boldsymbol{\cal R}}_1\esc\left[\hat{\bf S}-\hat{\bf I}\esc {\hat{\boldsymbol{\omega}}}_1\right]
\end{align}
where ${\bf r}_{1i},{\bf v}_{1i}$ are the position and velocity of particle $i$ in ${\cal S}^1$, and $\hat{\boldsymbol{\cal  R}}_1$ is  the  rotation  matrix that  brings
$\cal S$ to ${\cal S}^1$.     If     we     choose
$\hat{\boldsymbol{\omega}}_1=\hat{\boldsymbol{\Omega}}$,   then   from
(\ref{vel}) we have that the  angular momentum in
${\cal  S}^1$ vanishes,  $  \hat{\bf S}_1=0$.   Therefore, the  spin
velocity $\hat{\boldsymbol{\Omega}}$ introduced  in (\ref{vel}) is the
angular  velocity $\hat{\boldsymbol{\omega}}_1$  (with respect  to the
lab frame) of  the reference system ${\cal S}^1$ in  which the angular
momentum  of the  body  vanishes.   Note that  we  can  always find  a
rotation   matrix   $\hat{\boldsymbol{\cal   R}}_1$  for   which   the
corresponding   angular   velocity  $\hat{\boldsymbol{\omega}}_1$   is
prescribed  to be $\hat{\boldsymbol{\Omega}}$ which is given in  terms   of
$\hat{\bf S},\hat{\bf  I}$ through (\ref{vel}).  In fact, from  definition (\ref{omega1})
we have
\begin{align}
  \label{eq:109}
  \frac{d\hat{\boldsymbol{\cal R}}_1^{T}}{dt}(t)
  &=[\hat{\boldsymbol{\Omega}}(t)]_\times\esc\hat{\boldsymbol{\cal R}}_1^{{T}}(t)
\end{align}
which can be integrated with the help of the time-ordered exponential to give
\begin{align}
  \label{eq:110a}
  \hat{\boldsymbol{\cal R}}_1^{{T}}(t)
  &=\exp_+\left\{\int_0^t [\hat{\boldsymbol{\Omega}}(t')]_\times dt'\right\}
    \hat{\boldsymbol{\cal R}}_1^{{T}}(0)
\end{align}
where     we      assume     that     at     the      initial     time
$\hat{\boldsymbol{\cal    R}}_1^{T}(0)$     is    known.     Equation
(\ref{eq:110a})  gives   the  rotation   matrix  from  ${\cal   S}$  to
${\cal        S}^1$,       given        the       spin        velocity
$\hat{\boldsymbol{\Omega}}(t)$.

From  the  general  transformation  rule  (\ref{eq:111})  for  angular
momentum we also  have the connection between the  angular momentum in
frames ${\cal S}^0$ and ${\cal S}$, that is
\begin{align}
  \label{eq:270}
  \hat{\bf S}
  =\hat{\bf I}\esc\hat{\boldsymbol{\omega}}+\hat{\boldsymbol{\cal R}}^{T}\esc\hat{\bf S}_0
\end{align}
where the angular momentum in the principal axis system is
  \begin{align}
    \label{eq:237}
    \hat{\bf S}_0&\equiv\sum_im_i{\bf r}_{0i}\times{\bf v}_{0i}
  \end{align}
where ${\bf r}_{0i},{\bf v}_{0i}$ are the position and velocity of particle $i$ in ${\cal S}^0$.
Therefore, from  (\ref{vel}) and  (\ref{eq:270}) we
have the following relationship between the spin
velocity $\hat{\boldsymbol{\Omega}}$ and the 
angular velocity $\hat{\boldsymbol{\omega}}$
\begin{align}
  \label{eq:123a}
\hat{\boldsymbol{\Omega}}=\hat{\boldsymbol{\omega}}+\hat{\bf I}^{-1}\esc\hat{\boldsymbol{\cal R}}^{T}\esc\hat{\bf S}_0  
\end{align}
Because  the angular  momentum in  the principal  axis frame  does not
vanish in general, i.e. $\hat{\bf  S}_0 \neq 0$, (\ref{eq:123a}) shows
 the principal  axis  frame  ${\cal S}^0$  and  the zero  angular
momentum frame  ${\cal S}^1$ rotate with  different angular velocities
$\hat{\boldsymbol{\omega}},\hat{\boldsymbol{\Omega}}$ with  respect to
the  lab  frame  ${\cal  S}$,  respectively.   As  we  will  see,  the
difference between  these two  reference frames and  the corresponding
``angular                                                 velocities''
$\hat{\boldsymbol{\omega}},\hat{\boldsymbol{\Omega}}$ is one source of
dissipation.

It proves  convenient to introduce  the spin velocity relative  to the
principal  axis  frame  $\hat{\boldsymbol{\Omega}}_0$  by  analogy  to
(\ref{vel})
  \begin{align}
    \label{eq:86}
    \hat{\boldsymbol{\Omega}}_0&\equiv \hat{\mathbb{I}}^{-1}\esc\hat{\bf S}_0
  \end{align}
  {with the inertia tensor and the angular momentum referred to the principal axis frame.}
Using (\ref{eq:86}) and (\ref{ID}), (\ref{eq:123a}) may be written as
\begin{align}
  \label{eq:160}
  \hat{\boldsymbol{\Omega}}&=\hat{\boldsymbol{\omega}}+\hat{\boldsymbol{\cal R}}^{T}\esc\hat{\boldsymbol{\Omega}}_0
\end{align}
It is also convenient to introduce the spin velocity rotated to the principal axis frame
\begin{align}
  \label{eq:160a}
  \hat{\boldsymbol{\Omega}}_p
  &\equiv  \hat{\boldsymbol{\cal R}}\esc\hat{\boldsymbol{\Omega}}
\end{align}
which, by using (\ref{eq:4}) and (\ref{eq:160}) implies
\begin{align}
  \label{eq:12}
\hat{\boldsymbol{\Omega}}_p
&=\hat{\boldsymbol{\omega}}_0+\hat{\boldsymbol{\Omega}}_0 
\end{align}
In turn, (\ref{eq:160a}) renders  (\ref{eq:270}) into the form
\begin{align}
  \label{eq:200}
  \hat{\bf S}=\hat{\boldsymbol{\cal R}}^T\esc \hat{\mathbb{I}}\esc\hat{\boldsymbol{\Omega}}_p
\end{align}
This equation  forms  the basis  of Euler's  equations for  rigid body
dynamics.  The  usual  derivation  of  Euler's  equation  is  just  a
statement  of  angular  momentum  conservation.  By  taking  the  time
derivative of (\ref{eq:200}), and noting $\hat{\bf S}$ is conserved we
obtain
\begin{align}
  \label{eq:13b}
  0&= \left[\hat{\boldsymbol{\omega}}_0\right]_\times\esc
     \hat{\mathbb{I}}\esc\hat{\boldsymbol{\Omega}}_p 
     +\frac{d\hat{\mathbb{I}}}{dt}\esc\hat{\boldsymbol{\Omega}}_p 
     +  \hat{\mathbb{I}}\esc \frac{d     \hat{\boldsymbol{\Omega}}_p }{dt}
\end{align}

Equation (\ref{eq:13b}) is not yet  Euler's equation.  It  looks like  a differential  equation for 
$\hat{\boldsymbol{\Omega}}_p$ but it is,  in fact, just a relationship
between         three        different         phase        functions:
$\hat{\boldsymbol{\Omega}}_p,     \hat{\boldsymbol{\omega}}_0$,    and
$  \hat{\mathbb{I}}$.  It  is,  therefore, not  a closed  differential
equation and  is of limited value  unless a closure is  proposed.  The
rigid  body idealization  is  such   a  closure,  and  involves  two
assumptions.

The first assumption  ${\cal H}1$ is that the  principal moment matrix
$\hat{\mathbb{I}}$,  which  is  a  phase  function  depending  on  the
configuration of the  body, can be approximated by  a constant matrix $\overline{\mathbb{I}}$,
no longer dependent on the configuration,
\begin{align}
  \label{eq:95a}
{\cal H}1 \Rightarrow\quad\hat{\mathbb{I}}(z)\simeq \overline{\mathbb{I}}
\end{align}
The  second  assumption  ${\cal  H}2$ is  that  the  angular  momentum
$\hat{\bf S}_0$  with respect to  the principal axis  reference system
${\cal S}^0$ is  zero, $\hat{\bf S}_0=0$.  The usual  argument is that
for  a rigid  body  ``particles do  not move  in  the principal  axis
system''  and therefore,  $\hat{\bf  S}_0=0$ and $\hat{\boldsymbol{\Omega}}_0=0$.   This implies,  through
(\ref{eq:123a}),                                                   that
\begin{align}
  \label{eq:14}
{\cal H}2 \Rightarrow \quad \hat{\boldsymbol{\Omega}}=\hat{\boldsymbol{\omega}}  
\end{align}
 and that the two
reference systems ${\cal S}^0$ and ${\cal S}^1$ coincide, up to a time
independent   rotation   matrix.     With   these   two   assumptions,
(\ref{eq:13b}) becomes
\begin{align}
\label{eq:148a}
\overline{\mathbb{I}}\esc \frac{d\hat{\boldsymbol{\omega}}_0}{dt}
&= -\left[\hat{\boldsymbol{\omega}}_0\right]_\times\esc
                             \overline{\mathbb{I}}
                             \esc\hat{\boldsymbol{\omega}}_0
\end{align}
which  are  precisely   the  equations  obtained  by   Euler  for  the
description of  the free rigid  body.  This  vector equation is  now a
closed  differential equation  for $\hat{\boldsymbol{\omega}}_0$  that
can  be solved  with  appropriate initial  conditions.   Once we  know
$\hat{\boldsymbol{\omega}}_0$,  we  may   use   (\ref{omega0})  as   a
differential equation for  the rotation matrix which in turn  can be integrated
in a way similar to (\ref{eq:110a}), thus fully solving the problem of
the motion of a rigid  body.

An alternative path to Euler's equations 
formulates a differential  equation not for the  angular velocity, but
for  the  orientation itself,  bypassing  the  need for  an  additional
integration  (see the  discussion  on pg.   548 of  \cite{Gregory2006}
about  the ``deficiency''  of  Euler's  equations).  Under  assumption
(\ref{eq:14}), the kinematic condition (\ref{eq:29b}) takes the form
\begin{align}
  \label{eq:103}
  \frac{d\boldsymbol{\Lambda}}{dt}
  &={\bf B}(\boldsymbol{\Lambda})\esc\boldsymbol{\Omega}
\end{align}
Use  of (\ref{vel}),  (\ref{ID}), and assumption ${\cal H}1$ (\ref{eq:95a}),
allows us  to write (\ref{eq:103})  as a closed  differential equation
for the orientation
\begin{align}
  \label{eq:113}
  \frac{d\boldsymbol{\Lambda}}{dt}
  &={\bf B}(\boldsymbol{\Lambda})\esc e^{[{\boldsymbol{\Lambda}}]_\times}
    \esc\overline{\mathbb{I}}^{-1}\esc e^{-[{\boldsymbol{\Lambda}}]_\times}\esc {\bf S}
\end{align}
which is  entirely equivalent  to Euler's equation  of motion  for a
rigid body, but  whose solution gives directly the  orientation as a
function of time. 
\section{Dissipative Euler's equations}
While,  intuitively, assumption  ${\cal  H}1$ seems  reasonable for  a
solid  object, it  is  difficult  to reconcile  ${\cal  H}2$ with  the
obvious fact that particles actually move in the principal axis system
due to  thermal fluctuations.  The  rigid body limit is  expected when
the stiffness of the interactions connecting the atoms increases.  But
this results in high frequency,  small amplitude motions, for which it
is  not obvious  the  angular  momentum in  the  principal axis  frame
vanishes.

The main  objective of the  present work  is to go  beyond assumptions
${\cal H}1,{\cal H}2$ in order  to formulate from first principles the
dynamics of a quasi-rigid body in free motion. Advancing some results,
for a realistic  body large enough for the thermal  fluctuations to be
neglected,  the   orientation  dynamics  (\ref{eq:103})   contains  an
additional dissipative mechanism
\begin{align}
  \label{eq:17}
   \frac{d\boldsymbol{\Lambda}}{dt}
    & = {\bf B}({\boldsymbol{\Lambda}})\esc\left[
\boldsymbol{\Omega}
      -      
\boldsymbol{\cal D}\esc(\boldsymbol{\Omega}\times{\bf S})\right]
\end{align}
where  $\boldsymbol{\cal D}$  is  an orientational diffusion tensor,  whose
explicit form in  terms of orientation and principal  moments is given
in (\ref{eq:9}) and (\ref{eq:476}).  As we will discuss,  this dissipative
term governs the dynamics of the process by which a spinning body will
align its major principal axis  with the angular momentum vector.
Observe that (\ref{eq:17}) and (\ref{eq:29}) imply  the angular velocity $ \boldsymbol{\omega}$  is
\begin{align}
  \label{eq:146}
  \boldsymbol{\omega}=\boldsymbol{\Omega}-\boldsymbol{\cal D}\esc(\boldsymbol{\Omega}\times{\bf S})
\end{align}
which is not  equal to the spin  velocity $\boldsymbol{\Omega}$, the
difference being determined by the dissipative term.  By multiplying
both        sides        of         this        equation        with
$e^{-[\boldsymbol{\Lambda}]_\times}$,    and    using    definitions
(\ref{eq:4}), (\ref{eq:160a}), (\ref{eq:9}), we get
the corresponding equation in the principal axis frame
\begin{align}
  \label{eq:299}
  \boldsymbol{\omega}_0
  &=     \boldsymbol{\Omega}_p
    -\boldsymbol{\cal D}_0\esc (\boldsymbol{\Omega}_p\times{\bf S}_p)
\end{align}
By   using   (\ref{eq:299})   in  (\ref{eq:13b})   we   obtain   the
generalization of Euler's equations in the presence of dissipation
\begin{align}
\label{o-eq:34}
\mathbb{I}\esc \frac{d}{dt}\boldsymbol{\Omega}_p
&= -\left[\boldsymbol{\Omega}_p\right]_\times
\esc\mathbb{I}\esc \boldsymbol{\Omega}_p -
\frac{d\mathbb{I}}{dt}\esc \boldsymbol{\Omega}_p
\nonumber\\
&\quad +\left[ \boldsymbol{\cal D}_0  \esc
\left [\boldsymbol{\Omega}_p\right ]_\times\esc
\mathbb{I}\esc\boldsymbol{\Omega}_p\right]_\times
\esc\mathbb{I}\esc \boldsymbol{\Omega}_p
\end{align}
The first term on the right  hand side is the usual (reversible) Euler
equation (\ref{eq:148a}).  The second term takes into account the time
evolution of  the principal moments  of inertia, and  appears whenever
assumption  ${\cal H}1$  does not  hold.  In  order to  have a  closed
equation, in this  work we will also provide  the explicit ``viscoelastic''
dynamics  of the  principal moments.  Finally, the  third term  is the
dissipative  contribution that  emerges from  the dissipative  term in
(\ref{eq:17}), which is due to the failure of assumption ${\cal H}2$.
In the rest of the article, we provide the derivation of (\ref{eq:17}).
\section{The theory of coarse graining}
\label{sec:theo}
In this  section, we review the  theory of coarse graining as  put
forward by Green and Zwanzig \cite{Green1952,Zwanzig1961,Grabert1982}.
At a  CG level  of description, the  system is described  by a  set of
{$N_{\rm  CG}$} functions  $\hat{A}(z)$  which depend upon the set  of
positions  and  momenta $z$  of  the  atoms  of  the system.   The  CG
variables will  be assembled into  a column vector  $\hat{A}(z)$ {with
  components $\hat{A}_\mu(z), \mu=1\cdots,N_{\rm  CG}$}. The selection
of  the  {CG}  variables  $\hat{A}(z)$   is  a  crucial  step  in  the
description of a  non-equilibrium system.  In any case, the  set of CG
variables should include the dynamic invariants of the system, as they
determine the equilibrium state of the system.  Green \cite{Green1952}
and  later Zwanzig  \cite{Zwanzig1961}  derived  from the  microscopic
Hamiltonian dynamics  of the  system a general  Fokker-Planck equation
for the probability distribution $P(a,t)$ that the set of CG variables
$\hat{A}(z)$ take the values $a$
\begin{align}
\partial_t P(a,t)&=
-\frac{\partial}{\partial a}\esc \left[ V(a)
+M(a)\esc\frac{\partial S_B}{\partial {a}}(a) \right]P(a,t) 
\nonumber\\
&+
k_B\frac{\partial}{\partial a}  \esc M(a) \esc
\frac{\partial}{\partial {a}} P(a,t) \label{ZFPE}
\end{align}
Only  two assumptions  are invoked  in  deriving the  FPE.  The  first
assumption is the separation of time-scales, that is, the CG variables
evolve with  two distinct time  scales, an overall smooth  mean motion
plus a superimposed rapid variation modelled as a white noise. This is
possible  if the  dynamics  of the  CG  variable arise from the
cumulative effect  of a large number of minute contributions  (like atomic
collisions  or   vibrations).  The  second  assumption   concerns  the
statistics of initial  conditions which are assumed  to be distributed
in such a way that all  microstates corresponding to the initial value
of the CG variables are equiprobable \cite{Grabert1982}.

The different objects in (\ref{ZFPE}) have well-defined microscopic
definitions.  For  example, the  reversible  drift  is the  conditional
expectation of the time derivative of the CG variables and is given by
\begin{equation}
V(a) = \langle i{\cal L} \hat{A}\rangle^a 
\label{Aa}
\end{equation}
where  $i{\cal  L}$ is  the  Liouville  operator and  the  conditional
expectation is defined by
\begin{equation}
\langle\ldots\rangle^{a}=\frac{1}{\Omega(a)} \int
dz\delta(\hat{A}(z)-{a})\cdots 
\label{ca}
\end{equation}
where $\delta(\hat{A}(z)-{a})$ is a  product of Dirac delta functions,
one for  every component  of the  vector function  $\hat{A}(z)$.  With
this  definition, the  conditional  expectation acting  on a  function
$f(\hat{A}(z))$ of the CG variables gives
\begin{align}
  \label{eq:76}
\langle f(\hat{A})\rangle^{a}=f(a)\ .  
\end{align}
The ``volume'' of  phase space compatible with a  prescribed value $a$
of the {CG} variables is
\begin{equation}
\Omega(a)=\int dz\delta(\hat{A}(z)-{a}) 
\label{omega}
\end{equation}
and is  closely related  to the  entropy at  the level  of description
given by the CG variables $a$, which is defined through
\begin{equation}
S_B(a)= k_B\ln\Omega(a) 
\label{Entropy}
\end{equation}
where $k_B$ is Boltzmann's  constant. 

The dissipative matrix  $M(a)$   is  the  matrix   of  transport
coefficients expressed in the form of Green-Kubo formulae,
\begin{equation}
M(a)=\frac{1}{k_B}\int_0^\infty\langle {\cal Q} i{\cal L}  \hat{A}
\exp\{Qi{\cal L}t'\} {\cal Q} i{\cal L} \hat{A}^T \rangle^{a} dt' 
\label{M}
\end{equation}
where
${\cal Q}  i{\cal  L} \hat{A}=  i{\cal  L}  \hat{A}- \langle  i{\cal  L}
\hat{A}\rangle^{a}$  is  the  so  called  projected  current.    The projection  operator ${\cal Q}$ is  defined from
its action on any phase function $\hat{F}(z)$ \cite{Zwanzig1961}
\begin{equation}
{\cal Q}\hat{F}(z) = \hat{F}(z) - \langle \hat{F}\rangle^{\hat{A}(z)} 
\label{qop}
\end{equation}
and describes the  fluctuations of the phase function  with respect to
the conditional  expectation. The   dynamic  operator
$\exp\{Qi{\cal L}t'\}$ is usually named the projected dynamics, which
is,   strictly    speaking   different   from   the    real   dynamics
$\exp\{i{\cal L}t'\}$.  The  projected dynamics can  be usually
approximated by the  real dynamics, but then the  upper infinite limit
of integration in Eq.  (\ref{M}) has  to be replaced by $\tau$, a time
which is long compared with the correlation time of the integrand, but
short  compared  with  the  time   scale  for  the  evolution  of  the
macroscopic  variables.   This  is   the  well-known  plateau  problem
\cite{Grabert1982},\cite{Kirkwood1946},\cite{Espanol1993}.         The
symmetric  part of  the dissipative  matrix $M$  is positive  definite
\cite{Grabert1982}. Time reversibility leads to Onsager reciprocity in
the form \cite{Grabert1982}
  \begin{align}
    \label{eq:21}
    M_{\mu\nu}(\varepsilon a)&=\varepsilon_\mu\varepsilon_\nu M_{\nu\mu}(a)
  \end{align}
  where $\varepsilon_\mu=\pm 1$ depending on the time reversible character of the
CG  variable $a_\mu$.

The  Ito  stochastic  differential  equation  that  is  mathematically
equivalent to the Fokker-Planck equation (\ref{ZFPE}) is given by
\begin{equation}
da = \left[V(a) +M(a)\esc \frac{\partial S_B}{\partial {a}}(a) 
+V^{\rm sto}(a) \right]dt + d\tilde{a}
\label{sde}
\end{equation}
where $d\tilde{a}$  is a linear  combination of independent increments  of the
Wiener  process. Their covariance is given by the   Fluctuation-Dissipation 
theorem  (FDT)
\begin{eqnarray}
d\tilde{a}d\tilde{a}^T&=&2k_BM^S(a)dt  
\label{FD}
\end{eqnarray}
where $M^S$ is the symmetric  part of the dissipative matrix. {The
  stochastic drift term $V^{\rm sto}(a)$ is given, in component form, by
  \begin{align}
    \label{eq:84}
    V_\mu^{\rm sto}(a)&=k_B\frac{\partial M_{\mu\nu}}{\partial  a_\nu}
  \end{align}
  where  Einstein  summation  convention   over  repeated  indices  is
  assumed. The form of the  stochastic drift depends on the stochastic
  interpretation  of  the SDE.  The  present  form considers  the  Ito
  interpretation.}

A comparatively recent development in the theory of coarse-graining is
the               {\sc              generic}               formulation
\cite{Oettinger1997c},\cite{Ottinger2005} that  fully acknowledges the
presence of dynamic invariants (in particular the total energy) in the
system.   A  dynamic  invariant   $\hat{I}(z)$  is  a  phase  function
satisfying  $\hat{I}(z_t)=\hat{I}(z_0)$, with  $z_t$  the solution  of
Hamilton's equation with initial condition  $z_0$.  We will assume the
dynamic  invariants  can  be  fully  expressed  in  terms  of  the  CG
variables,  $\hat{I}(z)=I(\hat{A}(z))$  where  $I(a)$ is  the  dynamic
invariant at the CG level of description.
When there are additional dynamic invariants it is easy to show 
the dissipative matrix $M(a)$ in (\ref{M}) satisfies
\begin{align}
  M(a)\esc\frac{\partial I}{\partial a}(a)=0
\label{M-Degeneracy}
\end{align}
According to the FDT (\ref{FD}), in order
to  fulfill this  identity  the  random forces  need  to satisfy  the
orthogonality conditions
\begin{align}
  d\tilde{a}^T\esc\frac{\partial I}{\partial a}(a)=0
\label{Orthog}
\end{align}
The equilibrium solution  of the FPE is given by  the Einstein formula
for equilibrium  fluctuations suitably  modified to take  into account
the presence of dynamic invariants \cite{Espanol1990}, this is
\begin{align}
  \label{eq:207}
    P^{\rm eq}(a)&=\frac{1}{Z}P_0(I(a))\exp\left\{\frac{S_B(a)}{k_B}\right\}
\nonumber\\
Z&=\int da P_0(I(a))\exp\left\{\frac{S_B(a)}{k_B}\right\}
\end{align}
where $P_0(I)$  is the probability distribution  of dynamic invariants
at the initial time.  If we know with certainty the value $I_0$ of the
dynamic   invariants,    as it happens   in   MD    simulations,   then
$P_0(I)=\delta(I(a)-I_0)$.   Note    that   like   any    other   FPE,
Eq.    (\ref{ZFPE})   has    the    following   Liapunov    functional
\cite{Gardiner1983}
\begin{align}
  \label{eq:320}
  S[P(t)]&=k_B\int da P(a,t)\ln \frac{P(a,t)}{P^{\rm eq}(a)}
\end{align}
that satisfies
\begin{align}
  \label{eq:322}
\frac{  \partial}{\partial t}  S[P(t)]\ge 0
\end{align}
and ensures any initial  distribution $P(a,0)$ evolves,  as time
proceeds,    towards    the     equilibrium    distribution    $P^{\rm
  eq}(a)$.  $S[P(t)]$ is  the  entropy at  the  level of  distribution
functions and (\ref{eq:322}) is the Second Law.

Finally, an  important identity known as  the reversibility condition,
is  obtained  from  the   statement  that  the  Einstein  distribution
probability is  the actual  equilibrium solution of  the Fokker-Planck
equation. This translates into the identity
\begin{align}
V^T(a)\esc\frac{\partial S_B}{\partial a}(a)+  k_B\frac{\partial \esc V}{\partial a}(a)=0
\label{RevCond}
\end{align}
The  reversibility relation  is  extremely useful  because it  imposes
rather stringent conditions  on any approximate model  for the entropy
and reversible drift. In the GENERIC framework, the reversibility
  condition is captured by the  degeneracy condition of the reversible
  operator \cite{Ottinger2005}.

In situations where thermal fluctuations can be neglected (formally in
the limit $k_B\to0$ \cite{Grabert1982}),  one obtains from (\ref{sde})
a set of deterministic equations of the form
\begin{equation}
\frac{da}{dt}= V(a)+M(a) \esc\frac{\partial S_B}{\partial {a}}(a)
\label{z1}
\end{equation}
Using (\ref{z1}), the entropy production is
\begin{align}
  \label{eq:294}
  \frac{d}{dt}S_B&=\left(\frac{\partial S_B}{\partial {a}}\right)^T\esc
\frac{da}{dt}\cr
  &=\left(\frac{\partial S_B}{\partial {a}}\right)^T\esc V
+
  \left(\frac{\partial S_B}{\partial {a}}\right)^T\esc M \esc\frac{\partial S_B}{\partial {a}}
\ge0\end{align}
The first  term is zero as  a result of (\ref{RevCond})  with $k_B=0$,
showing that  the reversible  part of  the dynamics  does not  lead to
entropy  production.  This justifies referring to $V(a)$ as the reversible drift.  The second term is greater than or equal to zero owing to the
positive  definiteness of  the symmetric  part of  $M(a)$.  Thus,  the
deterministic  dynamics  satisfy  the  Second  Law  of  Thermodynamics
automatically.   The evolution  of $a(t)$  will be  such that  entropy
increases, while  conserving the dynamic invariants,  until it reaches
the value $a^{\rm eq}$ that maximizes the entropy  conditional to the
dynamic invariants.
The grand objective of  the  theory of  coarse-graining  is to
obtain the FPE  (\ref{ZFPE}) or SDE (\ref{sde}) governing  the
stochastic dynamics of the CG variables, by producing explicit
expressions for their building blocks: the entropy
$S_B(a)$,  the reversible  drift  $V(a)$, and  the dissipative  matrix
$M(a)$.   Usually,   symmetries  need   to  be  exploited   for  these
calculations,  and  some  modelling  is required  in  order  to  obtain
explicit  expressions.

In the  rest of the  paper, we derive  the CG stochastic  equations of
motion of a free  solid body by starting from Hamilton's
equations.  This is  achieved by computing the building  blocks of the
general  FPE  (\ref{ZFPE}) for  the  particular  set of  CG  variables
discussed in the next section.

\section{The CG variables}
\label{Sec:free}
The  most important  step  in  the theory  of  coarse-graining is  the
selection of the CG variables.  For example, we could take a continuum
description of the body and include elasticity variables (for example,
displacement and  velocity \textit{fields}).   Such a  formulation may
allow one to include dissipation and, in principle, also fluctuations.
However,  if  one  is  only  interested in  ``the  overall  shape  and
orientation''   of  the   body,   a  continuum   description  is   too
detailed. Simulating  a large number of  interacting quasi-rigid bodies
described  with elastic  field  variables  may be  too  costly from  a
computational  point of  view.  For  small systems  composed of a small
number of particles, such a field description may also be questionable.
Instead, in  this work we  select as CG  variables the center  of mass
position  $\hat{\bf R}$  and  gyration tensor  $\hat{\bf G}$,  closely
related  to the  inertia  tensor $\hat{\bf  I}$.   These CG  variables
capture how the  particles of the body distribute in  space, the first
giving the ``location''  of the body and the latter  giving a sense of
its ``shape and  orientation''.  These variables are the  ones used to
describe  a  rigid  body  in   Classical  Mechanics  under  the  rigid
constraint  assumption,  and it  is  natural  to include  these  phase
functions in the list of CG variables.  The gyration tensor is defined
as
\begin{align}
  \label{eq:311}
  \hat{\bf G}&\equiv  \frac{1}{4}\sum_{i}m_{i}\left({\bf r}_{i}-\hat{\bf R}\right)\left({\bf r}_{i}-\hat{\bf R}\right)^T
\end{align}
and  receives  its  name  by  analogy to  the  usual  gyration  tensor
introduced in polymer physics  \cite{Kroger2005}. In  the  pseudo-rigid  body
literature,  $4\hat{\bf  G}$  is  referred  to  as  the  Euler  tensor
\cite{OReilly2003}.  The prefactor $1/4$
in the definition  (\ref{eq:311}) will allow us  to interpret directly
the eigenvalues $\hat{ M}_\alpha$  of $\hat{\bf G}$ as ``dilational
masses'',  or  inertia to  dilations.   The  tensor $\hat{\bf  G}$  is
symmetric  and  positive  semidefinite.   
The inertia tensor (\ref{Geomoments}) can be expressed in terms of the
 gyration tensor (\ref{eq:311})  in a linear way
\begin{align}
  \label{eq:190}
  \hat{\bf I}&={\black  4}\left({\rm Tr}[\hat{\bf G}]\mathbb{1}-\hat{\bf G}\right)
\end{align}
where ${\rm Tr}[\cdots]$ denotes the trace of the matrix.  The tensors
$\hat{\bf  G},\hat{\bf I}$  commute, and  therefore, they  diagonalize
simultaneously         with         the        rotation         matrix
$e^{-[\hat{\boldsymbol{\Lambda}}]_\times}$ in the  same principal axis
system. Similarly  to (\ref{diagonalization}) and (\ref{ID}),  we have
the diagonalization of the gyration tensor $\hat{\bf G}$ as
\begin{align}
  \label{Gdiag}
\hat{\mathbb{G}}
  &=   e^{-[\hat{\boldsymbol{\Lambda}}]_\times}\esc\hat{\bf G} \esc e^{[\hat{\boldsymbol{\Lambda}}]_\times}
    \nonumber\\
    \hat{\bf G}&=e^{[\hat{\boldsymbol{\Lambda}}]_\times}\esc
\hat{\mathbb{G}}\esc e^{-[\hat{\boldsymbol{\Lambda}}]_\times}
\end{align}
where the elements of the diagonal matrix $\hat{\mathbb{G}}$ are the
central moments $\hat{M}_1,\hat{M}_2,\hat{M}_3$ which we write compactly as
a list $\hat{{\bf M}}=(\hat{M}_1,\hat{M}_2,\hat{M}_3)$.
The  eigenvalues   $\hat{I}_\alpha$   of  the   inertia  tensor
$\hat{\bf  I}$  are  the \textit{principal}  moments,  while  the
eigenvalues $\hat{M}_\alpha$ of the gyration  tensor $\hat{\bf G}$ are the
\textit{central} moments. They are related through
\begin{align}
  \label{dj}
\hat{I}_\alpha
  &={\black  4}\left(\hat{M}_1+\hat{M}_2+\hat{M}_3-\hat{M}_\alpha\right)
\end{align}
For example, $\hat{I}_1=4(\hat{M}_2+\hat{ M}_3)$, etc.
The gyration  tensor of  a homogeneous  ellipsoid of  semi-axis $a,b,c$
oriented   along   the   Cartesian   axis  has   the   diagonal   form
$\mathbb{G}=\frac{M}{20}{\rm Diag}[a^2,b^2,c^2]$, so ${\bf G}$ gives a
more intuitive idea of the shape  than the inertia tensor.  We want to
contemplate situations in which the body may \textit{deform}, and this
deformation is captured  by the evolution of the  central moments.  In
addition,  simpler expressions  for the  dilation kinetic  energy (see
below) are obtained when selecting  the gyration tensor instead of the
inertia  tensor, and  this is  the fundamental  reason for  preferring
$\hat{\bf G}$ instead  of $\hat{\bf I}$ as a way  to represent changes
in shape of the body.  Choosing $\hat{\bf G}$ itself as a CG variable,
however,  is not  entirely  convenient, because  this  is a  symmetric
matrix that  has redundant information.   Only six of the  nine matrix
elements are  independent, leading to somewhat  cumbersome expressions
when taking derivatives with respect to its elements. For this reason,
we will  choose directly the  orientation $\hat{\boldsymbol{\Lambda}}$
of the body and the central moments $\hat{{\bf M}}$ of the body as the
CG  variables.
We expect the  shape  of the  body
to evolve  on  time  scales  which  are comparable  with  those  of  the
orientation.  It   turns  out  that   in  the  principal   axis  frame
${\cal  S}^0$ it  is easier  to separate  physical effects,  like pure
rotations  through the  orientation and  pure deformation  through the
central  moments.  Alternatively,  we could  use quaternion  variables
instead of the orientation as a, perhaps, more convenient way from the
computational     point     of      view     (see     for     example,
Ref.~\cite{Silveira2017}).     However,    somewhat   more    physical
expressions   are   obtained   with  the   orientation   and   central
moments. Once  we have the equations  for the orientation, the  use of
quaternions implies just a change of variables.

The          9          ``geometric''         phase          functions
$\hat{\bf  R},\hat{\boldsymbol{\Lambda}},\hat{{\bf  M}}$  depend  only
upon the  positions of  the particles.  We  also include  the momentum
$\hat{\bf P}$, the angular momentum with respect to the center of mass
$\hat{\bf  S}$,  as  well  as  the Hamiltonian,  in  the  list  of  CG
variables, because they are conserved  by the dynamics (i.e.  they are
infinitely slow),  and determine  the equilibrium  state of  the body.
The ``dynamic''  CG variables  $\hat{\bf P},\hat{\bf S}$  (6 variables
per   body)    play   the    role   of   ``conjugate    momenta''   to
$\hat{\bf R},\hat{\boldsymbol{\Lambda}}$.  Just  from counting, we are
missing a  ``conjugate momenta''  associated with the  central moments
$\hat{{\bf  M}}$. For  this  reason,  we include  in  the  list of  CG
variables  the dilation  momentum $\hat{\boldsymbol{\Pi}}$  defined as
the time derivative of the central moments, this is
\begin{align}
  \label{eq:127}
  \hat{\boldsymbol{\Pi}}\equiv i{\cal L}\hat{{\bf M}}
\end{align}
In summary, our selected set of the CG variables $\hat{A}(z)$ is
\begin{align}
  \label{eq:21CG}
  \hat{A}(z)
  &=\{\hat{\bf     R},\hat{\boldsymbol{\Lambda}},\hat{\bf M}, \hat{\bf  P},\hat{\bf S},\hat{\boldsymbol{\Pi}},     \hat{H}\}
\end{align}
The numerical values of the phase functions $\hat{A}(z)$ are denoted with $a$
\begin{align}
      \label{eq:232}
  a&=\{{\bf R},{\boldsymbol{\Lambda}},{\bf M},{\bf  P},{\bf S},{\boldsymbol{\Pi}},E\}
\end{align}
with no circumflexed symbols.  Even though the dynamic invariants have
a trivial  evolution they in  fact determine the equilibrium  state of
the body, and it is important to retain them in the description.
Implicit in the selection of the orientation, the central moments, and
the  dilational momentum  as  CG variables,  is  the assumption  these
quantities change  as a result  of many small and  rapid contributions
due to vibrations  of the particles.  In this way,  memory effects can
be eliminated and short time scales can be described with white noise,
giving rise  to a diffusion  process described with  the Fokker-Planck
equation (\ref{ZFPE}).

The time  derivatives of the CG  variables play a fundamental  role in
the theory  as they  determine the structure  of both the reversible
drift and the dissipative matrix.   The time derivatives are obtained
from  the action  of  the Liouville  operator $i{\cal  L}$  on the  CG
variables, with the result
\begin{align}
  i{\cal L}\hat{\bf R} 
&= \hat{\bf V},
\label{iLR}\\
  i{\cal L}\hat{\boldsymbol{\Lambda}}
  &= \sum_{i}\frac{\partial \hat{\boldsymbol{\Lambda}}}{\partial {\bf r}_{i}}\esc{\bf v}_{i},
\label{iLLambda}\\
  i{\cal L}\hat{{\bf M}}
  &=
\sum_{i}\frac{\partial \hat{{{\bf M}}}}{\partial {\bf r}_{i}}\esc{\bf v}_{i}    =\hat{\boldsymbol{\Pi}},
    \label{iLd}
\\
  i{\cal L}\hat{\bf P}
  &= 0,
\label{iLP}\\
  i{\cal L}\hat{\bf S}
  &=  0,
\label{iLS}\\
  i{\cal L}\hat{\boldsymbol{\Pi}}
  &= \hat{\boldsymbol{\cal K}},
\label{K}\\
  i{\cal L}\hat{H}
  & =0,
    \label{iLE}
\end{align}
where     we     have      introduced     the     dilational     force
$  \hat{\boldsymbol{\cal  K}}$ as  the  time  rate  of change  of  the
dilational     momentum.      Because      the     phase     functions
$\hat{\boldsymbol{\Lambda}},\hat{{{\bf M}}}$ are not explicit in terms
of  the  positions of  the  particles,  the  calculation of  the  time
derivatives above is not trivial and requires the lengthy calculations
presented in the Supplemental Material.  The time derivative
$ i{\cal L}\hat{\boldsymbol{\Lambda}}, i{\cal  L}\hat{\bf M}$, and the
explicit form  of $\hat{\boldsymbol{\cal K}}$  are given in Sec.  M of
the Supplemental Material.

In the  following sections we  consider the different  building blocks
entering the SDE (\ref{sde}) governing  the stochastic dynamics of the
CG variables as provided by the theory of CG.

Concerning  notation, we  use the  following convention.   For vectors
like    ${\bf   r}_i,{\bf    R}$    the    components   are    denoted
${\bf  r}_{i\alpha},{\bf   R}_\alpha$.   For  non-tensor   lists  like
$\boldsymbol{\Lambda},{\bf   M},\boldsymbol{\Pi}$,  we   do  not   use
bold-faced    symbols   to    denote   the    components,   that    is
$\Lambda_\alpha,M_\alpha,\Pi_\alpha$.
\section{Block 1: The entropy $S_B(a)$}
\label{Sec:Entropy}
The entropy  $S_B(a)$ defined generically in (\ref{Entropy}) takes the following
form for the present level of description
\begin{align}
  &S_B({\bf    R},\boldsymbol{\Lambda},{{\bf M}},{\bf
    P},{\bf S},\boldsymbol{\Pi},E)
  =k_B\ln \int dz
    \nonumber\\
  \times&  \delta\left(\hat{\bf R}(z)-{\bf R} \right)
    \delta\left(\hat{\boldsymbol{\Lambda}}(z)-\boldsymbol{\Lambda}\right)
    \delta\left(\hat{{\bf M}}(z)-{{\bf M}} \right)
    \nonumber\\
  \times&\delta\left(\hat{\bf P}(z)-{\bf P} \right)
          \delta\left(\hat{\bf S}(z)-{\bf S} \right)
          \delta\left(\hat{\boldsymbol{\Pi}}(z)-\boldsymbol{\Pi}\right)
\nonumber\\
\times&              \delta\left(\hat{H}(z)-E \right)
\label{Trembling:Sa}
\end{align}
It  is quite  remarkable  this complicated  integral  can be  computed
exactly in  terms of simpler objects.   In Sec. A of  the Supplemental
Material, we show this entropy has the following form,
(no approximations are needed to arrive at this exact result)
\begin{align}
  \label{eq:104}
  &S_B({\bf R},\boldsymbol{\Lambda},{{\bf M}},{\bf P},{\bf S},\boldsymbol{\Pi},E)
    \nonumber\\
  & =     S^{\rm MT}({\cal E})+k_B\ln P^{\rm rest}_{\cal E}(\boldsymbol{\Lambda},{{\bf M}},{\bf 0})    
\end{align}
The first term in (\ref{eq:104}) is the entropy $ S_B^{\rm MT}({\cal E})$ 
defined as
\begin{align}
  \label{eq:122}
  S_B^{\rm MT}({\cal E})
  =& \:k_B \ln \Omega^{\rm MT}({\cal E})
    \nonumber\\
  \Omega^{\rm MT}({\cal E})
  =& \int dz    \delta\left(\hat{\bf R}(z) \right)
    \delta\left(\hat{\bf P}(z) \right)
    \delta\left(\hat{\bf S}(z) \right)
    \nonumber\\
  &\times
    \delta\left(\hat{H}(z)-{\cal E} \right)
\end{align}
This  is the  usual  definition of  the  microcanonical entropy  given
originally by Boltzmann, when all dynamic invariants of the system are
properly  accounted  for  \cite{Gross2001}. The  notation  ${\rm  MT}$
stands for the level of  description of Macroscopic Thermodynamics, in
which the only selected CG variables are the dynamic invariants of the
system.   Obviously, the  entropy  at the  present  detailed level  of
description  is   different  from  the  entropy   at  the  Macroscopic
Thermodynamics level of description. The difference is the second term
in  (\ref{eq:104}), and  involves  the logarithm  of the  probability
density
$P^{\rm rest}_{\cal E}(\boldsymbol{\Lambda},{{\bf M}},{\bf 0})$, discussed below.

Observe that most of the  arguments of the entropy are combined  within the thermal
energy ${\cal E}$.   The thermal energy is the  result of subtracting
from the total energy $E$ the ``organized forms of kinetic energy''
\begin{align}
  \label{Thermal}
    {\cal E}&\equiv E-K^{\rm trans}-    K^{\rm rot}-K^{\rm dil}
\end{align}
where the translational, rotational, and dilational kinetic energies are
\begin{align}
  \label{Kinetic}
  K^{\rm trans}
  &\equiv  \frac{{\bf P}^2}{2{\rm M}}
    \nonumber\\
  K^{\rm rot}
  &\equiv\frac{1}{2}{\bf S}^T\esc{\bf I}^{-1}\esc{\bf S}
    \nonumber\\
    K^{\rm dil}
  &\equiv \frac{1}{2}{\boldsymbol{\Pi}}^T\esc\mathbb{G}^{-1}\esc{\boldsymbol{\Pi}}
    = \sum_\alpha\frac{\Pi_\alpha^2}{2{M}_\alpha}
\end{align}
Here,  the inertia  tensor  is  given by  the  matrix valued  function
(\ref{eq:403}).    The  kinetic   energies   $   K^{\rm  trans}$   and
$K^{\rm  rot}$  are  the   usual  expressions  for  translational  and
rotational kinetic energies, while $K^{\rm dil}$ is the kinetic energy
associated with  changes in  the shape  of the  body.  Observe  that the
central moments $M_\alpha$ play the role of  a dilational mass.
As mentioned,  this is a  consequence of defining the  central moments
tensor  $\hat{\bf  G}$ in  (\ref{eq:311})  with  the prefactor  $1/4$.
Also,  the  simple  form  of   the  dilational  kinetic  energy  is  a
consequence  of selecting  the gyration tensor  $\hat{\bf  G}$ as a CG
variable.  In  terms of  the inertia tensor  $\hat{\bf I}$,  much more
involved expressions are found.

In                                                     (\ref{eq:104}),
$P^{\rm rest}_{\cal E}(\boldsymbol{\Lambda},{{\bf M}},{\bf 0})$ is the
equilibrium  probability density  that  the system  at  rest and  with
energy  ${\cal E}$   has  a   particular  realization  of   the  orientation
$\boldsymbol{\Lambda}$,  central  moments  ${{\bf M}}$,  and  dilation
momentum  $\boldsymbol{\Pi}$,   the  latter   one  set  to   zero.  By
definition, this probability is given by
\begin{align}
  \label{eq:123}
  &  P^{\rm rest}_{{\cal E}}(\boldsymbol{\Lambda},{{\bf M}},\boldsymbol{\Pi})
  \equiv\int dz \rho^{\rm rest}_{{\cal E}}(z) 
    \nonumber\\
  &\times\delta\left(\hat{\boldsymbol{\Lambda}}(z)-\boldsymbol{\Lambda}\right)
    \delta\left(\hat{{\bf M}}(z)-{{\bf M}}\right)
    \delta(\hat{\boldsymbol{\Pi}}(z)-\boldsymbol{\Pi})
\end{align}
where the rest microcanonical ensemble is
\begin{align}
  \label{eq:125mic}
      \rho^{\rm rest}_{{\cal E}}(z)  &\equiv\frac{
                  \delta\left(\hat{\bf R}(z) \right)
                    \delta\left(\hat{\bf P}(z) \right)
                    \delta\left(\hat{\bf S}(z) \right)
                    \delta\left(\hat{H}(z)-{\cal E} \right)}{\Omega^{\rm MT}({\cal E})}
\end{align}
For a Hamiltonian dynamics of  the mixing type \cite{Dorfman1999}, the
rest microcanonical  ensemble is the  ensemble sampled in  a molecular
dynamics simulation with  initial conditions such that  the system has
zero linear and angular momentum  (it is macroscopically at rest), and
with energy  ${\cal E}$.  As  shown in Sec. B  of the
Supplemental Material, the rotational  invariance of the
microcanonical ensemble implies the probability (\ref{eq:123}) has the
following factorized form
\begin{align}
  \label{eq:136a}
  P^{\rm rest}_{\cal E}(\boldsymbol{\Lambda},{{\bf M}},\boldsymbol{\Pi})
  =P^{\rm Haar}(\boldsymbol{\Lambda})P^{\rm rest}_{\cal E}({{\bf M}},\boldsymbol{\Pi})
\end{align}
and, therefore,  for a body  at rest the orientation  is statistically
independent of the  central moments and dilational  momentum.  A quite
remarkable result, shown in Sec. B  of the Supplemental
Material,  is  that the  marginal  probability  density
$P^{\rm  Haar}(\boldsymbol{\Lambda})$ of  a particular  orientation at
equilibrium is given by a universal function independent of the energy
content of the body, and depending only on the orientation through its
modulus $\Lambda=|\boldsymbol{\Lambda}|$, this is
\begin{align}
  \label{eq:254}
    P^{\rm Haar}(\boldsymbol{\Lambda})=\frac{1}{4\pi^2}\frac{1-\cos\Lambda}{\Lambda^2}
\end{align}
The  support of  this  probability  is given  by  the  ``space of  all
possible   orientations''   that,   as   we   have   discussed   after
(\ref{eq:143}),  is  a  sphere  of   radius  $\pi$  in  the  space  of
orientations   $\boldsymbol{\Lambda}$.    With   that   support,   the
probability  density  (\ref{eq:254})  is  normalized  to  unity.   The
interpretation  of the  functional form  of (\ref{eq:254})  requires 
consideration  of  what  a  ``uniformly  distributed  random  rotation
matrix''  means.  The  rotation  matrices are  elements  of the  group
SO(3).  In  this group there  is a natural  measure given by  the Haar
measure,  which  is  the unique measure invariant  under
rotations, and  allows one to  speak about ``equally  probable'' random
rotation  matrices.  Miles  in  Ref.   \cite{Miles1965} discussed  this
uniform  measure in  SO(3)  in  terms of  two  representations of  the
attitude  parameters, the  Euler angles  $(\alpha,\beta,\gamma)$ and
the angle/axis parameters,  described by $(\Lambda,\theta,\phi)$, with
$0\le\Lambda\le\pi,  0\le\theta\le\pi,0\le\phi\le\pi$.  In  the latter
representation,  the   uniform  measure  has  the   following  density
\cite{Miles1965}
\begin{align}
  \label{eq:93a}
  P(\Lambda,\theta,\phi)&=\frac{1}{2\pi^2}\left(\sin\frac{\Lambda}{2}\right)^2\sin\theta
\end{align}
The orientation parameters  $\boldsymbol{\Lambda}=(\Lambda_1,\Lambda_2,\Lambda_3)$ of our work are related
to  the  angle/axis  parameters $(\Lambda,\theta,\phi)$,  through  the
spherical                     coordinate                     transform
$(\Lambda_1,\Lambda_2,\Lambda_3)=(\Lambda\sin\phi\sin\theta,\Lambda\cos\phi\sin\theta,\Lambda\cos\theta)$.
The rule for relating the two probability densities is
\begin{align}
  \label{eq:112a}
  P^{\rm Haar}(\boldsymbol{\Lambda})= \left| \frac{\partial (\Lambda_1,\Lambda_2,\Lambda_3)}{\partial (\Lambda\theta\phi)}\right| P(\Lambda,\theta,\phi)
\end{align}
i.e   they  are   proportional  with   the  Jacobian   determinant  as
proportionality    factor.    This    Jacobian    takes   the    value
$\Lambda^2\sin\theta$.   Inserting  (\ref{eq:93a}) in  (\ref{eq:112a})
gives the density of the uniform Haar measure in SO(3) in terms of the
attitude  parameters  $\boldsymbol{\Lambda}$.   The  result  coincides
precisely       with        (\ref{eq:254})       and,       therefore,
$P^{\rm Haar}(\boldsymbol{\Lambda})$ is the probability density of the
Haar  measure  in  terms  of the  attitude  parameters.   Recall  that
(\ref{eq:254}) is  obtained in  Sec.  B  of the  Supplemental Material
from  the  observation  the microcanonical  ensemble  is  rotationally
invariant.  That this  leads precisely to the  uniform distribution in
SO(3) should then come as no surprise because, for a body at rest, all
possible  rotations of  the body  are equally  probable, as  one would
reasonably expect because there is no privileged frame.

In summary,  we have an  \textit{exact} expression for the  entropy at
the  present level  of  description obtained  from (\ref{eq:104})  and
(\ref{eq:136a})
\begin{align}
  \label{W:eq:6}
&  S_B({\bf R},\boldsymbol{\Lambda},{{\bf M}},{\bf P},{\bf S},\boldsymbol{\Pi},E)=
    \nonumber\\
  &     S^{\rm MT}({\cal E})+k_B\ln P^{\rm Haar}(\boldsymbol{\Lambda})
     +k_B\ln P^{\rm rest}_{\cal E}({{\bf M}},{\bf 0})    
\end{align}
where  $P^{\rm  rest}_{\cal  E}({\bf  M},{\bf 0})$  is  the  value  of
$P^{\rm      rest}_{\cal     E}({\bf    M},\boldsymbol{\Pi})$     at
$\boldsymbol{\Pi}={\bf 0}$. The result  (\ref{W:eq:6}) involves no approximations.
\subsection{The equilibrium probability}

  The   entropy   determines,    according   to   Einstein's   formula
  (\ref{eq:207}) the  probability of finding particular  values of the
  CG  variables  at  equilibrium.  In   the  present  case,  by  using
  (\ref{eq:207}), and (\ref{W:eq:6}), this probability is,
\begin{align}
\label{exact}  P^{\rm eq}(\boldsymbol{\Lambda},{{\bf M}},\boldsymbol{\Pi})&=
  P^{\rm Haar}(\boldsymbol{\Lambda})P_{\cal E}^{\rm rest}({{\bf M}},{\bf 0})
  \exp\left\{k_B^{-1}S^{\rm MT}({\cal E})\right\}
\end{align}
This result is  exact. Observe that for a non-spinning  body with zero
angular          momentum          the         thermal          energy
$   {\cal    E}=E-\sum_\alpha\frac{\Pi_\alpha^2}{2{M}_\alpha}$     is
  independent of the orientation. In this case, (\ref{exact}) predicts
  that  orientation  and  central  moments  are  independent  and,  in
  particular, that  the marginal  probability of finding  a particular
  orientation is given by the density of the uniform Haar measure.

\subsection{Approximate explicit model for the entropy}
The exact  result (\ref{W:eq:6}) is  not explicit, because  the actual
functional  forms of  $S^{\rm MT}(E)$  and $\ln P^{\rm rest}_{\cal E}({{\bf M}},{\bf 0}) $ are  not yet
known. We consider now simple models that render the entropy explicit.

For an harmonic  solid in which the particles interact  in a pair-wise
form  with linear  Hookean springs,   $S^{\rm MT}(E)$ can be evaluated analytically, yielding a
constant  heat  capacity  given  by the  well-known  Dulong-Petit  law
$C^{\rm   MT}=3Nk_B$.    This   implies  a   macroscopic   temperature
$T^{\rm MT}(E)=E/(3Nk_B)$ which  is linear in the  energy content. The
entropy  is obtained by integrating the usual  definition for  the
temperature $T^{\rm MT}(E)$ in Macroscopic Thermodynamics,
\begin{align}
  \label{eq:97}
  \frac{\partial S_B^{\rm MT}}{\partial E}(E)=\frac{1}{T^{\rm MT}(E)}
\end{align}
leading to
\begin{align}
  \label{eq:175}
  S^{\rm MT}(E)&=3Nk_B\ln E+{\rm ctn}
\end{align}
where the  constant is irrelevant  as only derivatives of  the entropy
appear  in the  dynamics.   In more  general  models, the  macroscopic
entropy  $S^{\rm   MT}(E)$  can  be   obtained  in  a   simulation  by
thermodynamic  integration  of  the   temperature  measured  from  the
equipartition  of  energy  (see Sec. N.6  of  the
Supplemental  Material  for the  demonstration  of  the
equipartition theorem for  a free body).  In  Ref. \cite{Faure2017} we
followed this procedure in a model for a star polymer in vacuum, which
is a rather fluffy body, and still  we got a fairly constant value for
the heat capacity.   Therefore, such a Dulong-Petit law seems  to be a
rather  robust   result,  possibly   extending  beyond   the  harmonic
approximation.

Because   the  probability   density  (\ref{eq:254})   is  normalized,
$P^{\rm  rest}_{\cal E}({\bf M},\boldsymbol{\Pi})$  is just  the marginal  of
$  P^{\rm  rest}_{\cal E}(\boldsymbol{\Lambda},{{\bf  M}},\boldsymbol{\Pi})$ in (\ref{eq:123}),
that is
\begin{align}
  \label{eq:131}
    P^{\rm rest}_{{\cal E}}({{\bf M}},\boldsymbol{\Pi})
  &=\int dz \rho^{\rm rest}_{{\cal E}}(z)
    \delta\left(\hat{{\bf M}}(z)-{{\bf M}}\right)
    \delta(\hat{\boldsymbol{\Pi}}(z)-\boldsymbol{\Pi}) 
\end{align}
The probability  (\ref{eq:131}) admits a simple  analytical expression
once we  use the  approximation of equivalence  of ensembles.  In this
approximation,  valid for  sufficiently large  bodies (say,  with more
than  ten  particles),  averages   computed  with  the  microcanonical
ensemble (\ref{eq:125mic}) may be  approximated with averages computed
with the canonical ensemble ``at rest''
\begin{align}
  \label{eq:289}
    \rho^{\rm rest}_{\beta}(z)
  &=\frac{1}{Z(\beta)}
    \delta\left(\hat{\bf R}(z) \right)
    \delta\left(\hat{\bf P}(z) \right)
    \delta\left(\hat{\bf S}(z) \right)
    e^{-\beta\hat{H}(z)}
\end{align}
where the  partition function  $Z(\beta)$ normalizes the  ensemble and
the  parameter  $\beta$   is  related  to  the  energy   $E'$  of  the
microcanonical  ensemble   (\ref{eq:125mic})  through   (see Sec. 
C in the Supplemental  Material)
\begin{align}
  \label{eq:336}
\beta&=\frac{1}{k_BT^{\rm MT}({\cal E})}
\end{align}
Under equivalence of ensembles we have that
\begin{align}
  \label{eq:350}
    P^{\rm rest}_{{\cal E}}({{\bf M}},\boldsymbol{\Pi})
  &\simeq    P^{\rm rest}_{\beta({\cal E})}({{\bf M}},\boldsymbol{\Pi})
\end{align}
where the right hand side is  identical to (\ref{eq:131}) but with the
canonical ensemble  instead of the microcanonical  ensemble.  As shown
 in  Sec. C  of the Supplemental  Material,  the  canonical
probability can be explicitly computed to give
\begin{align}
  \label{eq:139a}
  P^{\rm rest}_{\beta}({{\bf M}},\boldsymbol{\Pi})
  &=
  P^{\rm rest}_{\beta}({{\bf M}})  G^{\rm rest}_{\beta}({{\bf M}},\boldsymbol{\Pi})
\end{align}
where the dilational momentum is distributed according to a
normalized Gaussian
\begin{align}
  \label{eq:148}
    G^{\rm rest}_{\beta}({{\bf M}},\boldsymbol{\Pi})
  &=\prod_{\alpha'}\left(\frac{\beta}{2\pi{M}_{\alpha'}}\right)^{1/2}
\exp\left\{-\beta
    \sum_\alpha\frac{\Pi^2_\alpha}{2M_\alpha}\right\}
\end{align}
and  where  $  P^{\rm  rest}_{\beta}(
{\bf M})$  is  the  marginal  of
$  P^{\rm  rest}_{\beta}({\bf M},\boldsymbol{\Pi})$  and  is  given
microscopically by
\begin{align}
  \label{eq:95}
  P^{\rm rest}_{\beta}({{\bf M}})
  &\equiv\int dz \rho^{\rm rest}_{\beta}(z)
    \delta\left(\hat{{\bf M}}(z)-{{\bf M}}\right)                                       
\end{align}
The functional form of this marginal  is still unknown and needs to be
either  measured  in  a  simulation, or  modelled.  We  choose a simple
Gaussian model  for $P^{\rm rest}_\beta({\bf M})$  which is determined
by  the   average  and   covariance  of   central  moments,   that  is
\begin{align}
  \label{eq:149a}
    P^{\rm rest}_\beta({\bf M})=
    \frac{\beta^{3/2}}{\det(2\pi \boldsymbol{\Sigma})^{1/2}} e^{-\frac{\beta}{2}({{\bf M}}-{\bf M}^{\rm rest})^T\esc\boldsymbol{\Sigma}^{-1}\esc({{\bf M}}-{\bf M}^{\rm rest})}
\end{align}
where ${{\bf M}}^{\rm rest}$ is the equilibrium average of the central
moments for  the body at  rest, and $\boldsymbol{\Sigma}$ is proportional  to the
covariance of central moments fluctuations, given by
\begin{align}
  \label{eq:138a}
  \boldsymbol{\Sigma}&=\beta\llangle  (\hat{\bf M}-{\bf M}^{\rm rest})(\hat{\bf M}-{\bf M}^{\rm rest})^T\rrangle^{\cal E}
\end{align}
where $\llangle\cdots\rrangle^{\cal  E}$ is  an average with  the rest
microcanonical ensemble  (\ref{eq:125mic}).  We expect  
$\llangle  ({{\bf  M}}-{\bf   M}^{\rm  rest})({{\bf  M}}-{\bf  M}^{\rm
  rest})^T\rrangle^{\cal  E}$ to  be roughly  proportional to  the
temperature  and,   therefore,  the  matrix  $\boldsymbol{\Sigma}$ to be  roughly
independent of the temperature.

The Gaussian  model (\ref{eq:149a}) is  expected to be  reasonable for
values of ${\bf  M}$ near ${\bf M}^{\rm rest}$, but  most probably the
``tails''  of  the  real  distribution are  not  well  represented  by
Gaussian tails. In particular, note that the central moments cannot be
negative, but the Gaussian probability gives a non-zero (although very
small) probability of finding negative values of $\bf M$.

By   collecting    the   results    (\ref{eq:350}),   (\ref{eq:139a}),
(\ref{eq:148}),           and            (\ref{eq:149a})           for
$   P^{\rm  rest}_{\cal   E}({{\bf   M}},{\bf   0})$,  together   with
(\ref{eq:175}) for  the macroscopic entropy,  we get a  \textit{fully explicit}
model  for  the  entropy  (\ref{W:eq:6})   at  the  present  level  of
description, up to irrelevant constants
\begin{align}
  \label{eq:195}
  &  S_B({\bf R},\boldsymbol{\Lambda},{{\bf M}},{\bf P},{\bf S},\boldsymbol{\Pi},E)
  \nonumber\\
  &=S^{\rm MT}({\cal E})    +k_B\ln P^{\rm Haar}(\boldsymbol{\Lambda})
    \nonumber\\
  &  -\frac{1}{2T^{\rm MT}({\cal E})}({{\bf M}}-{\bf M}^{\rm rest})^T\esc\boldsymbol{\Sigma}^{-1}\esc({{\bf M}}-{\bf M}^{\rm rest})
       \nonumber\\
  & -\frac{k_B}{2}\ln\left(M_1 M_2 M_3\right)
-\frac{k_B}{2}\ln\det(\boldsymbol{\Sigma})
\end{align}
where  the macroscopic  entropy is  given by  (\ref{eq:175}), and  the
thermal   energy    ${\cal   E}$   is   given    in   (\ref{Thermal}).

\subsection{The thermodynamic forces}
The thermodynamic  forces are, by  definition, the derivatives  of the
entropy. By taking the derivatives of the exact result for the entropy
in (\ref{W:eq:6}) we obtain
\begin{align}
  \label{eq:63a}
  \frac{\partial S_B}{\partial a}(a)
  &    \to\left(
    \begin{array}{r}
      \frac{\partial S_B}{\partial{\bf R}}
      \\
      \\
      \frac{\partial S_B}{\partial{\boldsymbol{\Lambda}}}
      \\
      \\
      \frac{\partial S_B}{\partial{{\bf M}}}
      \\
      \\
      \frac{\partial S_B}{\partial{\bf P}}
      \\
      \\
      \frac{\partial S_B}{\partial{\bf S}}
      \\
      \\
      \frac{\partial S_B}{\partial{\boldsymbol{\Pi}}}
      \\
      \\
      \frac{\partial S_B}{\partial E}
    \end{array}
  \right)
  =
  \left(
  \begin{array}{c}
    0
    \\
    \\
    \frac{\partial S_B}{\partial{\boldsymbol{\Lambda}}}
    \\
    \\
    \frac{\partial S_B}{\partial{{\bf M}}}
    \\
    \\
    -\frac{{\bf V}}{T}
    \\
    \\
    -\frac{\boldsymbol{\Omega}}{T}
    \\
    \\
    -\frac{\boldsymbol{\nu}}{T}
    \\
    \\
    \frac{1}{T}
  \end{array}
  \right)
\end{align}
Let us discuss in detail how these derivatives are obtained.

The   first  entry    reflects   that  the   entropy
(\ref{W:eq:6}) does not depend on the  position of the center of mass.
The last  entry defines  the temperature  $T(a)$ of  the body  at this
level of description according to
\begin{align}
  \frac{1}{T(a)}&\equiv\frac{\partial S_B}{\partial E}(a)
\label{Temperature}
\end{align}
From the exact expression (\ref{W:eq:6}) for the entropy we have 
\begin{align}
  \label{App:eq:182}
  \frac{1}{T(a)}
  &=\frac{1}{T^{\rm MT}({\cal E})}+
k_B    \frac{\partial}{\partial {\cal E}}
    \ln P^{\rm rest}_{{\cal E}}({{\bf M}},{\bf 0})
\end{align}
where  the  macroscopic  temperature   is  defined  in  (\ref{eq:97}).
Observe  that  the  temperatures  $T(a)$  in  (\ref{Temperature})  and
$T^{\rm MT}({\cal E})$  in (\ref{eq:97}) are different,  as an obvious
consequence of the fact that the entropy $S_B(a)$ at the present level
of  description and  the entropy  $S^{\rm MT}(E)$  at the  Macroscopic
Thermodynamics level of description are also different functions.  For
the  Gaussian  model  (\ref{eq:139a}) and (\ref{eq:149a}),  the  explicit
relation between the two temperatures is
\begin{align}
  \label{eq:357}
&\frac{T^{\rm MT}{({\cal E})}}{T(a)}
  =1
    \nonumber\\
  &+\frac{k_B}{C^{\rm MT}}\left(\beta
\frac{({{\bf M}}-{\bf M}^{\rm rest})^T\esc\boldsymbol{\Sigma}^{-1}\esc({{\bf M}}-{\bf M}^{\rm rest})}{2}
 -3  \right)
\end{align}
where we have introduced the heat capacity according to the usual definition
\begin{align}
  \label{eq:354}
\frac{1}{C^{\rm MT}}&\equiv  \frac{\partial T^{\rm MT}}{\partial {\cal E}}
\end{align}
The heat capacity $C^{\rm MT}$  is extensive ($C^{\rm MT}=3Nk_B$ for a
harmonic solid), and the prefactor  $k_B/C^{\rm MT}$ is of order $1/N$.
We expect that  the  difference between  the
temperature     $T(a)$     and     the     macroscopic     temperature
$T^{\rm MT}({\cal E})$ is {of order $k_B/C^{\rm MT}$, and }negligible in the thermodynamic limit.

We move  now to  the derivatives  of the entropy  with respect  to the
momenta in (\ref{eq:63a}).  All these  terms have the same structure in
terms of ``velocities''.  By  using the definition (\ref{Temperature})
for  the temperature  and the  fact the only  dependence of  the
entropy on momenta is through the thermal energy ${\cal E}$ defined in
(\ref{Thermal}) and (\ref{Kinetic}) we obtain
\begin{align}
  \label{eq:189}
  \frac{\partial S_B}{\partial{\bf P}}
  &= \frac{1}{T}\frac{\partial {\cal E}}{\partial{\bf P}}
    = -\frac{{\bf V}}{T}
    \nonumber\\
  \frac{\partial S_B}{\partial{\bf S}}
  &= \frac{1}{T} \frac{\partial {\cal E}}
    {\partial{\bf S}} = -\frac{\boldsymbol{\Omega}}{T}
    \nonumber\\
  \frac{\partial S_B}{\partial\boldsymbol{\Pi}}
  &= \frac{1}{T}\frac{\partial {\cal E}}{\partial\boldsymbol{\Pi}}
    = -\frac{\boldsymbol{\nu}}{T}
\end{align}
where the linear ${\bf V}$, spin $    \boldsymbol{\Omega}$, and dilational
$   \boldsymbol{\nu}$ velocities are 
\begin{align}
  \label{eq:117}
  {\bf V}&\equiv {\rm M}^{-1}{\bf P}
           \nonumber\\
  \boldsymbol{\Omega}&\equiv{\bf I}^{-1}\esc{\bf S}
                       \nonumber\\
  \boldsymbol{\nu}
  &\equiv\mathbb{G}^{-1}\esc\boldsymbol{\Pi}
\end{align}
We appreciate  these  ``velocities'' are  of the  form ``momentum
divided by (linear, angular,  dilational) mass'' \footnote{In spite of
  the assigned  name, $  \boldsymbol{\Omega}, \boldsymbol{\nu}  $ have
  physical  dimensions of  frequency.}. The explicit form of the spin velocity
in terms of the selected CG variables is
\begin{align}
  \label{eq:82}
  \boldsymbol{\Omega}
  &=e^{[\boldsymbol{\Lambda}]_\times}\esc\mathbb{I}^{-1}\esc
    e^{-[\boldsymbol{\Lambda}]_\times}\esc{\bf S}
\end{align}
where the diagonalized inertia tensor $\mathbb{I}$ (\ref{diagonalization}) is the function
(\ref{dj}) of the central moments.

Finally,  the  derivatives   of  the  entropy  with   respect  to  the
orientation   and  central   moments   are   computed  in Sec.
D  of Supplemental  Material, with the explicit results
\begin{align}
  \label{eq:184}
{T}  \frac{\partial S_B}{\partial{\boldsymbol{\Lambda}}}
  &=-{\bf B}^{-T}\esc(\boldsymbol{\Omega}\times{\bf S}) +k_BT
    \frac{\partial}{\partial \boldsymbol{\Lambda}}\ln P^{\rm Haar}(\boldsymbol{\Lambda})
\\
  \label{eq:380}
  T    \frac{\partial S_B}{\partial M_\alpha}
  &=\frac{1}{2}\nu_\alpha^2
    +2\left(    \boldsymbol{\Omega}_p^T\esc{\boldsymbol{\Omega}_p}-( {\Omega_p^{\alpha}})^2 \right)
    \nonumber\\
  &+    k_BT\frac{\partial}{\partial {
    M}_\alpha}\ln P^{\rm rest}_{\cal E}({{\bf M}},{\bf 0})
\end{align}
where, {from (\ref{eq:254}) we have
\begin{align}
  \label{eq:58}
  \frac{\partial}{\partial \boldsymbol{\Lambda}}\ln P^{\rm Haar}(\boldsymbol{\Lambda})
  &=-2f(\Lambda)\boldsymbol{\Lambda}
\end{align}
with the function defined as
\begin{align}
      \label{eq:5}
       f(\Lambda)&\equiv\frac{1}{\Lambda^2}\left(1-\frac{\Lambda}{2}\cot\frac{\Lambda}{2}\right)
\end{align} and,}
analogously to (\ref{eq:160a}) but with no circumflexed symbols,
\begin{align}
  \label{eq:453}
{\boldsymbol{\Omega}_p}&\equiv  e^{-[\boldsymbol{\Lambda}]_\times}\esc\boldsymbol{\Omega}
= \mathbb{I}^{-1}\esc e^{-[\boldsymbol{\Lambda}]_\times}\esc{\bf S} 
\end{align}
Observe that the component $\alpha=1$ of the  ``centrifugal term'' 
in (\ref{eq:380}) involves the $2,3$ components, that is
\begin{align}
  \label{eq:452}
  \boldsymbol{\Omega}^T_p\esc\boldsymbol{\Omega}_p
  -(\boldsymbol{\Omega}_p^1)^2    
  &=    (\boldsymbol{\Omega}_p^2)^2+(\boldsymbol{\Omega}_p^3)^2
\end{align}

For the  Gaussian model (\ref{eq:139a}) and  (\ref{eq:149a}), the last
term in (\ref{eq:380}) is explicit in the CG variables
\begin{align}
  \label{eq:65}
    k_BT\frac{\partial}{\partial M_\alpha}\ln P^{\rm rest}_{\cal E}({{\bf M}},{\bf 0})
  &=-\frac{T}{T^{MT}}[\Sigma^{-1}]_{\alpha\beta}(M_\beta-M_\beta^{\rm rest})
    \nonumber\\
  &-\frac{k_BT}{2 M
    _\alpha}
\end{align}
where repeated indices are summed over by following Einstein's convention.

\section{Block 2: The reversible drift}
\label{Sec:Rev}
The reversible part  of the dynamics is given by  the reversible drift
(\ref{Aa}), specified by the time-derivatives (\ref{iLR})-(\ref{iLE}).  The  time
derivatives of $\hat{\bf R}$ and $\hat{{\bf M}}$ are given in terms of
the   linear   momentum   $\hat{\bf    P}$   and   dilation   momentum
$\hat{\boldsymbol{\Pi}}$ so their conditional expectations are, from (\ref{eq:76}),
\begin{align}
  \label{W:eq:58}
  \llangle i{\cal L}\hat{\bf R}\rrangle^a
  &={\bf V}
    \nonumber\\
  \llangle i{\cal L}\hat{{\bf M}}\rrangle^a
  &=\boldsymbol{\Pi}
\end{align}
Other trivial terms are those corresponding to the conserved variables
\begin{align}
  \label{eq:199}
  \llangle i{\cal L}\hat{\bf P}\rrangle^a
  &=0
    \nonumber\\
  \llangle i{\cal L}\hat{\bf S}\rrangle^a
  &=0
    \nonumber\\    \llangle i{\cal L}\hat{H}\rrangle^a
  &=0
\end{align}
The only non-trivial  reversible terms are those  corresponding to the
orientation    and    the    dilation   momentum.     In    Sec.     E
 of the Supplemental Material, we
show the reversible dynamics of the orientation is given by
\begin{align}
  \label{ConExp}
        \llangle i{\cal L}\hat{\boldsymbol{\Lambda}}\rrangle^a&=  {\bf B}(\boldsymbol{\Lambda})\esc{\boldsymbol{\Omega}}
\end{align}
where the matrix valued function ${\bf B}(\boldsymbol{\Lambda})$ is given in (\ref{eq:53}).

The remaining non-trivial reversible part is
\begin{align}
  \label{eq:61}
  \llangle i{\cal L}\hat{\boldsymbol{\Pi}}\rrangle^a&=  \llangle \hat{\boldsymbol{\cal K}}\rrangle^a \equiv {\boldsymbol{\cal K}}
\end{align}
where    the    microscopic    form     of    the    dilation    force
$\hat{\boldsymbol{\cal K}}$ is  given in Sec. M.3  of the Supplemental
Material, and the dilation mean force ${\boldsymbol{\cal K}}$ (without
circumflex) is  defined in  (\ref{eq:61}) as its  conditional average.
In    order   to    find    the   explicit    functional   form    for
${\boldsymbol{\cal  K}}$, we  resort  to  the reversibility  condition
(\ref{RevCond})  that ensures  the  Einstein equilibrium  distribution
function  is the  stationary solution  of the  Fokker-Planck equation.
The reversibility condition represents a very strong constraint on the
functional form of the reversible  drift and the entropy function.  In
the  present  case,  it  allows  us to  obtain  the  dilational  force
explicitly in terms of derivatives of the entropy.  As shown in Sec. F
  of the  Supplemental  Material, the  $\alpha$
component of the mean force is given by
\begin{align}
  \label{eq:101}
{\cal K}_\alpha
  &= M_{\underline{\alpha}}\left( \frac{1}{2}\nu_{\underline{\alpha}}^2
    +2\left(    \boldsymbol{\Omega}_p^T\esc\boldsymbol{\Omega}_p - \boldsymbol{\Omega}_{p\underline\alpha}^2 \right)\right )
    \nonumber\\
  &+   k_BT^{\rm MT}\left( M_{\underline{\alpha}}  \frac{\partial}{\partial{
    M}_{\underline{\alpha}}}\ln P^{\rm rest}_{\cal E}({{\bf M}},{\bf 0})+1\right)
\end{align}
where underlined repeated indices  do not follow Einstein's convention
and  are not  summed  over.   In particular,  for  the Gaussian  model
(\ref{eq:65}) we obtain the fully explicit result
\begin{align}
  \label{eq:137}
  {\cal K}_\alpha
  &=M_{\underline{\alpha}}\left( \frac{1}{2}{\nu}_{\underline{\alpha}}^2
    +2\left(    \boldsymbol{\Omega}_p^T\esc\boldsymbol{\Omega}_p -\boldsymbol{\Omega}_{p\underline\alpha}^2 \right)\right.
    \nonumber\\
&\quad\left . +\frac{k_BT^{\rm MT}}{2{M}_{\underline{\alpha}}}
- [\Sigma^{-1}]_{\underline{\alpha}\beta}(M_\beta-M_\beta^{\rm rest})\right )
\end{align}
We  refer  to  the  contribution  to ${\boldsymbol{\cal  K}}$ quadratic in
 $\boldsymbol{\nu}$  as  the convective term,  the contribution quadratic
in $\boldsymbol{\Omega}$ as  the  centrifugal term,  and  the  last  term
proportional  to  the covariance  matrix  as  the  elastic term.

\begin{widetext}

\section{Block 3: The dissipative matrix }
\label{Sec:Diss}
The friction  matrix $M$  given in general  by the  Green-Kubo formula
(\ref{M}) has,  at the  present level  of description,  the following
entries
\begin{align}
\label{Ma}  
  &M(a) =
\small{\left(
    \begin{array}{ccccccc}
      \GK {\dot{\bf R}}{\dot{\bf R}}
      & \GK {\dot{\bf R}}{\dot{\boldsymbol{\Lambda}}}
      & \GK {\dot{\bf R}}{\dot{{\bf M}}}
      & \GK {\dot{\bf R}}{\dot{\bf P}}
      & \GK {\dot{\bf R}}{\dot{\bf S}}
      & \GK {\dot{\bf R}}{\dot{\boldsymbol{\Pi}}}
      & \GK {\dot{\bf R}}{\dot{H}}\\
      \\
      \GK {\dot{\boldsymbol{\Lambda}}}{\dot{\bf R}}
      & \GK {\dot{\boldsymbol{\Lambda}}}{\dot{\boldsymbol{\Lambda}}}
      & \GK {\dot{\boldsymbol{\Lambda}}}{\dot{{\bf M}}}
      & \GK {\dot{\boldsymbol{\Lambda}}}{\dot{\bf P}}
      & \GK {\dot{\boldsymbol{\Lambda}}}{\dot{\bf S}}
      & \GK {\dot{\boldsymbol{\Lambda}}}{\dot{\boldsymbol{\Pi}}}
      & \GK {\dot{\boldsymbol{\Lambda}}}{\dot{H}}\\
      \\
      \GK {\dot{\bf P}}{\dot{\bf R}}
      & \GK {\dot{\bf P}}{\dot{\boldsymbol{\Lambda}}}
      & \GK {\dot{\bf P}}{\dot{{\bf M}}}
      & \GK {\dot{\bf P}}{\dot{\bf P}}
      & \GK {\dot{\bf P}}{\dot{\bf S}}
      & \GK {\dot{\bf P}}{\dot{\boldsymbol{\Pi}}}
      & \GK {\dot{\bf P}}{\dot{H}}\\
      \\
      \GK {\dot{\bf S}}{\dot{\bf R}}
      & \GK {\dot{\bf S}}{\dot{\boldsymbol{\Lambda}}}
      & \GK {\dot{\bf S}}{\dot{{\bf M}}}
      & \GK {\dot{\bf S}}{\dot{\bf P}}
      & \GK {\dot{\bf S}}{\dot{\bf S}}
      & \GK {\dot{\bf S}}{\dot{\boldsymbol{\Pi}}}
      & \GK {\dot{\bf S}}{\dot{H}}\\
      \\
      \GK {\dot{\boldsymbol{\Pi}}}{\dot{\bf R}}
      & \GK {\dot{\boldsymbol{\Pi}}}{\dot{\boldsymbol{\Lambda}}}
      & \GK {\dot{\boldsymbol{\Pi}}}{\dot{{\bf M}}}
      & \GK {\dot{\boldsymbol{\Pi}}}{\dot{\bf P}}
      & \GK {\dot{\boldsymbol{\Pi}}}{\dot{\bf S}}
      & \GK {\dot{\boldsymbol{\Pi}}}{\dot{\boldsymbol{\Pi}}}
      & \GK {\dot{\boldsymbol{\Pi}}}{\dot{H}}\\
      \\
      \GK {\dot{H}}{\dot{\bf R}}
      & \GK {\dot{H}}{\dot{\boldsymbol{\Lambda}}}
      & \GK {\dot{H}}{\dot{{\bf M}}}
      & \GK {\dot{H}}{\dot{\bf P}}
      & \GK {\dot{H}}{\dot{\bf S}}
      & \GK {\dot{H}}{\dot{\boldsymbol{\Pi}}}
      & \GK {\dot{H}}{\dot{H}}
    \end{array}\right)}
\end{align}
where we use the shorthand notation for the Green-Kubo formulae
\begin{align}
  \label{GKShorthand}
  \GK {\dot{\bf A}}{\dot{\bf B}}&=\frac{1}{k_B}  \int_0^\infty dt\llangle \left({\cal Q}i{\cal L}\hat{\bf A}\right)
\left(\exp\{i{\cal Q}{\cal L}t\}{\cal Q}i{\cal L}\hat{\bf B}\right)^T\rrangle^{a}
\end{align}
\end{widetext}
The projection operator $\cal Q$ acting on any CG variable gives zero,
then       ${\cal       Q}i{\cal      L}\hat{\bf       R}=0$       and
${\cal Q}i{\cal L}\hat{{\bf M}}=0$.  Also, linear and angular momenta, and   the     energy are    conserved,    implying
$i{\cal    L}\hat{\bf   P}=0$,    $i{\cal   L}\hat{\bf    S}=0$,   and
$i{\cal   L}\hat{H}=0$.  The   dissipative   matrix  (\ref{Ma})   thus
simplifies to
\begin{align}
  \label{eq:36}
M(a)
  &=T
\left(
    \begin{array}{ccccccc}
 0
      & 0
      & 0
      & 0
      & 0
      & 0
      & 0
      \\
0
      & \boldsymbol{\Gamma}_{_\Lambda}
      & 0
      & 0
      & 0
      & \boldsymbol{\Gamma}_{{_\Lambda} {_\Pi}}
      & 0\\
0
      &0
      & 0
      & 0
      & 0
      & 0
      & 0\\
0
      &0
      & 0
      & 0
      & 0
      & 0
      & 0\\
0
      & 0
      & 0
      & 0
      & 0
      & 0
      & 0\\
0
      & \boldsymbol{\Gamma}_{{_\Pi}{_\Lambda} }
      & 0
      & 0
      & 0
      & \boldsymbol{\Gamma}_{_\Pi}
      & 0\\
0
      & 0
      & 0
      & 0
      & 0
      & 0
      & 0
    \end{array}\right)
\end{align}
where
\begin{align}
  \label{eq:39}
    {\boldsymbol{\Gamma}}_{_\Lambda}(a)
  &=  \frac{1}{k_BT}  \int_0^\infty dt
\llangle \left({\cal Q}i{\cal L}\hat{\boldsymbol{\Lambda}}\right)
\left(\exp\{i{\cal Q}{\cal L}t\}{\cal Q}i{\cal L}\hat{\boldsymbol{\Lambda}}\right)^T\rrangle^{a}
    \nonumber\\
    {\boldsymbol{\Gamma}}_{{_\Lambda} {_\Pi}}(a)
  &=  \frac{1}{k_BT}  \int_0^\infty dt
\llangle \left({\cal Q}i{\cal L}\hat{\boldsymbol{\Lambda}}\right)
\left(\exp\{i{\cal Q}{\cal L}t\}{\cal Q}i{\cal L}\hat{\boldsymbol{\Pi}}\right)^T\rrangle^{a}
    \nonumber\\
    {\boldsymbol{\Gamma}}_{{_\Pi}{_\Lambda} }(a)
  &=  \frac{1}{k_BT}  \int_0^\infty dt
\llangle \left({\cal Q}i{\cal L}\hat{\boldsymbol{\Pi}}\right)
\left(\exp\{i{\cal Q}{\cal L}t\}{\cal Q}i{\cal L}\hat{\boldsymbol{\Lambda}}\right)^T\rrangle^{a}
    \nonumber\\
    {\boldsymbol{\Gamma}}_{_\Pi}(a)
  &=  \frac{1}{k_BT}  \int_0^\infty dt
\llangle \left({\cal Q}i{\cal L}\hat{\boldsymbol{\Pi}}\right)
\left(\exp\{i{\cal Q}{\cal L}t\}{\cal Q}i{\cal L}\hat{\boldsymbol{\Pi}}\right)^T\rrangle^{a}
\end{align}
This matrix of transport coefficients  depends, in general, on the
CG variables $a$, because the  Green-Kubo expression is given in terms
of a  conditional expectation. {The definitions (\ref{eq:39}) of the
  transport coefficients with the prefactor $1/k_BT$ is conventional, and
  motivated by the expectation that, defined in this way, the transport coefficients
  are roughly independent on temperature.}

Time reversibility  leads to Onsager's
reciprocity in the  form given in (\ref{eq:21}). At  the present level
of description  the time reversed CG variables are
\begin{align}
  \label{eq:197}
  \varepsilon a= ({\bf R},\boldsymbol{\Lambda},{{\bf M}},-{\bf P},-{\bf S},-\boldsymbol{\Pi},E)
\end{align}
On the other hand, we expect   the dependence of the dissipative
matrix  on the  momenta ${\bf P},{\bf S},\boldsymbol{\Pi}$ to be  only through  the thermal  energy
(\ref{Thermal}),   which  is   quadratic   in   momenta  thus giving
$M_{\mu\nu}(\varepsilon  a)=M_{\mu\nu}(a)$.   Onsager's  reciprocity
(\ref{eq:21}) implies  the elements (\ref{eq:39}) satisfy
\begin{align}
  \label{eq:204}
  {\boldsymbol{\Gamma}}_{_\Lambda}(a)
  &=      \left[{\boldsymbol{\Gamma}}_{_\Lambda}(a)\right]^T
    \nonumber\\
  {\boldsymbol{\Gamma}}_{{_\Lambda} {_\Pi}}(a)
  &=-   \left[ {\boldsymbol{\Gamma}}_{{_\Pi}{_\Lambda} }(a)\right]^T
    \nonumber\\
  {\boldsymbol{\Gamma}}_{_\Pi}(a)
  &= \left[    {\boldsymbol{\Gamma}}_{_\Pi}(a)\right]^T
\end{align}
 The symmetric
part of   $M(a)$ in  (\ref{eq:36}) then has 
off-diagonal                       block                      elements
$\frac{1}{2}({\boldsymbol{\Gamma}}_{{_\Lambda} {_\Pi}}(a)+({\boldsymbol{\Gamma}}_{{_\Pi}{_\Lambda} }(a))^T)=0$, which
vanish. Note that  the symmetric part of $M(a)$  is what appears
in the  Fluctuation-Dissipation theorem  and, therefore,  this Onsager
symmetry  implies   the noise  terms  in the  orientation and  the
dilational momentum are statistically independent.  For simplicity, we
will  assume the cross  correlation blocks  (and not  only their
symmetric parts) vanish in general, this is
\begin{align}
  \label{eq:205}
  {\boldsymbol{\Gamma}}_{{_\Lambda} {_\Pi}}(a)&=
 {\boldsymbol{\Gamma}}_{{_\Pi}{_\Lambda} }(a)  =0  
\end{align}
Therefore, in  the matrix  (\ref{eq:36}) only  the two  block elements
$\boldsymbol{\Gamma}_{_\Lambda},\boldsymbol{\Gamma}_{_\Pi}$        are
different from zero. Let us analyze both terms.
\subsection{The element $\boldsymbol{\Gamma}_{_\Lambda}$}
The projected current  associated with the orientation  can be expressed
using (\ref{eq:29}) and (\ref{ConExp}) as
\begin{align}
  \label{eq:54}
  {\cal Q}i{\cal L}\hat{\boldsymbol{\Lambda}}
  &\overset{(\ref{qop})}{=}i{\cal L}\hat{\boldsymbol{\Lambda}}
    -\llangle i{\cal L}\hat{\boldsymbol{\Lambda}}\rrangle^{\hat{A}}
  ={\bf B}(\boldsymbol{\Lambda})\esc\Delta\hat{\boldsymbol{\omega}}
\end{align}
where $  \Delta\hat{\boldsymbol{\omega}}$ 
\begin{align}
      \label{eq:265}
  \Delta\hat{\boldsymbol{\omega}}
  &\equiv  \hat{\boldsymbol{\omega}}-\hat{\boldsymbol{\Omega}}
\end{align}
is the difference between angular $\hat{\boldsymbol{\omega}}$ and spin
$\hat{\boldsymbol{\Omega}}$  velocities. Observe  that the  difference
between these two velocities, in  violation of assumption ${\cal H}2$,
produces a non-vanishing  element in the dissipative matrix.
Using (\ref{eq:54}) in (\ref{eq:39}) gives
\begin{align}
  \label{eq:388}
    \boldsymbol{\Gamma}_{_\Lambda}(a)
  &=
    {\bf B}({\boldsymbol{\Lambda}})\esc
    \boldsymbol{\cal D}(a)
    \esc{\bf B}^T({\boldsymbol{\Lambda}}) 
\end{align}
where we have introduced the orientational diffusion tensor
\begin{align}
  \label{eq:323}
  \boldsymbol{\cal D}(a)
  &    \equiv \frac{1}{k_BT}\int_0^\infty dt
    \llangle\Delta\hat{\boldsymbol{\omega}}(0)\Delta\hat{\boldsymbol{\omega}}^T(t)
    \rrangle^a
\end{align}
We  show  in  Sec. G of the Supplemental  Material
this takes the form
  \begin{align}
    \label{eq:9}
      \boldsymbol{\cal D}(a)=    e^{[{{\boldsymbol{\Lambda}}}]_\times}\esc
    \boldsymbol{\cal D}_0({{\bf M}},{\cal E})\esc
     e^{-[{{\boldsymbol{\Lambda}}}]_\times}
  \end{align}
where the orientational  diffusion tensor in the rest frame  is given by the
Green-Kubo expression
\begin{align}
  \label{eq:476}
  \boldsymbol{\cal D}_0({{\bf M}},{\cal E})
  &\equiv \frac{1}{k_BT}\int_0^\infty dt
    \llangle\hat{\boldsymbol{\omega}}_{0}(0)\hat{\boldsymbol{\omega}}_{0}^{T}(t)
    \rrangle^{\boldsymbol{\Lambda}{{\bf M}}{\cal E}}_{\rm rest}
\end{align}
The  microscopic expression  for the  components $x=1,y=2,z=3$  of the
angular velocity $\hat{\boldsymbol{\omega}}_{0}$ in the principal axis
frame is given by (see Sec. M of the Supplemental Material)
\begin{align}
  \label{eq:478}
  \hat{\boldsymbol{\omega}}_0^{x}
  &=    \frac{1}{4\left(\hat{M}_2-\hat{M}_3\right)}\sum_im_i({\bf v}_{pi}^{y} {\bf r}_{0i}^{z}+{\bf r}_{0i}^{y} {\bf v}_{pi}^{z})
    \nonumber\\
  \hat{\boldsymbol{\omega}}_0^{y}
  &=    \frac{1}{4\left(\hat{M}_3-\hat{M}_1\right)}    \sum_im_i({\bf v}_{pi}^{x} {\bf r}_{0i}^{z}+{\bf r}_{0i}^{x} {\bf v}_{pi}^{z})
    \nonumber\\
  \hat{\boldsymbol{\omega}}_0^{z}
  &=    \frac{1}{4\left(\hat{M}_1-\hat{M}_2\right)}\sum_im_i({\bf v}_{pi}^{x} {\bf r}_{0i}^{y}+{\bf r}_{0i}^{x} {\bf v}_{pi}^{y})
\end{align}
where ${\bf r}_{0i}$ from (\ref{eq:74}) are
the particle positions in the principal axis frame, and
\begin{align}
\label{eq:477}
{\bf v}_{pi} = e^{-[{\hat{\boldsymbol{\Lambda}}}]_\times}\esc
( {\bf v}_i -\hat{\bf V} )
\end{align}
are particle velocities projected to the principal axis frame.
Observe that ${\bf v}_{pi}$ is different from the velocity
  of the particles with respect to the principal axis ${\bf v}_{0i}$ given in (\ref{eq:41}).

  As shown  in Sec. B.3  of the Supplemental  Material, the  orientational diffusion tensor in  (\ref{eq:476}) is
  independent of the orientation, in  spite of the average conditional
  on  $\boldsymbol{\Lambda}$.  Therefore,  the  dissipative matrix  is
  explicit in the orientation variables
\begin{align}
  \label{eq:475}
        \boldsymbol{\Gamma}_{_\Lambda}({\boldsymbol{\Lambda}},{\bf M},{\cal E})
  &=    {\bf B}({\boldsymbol{\Lambda}})\esc
    e^{[{{\boldsymbol{\Lambda}}}]_\times}\esc
    \boldsymbol{\cal D}_0({{\bf M}},{\cal E})\esc
     e^{-[{{\boldsymbol{\Lambda}}}]_\times}
    \esc{\bf B}^T({\boldsymbol{\Lambda}}) 
\end{align}
No approximations were used in arriving at this result. The functional
dependence  of   $  \boldsymbol{\cal  D}_0({{\bf  M}},{\cal   E})$  on
${\bf  M},  {\cal E}$  generated  by  the conditional  expectation  in
(\ref{eq:476}) is  not trivial.  A simple  modelling assumption  is to
consider that  the conditional expectation involved  in the Green-Kubo
formula can be well approximated with an ordinary equilibrium average,
in              such              a              way              that
$  \boldsymbol{\cal  D}_0({{\bf M}},{\cal  E})=\boldsymbol{\cal  D}_0$
independent on the instantaneous state.  \color{black}
\subsection{The element $\boldsymbol{\Gamma}_{_\Pi}$}
In Sec. G Supplemental  Material we  have computed  the
dissipative matrix $\boldsymbol{\Gamma}_{_\Pi}$ with the result
\begin{align}
  \label{eq:152}
  \boldsymbol{\Gamma}_{_\Pi}({\cal E})&\simeq\frac{1}{k_BT}\int_0^\infty dt
  \nonumber\\
  &\times\llangle
  \left(\hat{\boldsymbol{\cal K}}(0)-\llangle \hat{\boldsymbol{\cal K}}\rrangle^{{\cal E}}\right)
  \left(\hat{\boldsymbol{\cal K}}^T(t)-\llangle \hat{\boldsymbol{\cal K}}^T\rrangle^{{\cal E}}\right)\rrangle^{{\cal E}}    
\end{align}
Therefore, this ``dilation friction'' matrix  is given in terms of the
time   integral  of   the  autocorrelation   of  the   dilation  force
$\hat{\boldsymbol{\cal           K}}$.           The           average
$\llangle\cdots\rrangle^{\cal  E}$   is  an   ordinary  microcanonical
equilibrium  average computed  with the  rest microcanonical  ensemble
(\ref{eq:125mic}) at  the thermal energy ${\cal  E}$.  The microscopic
expression of the dilation force  in terms of positions and velocities
of  the   particles  is  given in Sec. M of the Supplemental Material.
\section{Block 4: The noise and stochastic drift }
\label{Sec:Noise}
The  noise  in  the  SDE (\ref{sde})  satisfies  the  FDT  (\ref{FD}).
Comparison  of  the  FDT   and  the  Green-Kubo  expression  (\ref{M})
indicates that  one way to construct  the noise term is  by looking at
the structure  of the projected current,  and modeling the noise  as a
linear  combination of  independent increments  of the  Wiener process
respecting this structure. In the present case, the FDT is given by
\begin{align}
  \label{eq:268}
  d\tilde{\boldsymbol{\Lambda}}  d\tilde{\boldsymbol{\Lambda}}^T
  &=2k_BT\boldsymbol{\Gamma}_{_\Lambda}dt
    \nonumber\\
  d\tilde{\boldsymbol{\Lambda}}  d\tilde{\boldsymbol{\Pi}}^T
  &=0
    \nonumber\\
  d\tilde{\boldsymbol{\Pi}}  d\tilde{\boldsymbol{\Pi}}^T
  &=2k_BT\boldsymbol{\Gamma}_{_\Pi}dt
\end{align}
 By  inspection  of the  form  (\ref{eq:475}),  we
propose the following noise for the orientation
\begin{align}
  \label{eq:68}
  d\tilde{\boldsymbol{\Lambda}}&=(2k_BT)^{1/2}
    {\bf B}({\boldsymbol{\Lambda}})\esc
    e^{[{{\boldsymbol{\Lambda}}}]_\times}\esc
\boldsymbol{\cal D}_0^{1/2}\esc d\tilde{\bf W}
\end{align}
where $d\tilde{\bf W}$  is a vector of independent increments  of the Wiener
process satisfying the mnemotechnical Ito rule
\begin{align}
  \label{eq:293}
  d\tilde{\bf W}d\tilde{\bf W}^T=\mathbb{1}dt
\end{align}
and the matrix $\boldsymbol{\cal D}_0^{1/2}$ satisfies
\begin{align}
  \label{eq:468}
  \boldsymbol{\cal D}_0^{1/2}\esc\left(\boldsymbol{\cal D}_0^{1/2}\right)^T&=\boldsymbol{\cal D}_0
\end{align}
The     noise     (\ref{eq:68})     automatically     fulfills     the
fluctuation-dissipation theorem (first equation in (\ref{eq:268})) at the present level of description.

On the other hand, we propose the following form for the noise $d\tilde{\boldsymbol{\Pi}}$
\begin{align}
  \label{eq:157}
  d\tilde{\boldsymbol{\Pi}}&=(2k_BT)^{1/2} \boldsymbol{\Gamma}_{_\Pi}^{1/2}d\tilde{\bf V}
\end{align}
where the matrix satisfies
\begin{align}
  \label{eq:165}
    \boldsymbol{\Gamma}_{_\Pi}^{1/2}\esc\left(\boldsymbol{\Gamma}_{_\Pi}^{1/2}\right)^T&=\boldsymbol{\Gamma}_{_\Pi}
\end{align}
and  $d\tilde{\bf V}$  is a vector of independent increments  of the Wiener
process satisfying the mnemotechnical Ito rule
\begin{align}
  \label{eq:293b}
  d\tilde{\bf V}d\tilde{\bf V}^T=\mathbb{1}dt
\end{align}

The stochastic drift term $V^{\rm sto}(a)$ given by (\ref{eq:84})
produces    two    components    for    the    drift,
$\boldsymbol{V}^{\rm  sto}_{_\Lambda}$  that  will  go  in  the
orientation  dynamics,  and  $\boldsymbol{V}^{\rm  sto}_{_\Pi}$
that  will go  in  the dilational  momentum  dynamics.  By
recalling the cross terms (\ref{eq:205}) vanish, we have
\begin{align}
  \label{eq:269}
 \boldsymbol{V}^{{\rm sto }}_{_\Lambda}
  &=k_B\frac{\partial}{\partial\boldsymbol{\Lambda}}\esc
    T\boldsymbol{\Gamma}_{_\Lambda}
    \nonumber\\
 \boldsymbol{V}^{\rm sto}_{_\Pi}
  &=k_B\frac{\partial}{\partial\boldsymbol{\Pi}}\esc T\boldsymbol{\Gamma}_{_\Pi}
\end{align}
The stochastic drift is computed in Sec. H of Supplemental  Material
with the result
  \begin{align}
    \label{eq:290}
      \boldsymbol{V}^{{\rm sto }}_{_\Lambda}
  &{=}  -\frac{k_B}{C^{\rm MT}}\frac{\partial T\boldsymbol{\Gamma}_{_\Lambda}}{\partial T}\esc
    {\bf B}^{-T}\esc(\boldsymbol{\Omega}\times{\bf S})+ k_BT\frac{\partial\esc \boldsymbol{\Gamma}_{_\Lambda}} {\partial\boldsymbol{\Lambda}}
\end{align}
The first term arises from the dependence of the dissipative matrix on
the   temperature,  while   the  second   term  originates   from  the
explicit   dependence  of   the   dissipative   matrix  on   the
orientation.
The other component of the stochastic drift in (\ref{eq:269})
gives the result
\begin{align}
  \label{eq:312}
    \boldsymbol{V}^{\rm sto}_{_\Pi}
  &=  -\frac{k_B}{C^{\rm MT}}\frac{\partial T \boldsymbol{\Gamma}_{_\Pi}^{\rm eq}}{\partial T}
\esc    \boldsymbol{\nu}
\end{align}

\section{The final SDE and FPE}
\label{Sec:FinSDE}
We  have now  all the  necessary building  blocks to  specify the  SDE
(\ref{sde})  of the  present level  of description.   Without loss  of
generality, we may select the inertial reference frame ${\cal S}$ with
the origin at the center of mass, for which ${\bf R}=0,{\bf P}=0$. The
CG  variables ${\bf  R},{\bf P},{\bf  S},E$ are  conserved and  have a
trivial evolution.   The three evolving variables  are the orientation
$\boldsymbol{\Lambda}$, the central moments ${\bf M}$ and the dilation
momentum $\boldsymbol{\Pi}$.   By collecting  all the  building blocks
presented  in  the  previous  sections, the  general  SDE  (\ref{sde})
becomes
\begin{align}
  \label{eq:115}
d\boldsymbol{\Lambda}
  &= {\bf B}\esc\boldsymbol{\Omega}dt +{\boldsymbol{\Gamma}}_{_\Lambda}
    \esc T\frac{\partial S_B}{\partial \boldsymbol{\Lambda}}dt
    +   \boldsymbol{V}^{\rm sto}_{_\Lambda} dt+d\tilde{\boldsymbol{\Lambda}}
    \nonumber\\
  d{{\bf M}}&=\boldsymbol{\Pi}dt
            \nonumber\\
  d\boldsymbol{\Pi}&={\boldsymbol{\cal K}}dt 
{+{\boldsymbol{\Gamma}}_{_\Pi}
                     \esc T\frac{\partial S_B}{\partial \boldsymbol{\Pi}}dt}
                     +    \boldsymbol{V}^{\rm sto}_{_\Pi} dt+d\tilde{\boldsymbol{\Pi}}
\end{align}
where we clearly distinguish the reversible part, the dissipative
  part proportional to the thermodynamic forces, the stochastic drift,
  and       the      random       noise      terms. The corresponding
  FPE (\ref{ZFPE}) governing the evolution of the probability of the CG variables
  is given now by 
\begin{align}
  \label{eq:314}
  \frac{\partial}{\partial t} P(a,t)
  &=-\frac{\partial}{\partial \boldsymbol{\Lambda}}\esc 
    \left({\bf B}\esc\boldsymbol{\Omega} +{\boldsymbol{\Gamma}}_{_\Lambda}
    \esc T\frac{\partial S_B}{\partial \boldsymbol{\Lambda}}\right)
    P(a,t)
    \nonumber\\
  &-\frac{\partial}{\partial {\bf M}}\esc\boldsymbol{\Pi}P(t)
    -\frac{\partial}{\partial\boldsymbol{\Pi}}
    \esc \left({\boldsymbol{\cal K}} +\boldsymbol{\Gamma}_{_\Pi}
                     \esc T\frac{\partial S_B}{\partial \boldsymbol{\Pi}}
\right)P(a,t)
    \nonumber\\
  &+k_B  \frac{\partial}{\partial \boldsymbol{\Lambda}}
    \esc T\boldsymbol{\Gamma}_{_\Lambda}\esc
    \frac{\partial}{\partial \boldsymbol{\Lambda}}
  P(a,t)                               
    \nonumber\\
  &+k_B  \frac{\partial}{\partial \boldsymbol{\Pi}}
    \esc T\boldsymbol{\Gamma}_{_\Pi}\esc
    \frac{\partial}{\partial \boldsymbol{\Pi}}
  P(a,t)                               
\end{align}
Here      $P(a,t)$      is      short      hand      notation      for
$P(\boldsymbol{\Lambda},{\bf M},\boldsymbol{\Pi},t)$, as the arguments
${\bf  R},{\bf  P},{\bf  S},E$  are  constants  of  motion.   The  FPE
(\ref{eq:314}) is less useful in  practice than the SDE (\ref{eq:115})
because  it is  a partial  differential  equation in  a 9  dimensional
space, whereas the SDE can be easily implemented numerically. However,
both  describe the  evolution  towards equilibrium  of the  stochastic
process of the CG variables.

We  obtain  a  more  explicit  form  of  the  SDEs  (\ref{eq:115})  by
introducing the  explicit forms (\ref{eq:189}) and (\ref{eq:184})  for the
thermodynamic   forces,   and   (\ref{eq:290}) and
(\ref{eq:312}) for the stochastic drift. The result is
  \begin{align}
    \label{eq:212}
    d\boldsymbol{\Lambda}
      &= {\bf B}\esc\boldsymbol{\Omega}dt -  {\boldsymbol{\Gamma}}'_{_\Lambda}
        \esc{\bf B}^{-T}\esc(\boldsymbol{\Omega}\times{\bf S})dt
     + k_BT{\bf F}^{\rm th} dt+d\tilde{\boldsymbol{\Lambda}}
        \nonumber\\
    d{{\bf M}}&=\boldsymbol{\Pi}dt
                \nonumber\\
    d\boldsymbol{\Pi}&={\boldsymbol{\cal K}}dt-\boldsymbol{\Gamma}'_{_\Pi}\esc\boldsymbol{\nu}dt
                       +   d\tilde{\boldsymbol{\Pi}}
  \end{align}
  where we have introduced the thermal drift 
  \begin{align}
    \label{eq:356}
   {\bf F}^{\rm th}
    &\equiv  \frac{\partial\esc   \boldsymbol{\Gamma}_{_\Lambda}}{\partial\boldsymbol{\Lambda}}
+  \boldsymbol{\Gamma}_{_\Lambda}\esc   \frac{\partial}{\partial \boldsymbol{\Lambda}}\ln P^{\rm Haar}(\boldsymbol{\Lambda})
  \end{align}
  that    collects     a    bit    from    the     stochastic    drift
  ${\bf V}^{\rm sto}_{_\Lambda}$  and a bit corresponding  to the Haar
  measure of the gradient of the entropy. In the absence of any better
  name, we will  refer to ${\bf F}^{\rm th}$ as  the thermal drift, as
  its overall effect is proportional to $k_BT$.  This term is computed
  in  Sec. I  of  the  Supplemental Material, with  the
  explicit result
\begin{align}
  \label{eq:454}
  {\bf F}^{\rm th}
  &=F_1{\rm Tr}\left[\left[\mathbb{1}-{\bf n}{\bf n}^T\right]\esc\boldsymbol{\cal D}_0\right]{\bf n}
+      F_2 \left[\mathbb{1}-{\bf n}{\bf n}^T\right]\esc\boldsymbol{\cal D}_0\esc{{\bf n}}
    \nonumber\\
  &+     F_3 [{{\bf n}}]_\times\esc    \boldsymbol{\cal D}_0\esc{{\bf n}}
\end{align}
where
\begin{align}
  \label{eq:288}
  F_1&=\frac{1}{2}\frac{\sin\Lambda -\Lambda}{1-\cos\Lambda}
       \nonumber\\
    F_2&=\cot\left(\frac{\Lambda}{2}\right)\left(1-\frac{\Lambda}{2}\cot\left(\frac{\Lambda}{2}\right)\right)
       \nonumber\\
  F_2&=\left(1-\frac{\Lambda}{2}\cot\left(\frac{\Lambda}{2}\right)\right)
\end{align}
The thermal drift  is a function of the orientation, and involves the
dissipative matrix $\boldsymbol{\cal D}_0({\bf M},{\cal E})$.
We have also introduced
\begin{align}
  \label{eq:2b}
  {\boldsymbol{\Gamma}}'_{_\Lambda}
  &\equiv {\boldsymbol{\Gamma}}_{_\Lambda}+\frac{k_B}{C^{\rm MT}}\frac{\partial T{\boldsymbol{\Gamma}}_{_\Lambda}}{\partial T}
    \nonumber\\
  \boldsymbol{\Gamma}'_{_\Pi}&\equiv   \boldsymbol{\Gamma}_{_\Pi}+\frac{k_B}{C^{\rm MT}}\frac{\partial T\boldsymbol{\Gamma}_{_\Pi}}{\partial T}
\end{align}

The  SDEs (\ref{eq:212})  constitute the  main result  of the  present
work, and determine the dynamics  of the orientation and dilation of
  the body.  These equations are  fully explicit in the CG variables:
${\bf B}(\boldsymbol{\Lambda})$  is given  by (\ref{eq:53}),  the spin
$\boldsymbol{\Omega}$ and dilational  $\boldsymbol{\nu}$ velocities by
(\ref{eq:117}),  the dilation  mean force  ${\boldsymbol{\cal K}}$  by
(\ref{eq:137}),            the           transport            matrices
$\boldsymbol{\Gamma}_{_\Lambda},\boldsymbol{\Gamma}_{_\Pi}$   by (\ref{eq:475})  and (\ref{eq:152}),  the thermal
drift ${\bf F}^{\rm th}$ by (\ref{eq:454}), and
the                             noise                            terms
$d\tilde{\boldsymbol{\Lambda}},   d\tilde{\boldsymbol{\Pi}}$ by (\ref{eq:268}), (\ref{eq:68}), and (\ref{eq:157}).
\section{Summary of approximations}
\label{Sec:Summary}
The   implicit  assumptions   made   in  deriving   the  main   result
(\ref{eq:212}) are  those inherent  to the theory  of coarse-graining,
that is,  the separation of  time scales, and the  uniform statistical
distribution  of  initial  conditions.  Because  the  gyration  tensor
involves a sum over  all the particles in the system,  we expect it to
evolve on  an overall  time scale  much larger  than the  typical time
scale for molecular oscillations.  Additional assumptions are required
to provide the explicit forms for the temperature $T$, the probability
$P^{\rm rest}_{\cal  E}({{\bf M}},\boldsymbol{\Pi})$  appearing inside
the  dilational force  $\boldsymbol{\cal  K}$,  and the  orientational
diffusion tensor $\boldsymbol{\cal  D}_0$.  The entropy (\ref{W:eq:6})
is  exact,  but the  explicit  form  requires  an expression  for  the
macroscopic  entropy  $S_{\rm  MT}({\cal   E})$  and  the  probability
$P^{\rm   rest}_{\cal   E}({{\bf  M}},\boldsymbol{\Pi})$.    For   the
macroscopic entropy  we have  assumed a  constant heat  capacity model
that leads to  a logarithmic equation of state $T({\cal  E})$. This is
probably a very robust approximation. The probability factorizes under
the  assumption of  equivalence  of ensembles  leading  to a  Gaussian
distribution  of  the dilational  momentum,  which  is also  a  robust
approximation.     The   probability    of    the   central    moments
$P^{\rm rest}_{\cal E}({{\bf M}})$ is  assumed to be Gaussian, leading
to a  ``linear elastic'' behaviour of  the system. This seems  to be a
good approximation  for molecular  systems, but for  large macroscopic
bodies more realistic non-linear elastic models are probably required.
The dilation mean force ${\boldsymbol{\cal  K}}$ is computed under the
assumption the  body is sufficiently  large (an approximation  akin to
the  equivalence of  ensembles).  As  it involves  the central  moment
probability $P^{\rm  rest}_{\cal E}({{\bf M}})$  it is subject  to the
above-mentioned linear  elasticity assumption.   The structure  of the
orientational  dissipative matrix  $\boldsymbol{\Gamma}_{_\Lambda}$ in
(\ref{eq:475}) is  robust, as it  has been obtained by  using symmetry
arguments.   However,  modelling  assumptions  are  required  for  the
orientational                     diffusion                     tensor
$   \boldsymbol{\cal   D}_0({{\bf    M}},{\cal   E})$   appearing   in
(\ref{eq:475}).   Cross elements  in the  dissipative matrix  have been
neglected      for     simplicity.       Finally,     the      element
$\boldsymbol{\Gamma}_{_\Pi}(\boldsymbol{\Lambda},{\bf M},{\cal E})$ of
the dissipative  matrix given by the  Green-Kubo formula (\ref{eq:39})
is  approximated  with a  rest  equilibrium  ensemble.  This  gives  a
damping of the central moments which is proportional to the dilational
velocity.  This  seems to be  a minimal approximation that  may render
reasonable  results.    The  combination  of  the   linear  elasticity
assumption and the linear damping with the dilational velocity is akin
to modelling the  rheology of the body through  a viscoelastic Maxwell
model.

\section{Deterministic equations and physical meaning of the different terms}
\label{Sec:Det}
The discussion of  the physical meaning of the different  terms in the
SDEs (\ref{eq:115}) or (\ref{eq:212}) is  facilitated when the body is
large, and thermal fluctuations can be neglected. Then, the Stochastic
Differential  Equations become  Ordinary Differential  Equations.  The
deterministic  limit is  obtained by formally taking $k_B\to0$\cite{Grabert1982}.

The orientation   dynamics    (\ref{eq:212})   has   the   following
deterministic limit for large bodies
\begin{align}
    \label{Lambda-Det}
 \frac{d\boldsymbol{\Lambda}}{dt}
    & = {\bf B}({\boldsymbol{\Lambda}})\esc\left[
\boldsymbol{\Omega}
      -      
\boldsymbol{\cal D}\esc(\boldsymbol{\Omega}\times{\bf S})\right]
\end{align}
The   term   ${\bf   B}\esc\boldsymbol{\Omega}$   in   (\ref{Lambda-Det})
corresponds  to  the  motion  of  a  rigid  body  in  the  absence  of
dissipation, as  discussed in  (\ref{eq:103}).  This term  couples the
orientation dynamics with the shape dynamics because the spin velocity
(\ref{eq:117}) is given in terms of the inertia tensor that depends on
the  central  moments ${\bf  M}$.   The  second   term  in   (\ref{Lambda-Det})  is
dissipative,  and has  its  origin  in the  term  that multiplies  the
gradient of the entropy. As we discuss in Sec.~\ref{Sec:2nL}, this  term essentially drives the orientation
towards the  maximum of the  entropy with respect to  the orientation.

Concerning the dilational  dynamics in (\ref{eq:212}) for  the case of
large bodies, the $k_B\to0$ limit leads to the deterministic equations
\begin{align}
  \label{eq:83}
   \frac{d{\bf M}}{dt}&=\boldsymbol{\Pi}
            \nonumber\\
 \frac{d\boldsymbol{\Pi}}{dt}  &={\boldsymbol{\cal K}} -\boldsymbol{\Gamma}_{_\Pi}\esc\boldsymbol{\nu}
\end{align}
These equations  are similar  to the equations  of a  particle with
position   ${\bf  M}$ and   momentum  $\boldsymbol{\Pi}$.   The  dilational  force $\boldsymbol{\cal K}$ is  given
explicitly  in (\ref{eq:137}), where the first convective term is non-linear,
and shows that the larger the dilational velocity $\boldsymbol{\nu}$, the
larger its  rate of  increase.  The physical meaning of this convective term
$\nu_\alpha^2/2M_\alpha=\Pi^2_\alpha/2M_\alpha$  can be
understood from a simple argument.  Imagine the body is undergoing pure
dilational motion of the form ${\bf r}_i(t)=g(t){\bf r}_i(0)$, for some
function $g(t)$.  This means the gyration tensor has the form
${\bf G}(t)=g^2(t){\bf G}(0)$ and, consequently the central moments evolve
according to $M_\alpha(t)=g^2(t)M_\alpha(0)$.  The dilational momentum then
takes the form $\Pi_\alpha(t)=\dot{M}_\alpha(t)=(2\dot{g}(t)/g(t))M_\alpha(t)$
and its time derivative takes the form
$\dot{\Pi}_\alpha(t)=\Pi_\alpha^2(t)/2M_\alpha(t)+(2\ddot{g}(t)/g(t))\Pi_\alpha(t)$.
Therefore, the convective term accounts for the dilational
momentum changes under constant rate dilations (for which $\ddot{g}(t)=0$).

In spite of this physical interpretation,
we may  worry about unstable, runaway behaviour in
the  equations due to the convective term.   A closer  look,  however,  shows  there is  no  such
runaway.   In fact,  if we  forget about  the coupling  with the  spin
velocity and  the ``elastic''  restoring force,  the structure  of any
single component of the two equations (\ref{eq:83}), with the dilation
force (\ref{eq:137}) is
\begin{align}
  \label{eq:439}
  \frac{d M}{dt}
  &=\Pi
    \nonumber\\
  \frac{d\Pi}{dt}
    &=\frac{\Pi^2}{2M}-\gamma \Pi
\end{align}
where $\gamma$ is a friction coefficient.
The solution of this system of equations is
\begin{align}
  \label{eq:442}
   M(t)
  &=\left(\frac{e^{\gamma (c_2 -t)}+c_1}{2\gamma}\right)^2
    \nonumber\\
{\Pi}(t)
  &=-2\gamma  M(t)+c_1M^{1/2}(t)
\end{align}
where $c_1,c_2$ are integration constants. At long times we have
\begin{align}
  \label{eq:443}
  \lim_{t\to\infty} M(t)
  &=\left(\frac{c_1}{2\gamma}\right)^2
    \nonumber\\
  \lim_{t\to\infty} {\Pi}(t)
  &=0
\end{align}
which is  perfectly behaved. Of  course, the inclusion of  the elastic
term will  produce oscillatory  motion and, if  the body  is spinning,
there will be  also a coupling with the spin  velocity, but no runaway
should occur.

The centrifugal  contribution in  the dilational  force (\ref{eq:137})
couples the  motion of the central  moments to the orientation  in the
SDE's  (\ref{eq:212})  because,  according to  (\ref{vel}),  the  spin
velocity contains the  inertia tensor (hence, the  central moments) in
its definition.  Finally, the  elastic contribution to  the dilational
force  gives a  restoring dilational  force  that tries  to bring  the
central moments  to their  value at rest.   Should the  convective and
centrifugal terms be absent, the equations would be isomorphic to that
of a  particle tethered with  an elastic spring. Therefore,  we expect
oscillatory damped dynamics  for the central moments  ${\bf M}$, which
will also produce oscillations in the orientation dynamics.

As a final remark,  SDE (\ref{eq:212}) contains three \textit{material
  properties}  given  by the  following  matrices:  the covariance  of
central moments  $\boldsymbol{\Sigma}$  specifying the  elastic properties  of the
body, the  orientational diffusion tensor $\boldsymbol{\cal  D}_0$ governing
the rate  at which  the orientation  evolves towards  equilibrium and,
finally,  the dissipative  matrix $\boldsymbol{\Gamma}_{_\Pi}$  governing
the  relaxation  of  the  central moments  towards  their  equilibrium
values.  These  material  properties  determine  the  time  scales  of
evolution of  the CG variables.  For  example, for a body  with a high
``bulk  viscosity''  (large  $\boldsymbol{\Gamma}_{_\Pi}$) that  is  very
``rigid''  (small  $\boldsymbol{\Sigma}$)  the  central  moments are  rapidly  and
strongly damped  towards their equilibrium values  ${\bf M}^{\rm eq}$.
In this  case, the central  moment dynamics  are fast compared  to the
orientational dynamics,  so we have  a separation of time  scales, and
the orientation dynamics decouple from the shape dynamics. On the time
scales of relaxation of the orientation towards the equilibrium state,
the central moments  adopt their equilibrium values.  In  this case, a
much simpler CG theory can be constructed  using the orientation as the only relevant CG
variable.  This theory leads precisely to
the  SDE for  the orientation  (\ref{eq:212}) with  the values  of the
central moments fixed at their equilibrium values.
\subsection{The Second Law}
  \label{Sec:2nL}
  Observe  that,  formally,  the  set of  coupled  ODEs (\ref{Lambda-Det}) and
(\ref{eq:83}) are obtained from  (\ref{eq:212}) by  setting $k_B\to0$.
In that limit, the entropy (\ref{eq:195}) becomes 
\begin{align}
  \label{eq:310}
S_B  &=S^{\rm MT}({\cal E})
    -\frac{1}{2T^{\rm MT}}({{\bf M}}-{\bf M}^{\rm rest})^T\esc\boldsymbol{\Sigma}^{-1}\esc({{\bf M}}-{\bf M}^{\rm rest})
\end{align}
The time derivative of the entropy (\ref{eq:310}) can be computed from
the chain rule, and  the dynamics (\ref{Lambda-Det}) and (\ref{eq:83}) of
the CG variables, giving
\begin{align}
  \label{eq:313}
\frac{dS_B}{dt}&=  \frac{1}{T^{\rm MT}}\left[
    \left(\boldsymbol{\Omega}\times{\bf S}\right)^T\esc\boldsymbol{\cal D}\esc    \left(\boldsymbol{\Omega}\times{\bf S}\right)
+     \boldsymbol{\nu}^T\esc\boldsymbol{\Gamma}_{_\Pi}\esc\boldsymbol{\nu}\right]\ge0
\end{align}
This  time derivative  is  always  positive, as  a  consequence of  the
positive   character   of    the   Green-Kubo   dissipative   matrices
$\boldsymbol{\cal    D},\boldsymbol{\Gamma}_{_\Pi}$.    Therefore,    the
deterministic  dynamic equations  satisfy  the  Second Law.   Of course,
(\ref{eq:313}) is  just the particular instance of the Second  Law (\ref{eq:294}) of  the general
theory of CG.

The state  $a(t)$ of  the system will  evolve in such  a way  that the
entropy  increases monotonically  while conserving  total energy,  and
linear and  angular momenta. This  implies  $a(t)\to  a^{\rm eq}$
for the equilibrium state $a^{\rm eq}$ that maximizes the
entropy, where the time derivative  (\ref{eq:313}) vanishes.
Due         to        the         positive        character         of
$\boldsymbol{\cal  D},\boldsymbol{\Gamma}_{_\Pi}$, this  can only  happen
if,
\begin{align}
  \label{eq:316}
  \boldsymbol{\Omega}^{\rm eq}\times{\bf S}
  &=0
    \nonumber\\
  \boldsymbol{\nu}^{\rm eq}
  &=0
\end{align}
The last condition  means the body does  not move dilatationally,
and has constant values ${\bf M}^{\rm eq}$ of the central moments. The
first  condition  means  the spin  velocity 
\begin{align}
  \label{eq:17b}
  \boldsymbol{\Omega}^{\rm eq}={\bf  I}^{-1}(\boldsymbol{\Lambda}^{\rm
  eq},{\bf M}^{\rm eq})\esc{\bf  S}  
\end{align}
should be parallel to  ${\bf  S}$.   This,  in turn,  is  only
possible if ${\bf S}$ becomes an eigenvector of the inertia tensor or,
in                             other                            words,
${\bf S}/I  = {\bf I}^{-1}(\boldsymbol{\Lambda}^{\rm  eq},{\bf M}^{\rm
  eq})\esc{\bf   S}$  with   $I$   one  of   the  principal   moments,
$I\in\{ I_1^{\rm  eq}, I_2^{\rm  eq}, I_3^{\rm eq}\}$.   The resulting
rotational  kinetic  energy is  $K^{\rm  rot}=S^2/(2I)$  with $S$  the
modulus   of   the   angular  momentum.    From   (\ref{eq:104})   and
(\ref{Thermal}), the maximum of the  entropy is reached at the minimum
of the kinetic energy, which  happens for the largest principal moment
$I_3^{\rm  eq}$.  Therefore,  at equilibrium  the major principal  axis with
{\em largest}  principal moment of  inertia aligns with  the conserved
angular   momentum   vector.     Observe   from   (\ref{eq:146})   and
(\ref{eq:316})                                                    that
$\boldsymbol{\Omega}^{\rm eq}=\boldsymbol{\omega}^{\rm   eq}$.   When
dissipation is taken  into account, the two vectors,  angular and spin
velocity, align with the angular momentum vector, that is accommodated
in the  major principal axis  of the  body. This evolution  is rather
different from  the prediction  of the  angular velocity  from Euler's
equations for  a free rigid body,  that follows a polhode  dictated by
the Poinsot construction.

We may study the rate of rotational kinetic energy transfer into thermal energy
during the evolution towards equilibrium by using
\begin{align}
  \label{eq:7}
  \frac{dK^{\rm rot}}{dt}&=\frac{\partial K^{\rm rot}}{\partial \boldsymbol{\Lambda}}\esc \frac{d\boldsymbol{\Lambda}}{dt}+
                           \frac{\partial K^{\rm rot}}{\partial {\bf M}}\esc \frac{d{\bf M}}{dt}
\end{align}
The derivatives of $K^{\rm rot}$ with respect to $\boldsymbol{\Lambda}$ is computed
in Sec K.4 of the Supplemental Material and with respect to ${\bf M}$ in
Sec D. Using (\ref{Lambda-Det}) and (\ref{ID}), (\ref{Kinetic}), we have
\begin{align}
  \label{eq:11-M}
  \frac{dK^{\rm rot}}{dt}
  &=
    -     \left(\boldsymbol{\Omega}\times{\bf S}\right)^T\esc\boldsymbol{\cal D}\esc\left(\boldsymbol{\Omega}\times {\bf S}\right)
    - \frac{1}{2}   \boldsymbol{\Omega}_p^T\esc \frac{d\mathbb{I}}{dt}\esc   \boldsymbol{\Omega}_p
\end{align}

The  first term  on the  right  hand side  is closely  related to  the
orientational  diffusion  contribution  to  the  dissipation  rate  in
(\ref{eq:313}), and is always  negative, because $\boldsymbol{\cal D}$
is positive definite.   The second term involving  the time derivative
of the principal  moments is eventually linked,  through the kinematic
condition $\dot{\bf M}=\boldsymbol{\Pi}$,  to the dilational momentum,
which we  expect to  follow a  damped oscillation.   The last  term in
(\ref{eq:11-M})  has no  definite  sign, and  may  either increase  or
decrease  the rotational  kinetic energy.   Therefore, (\ref{eq:11-M})
shows orientational  diffusion, and  not the  dynamics of  the central
moments,  is  the ultimate  responsible  for  reducing the  rotational
kinetic energy of the body.  Undoubtedly, both orientational diffusion
and   viscoelasticity    contribute   to   the    entropy   production
(\ref{eq:313}), but only the former produces the alignment of the body
by minimizing the rotational kinetic energy.

\subsection{Heating}   
For a spinning body, the orientation  will evolve in order to minimize
the  rotational kinetic  energy, while  conserving total  energy.  The
rotational kinetic energy ``lost'' in  the alignment process causes an
increase of the thermal energy  ${\cal E}$ defined in (\ref{Thermal}).
Therefore,  in this  alignment  process  the body  will  heat up,  and
increase its temperature.  For example, a flat  uniform disk initially
set rotating about one diameter will eventually end up rotating around
the  axis perpendicular  to the  disk, where  the moment  is twice  as
large.  In  the process, it will  convert half of the  initial kinetic
energy into heat, increasing the  temperature of the body accordingly.
This  dissipation arises  because at  the atomistic  level, atoms  can
vibrate and thus sample the more probable configurations corresponding
to smaller  rotational kinetic  energy.  In  time, this  will transfer
organized rotation kinetic energy  into disorganized thermal energy as
the system moves towards the lowest rotational kinetic energy state at
equilibrium.

\subsection{The equilibrium shape of the body}
We  predict  now   the  shape  of  the  body   at  equilibrium.  Using
(\ref{eq:316})   in   (\ref{eq:83}) implies $\boldsymbol{\cal K}=0$.  With   the   dilational   force
(\ref{eq:137}) we obtain the equilibrium condition
  \begin{align}
  \label{eq:55}
    2\left( \boldsymbol{\Omega}_p^T\esc{\boldsymbol{\Omega}_p}-\boldsymbol{\Omega}_{p\alpha}^2 \right)
  &=[{\Sigma}^{-1}]_{\alpha\beta}({M}^{\rm eq}_\beta-{M}^{\rm rest}_\beta)
\end{align}
Using (\ref{eq:453}) and (\ref{eq:17b}) then gives
\begin{align}
  \label{eq:208}
  \boldsymbol{\Omega}_p^{\rm eq}
  &=
e^{-[\boldsymbol{\Lambda}^{\rm eq}]_\times}\esc\boldsymbol{\Omega}^{\rm eq}=
     e^{-[\boldsymbol{\Lambda}^{\rm eq}]_\times}\esc\frac{\bf S}{I_3^{\rm eq}}
=\left(0,0,\frac{S}{I_3^{\rm eq}}\right)                         
\end{align}
where the last result
takes into account that the body at equilibrium rotates around the
major axis along the angular momentum vector $\bf S$, chosen to be in the  $z$-axis.  By
introducing the  distortion $\Delta^\alpha$ as the  difference between
the  equilibrium central  moment and its value at rest
\begin{align}
  \label{eq:259}
\Delta_\alpha&\equiv  M^{\rm eq}_\alpha-M_\alpha^{\rm rest}
\end{align}
we may write (\ref{eq:55}) as
\begin{align}
  \label{eq:233}
\Delta_\alpha
  &=2 \left( \Sigma_{\alpha 1}+ \Sigma_{\alpha 2} \right)\left(\frac{S}{I_3^{\rm eq}}\right)^2
\end{align}
This gives the  distortion of the body at equilibrium  in terms of the
covariance matrix and the angular momentum of the body.

\section{Discussion and Conclusions}
\label{Sec:Conclusions}

In this paper  we have constructed the equations of  motion for a free
deformable body composed of bonded  particles that is represented at a
coarse-grained level  in terms of ``shape  and orientation'' variables
determined by the gyration tensor, specifically by the central moments
and attitude parameters.  These equations generalize Euler's equations
for the  motion of a rigid  body by including dissipative  effects and
thermal fluctuations, both related through the Fluctuation-Dissipation
Theorem.  We  emphasize that the  dissipative and thermal  effects are
intrinsic  to  the  body  and   there  is  no  external  thermal  bath
present. This  departs from existing literature  on so-called Brownian
rotors.

We  have highlighted  two assumptions  implicit in  the derivation  of
Euler's equations  in Classical Mechanics: constancy  of the principal
moments (${\cal  H}1$), and the  vanishing of angular momentum  in the
principal   axis   system   (${\cal    H}2$).    As   seen   in   Sec.
\ref{Sec:Rig.Sol},  with these  two assumptions  one obtains  a closed
equation for  the orientation dynamics,  leading to the  usual Euler's
equations for a rigid body.  These two assumptions bypass the need for
a  statistical  treatment  in  the  usual  presentation  in  Classical
Mechanics,  but they  are clearly  insufficient. Rigid  bodies do  not
truly  exist,  and  the  inertia  tensor evolves  in  general  due  to
deformations, violating ${\cal H}1$, while thermal fluctuations induce
a non-zero  angular momentum  in the  principal axis  frame, violating
${\cal H}2$.

We  have  shown  how  it is  necessary  to  introduce  non-equilibrium
statistical mechanics concepts when these two assumptions do not hold.
The theory of coarse graining allows  us to construct the equations of
motion, with all  their building blocks defined  in microscopic terms.
These   microscopic   expressions  involve   non-trivial   conditional
expectations that  render the  equations of motion  highly non-linear.
However, we have been able to provide the explicit functional form for
all building blocks.   Some approximations have been made,  but not to
the point of blurring the  physical content.  The information required
in the  building blocks can  be obtained from ordinary  equilibrium MD
simulations  for  small bodies,  or  fitting  to observations  of  the
dynamics for macroscopic bodies.

The theory has  a number of fascinating  qualitative predictions.  Due
to  the  thermal  motion  of  the atoms,  the  orientation  follows  a
diffusive process with an associated dissipative mechanism.  Even when
the body does not spin (zero  angular momentum), it keeps changing its
orientation randomly, exploring all possible orientations according to
the uniform Haar measure.  For macroscopic bodies that conserve energy
(not  cats  or  trampolinists  that  transform  chemical  energy  into
mechanical energy to actuate their parts) this seems counterintuitive.
A torque-free asteroid at rest in free space is expected to stay still
in its orientation  forever.  \textit{Eppur si muove},  due to thermal
fluctuations, with a time scale that should grow with the size of the body.
For  a molecule  in  vacuum  with zero  angular  momentum, the  random
exploration  of   orientations  is   very  noticeable  in   actual  MD
simulations.   When the  body spins  with angular momentum along an
arbitrary  axis, the  dissipative  mechanism  associated with  thermal
fluctuations degrades  rotational kinetic energy into  thermal energy,
and the body  tends to a minimum of the  rotational kinetic energy, in
which the principal  axis of largest principal moment  aligns with the
conserved  angular momentum.   This is  a natural  consequence of  the
Second Law of thermodynamics, which is built into the theory of coarse
graining.   In the  process of  reorienting, the  body heats  up, and
increases its temperature.  The  reorientation and transfer of kinetic
energy  to  thermal  energy  in   a  spinning  body,  and  the  random
exploration  of  orientations in  a  non-spinning  body are  forbidden
processes in an ideal rigid body.   The distortion of the shape of the
body when it  spins along the principal axis in  the equilibrium state
is predicted quantitatively in the present theory.

While the alignment of spinning  bodies is well-known from equilibrium
statistical mechanics  \cite{Landau1980}, the present  theory predicts
quantitatively  how   this  process  occurs,  as   it  formulates  the
dissipative  Euler's   equations  governing   the  evolution   of  the
orientation  and  central  moments towards  the  equilibrium  spinning
state.   In addition,  we  have identified  two different  dissipative
mechanisms,  one  associated  with  the  random  fluctuations  of  the
orientation due to thermal molecular agitation, and another one due to
damping of the evolution of the  central moments, which plays the role
of  viscoelasticity in  the present  level of  description.  Only  the
first  mechanism  is  responsible  for precession  relaxation  in  our
theory.              As            current             interpretations
\cite{Lazarian2007,Efroimsky2000,Efroimsky2001,Warner2009,Breiter2012}
assume  viscoelastic dissipation  is  responsible  for the  precession
relaxation of astrophysical objects, we  should clarify our claim that
``viscoelastic dissipation does not  cause precession relaxation".  As
discussed  in the  Introduction, we  may use  two different  levels of
description when describing a body: a coarse one based on the gyration
tensor, as  done in  the present  work, or  a finer  one based  on the
elastic fields  of mass  density, displacement, and  momentum density,
for example.  This  field theory can describe the  complex rheology of
the body, i.e. viscoelasticity. For the  sake of discussion, it may be
simpler to consider a discretization of a viscoelasticity field theory
in  terms  of  mesoscopic   particles  interacting  with  springs  and
dashpots,  as  done  in   Ref.   \cite{Quillen2019}.  Each  mesoscopic
particle represents  a group of  microscopic atoms.  In  this particle
model, springs are  responsible for the ``elastic"  part, and dashpots
for  the ``visco"  part of  the dynamics.   The number  of degrees  of
freedom in each CG model  (gyration tensor-based or particle-based) is
very different, but  both models aim to represent  the same underlying
atomistic description  of the  body, albeit at  different resolutions.
The observation  of the present  low resolution theory is  that within
the many different modes in which the mesoscopic particles may move in
the  high  resolution  theory,  the  modes  involving  the  motion  of
particles  in  pure  dilation  should  not  contribute  to  precession
relaxation.  Only modes with non-zero angular velocity will contribute
to the  effect of changing  the orientation, which is  consistent with
(\ref{eq:29b}).

By construction, the theory is limited to bodies whose ``shape'' can
be reasonably  described with central moments.   Examples of systems
that can  be addressed with  these kind  of CG variables  range from
complex molecules in vacuum, to spinning liquid droplets, asteroids,
and other  large masses in free  space.  On the other  hand, a large
flexible plate or filament excited with high frequency modes may not
be  well described  with  the present  theory,  neither droplets  in
regimes  where breakup  due to  fast rotations  occurs, because  the
``shape''  of these  bodies requires  additional degrees  of freedom
beyond central moments.

The present theory can be  extended in several directions.  First, the
effect of  external fields on  a single  free body can  be considered.
When the  external field is due  to a central gravitational  field, we
will present in a future  publication a first principles derivation of
the governing equations for  tidal locking, the dissipative phenomenon
responsible for the  synchronization of the day and year  of bodies in
revolution.  Second, one may consider the formulation, from the theory
of  coarse graining,  of  the Brownian  rotor, in  which  the body  is
immersed  in  a  passive  fluid.    Finally,  the  extension  to  many
interacting  thermal and  deformable bodies  can be  considered.  This
points to  the possibility  of constructing  mesoscopic coarse-grained
models for complex soft matter from first principles, when intuitively
one expects  shape variables  are required.  A recent  example is
  given in Ref.  \cite{Goujon2020} where polymer chains are modelled
  with  anisotropic  beads.  Other  examples  are  deforming globular polymer
molecules under flow, granular  matter based on deformable ellipsoids,
and  liquid  crystals with  intrinsic  fluctuations.   We believe  the
introduction of shape as  a CG variable may be a  valuable tool in the
formulation of CG models of soft and granular matter, and that many of
the mathematical results presented in this work may be instrumental to
this goal.

\color{black}

\section{Acknowledgment}
P.E. deeply acknowledges Aleksandar  Donev and Hans Christian Ottinger
for  very useful  comments  on this  manuscript  and their  continuing
support and  encouragement. This  research has been  supported through
grants PDC2021-121441-C22, PID2020-117080RB-C54 funded  by MCIN,
and a grant from the Natural Sciences and Engineering Research Council
(NSERC) of Canada.

\end{document}


\title{Supplemental Material for ``The role of thermal fluctuations in the motion of a free body''}

\author{Pep Espa\~nol}
\affiliation{ Dept.   F\'{\i}sica Fundamental, Universidad Nacional
  de Educaci\'on a Distancia, Madrid, Spain}
\author{Mark Thachuk}
\affiliation{ Department of Chemistry, University of British Columbia, Vancouver, Canada}
\author{J.A. de la Torre}
\affiliation{ Dept.   F\'{\i}sica Fundamental, Universidad Nacional
  de Educaci\'on a Distancia, Madrid, Spain}

\date{\today}
\begin{abstract}
We present the explicit calculations required in the formulation of the stochastic differential equations governing the orientation and central moments of a free body, discussed in the main text.
\end{abstract}
\maketitle

  \color{black}
\section{A guide to the Supplemental Material}
In the  main text we have  presented the different building  blocks in
the SDE governing the dynamics of the CG variables, while the explicit
calculation of  these results has  been deferred to  this Supplemental
Material. The Sections  can be divided into groups.   The first group,
Secs.  \ref{App:Entropy}-\ref{App:cross}, presents  the calculation of
the building blocks entering the dynamic equations.  The second group,
Secs.   \ref{App:DerRot}-\ref{App:Sym}  presents   properties  of  the
rotation matrix, the angular velocity and its connection with the time
derivative  of the  orientation.  The  third group  collected in  Sec.
\ref{App:Mic} gives  different phase  functions expressed in  terms of
the coordinates  and velocities  of the  particles. In  particular, we
obtain the explicit microscopic form of the angular velocity, dilation
momentum,  dilational  force, as  well  as  a  number of  other  phase
functions related to  the derivatives of CG variables  with respect to
particles  coordinates.   The  explicit  form of  these  variables  is
required when their averages and correlations are to be measured in MD
simulations.   The  last  section \ref{App:Submanifold}  presents  the
calculation  of  special  Gaussian integrals  containing  Dirac  delta
functions of  conserved quantities that are  instrumental in computing
momentum integrals  in the evaluation  of the entropy  and conditional
expectations.  All the results  presented in the Supplemental Material
are used at some point in the calculations of the building blocks.

\setcounter{section}{0}
\renewcommand{\appendixname}{\large{Appendix}}
\renewcommand{\theequation}{\arabic{equation}}
\newcounter{my_equation_counter}
\renewcommand{\thesubsubsection}{\thesubsection.\alph{subsubsection}}
\renewcommand{\thesection}{\large{\Alph{section}}}
\renewcommand{\thesubsection}{\thesection.\arabic{subsection}}
\section{Calculation of the entropy}
\setcounter{equation}{\getrefnumber{finaleq}}
\label{App:Entropy}
In this section we evaluate the entropy function
at  the present  level of  description, which is given in terms of the integral
in phase space
\begin{align}
  \label{SM:1}
S_B({\bf    R},\boldsymbol{\Lambda},{{\bf M}},{\bf
    P},{\bf S},\boldsymbol{\Pi},E)
&  =k_B\ln \int dz
  \delta\left(\hat{\bf R}(z)-{\bf R} \right)
    \delta\left(\hat{\boldsymbol{\Lambda}}(z)-\boldsymbol{\Lambda}\right)
    \delta\left(\hat{{\bf M}}(z)-{{\bf M}} \right)
\nonumber\\
\times&\delta\left(\hat{\bf P}(z)-{\bf P} \right)
          \delta\left(\hat{\bf S}(z)-{\bf S} \right)
          \delta\left(\hat{\boldsymbol{\Pi}}(z)-\boldsymbol{\Pi}\right)
              \delta\left(\hat{H}(z)-E \right)
\end{align}
The evaluation  is possible  because the
momentum integrals involve the  intersection of a hypersphere, defined
by the  Dirac delta  on the  kinetic energy  in the  Hamiltonian, with
three   hyperplanes,   defined   by   the  three   Dirac   deltas   on
${\bf  P},{\bf   S},\boldsymbol{\Pi}$.   Motivated  by   the  Gaussian
integrals  discussed in Sec. \ref{App:Submanifold},  consider the
following   change  of   variables  from   $z$  to   \textit{co-moving
  coordinates} $z'(z)$
\begin{align}
{\bf r}'_{i}&={\bf r}_{i}-{\bf R}
\nonumber\\  
 {\bf p}_{i}'&
={\bf p}_{i} - m_{i}{\bf V} +  m_{i} ({\bf r}_{i}-{\bf R})\times\hat{\boldsymbol{\Omega}}
               -\frac{\partial \hat{{\bf M}}}{\partial {\bf r}_{i}}\esc\hat{\boldsymbol{\nu}}
\label{W:BodyFrame}
\end{align}
where we have introduced the linear, spin, and dilational velocities as
\begin{align}
  \label{eq:85}
    {\bf V}
  &=\frac{{\bf P}}{{\rm M}}
\nonumber\\
  \hat{\boldsymbol{\Omega}}
  &=\hat{\bf I}^{-1}\esc{\bf S}
    \nonumber\\
  \hat{\boldsymbol{\nu}}
  &=\hat{\bf T}^{-1}\esc\boldsymbol{\Pi}
\end{align}
Here,                            the                            values
${\bf R}, {\bf  P}, {\bf S}$, and $ \boldsymbol{\Pi}$ are
those appearing  in the Dirac  delta function of
(\ref{SM:1}).   They are  numbers,  not  phase functions.  The
total  mass of  the  body is  ${\rm  M}$ and  we  have introduced  the
dilation  inertia matrix  (with the  same physical  dimensions as  the
inertia tensor)
\begin{align}
  \label{eq:42b}
  \hat{\bf T}
  &\equiv\sum_i\frac{1}{m_i}\frac{\partial \hat{{\bf M}}}{\partial {\bf r}_i}
    \esc \frac{\partial \hat{{\bf M}}}{\partial {\bf r}_i}
\end{align}
where  the  dot  product  means   summation  over  the  components  of
${\bf r}_i$.  Even though we may  recognize in the change of variables
(\ref{W:BodyFrame}) a translation and rotation of the coordinates to a
rotating frame  in which the  linear and  spin momentum are  zero, the
last term  involving the dilational  velocity $\hat{\boldsymbol{\nu}}$
may  appear  intriguing.   As  we   show  below,  this  last  term  in
(\ref{W:BodyFrame})  ensures  \textit{all} momenta  (linear,  angular,
\textit{and}    dilational)    simultaneously    vanish    in    these
coordinates. Furthermore,  the Jacobian determinant of  this change of
variables   is    unity   because    $\hat{\boldsymbol{\Omega}}$   and
$\hat{\boldsymbol{\nu}}$  in  (\ref{eq:85})  do   not  depend  on  the
velocities ${\bf v}_i$.  The inverse transformation $z(z')$ is
\begin{align}
  {\bf r}_{i}
  &={\bf r}'_{i}+{\bf R} 
\nonumber\\  
  {\bf p}_{i}
  &={\bf p}'_{i} + m_{i}{\bf V} -  m_{i} {\bf r}'_{i}\times\hat{\boldsymbol{\Omega}}'(z')
    +\frac{\partial \hat{{\bf M}}(z')}{\partial {\bf r}'_{i}}\esc\hat{\boldsymbol{\nu}}(z')
    \label{W:BodyFrameInv}
\end{align}
where
\begin{align}
  \label{eq:217}
  \hat{\boldsymbol{\Omega}}'(z')
  &=  \hat{\boldsymbol{\Omega}}(z(z'))
    =\hat{\bf I}^{-1}(z(z'))\esc{\bf S}=
    [\hat{\bf I}']^{-1}(z')\esc{\bf S}
  \nonumber\\
  \hat{\boldsymbol{\nu}}'(z')
  &=\hat{\boldsymbol{\nu}}(z'(z))
    =\hat{\bf T}^{-1}(z(z'))\esc\boldsymbol{\Pi}
    =[\hat{\bf T}']^{-1}(z')\esc\boldsymbol{\Pi}
\end{align}
The inertia tensor in the new variables is
\begin{align}
  \label{eq:218}
\hat{\bf I}'(z')&\equiv\sum_im_i[{\bf r}_i']_\times^T \esc  [{\bf r}_i']_\times
\end{align}
and the tensor ${\bf T}'(z')$ is
\begin{align}
  \label{eq:228}
  \hat{\bf T}'(z')&\equiv  \hat{\bf T}(z(z'))
\end{align}
Define the operator ${\cal G}$
whose action on an arbitrary phase function is
\begin{align}
  \label{eq:35}
  {\cal G}\hat{F}(z')=\hat{F}(z(z'))
\end{align}
According to  the general  form for a change  of variables  (with unit
Jacobian)  (\ref{SM:1}) can be written as
\begin{align}
S_B({\bf    R},\boldsymbol{\Lambda},{{\bf M}},{\bf
  P},{\bf S},\boldsymbol{\Pi},E)
  &=k_B\ln \int dz'
\delta\left({\cal G}\hat{\bf R}(z')-{\bf R} \right)
    \delta\left({\cal G}\hat{\boldsymbol{\Lambda}}(z')-\boldsymbol{\Lambda}\right)
    \delta\left({\cal G}\hat{{\bf M}}(z')-{{\bf M}} \right)
\nonumber\\
&\quad\times\delta\left({\cal G}\hat{\bf P}(z')-{\bf P} \right)
              \delta\left({\cal G}\hat{\bf S}(z')-{\bf S} \right)
                  \delta\left({\cal G}\hat{\boldsymbol{\Pi}}(z')-\boldsymbol{\Pi}\right)
              \delta\left({\cal G}\hat{H}(z')-E \right)
\label{W:EntropyBody}
\end{align}
So we  only have to use (\ref{W:BodyFrameInv}) to determine the effect of the operator ${\cal G}$ on the CG  variables.
  The center of mass position 
transforms as
\begin{align}
{\cal G}\hat{\bf R}(z')&=  \hat{\bf R}(z(z'))=
\sum_{i}\frac{m_{i}}{{\rm M}}{\bf r}'_{i}+{\bf R} 
\equiv \hat{\bf R}(z') +{\bf R}   
\end{align}
The inertia tensor transforms as
\begin{align}
  \label{W:eq:208}
{\cal G}  \hat {\bf I}(z')
&=\sum_{i}m_{i}[{\bf r}'_{i}]_\times^T\esc[{\bf r}'_{i}]_\times 
\equiv \hat {\bf I}(z')+{\rm M}[\hat{\bf R}(z')]_\times\esc[\hat{\bf R}(z')]_\times
\end{align}
and the  gyration tensor as
\begin{align}
  \label{eq:287}
{\cal G}    \hat {\bf G}(z')
&=\frac{1}{4}\sum_{i}m_{i}{{\bf r}'}_{i}{{\bf r}'_{i}}^T
\equiv\hat{\bf G}(z')+\frac{{\rm M}}{4}\hat{\bf R}(z')\hat{\bf R}^T(z')
\end{align}
Because  the  inertia  tensor  only   depends  on  positions  and  the
transformation (\ref{W:BodyFrame}) is just  a translation,
the orientation and principal  moments in
the new configuration $r'$ are the same as in $r$, that is
\begin{align}
  \label{W:eq:209}
{\cal G}  \hat{\boldsymbol{\Lambda}}(z')
  &=\hat{\boldsymbol{\Lambda}}(z')
  \nonumber\\
{\cal G}  \hat{{\bf M}}(z')
  &=\hat{{\bf M}}(z')
\end{align}

The momentum transforms as 
\begin{align}
{\cal G}\hat{\bf P}(z') &=
\sum_{i}\left({\bf p}'_{i} -  m_{i} {\bf r}'_{i}\times{\hat{\boldsymbol{\Omega}}'}(z')+ m_{i}{\bf V} +\frac{\partial \hat{{\bf M}}(z')}{\partial {\bf r}'_{i}}\esc\hat{\boldsymbol{\nu}}'(z')\right)
\equiv     \hat{\bf P}(z') - {\rm M}\hat{\bf R}(z')\times{\hat{\boldsymbol{\Omega}}'(z')}+{\bf P}
\end{align}
where we have used  the translation invariance property (\ref{trans2}).
The  spin   transforms as
\begin{align}
{\cal G}  \hat{\bf S}(z')
  &=\sum_{i}{\bf r}'_{i}\times
    \left({\bf p}'_{i} -  m_{i} {\bf r}'_{i}\times{\hat{\boldsymbol{\Omega}}}(z')
    + m_{i}{\bf V} +\frac{\partial \hat{{\bf M}}(z')}{\partial {\bf r}'_{i}}\esc\hat{\boldsymbol{\nu}}(z')\right)
    \nonumber\\
  &\overset{(\ref{dsym})}{=}\sum_{i} {\bf r}'_{i}\times
    {\bf p}'_{i} 
    -\sum_{i} {\bf r}'_{i}\times
    \left(  m_{i} {\bf r}'_{i}\times{\hat{\boldsymbol{\Omega}}}(z') \right)
    +
    \hat{\bf R}(r')\times {\bf P} 
\nonumber\\
  &= \hat{\bf S}(z')+\hat{\bf I}(z')\esc {\hat{\boldsymbol{\Omega}}}(z')
    +\hat{\bf R}(z')\times{\bf P}
\end{align}
where the inertia tensor transforms as (\ref{W:eq:208}),
and (\ref{eq:85}) shows the second term on the right hand side is the constant vector ${\bf S}$.
The dilation momentum  transforms as
\begin{align}
  \label{eq:40}
{\cal G}  \hat{\boldsymbol{\Pi}}(z')
  &=\sum_i\frac{\partial \hat{{\bf M}}(z')}{\partial {\bf r}'_i}\esc
 \left({\bf v}'_{i} -   {\bf r}'_{i}\times{\hat{\boldsymbol{\Omega}}}(z')
    +{\bf V} +\frac{1}{m_i}\frac{\partial \hat{{\bf M}}(z')}{\partial {\bf r}'_{i}}\esc\hat{\boldsymbol{\nu}}(z')\right)
=
    \hat{\boldsymbol{\Pi}}(z')+\hat{\bf T}(z')\esc\hat{\boldsymbol{\nu}}(z') =
      \hat{\boldsymbol{\Pi}}(z')+\boldsymbol{\Pi}
\end{align}
where  we  have  used  the  translation  (\ref{trans2})  and  rotation
(\ref{dsym}) invariance  of the central moments,  and the definition
of the  dilational velocity  (\ref{eq:85}).  Finally,  the Hamiltonian
 transforms as
\begin{align}
{\cal G}\hat{H}(z') 
&= \sum_{i}\frac{m_{i}}{2}\left({\bf v}'_{i}+{\bf V}
                  -{\bf r}'_{i}\times\hat{\boldsymbol{\Omega}}(z')
                  +\frac{1}{m_i}\frac{\partial \hat{{\bf M}}(z')}
                  {\partial {\bf r}'_{i}}\esc\hat{\boldsymbol{\nu}}(z')
                  \right)^2+\hat{\Phi}(z')
                    \nonumber\\
    &= \hat{H}(z')
      + \hat{\bf P}(z')\esc{\bf V}
      + \hat{\bf S}(z')\esc\hat{\boldsymbol{\Omega}}(z')+\hat{\boldsymbol{\Pi}}(z')\esc\hat{\boldsymbol{\nu}}(z')
           -M (\hat{\bf R}(z')\times\hat{\boldsymbol{\Omega}}(z'))\esc{\bf V}
      \nonumber\\
  &+ \frac{{\rm M}{\bf V}^2}{2}
    + \frac{1}{2}\hat{\boldsymbol{\Omega}}^T(z')\esc\hat{\bf I}(z')\esc\hat{\boldsymbol{\Omega}}(z')
       + \frac{1}{2}\hat{\boldsymbol{\nu}}^T(z')\esc\hat{\bf T}(z')\esc\hat{\boldsymbol{\nu}}(z')
      + \frac{1}{2}{\rm M}(\hat{\bf R}(z')\times\hat{\boldsymbol{\Omega}}(z'))^2
\end{align}
where  we   have  used   the  fact   the  intramolecular  potential
$\hat{\Phi}(z)$ is translationally invariant, and (\ref{dsym}), (\ref{trans2}).

Substitution   of  the   above   transformed   phase  functions   into
(\ref{W:EntropyBody}) leads to
\begin{align}
  \label{W:eq:122}
S_B({\bf    R},\boldsymbol{\Lambda},{{\bf M}},{\bf
  P},{\bf S},\boldsymbol{\Pi},E)=
k_B\ln
\int dz'
  &\delta\left(\hat{\bf R}(z')\right)
    \delta\left(\hat{\boldsymbol{\Lambda}}(z')-\boldsymbol{\Lambda}\right)
    \delta\left( \hat{{\bf M}}(z')-{{\bf M}} \right)
  \delta\left(\hat{\bf S}(z')\right)
    \delta\left(\hat{\bf P}(z')\right)
    \delta\left(\hat{\boldsymbol{\Pi}}(z')\right)
  \nonumber\\
  &\times
    \delta\left(\hat{H}(z')-\left(E-\frac{{\bf P}^2}{2{\rm M}}
    -\frac{1}{2}{\bf S}^T\esc\hat{\bf I}^{-1}(z')\esc{\bf S}
        -\frac{1}{2}\boldsymbol{\Pi}^T\esc\hat{\bf T}^{-1}(z')\esc\boldsymbol{\Pi}
    \right)\right)
\end{align}
in which  the condition $\hat{\bf  R}(z')=0$, as imposed by  the Dirac
delta  functions in  (\ref{W:eq:122}), has  already been  applied.  We
show  in (\ref{TG})  that $\hat{\bf  T}(z)=\hat{\mathbb{G}}(z)$, where
$\hat{\mathbb{G}}(z)$  is  the  diagonalized gyration matrix, and  is a function  of the microstate $z$  only through
$\hat{\bf M}(z)$.  Similarly,  the $\hat{\bf I}$ is a  function of the
microstate  $z$   only  through   $\hat{\boldsymbol{\Lambda}}(z)$  and
$\hat{\bf  M}(z)$. Because  of  the  presence of  the  Dirac delta  on
$\hat{\bf    M}(z')$     and    $\hat{\boldsymbol{\Lambda}}(z')$    in
(\ref{W:eq:122}) we  may simply substitute these  phase functions with
the   numerical  values   ${{\bf   M}}$  and   $\boldsymbol{\Lambda}$,
respectively.  Therefore, we  may introduce the thermal  energy as the
total  energy  minus  the translational,  rotational,  and  dilational
kinetic energies, that is
\begin{align}
  \label{Thermal-App}
    {\cal E}&=E-K^{\rm trans}-    K^{\rm rot}-K^{\rm dil}
\end{align}
where the kinetic energies are
\begin{align}
  \label{W:eq:286}
  K^{\rm trans}
  &\equiv  \frac{{\bf P}^2}{2{\rm M}}
    \nonumber\\
  K^{\rm rot}
  &\equiv\frac{1}{2}{\bf S}^T\esc{\bf I}^{-1}\esc{\bf S}
    \nonumber\\
  K^{\rm dil}
  &\equiv  \frac{1}{2}{\boldsymbol{\Pi}}^T\esc\mathbb{G}^{-1}\esc{\boldsymbol{\Pi}}=
  \sum_\alpha \frac{{\Pi}_\alpha^2}{2{ M}_\alpha}
\end{align}
With these definitions we may write the entropy in (\ref{W:eq:122}) in the form
\begin{align}
  \label{W:eq:282}
  S_B({\bf R},\boldsymbol{\Lambda},{{\bf M}},{\bf P},{\bf S},\boldsymbol{\Pi},E)
  &= S^{\rm rest}(\boldsymbol{\Lambda},{{\bf M}}, {\cal E})
\end{align}
where the ``entropy at rest'' $S^{\rm rest}(\boldsymbol{\Lambda},{{\bf M}}, {\cal E})$
is the  entropy of a non-spinning and non-dilating body  of energy $ {\cal E}$ with its center of mass  fixed at the origin, that is
\begin{align}
  \label{W:eq:280}
  S^{\rm rest}(\boldsymbol{\Lambda},{{\bf M}},{\cal E})
  &\equiv  k_B\ln\Omega^{\rm rest}(\boldsymbol{\Lambda},{{\bf M}},{\cal E})
    \nonumber\\
  \Omega^{\rm rest}(\boldsymbol{\Lambda},{{\bf M}},{\cal E})
  &\equiv\int dz
    \delta\left(\hat{\bf R}(z)\right)
    \delta\left(\hat{\boldsymbol{\Lambda}}(z)-\boldsymbol{\Lambda}\right)
    \delta\left(\hat{{\bf M}}(z)-{{\bf M}}\right)
    \delta\left(\hat{\bf P}(z) \right)
    \delta\left(\hat{\bf S}(z) \right)
    \delta\left(\hat{\boldsymbol{\Pi}}(z)\right)
    \delta\left(\hat{H}(z)-{\cal E}  \right)
\end{align}
We   see  now   the  justification   for  the   change  of   variables
(\ref{W:BodyFrame}) as  it sets to zero  all the momenta in  the Dirac
delta  functions  that  define  $  S^{\rm  rest}$,  and  combines  the
arguments in terms of a  physically sensible thermal energy ${\cal E}$
defined in (\ref{Thermal-App}).

An  appealing  expression   is  obtained  for  the   rest  entropy  by
multiplying   and  dividing   the   argument  of   the  logarithm   in
(\ref{W:eq:280}) with the following measure
\begin{align}
  \label{W:eq:290}
    \Omega^{\rm MT}({\cal E})&= \int dz
             \delta\left(\hat{\bf R}(z) \right)
             \delta\left(\hat{\bf P}(z) \right)
             \delta\left(\hat{\bf S}(z) \right)
             \delta\left(\hat{H}(z)-{\cal E} \right)
\end{align}
In this way, we can express the rest entropy in the form
\begin{align}
  \label{W:eq:287}
  S^{\rm rest}(\boldsymbol{\Lambda},{{\bf M}},{\cal E})
  &=S_B^{\rm MT}({{\cal E}})+k_B\ln P^{\rm rest}_{{\cal E}}(\boldsymbol{\Lambda},{{\bf M}},{\bf 0})
\end{align}
where  the   entropy  at  the  macroscopic   thermodynamics  level  of
description  (with proper  account of  all dynamic  invariants of  the
system) is given by Boltzmann's original definition
\begin{align}
  \label{W:eq:291}
    S_B^{\rm MT}({\cal E})  &=k_B \ln \Omega^{\rm MT}({\cal E})
\end{align}
This  is the  entropy  of  a thermodynamic  system  when  it does  not
translate nor rotate.  The  probability introduced in (\ref{W:eq:287})
is defined as
\begin{align}
  \label{W:eq:288}
  P^{\rm rest}_{{\cal E}}(\boldsymbol{\Lambda},{{\bf M}},\boldsymbol{\Pi})
  &\equiv\int dz \rho^{\rm rest}_{{\cal E}}(z) 
    \delta\left(\hat{\boldsymbol{\Lambda}}(z)-\boldsymbol{\Lambda}\right)
    \delta\left(\hat{{\bf M}}(z)-{{\bf M}}\right)
    \delta(\hat{\boldsymbol{\Pi}}(z)-\boldsymbol{\Pi})
\end{align}
where the rest microcanonical ensemble is
\begin{align}
  \label{W:eq:289}
    \rho^{\rm rest}_{{\cal E}}(z)  &\equiv\frac{1}{\Omega^{\rm MT}({\cal E})}
                  \delta\left(\hat{\bf R}(z) \right)
                    \delta\left(\hat{\bf P}(z) \right)
                    \delta\left(\hat{\bf S}(z) \right)
                    \delta\left(\hat{H}(z)-{\cal E} \right),
\end{align}
We   assume    the  Hamiltonian   dynamics  is   of  the   mixing
type\footnote{This   is   not   guaranteed   for   bodies   that   are
  quasi-harmonic,    as    the   Fermi-Pasta-Ulam-Tsingou    numerical
  experiments revealed.}.  Therefore,  a molecular dynamics simulation
with  initial conditions  such that  the  system has  zero linear  and
angular momentum, i.e.  it is macroscopically ``at rest'', and with energy
${\cal E}$, will  sample $\rho^{\rm rest}_{{\cal E}}(z)$.
The  probability  $P_{{\cal E}}(\boldsymbol{\Lambda},{\bf  M},{\bf  0})$  in
(\ref{W:eq:287})  is  the  equilibrium   probability  of  finding  the
orientation $\boldsymbol{\Lambda}$,  central moments ${\bf  M}$, and
zero value of the dilational momentum $\boldsymbol{\Pi}$, when the
body is ``at rest''.

\subsection{Calculation of conditional expectations}
Consider the change of variables (\ref{W:BodyFrameInv}) when computing
conditional expectations, $\llangle\hat{F}\rrangle^a$, of an arbitrary
phase function $\hat{F}(z)$  at the
present level of description, that is
\begin{align}
  \llangle\hat{F}\rrangle^a
  \equiv\frac{1}{ \Omega(a)}
  \int dz\hat{F}(z)
  &\delta\left(\hat{\bf R}(z)-{\bf R}\right)
    \delta\left(\hat{\boldsymbol{\Lambda}}(z)-\boldsymbol{\Lambda}\right)
    \delta\left(\hat{{\bf M}}(z)-{{\bf M}}\right)
    \nonumber\\
\times  &
    \delta\left(\hat{\bf P}(z)-{\bf P}\right)
    \delta\left(\hat{\bf S}(z)-{\bf S}\right)
        \delta\left(\hat{\boldsymbol{\Pi}}(z)-\boldsymbol{\Pi}\right)
    \delta\left(\hat{H}(z)-E\right)
    \label{W:Fav}
\end{align}
Using the operator $\cal G$ to change variables $z\to z' $ gives
\begin{align}
  \label{eq:43}
  \llangle \hat{F}\rrangle^a
  &=\llangle {\cal G}\hat{F}\rrangle^{\boldsymbol{\Lambda}{\bf M}{\cal E}  }_{\rm rest}
\end{align}
where          the          rest          conditional          average
$\llangle  \cdots\rrangle^{\boldsymbol{\Lambda}{\bf M}{\cal  E} }_{\rm
  rest}$ is defined as
\begin{align}
  \label{W:Fav2}
  \llangle\hat{F}\rrangle^{\boldsymbol{\Lambda}{{\bf M}}{\cal E}}_{\rm rest}
  \equiv \frac{1}{ \Omega^{\rm rest}(\boldsymbol{\Lambda},{{\bf M}},{\cal E})}
  \int dz\hat{F}(z)
  &\delta\left(\hat{\bf R}(z)\right)
    \delta\left(\hat{\boldsymbol{\Lambda}}(z)-\boldsymbol{\Lambda}\right)
    \delta\left(\hat{{\bf M}}(z)-{{\bf M}}\right)
    \delta\left(\hat{\bf P}(z)\right)
    \delta\left(\hat{\bf S}(z)\right)
    \delta\left(\hat{\boldsymbol{\Pi}}(z)\right)
    \delta\left(\hat{H}(z)-{\cal E}\right)
\end{align}
with  $\Omega^{\rm   rest}(\boldsymbol{\Lambda},{{\bf  M}},{\cal  E})$
defined  in (\ref{W:eq:280}).   The rest  conditional ensemble  is the
conditional  average $\llangle\cdots\rrangle^a$  in (\ref{W:Fav})  for
the                particular                 values                of
$a=\{{\bf R}={\bf  0},\boldsymbol{\Lambda},{\bf M},{\bf  P}={\bf 0},{\bf
  S}={\bf 0},\boldsymbol{\Pi}={\bf  0}, E={\cal  E}\}$, and can  also be
expressed  in   the  following  convenient   form  in  terms   of  the
microcanonical ensemble (\ref{W:eq:289}) as
\begin{align}
  \label{eq:176}
    \llangle\hat{F}\rrangle^{\boldsymbol{\Lambda}{{\bf M}}{\cal E}}_{\rm rest}
&= \frac{1}{ P_{\cal E}^{\rm rest}(\boldsymbol{\Lambda},{{\bf M}},{\bf 0})}
  \int dz\hat{F}(z)\rho^{\rm rest}_{\cal E}(z)
    \delta\left(\hat{\boldsymbol{\Lambda}}(z)-\boldsymbol{\Lambda}\right)
    \delta\left(\hat{{\bf M}}(z)-{{\bf M}}\right)
    \delta\left(\hat{\boldsymbol{\Pi}}(z)\right)
\end{align}

\setcounter{my_equation_counter}{\value{equation}}
\section{Consequences of rotational symmetry}
\setcounter{equation}{\value{my_equation_counter}}
\label{App:W:RotInv}
In this section,  we examine the effect  of rotational symmetry on
the   functional  forms  for  the  entropy,   the  rest   probability
distribution, and the conditional expectations of tensorial quantities.

\subsection{Rotational symmetry of the entropy}
Because,  as  shown in  (\ref{W:eq:282}),  the  entropy can  be  fully
expressed in  terms of  the rest entropy,  we consider  the rotational
invariance of the latter.  Consider  the following change of variables
$z=Qz'$
\begin{align}
  \label{W:eq:182}
  {\bf r}_i
  &=\boldsymbol{\cal Q}\esc{\bf r}_i'=e^{[{\bf A}]_\times}\esc{\bf r}_i'
  \nonumber\\
  {\bf p}_i
  &=\boldsymbol{\cal Q}\esc{\bf p}_i'=e^{[{\bf A}]_\times}\esc{\bf p}_i'
\end{align}
where  $\boldsymbol{\cal  Q}=e^{[{\bf  A}]_\times}$  is  an  arbitrary
rotation  with orientation  ${\bf A}$. This change
of variables in the phase  space integral (\ref{W:eq:280}) gives (with
a renaming of $z'$ with $z$ for easier notation)
\begin{align}
  \label{W:eq:210}
&  S^{\rm rest}(\boldsymbol{\Lambda},{{\bf M}},{\cal E})
                     \nonumber\\
  &=  k_B\ln
    \int dz
    \delta\left(\hat{\bf R}(Qz)\right)
    \delta\left(\hat{\boldsymbol{\Lambda}}(Qz)-\boldsymbol{\Lambda}\right)
    \delta\left(\hat{{\bf M}}(Qz)-{{\bf M}}\right)
    \delta\left(\hat{\bf P}(Qz) \right)
    \delta\left(\hat{\bf S}(Qz) \right)
    \delta\left(\hat{\boldsymbol{\Pi}}(Qz)\right)
    \delta\left(\hat{H}(Qz)-{\cal E}  \right)
    \nonumber\\
  &=  k_B\ln
    \int dz
    \delta\left(\boldsymbol{\cal Q}\esc\hat{\bf R}(z)\right)
    \delta\left(\hat{\boldsymbol{\Lambda}}(Qz)-\boldsymbol{\Lambda}\right)
    \delta\left(\hat{{\bf M}}(z)-{{\bf M}}\right)
    \delta\left(\boldsymbol{\cal Q}\esc\hat{\bf P}(z) \right)
    \delta\left(\boldsymbol{\cal Q}\esc\hat{\bf S}(z) \right)
    \delta\left(\hat{\boldsymbol{\Pi}}(z)\right)
    \delta\left(\hat{H}(z)-{\cal E}  \right)
    \nonumber\\
  &=  k_B\ln
    \int dz
    \delta\left(\hat{\bf R}(z)\right)
    \delta\left(\hat{\boldsymbol{\Lambda}}(Qz)-\boldsymbol{\Lambda}\right)
    \delta\left(\hat{{\bf M}}(z)-{{\bf M}}\right)
    \delta\left(\hat{\bf P}(z) \right)
    \delta\left(\hat{\bf S}(z) \right)
    \delta\left(\hat{\boldsymbol{\Pi}}(z)\right)
    \delta\left(\hat{H}(z)-{\cal E}  \right)
\end{align}
where    in   the    second    line   we    have    used   the    fact
$\hat{\bf  R},\hat{\bf  P},\hat{\bf  S}$   transform  as  vectors  and
$\hat{{\bf  M}},\hat{\boldsymbol{\Pi}},\hat{H}$  are  invariant  under
rotations (as discussed in  Sec.  \ref{App:Orientation}). In the third
line we used the Fourier representation of the Dirac delta function to
show that
  \begin{align}
    \label{eq:202}
    \delta\left(\boldsymbol{\cal Q}\esc\hat{\bf P}(z) \right)
    &=\frac{1}{(2\pi)^3}\int d{\bf k}^T\exp\{i{\bf k}^T\esc\boldsymbol{\cal Q}\esc\hat{\bf P}(z) \}
      =\frac{1}{(2\pi)^3}\int d{{\bf k}^T}'\exp\{{i{\bf k}^T}'\esc\hat{\bf P}(z) \}=
      \delta\left(\hat{\bf P}(z) \right)
  \end{align}
  where the change of variables  ${\bf k}'=\boldsymbol{\cal Q}^T\esc{\bf k}$ with unit
  Jacobian determinant has been used.
Note  the  left  hand  side  of  (\ref{W:eq:210})  is  independent  of
${\bf  A}$ so  taking the  derivative of  both sides  with respect  to
${\bf A}$ and evaluating at ${\bf A}=0$ gives in component notation
\begin{align}
\label{W:eq:259}
  0&=\left . \frac{\partial}{\partial{\bf A}^\gamma} \int dz
     \delta\left(\hat{\bf R}(z)\right)
     \delta\left(\hat{\boldsymbol{\Lambda}}(Qz)-\boldsymbol{\Lambda}\right)
     \delta\left(\hat{{\bf M}}(z)-{{\bf M}}\right)
     \delta\left(\hat{\bf P}(z) \right)
     \delta\left(\hat{\bf S}(z) \right)
     \delta\left(\hat{\boldsymbol{\Pi}}(z)\right)
     \delta\left(\hat{H}(z)-{\cal E} \right)\right |_{{\bf A}=0}\cr
  &
=\int dz \delta\left(\hat{\bf R}(z)\right) \left .\left (
     \frac{\partial}{\partial {\Lambda}_{\gamma'}(Qz)}
     \delta\left(\hat{\boldsymbol{\Lambda}}(Qz)-\boldsymbol{\Lambda}\right)
    \frac{\partial}{\partial{\bf A}^\gamma}\hat{\Lambda}_{\gamma'}(Qz)\right )\right |_{{\bf A}=0}
    \nonumber\\
  &\times
    \delta\left(\hat{{\bf M}}(z)-{{\bf M}}\right)
    \delta\left(\hat{\bf P}(z) \right)
    \delta\left(\hat{\bf S}(z) \right)
    \delta\left(\hat{\boldsymbol{\Pi}}(z)\right)
    \delta\left(\hat{H}(z)-{\cal E} \right)\cr
  &=-\frac{\partial}{\partial {\Lambda}_{\gamma'}}
    \left .  \int dz \delta\left(\hat{\bf R}(z)\right)
  \frac{\partial }{\partial{\bf A}^\gamma}\hat{\Lambda}_\alpha(Qr)\right |_{{\bf A}=0}
     \delta\left(\hat{\boldsymbol{\Lambda}}(z)-\boldsymbol{\Lambda}\right)
    \delta\left(\hat{{\bf M}}(z)-{{\bf M}}\right)
    \delta\left(\hat{\bf P}(z) \right)
    \delta\left(\hat{\bf S}(z) \right)
    \delta\left(\hat{\boldsymbol{\Pi}}(z)\right)
    \delta\left(\hat{H}(z)-{\cal E} \right)
   \end{align}
   where we have used  the chain rule and noted in  the last line that
   $\hat{\boldsymbol{\Lambda}}$ is  a function  of position  only. Now
   make use of the following identity
\begin{align}
  \label{eq:24}
  \left.  \frac{\partial }{\partial{\bf A}^\gamma}\hat{\Lambda}_{\gamma'}(Qr)\right |_{{\bf A}=0}
  &\overset{\mbox{chain rule}}{=} \left.\sum_i \frac{\partial \hat{\Lambda}_{\gamma'}}
    {\partial{\bf r}_i^{\alpha}}(Qr)\frac{\partial \boldsymbol{\cal Q}^{\alpha\beta}{\bf r}_i^\beta}{\partial{\bf A}^\gamma}\right |_{{\bf A}=0}
    \overset{(\ref{eq:711})}{=}
    \sum_i        \frac{\partial\hat{\Lambda}_{\gamma'}}{\partial{\bf r}^{\alpha}_i}
    \epsilon_{\alpha\gamma\beta} {\bf r}^\beta_i
    =  -\sum_i \frac{\partial\hat{\Lambda}_{\gamma'}}{\partial{\bf r}^{\alpha'}}
    [{\bf r}_i]_\times^{\alpha\gamma}
    \overset{(\ref{eq:255})}{=}{B}_{\gamma'\gamma}(\hat{\boldsymbol{\Lambda}}(r))
\end{align}
Because ${B}_{\gamma'\gamma}(\hat{\boldsymbol{\Lambda}}(r))$
depends on the configuration $r$ only through the orientation, the
Dirac delta function on the orientation in (\ref{W:eq:259}) allows us to move
this term outside the integral giving
\begin{align}
  \label{eq:37}
0   = \frac{\partial}{\partial {\Lambda}_{\gamma'}}
    {B}_{\gamma'\gamma}(\boldsymbol{\Lambda}) \Omega^{\rm rest}(\boldsymbol{\Lambda},{{\bf M}},{\cal E})
    \end{align}
Performing this
derivative, using (\ref{W:eq:280}), and simplifying gives
\begin{align}
  \label{W:eq:266}
  \frac{\partial S^{\rm rest}}{\partial \boldsymbol{\Lambda}}(\boldsymbol{\Lambda},{{\bf M}},{\cal E})
  \esc  {\bf B}(\boldsymbol{\Lambda})
  + k_B \frac{\partial}{\partial \boldsymbol{\Lambda}}\esc{\bf B}(\boldsymbol{\Lambda})
  =& 0
\end{align}
This identity reflects the rotational invariance of the entropy.

\subsection{Rotational      symmetry      of      the      probability
  $ P^{\rm rest}_{\cal E}(\boldsymbol{\Lambda},{{\bf M}},\boldsymbol{\Pi})$}
\label{W:App:RotProb}

The rest  microcanonical ensemble (\ref{W:eq:289}) is  invariant under
the  rotation (\ref{W:eq:182}),  as are the  two phase  functions
$\hat{{\bf M}}(z),\hat{\boldsymbol{\Pi}}(z)$, according to (\ref{eq:47}),
  (\ref{eq:51}).   Therefore,   performing    the   change   of   variables
  (\ref{W:eq:182}) gives
  \begin{align}
    \label{eq:327}
      P^{\rm rest}_{\cal E}(\boldsymbol{\Lambda},{{\bf M}},\boldsymbol{\Pi})
  &=\int dz \rho^{\rm rest}_{\cal E}(z) \delta\left(\hat{\boldsymbol{\Lambda}}(Qz)-\boldsymbol{\Lambda}\right)
\delta\left(\hat{{\bf M}}(z)-{{\bf M}}\right)                                       \delta(\hat{\boldsymbol{\Pi}}(z)-\boldsymbol{\Pi}) 
\end{align}
 Analogous  to the procedure used in obtaining
(\ref{W:eq:266}),   by   taking   the   derivative   with   respect   to
${\bf A}^\gamma$ and evaluating the result at ${\bf A}=0$ on both sides we obtain
\begin{align}
  \label{W:eq:450}
  0
  &=\left .\int dz \rho_{\cal E}(z)  \frac{\partial}{\partial {\bf A}^\gamma}
    \delta\left(\hat{\boldsymbol{\Lambda}}(Qz)-\boldsymbol{\Lambda}\right)
    \right |_{{ {\bf A}=0 }}
    \delta\left(\hat{{\bf M}}(z)-{{\bf M}}\right)
    \delta(\hat{\boldsymbol{\Pi}}(z)-\boldsymbol{\Pi}) 
= -\frac{\partial}{\partial {\Lambda}_{\gamma'}}
    {B}_{\gamma'\gamma}(\boldsymbol{\Lambda})P^{\rm rest}_E(\boldsymbol{\Lambda},{{\bf M}},\boldsymbol{\Pi})
    \end{align}
Using the chain rule and some rearranging then gives
\begin{align}
  \label{W:eq:113}
  \left(\frac{\partial}{\partial {\Lambda}_{\gamma'}}
{B}_{\gamma'\gamma}(\boldsymbol{\Lambda})\right)
{B}^{-1}_{\gamma\alpha} +
  \frac{\partial}{\partial {\Lambda}_{{\alpha}}}
\ln P^{\rm rest}_E(\boldsymbol{\Lambda},{{\bf M}},\boldsymbol{\Pi})=0
\end{align}
Further use of (\ref{eq:173}) gives
\begin{align}
  \label{W:eq:128}
-\frac{\partial }{\partial {\Lambda}_\alpha}
\ln\left(\frac{1-\cos\Lambda}{\Lambda^2}\right) +
  \frac{\partial}{\partial {\Lambda}_{{\alpha}}}\ln
  P^{\rm rest}_E(\boldsymbol{\Lambda},{{\bf M}},\boldsymbol{\Pi})=0
\end{align}
implying
\begin{align}
  \label{eq:57}
  \frac{\partial}{\partial {\Lambda}_{{\alpha}}}\ln
\frac{  P^{\rm rest}_E(\boldsymbol{\Lambda},{{\bf M}},\boldsymbol{\Pi})}{\left(\frac{1-\cos\Lambda}{\Lambda^2}\right)}=0
\end{align}
which shows  the  argument of  the logarithm  is a  constant with
respect to $\boldsymbol{\Lambda}$ and therefore, must be a function of
${{\bf    M}},\boldsymbol{\Pi},E$   only,    that    we   denote    as
$P^{\rm rest}_E({\bf M},\boldsymbol{\Pi})$. We conclude
\begin{align}
  \label{eq:119}
  P^{\rm rest}_E(\boldsymbol{\Lambda},{{\bf M}},\boldsymbol{\Pi})=\frac{1-\cos\Lambda}{4\pi^2\Lambda^2}P^{\rm rest}_E({{\bf M}},\boldsymbol{\Pi})
 =P^{\rm Haar}(\boldsymbol{\Lambda})P^{\rm rest}_E({{\bf M}},\boldsymbol{\Pi})
\end{align}
where we have  introduced the probability density $P^{\rm Haar}(\boldsymbol{\Lambda})$   in the orientation
space
\begin{align}
  \label{eq:150}
P^{\rm Haar}(\boldsymbol{\Lambda})&\equiv  \frac{1-\cos\Lambda}{4\pi^2\Lambda^2}
\end{align}
We  have discussed  in  the main  text the  probability
density   of   the   Haar   measure  in   the   parameter   space   of
$\boldsymbol{\Lambda}$.    Because   this   probability   density   is
normalized  on   the  sphere  of   radius  $\pi$  in  the   space  of
orientations, the  factor $P^{\rm  rest}_E({\bf M},\boldsymbol{\Pi})$
 in    (\ref{eq:119})   is   just   the    marginal   of
$  P^{\rm  rest}_E(\boldsymbol{\Lambda},{{\bf  M}},\boldsymbol{\Pi})$,
that is
\begin{align}
  \label{eq:151}
  P^{\rm rest}_{\cal E}({{\bf M}},\boldsymbol{\Pi})
  &=\int dz \rho^{\rm rest}_{\cal E}(z)
    \delta\left(\hat{{\bf M}}(z)-{{\bf M}}\right)
    \delta(\hat{\boldsymbol{\Pi}}(z)-\boldsymbol{\Pi}) 
\end{align}
The rotational invariance expressed  in (\ref{eq:119}) implies, therefore,
that   the  orientation   $\boldsymbol{\Lambda}$  is   distributed  at
equilibrium according  to the  Haar measure,  and  is statistically
independent  of  the values  of  the  central moments  and  dilation
momentum.
{The Haar density probability is defined microscopically as
  \begin{align}
    \label{eq:33}
    P^{\rm Haar}(\boldsymbol{\Lambda})&\equiv \int dz \rho^{\rm rest}_E(z)\delta(\hat{\boldsymbol{\Lambda}}(z)- \boldsymbol{\Lambda})
  \end{align}
  Quite remarkably,  by recourse  to rotational invariance,  in this
  section  we   have  computed   explicitly  the  integral   in  this
  probability  density, with  the result  (\ref{eq:150}), without  the
  need of knowing  the explicit functional form of  the phase function
  $\hat{\boldsymbol{\Lambda}}(z)$.}

\subsection{Rotational symmetry of conditional expectations}
\label{App:RotInv}
In   this    section   we   derive   the    explicit   dependence   on
$\boldsymbol{\Lambda}$ for  the conditional expectations  of tensorial
quantities. Consider  the transformation  of three different  types of
phase functions:  scalar, vectorial,  or tensorial under  an arbitrary
rotation $\boldsymbol{\cal Q}$, respectively
\begin{align}
  \label{W:eq:52}
  \hat{F}\left(Qz\right)
  &=  \hat{F}\left(z\right)
    \nonumber\\
    \hat{\bf V}\left(Qz\right)
  &=   \boldsymbol{\cal Q}\esc\hat{\bf V}\left(z\right)
    \nonumber\\
    \hat{\bf W}\left(Qz\right)
  &=   \boldsymbol{\cal Q}\esc\hat{\bf W}\left(z\right)\esc\boldsymbol{\cal Q}^T
\end{align}
We  focus  our  discussion  on the  ``rest''  conditional  expectation
(\ref{W:Fav2})  as  we  will  consider almost  always  the  change  of
variables (\ref{W:BodyFrame}). The explicit result we will demonstrate in
this section is
\begin{align}
  \label{W:eq:469}
  \llangle \hat{F}\rrangle^{\boldsymbol{\Lambda}{{\bf M}}{\cal E}}_{\rm rest}
  &=F_0({\bf M},{\cal E})
    \nonumber\\
    \llangle \hat{\bf V}\rrangle^{\boldsymbol{\Lambda}{{\bf M}}{\cal E}}_{\rm rest}
  &= e^{[\boldsymbol{\Lambda}]_\times}\esc
    {\bf V}_0({{\bf M}},{\cal E})
    \nonumber\\
    \llangle \hat{\bf W}\rrangle^{\boldsymbol{\Lambda}{{\bf M}}{\cal E}}_{\rm rest}
  &= e^{[\boldsymbol{\Lambda}]_\times}\esc
    {\bf W}_0({{\bf M}},{\cal E})\esc e^{-[\boldsymbol{\Lambda}]_\times}
\end{align}
where $F_0({\bf  M},{\cal E})$,  ${\bf V}_0({{\bf M}},{\cal  E})$, and
${\bf W}_0({{\bf  M}},{\cal E})$  are independent of  the orientation.
Performing  the change  of variables  (\ref{W:eq:182}), and  noting the
Dirac                delta                 functions                of
${\bf R},{{\bf M}},{\bf P},{\bf S},\boldsymbol{\Pi}$, and $E$ are left
invariant  under this  change, the  conditional expectations  of these
three types of functions become
\begin{align}
  \label{W:eq:279}
\llangle\hat{F}\rrangle^{\boldsymbol{\Lambda}{{\bf M}}{\cal E}}_{\rm rest}
  =\frac{1}{ \Omega^{\rm rest}}
  \int dz\hat{F}(z)
  &\delta\left(\hat{\bf R}(z)\right)
    \delta\left(\hat{\boldsymbol{\Lambda}}(Qz)-\boldsymbol{\Lambda}\right)
    \delta\left(\hat{{\bf M}}(z)-{{\bf M}}\right)
    \delta\left(\hat{\bf P}(z)\right)
    \delta\left(\hat{\bf S}(z)\right)
    \delta\left(\hat{\boldsymbol{\Pi}}(z)\right)
    \delta\left(\hat{H}(z)-{\cal E}\right)
    \nonumber\\
\llangle\hat{\bf V}\rrangle^{\boldsymbol{\Lambda}{\bf M}{\cal E}}_{\rm rest}
  =\boldsymbol{\cal Q} \esc \frac{1}{ \Omega^{\rm rest}}
  \int dz\hat{\bf V}(z)
  &\delta\left(\hat{\bf R}(z)\right)
    \delta\left(\hat{\boldsymbol{\Lambda}}(Qz)-\boldsymbol{\Lambda}\right)
    \delta\left(\hat{{\bf M}}(z)-{{\bf M}}\right)
    \delta\left(\hat{\bf P}(z)\right)
    \delta\left(\hat{\bf S}(z)\right)
    \delta\left(\hat{\boldsymbol{\Pi}}(z)\right)
    \delta\left(\hat{H}(z)-{\cal E}\right)
    \nonumber\\
\llangle\hat{\bf W}\rrangle^{\boldsymbol{\Lambda}{\bf M}{\cal E}}_{\rm rest}
  =
\boldsymbol{\cal Q} \esc \frac{1}{ \Omega^{\rm rest}} \int dz\hat{\bf W}(z)
  &\delta\left(\hat{\bf R}(z)\right)
    \delta\left(\hat{\boldsymbol{\Lambda}}(Qz)-\boldsymbol{\Lambda}\right)
    \delta\left(\hat{{\bf M}}(z)-{{\bf M}}\right)
    \delta\left(\hat{\bf P}(z)\right)
    \delta\left(\hat{\bf S}(z)\right)
    \delta\left(\hat{\boldsymbol{\Pi}}(z)\right)
    \delta\left(\hat{H}(z)-{\cal E}\right)\esc\boldsymbol{\cal Q}^T
\end{align}
Now  take  the derivative  of  these  equations  with respect  to  the
arbitrary orientation vector  ${\bf A}$, noting the left  hand side is
independent of ${\bf A}$, to give component-wise
\begin{align}
  \label{W:eq:443}
0
  &=\frac{1}{ \Omega^{\rm rest}}
  \int dz\hat{F}(z)
  \delta\left(\hat{\bf R}(z)\right)\frac{\partial}{\partial {\bf A}^\gamma}
    \delta\left(\hat{\boldsymbol{\Lambda}}(Qz)-\boldsymbol{\Lambda}\right)
    \delta\left(\hat{{\bf M}}(z)-{{\bf M}}\right)
    \delta\left(\hat{\bf P}(z)\right)
    \delta\left(\hat{\bf S}(z)\right)
    \delta\left(\hat{\boldsymbol{\Pi}}(z)\right)
    \delta\left(\hat{H}(z)-{\cal E}\right)
    \nonumber\\
0
  &= \frac{1}{ \Omega^{\rm rest}}
\frac{\partial}{\partial {\bf A}^\gamma} \boldsymbol{\cal Q}^{\alpha\alpha'}
  \int dz\hat{\bf V}^{\alpha'}(z)
  \delta\left(\hat{\bf R}(z)\right)
    \delta\left(\hat{\boldsymbol{\Lambda}}(Qz)-\boldsymbol{\Lambda}\right)
    \delta\left(\hat{{\bf M}}(z)-{{\bf M}}\right)
    \delta\left(\hat{\bf P}(z)\right)
    \delta\left(\hat{\bf S}(z)\right)
    \delta\left(\hat{\boldsymbol{\Pi}}(z)\right)
    \delta\left(\hat{H}(z)-{\cal E}\right)
    \nonumber\\
0
  &=\frac{1}{ \Omega^{\rm rest}}\frac{\partial}{\partial {\bf A}^\gamma}
    {\boldsymbol{\cal Q}}^{\alpha\alpha'} {\boldsymbol{\cal Q}}^{\beta\beta'}
    \nonumber\\
  &\times\int dz\hat{\bf W}^{\alpha'\beta'}(z)
    \delta\left(\hat{\bf R}(z)\right)
    \delta\left(\hat{\boldsymbol{\Lambda}}(Qz)-\boldsymbol{\Lambda}\right)
    \delta\left(\hat{{\bf M}}(z)-{{\bf M}}\right)
    \delta\left(\hat{\bf P}(z)\right)
    \delta\left(\hat{\bf S}(z)\right)
    \delta\left(\hat{\boldsymbol{\Pi}}(z)\right)
    \delta\left(\hat{H}(z)-{\cal E}\right)
\end{align}
For the scalar function, by evaluating the derivative at ${\bf A}=0$, and using (\ref{eq:24}), and (\ref{eq:37}) we have
\begin{align}
  \label{W:eq:444}
  0  &=-\frac{1}{\Omega^{\rm rest}(\boldsymbol{\Lambda},{\bf M},{\cal E})}
    \frac{\partial}{\partial {\Lambda}_{\gamma'}}  {B}_{\gamma'\gamma}(\boldsymbol{\Lambda})
    \Omega^{\rm rest}(\boldsymbol{\Lambda},{{\bf M}},{\cal E})\llangle \hat{F}\rrangle^{\boldsymbol{\Lambda}{{\bf M}}{\cal E}}_{\rm rest}
       =-{B}_{\gamma'\gamma}(\boldsymbol{\Lambda})    \frac{\partial}{\partial {\Lambda}_{\gamma'}}  
  \llangle \hat{F}\rrangle^{\boldsymbol{\Lambda}{{\bf M}}{\cal E}}_{\rm rest}
\end{align}
Because ${\bf  B}(\boldsymbol{\Lambda})$ is non-singular it must be that
\begin{align}
  \label{W:eq:446}
    \frac{\partial}{\partial {\Lambda}_{\gamma'}}  
  \llangle \hat{F}\rrangle^{\boldsymbol{\Lambda}{{\bf M}}{\cal E}}_{\rm rest}
  &=0
\end{align}
which  means  the  conditional  average does  not  depend on  the
orientation, so that
\begin{align}
  \label{eq:321}
    \llangle \hat{F}\rrangle^{\boldsymbol{\Lambda}{{\bf M}}{\cal E}}_{\rm rest}&=F_0({\bf M},{\cal E})
\end{align}
For vectorial functions, evaluating the  derivative at ${\bf A}=0$ for
the second equation of (\ref{W:eq:443})  and using the first relation in
(\ref{eq:457}) gives
\begin{align}
  \label{W:eq:448}
  0
  &=\epsilon_{\alpha\gamma\alpha'}
\llangle \hat{\bf V}^{\alpha'}\rrangle^{\boldsymbol{\Lambda}{{\bf M}}{\cal E}}_{\rm rest}
-    
\frac{1}{ \Omega^{\rm rest}(\boldsymbol{\Lambda},{{\bf M}},{\cal E})}
    \frac{\partial{B}_{\gamma'\gamma}}{\partial {\Lambda}_{\gamma'}}(\boldsymbol{\Lambda})
    \Omega^{\rm rest}(\boldsymbol{\Lambda},{{\bf M}},{\cal E})\llangle \hat{\bf V}^{\alpha}\rrangle^{\boldsymbol{\Lambda}{{\bf M}}{\cal E}}_{\rm rest}
\end{align}
which with the use of (\ref{eq:37}) gives
\begin{align}
  \label{W:eq:449}
  0 &= -\llangle [\hat{\bf V}]_\times^{\alpha\gamma}
    \rrangle^{\boldsymbol{\Lambda}{{\bf M}}{\cal E}}_{\rm rest} -
    {B}_{\gamma'\gamma}(\boldsymbol{\Lambda})
    \frac{\partial}{\partial {\Lambda}_{\gamma'}}
    \llangle \hat{\bf V}^{\alpha}\rrangle^{\boldsymbol{\Lambda}{{\bf M}}{\cal E}}_{\rm rest}
\end{align}
Let us define the vector ${\bf V}_0(\boldsymbol{\Lambda},{{\bf M}},{\cal E})$ through
\begin{align}
  \label{W:eq:466}
  \llangle \hat{\bf V}\rrangle^{\boldsymbol{\Lambda}{{\bf M}}{\cal E}}_{\rm rest}
  &=   e^{[\boldsymbol{\Lambda}]_\times}\esc{\bf V}_0(\boldsymbol{\Lambda},{{\bf M}},{\cal E})
\end{align}
Using this definition in (\ref{W:eq:449}) along with 
(\ref{eq:99}) gives
\begin{align}
0&=-\llangle [\hat{\bf V}]_\times^{\alpha\gamma}
\rrangle^{\boldsymbol{\Lambda}{{\bf M}}{\cal E}}_{\rm rest} -
{B}_{\gamma'\gamma}(\boldsymbol{\Lambda})
\left ( \frac{\partial}{\partial {\Lambda}_{\gamma'}} \left[
e^{[\boldsymbol{\Lambda}]_\times}\right]_{\alpha\alpha'}\right )
{\bf V}^{\alpha'}_0(\boldsymbol{\Lambda},{{\bf M}},{\cal E}) -
{B}_{\gamma'\gamma}(\boldsymbol{\Lambda}) \left[
e^{[\boldsymbol{\Lambda}]_\times}\right]_{\alpha\alpha'}
\frac{\partial}{\partial {\Lambda}_{\gamma'}}
{\bf V}^{\alpha'}_0(\boldsymbol{\Lambda},{{\bf M}},{\cal E})\cr
0&=-\llangle [\hat{\bf V}]_\times^{\alpha\gamma}
\rrangle^{\boldsymbol{\Lambda}{{\bf M}}{\cal E}}_{\rm rest} -
{B}_{\gamma'\gamma}(\boldsymbol{\Lambda}) \epsilon_{\alpha\gamma''\beta'}
\left[e^{[\hat{\boldsymbol{\Lambda}}]_\times}\right]_{\beta'\alpha'}
{B}^{-1}_{\gamma''\gamma'}
{\bf V}^{\alpha'}_0(\boldsymbol{\Lambda},{{\bf M}},{\cal E}) -
{B}_{\gamma'\gamma}(\boldsymbol{\Lambda}) \left[
e^{[\boldsymbol{\Lambda}]_\times}\right]_{\alpha\alpha'}
\frac{\partial}{\partial {\Lambda}_{\gamma'}}
{\bf V}^{\alpha'}_0(\boldsymbol{\Lambda},{{\bf M}},{\cal E})\cr
0&=-\llangle [\hat{\bf V}]_\times^{\alpha\gamma}
\rrangle^{\boldsymbol{\Lambda}{{\bf M}}{\cal E}}_{\rm rest} -
\epsilon_{\alpha\gamma\beta'}
\llangle {\bf V}^{\beta'}\rrangle^{\boldsymbol{\Lambda}{{\bf M}}{\cal E}}_{\rm rest} -
{B}_{\gamma'\gamma}(\boldsymbol{\Lambda}) \left[
e^{[\boldsymbol{\Lambda}]_\times}\right]_{\alpha\alpha'}
\frac{\partial}{\partial {\Lambda}_{\gamma'}}
{\bf V}^{\alpha'}_0(\boldsymbol{\Lambda},{{\bf M}},{\cal E})\cr
0&=-{B}_{\gamma'\gamma}(\boldsymbol{\Lambda}) \left[
e^{[\boldsymbol{\Lambda}]_\times}\right]_{\alpha\alpha'}
\frac{\partial}{\partial {\Lambda}_{\gamma'}}
{\bf V}^{\alpha'}_0(\boldsymbol{\Lambda},{{\bf M}},{\cal E})
\end{align}
from which we conclude the vector ${\bf  V}_0$ is independent of the orientation.

Proceeding in the same manner for the average of the tensor valued phase
function in (\ref{W:eq:279}) using (\ref{eq:457}) and noting
${\boldsymbol{\cal Q}}^{\alpha\alpha'}|_{{\bf A}=0}=\delta_{\alpha\alpha'}$ gives 
\begin{align}
  \label{W:eq:454}
0
  &=\frac{1}{ \Omega^{\rm rest}}\frac{\partial}{\partial {\bf A}^\gamma}
{\boldsymbol{\cal Q}}^{\alpha\alpha'}{\boldsymbol{\cal Q}}^{\beta\beta'}  \left . \int dz\hat{\bf W}^{\alpha'\beta'}(z)
  \delta\left(\hat{\bf R}(z)\right)
    \delta\left(\hat{\boldsymbol{\Lambda}}(Qz)-\boldsymbol{\Lambda}\right)
    \delta\left(\hat{{\bf M}}(z)-{{\bf M}}\right)
    \delta\left(\hat{\bf P}(z)\right)
    \delta\left(\hat{\bf S}(z)\right)
    \delta\left(\hat{\boldsymbol{\Pi}}(z)\right)
    \delta\left(\hat{H}(z)-{\cal E}\right)\right |_{{\bf A}=0}
    \nonumber\\
  &=\left .\left(\frac{\partial}{\partial {\bf A}^\gamma}
     {\boldsymbol{\cal Q}}^{\alpha\alpha'}{\boldsymbol{\cal Q}}^{\beta\beta'}\right )\right |_{{\bf A}=0}
    \llangle \hat{\bf W}^{\alpha'\beta'}\rrangle^{\boldsymbol{\Lambda}{{\bf M}}{\cal E}}_{\rm rest}
 -
    {B}_{\gamma'\gamma}(\boldsymbol{\Lambda})
    \frac{\partial}{\partial {\Lambda}_{\gamma'}}
\llangle \hat{\bf W}^{\alpha\beta}\rrangle^{\boldsymbol{\Lambda}{{\bf M}}{\cal E}}_{\rm rest}
    \nonumber\\
  &=\left(\epsilon_{\alpha\gamma\alpha'}\delta_{\beta\beta'}
    +\delta_{\alpha\alpha'}\epsilon_{\beta\gamma\beta'}\right)
    \llangle \hat{\bf W}^{\alpha'\beta'}\rrangle^{\boldsymbol{\Lambda}{{\bf M}}{\cal E}}_{\rm rest}
  -
    {B}_{\gamma'\gamma}(\boldsymbol{\Lambda})
    \frac{\partial}{\partial {\Lambda}_{\gamma'}}
\llangle \hat{\bf W}^{\alpha\beta}\rrangle^{\boldsymbol{\Lambda}{{\bf M}}{\cal E}}_{\rm rest}
    \nonumber\\
  &=\epsilon_{\alpha\gamma\alpha'}
    \llangle \hat{\bf W}^{\alpha'\beta}\rrangle^{\boldsymbol{\Lambda}{{\bf M}}{\cal E}}_{\rm rest}
    +\epsilon_{\beta\gamma\beta'}
    \llangle \hat{\bf W}^{\alpha\beta'}\rrangle^{\boldsymbol{\Lambda}{{\bf M}}{\cal E}}_{\rm rest}
-
    {B}_{\gamma'\gamma}(\boldsymbol{\Lambda})
    \frac{\partial}{\partial {\Lambda}_{\gamma'}}
\llangle \hat{\bf W}^{\alpha\beta}\rrangle^{\boldsymbol{\Lambda}{{\bf M}}{\cal E}}_{\rm rest}
\end{align}
Let us introduce the tensor  ${\bf W}_0(\boldsymbol{\Lambda},{{\bf M}},{\cal E})$ through
\begin{align}
  \label{W:eq:455}
  \llangle \hat{\bf W}\rrangle^{\boldsymbol{\Lambda}{{\bf M}}{\cal E}}_{\rm rest}
  &\equiv e^{[\boldsymbol{\Lambda}]_\times}\esc
    {\bf W}_0(\boldsymbol{\Lambda},{{\bf M}},{\cal E})\esc e^{-[\boldsymbol{\Lambda}]_\times}
\end{align}

The last term of (\ref{W:eq:454}) along with
relations (\ref{eq:99}) and (\ref{eq:183}) gives
\begin{align}
\label{W:eq:456}
&{B}_{\gamma'\gamma}(\boldsymbol{\Lambda})
 \frac{\partial}{\partial {\Lambda}_{\gamma'}}
\llangle \hat{\bf W}^{\alpha\beta}\rrangle^{\boldsymbol{\Lambda}{{\bf M}}{\cal E}}_{\rm rest}\nonumber\\
&=\left ( \frac{\partial}{\partial {\Lambda}_{\gamma'}}
\left[ e^{[\boldsymbol{\Lambda}]_\times}\right]_{\alpha\alpha''}\right )
{B}_{\gamma'\gamma}(\boldsymbol{\Lambda}) {\bf W}^{\alpha''\beta''}_0
\left[ e^{-[\boldsymbol{\Lambda}]_\times}\right ]^{\beta''\beta}
+ \left[ e^{[\boldsymbol{\Lambda}]_\times}\right]_{\alpha\alpha''}
{\bf W}^{\alpha''\beta''}_0
\left ( \frac{\partial}{\partial {\Lambda}_{\gamma'}}
\left[e^{-[\boldsymbol{\Lambda}]_\times}\right]_{\beta''\beta}\right )
{B}_{\gamma'\gamma}(\boldsymbol{\Lambda})\nonumber\\
&\quad+{B}_{\gamma'\gamma}(\boldsymbol{\Lambda})
\left [ e^{[\boldsymbol{\Lambda}]_\times}\right ]^{\alpha\alpha''}
    \left(\frac{\partial}{\partial {\Lambda}_{\gamma'}}
    {\bf W}^{\alpha''\beta''}_0\right)
\left[e^{-[\boldsymbol{\Lambda}]_\times}\right]_{\beta''\beta}\nonumber\\
&=\epsilon_{\alpha\gamma''\alpha'}
\left[e^{[\hat{\boldsymbol{\Lambda}}]_\times}\right]_{\alpha'\alpha''}
\left[{B}^{-1}(\boldsymbol{\Lambda})\right]_{\gamma''\gamma'}
{B}_{\gamma'\gamma}(\boldsymbol{\Lambda}) {\bf W}^{\alpha''\beta''}_0
\left[ e^{-[\boldsymbol{\Lambda}]_\times}\right ]^{\beta''\beta}\nonumber\\
&\quad+ \left[ e^{[\boldsymbol{\Lambda}]_\times}\right]_{\alpha\alpha''}
{\bf W}^{\alpha''\beta''}_0
\epsilon_{\beta\gamma''\beta'}
\left[e^{-[\hat{\boldsymbol{\Lambda}}]_\times}\right]_{\beta''\beta'}
\left[{B}^{-1}(\boldsymbol{\Lambda})\right]_{\gamma''\gamma'}
{B}_{\gamma'\gamma}(\boldsymbol{\Lambda})\nonumber\\
&\quad+{B}_{\gamma'\gamma}(\boldsymbol{\Lambda})
\left [ e^{[\boldsymbol{\Lambda}]_\times}\right ]^{\alpha\alpha''}
    \left(\frac{\partial}{\partial {\Lambda}_{\gamma'}}
    {\bf W}^{\alpha''\beta''}_0\right)
\left[e^{-[\boldsymbol{\Lambda}]_\times}\right]_{\beta''\beta}\nonumber\\
&=\epsilon_{\alpha\gamma\alpha'}
  \llangle{\bf W}^{\alpha'\beta}\rrangle^{\boldsymbol{\Lambda}{{\bf M}}{\cal E}}_{\rm rest}
+ \epsilon_{\beta\gamma\beta'}
\llangle {\bf W}^{\alpha\beta'}\rrangle^{\boldsymbol{\Lambda}{{\bf M}}{\cal E}}_{\rm rest}
 +{B}_{\gamma'\gamma}(\boldsymbol{\Lambda})
  \left[ e^{[\boldsymbol{\Lambda}]_\times}\right]_{\alpha\alpha''}
    \left(\frac{\partial}{\partial {\Lambda}_{\gamma'}}
    {\bf W}^{\alpha''\beta''}_0\right)
  \left[e^{-[\boldsymbol{\Lambda}]_\times}\right]_{\beta''\beta}
\end{align}
Inserting this result into (\ref{W:eq:454}) then gives
\begin{align}
  \label{W:eq:462}
  {B}_{\gamma'\gamma}(\boldsymbol{\Lambda})\left[ e^{[\boldsymbol{\Lambda}]_\times}\right]_{\alpha\alpha''}
    \left(\frac{\partial}{\partial {\Lambda}_{\gamma'}}
  {\bf W}^{\alpha''\beta''}_0\right)\left[e^{-[\boldsymbol{\Lambda}]_\times}\right]_{\beta''\beta}
  &=0
\end{align}
Because ${\bf B}$, $e^{[\boldsymbol{\Lambda}]_\times}$, and
$e^{-[\boldsymbol{\Lambda}]_\times}$ are non-singular matrices, this is
only possible if
\begin{align}
  \label{W:eq:465}
\frac{\partial}{\partial {\Lambda}_{\gamma'}}
    {\bf W}^{\alpha''\beta''}_0 &=0
\end{align}
which means the tensor ${\bf W}_0$ is independent of the orientation.

In conclusion,
the conditional expectation of scalar, vector, and tensor valued
phase functions is of the form (\ref{W:eq:469}).
Expressions for the quantities $F_0({\bf M},{\cal E})$, ${\bf  V}_0({{\bf M}},{\cal E})$,
and ${\bf W}_0({{\bf M}},{\cal  E})$ can be obtained by writing (\ref{W:eq:469}) as
\begin{align}
  \label{W:eq:470}
F_0({\bf M},{\cal E})
  &=   \llangle \hat{F}\rrangle^{\boldsymbol{\Lambda}{{\bf M}}{\cal E}}_{\rm rest}
    \nonumber\\
    {\bf V}_0({{\bf M}},{\cal E})
  &= 
    \llangle e^{-[\hat{\boldsymbol{\Lambda}}]_\times}\esc \hat{\bf V}\rrangle^{\boldsymbol{\Lambda}{{\bf M}}{\cal E}}_{\rm rest}
    \nonumber\\
    {\bf W}_0({{\bf M}},{\cal E})
  &= 
    \llangle e^{-[\hat{\boldsymbol{\Lambda}}]_\times}\esc\hat{\bf W}
\esc e^{[\hat{\boldsymbol{\Lambda}}]_\times}\rrangle^{\boldsymbol{\Lambda}{{\bf M}}{\cal E}}_{\rm rest}
\end{align}
where we  have introduced the  rotation matrix inside  the conditional
expectations by virtue of the Dirac delta function on the orientation.
The  main conclusion  of this  section is,  therefore, that  the rest
conditional  expectation of  a  tensor (of  any  order),  when
expressed  in  the   principal  axis  frame, is   independent  of  the
orientation.

\setcounter{my_equation_counter}{\value{equation}}
\section{Calculation       of      the      probability
  $P_{\cal E}^{\rm   rest}({\bf M},\boldsymbol{\Pi})$   in   the   Gaussian
  approximation}
\setcounter{equation}{\value{my_equation_counter}}
\label{App:PGauss}
Even  though  the  momentum   integrals  in  (\ref{eq:151})  with  the
microcanonical ensemble  (\ref{W:eq:289}) can be  explicitly computed,
simpler expressions are  obtained if we use a  canonical ensemble.  In
fact,   for  large   bodies   we  may   approximate  the   probability
(\ref{eq:151}) with
\begin{align}
  \label{eq:22}
  P^{\rm rest}_{\beta}({{\bf M}},\boldsymbol{\Pi})
  &=\int dz \rho^{\rm rest}_{\beta}(z)
    \delta\left(\hat{{\bf M}}(z)-{{\bf M}}\right)
    \delta(\hat{\boldsymbol{\Pi}}(z)-\boldsymbol{\Pi}) 
\end{align}
where the rest canonical ensemble is
\begin{align}
  \label{eq:16}
  \rho^{\rm rest}_{\beta}(z)
  &=\frac{1}{Z(\beta)}
    \delta\left(\hat{\bf R}(z) \right)
    \delta\left(\hat{\bf P}(z) \right)
    \delta\left(\hat{\bf S}(z) \right)
    e^{-\beta\hat{H}(z)}
\end{align}
To appreciate  the rigorous  connection between the  two probabilities
(\ref{eq:151}) and (\ref{eq:22}), insert      the      identity
\begin{align}
  \label{eq:450}
  \int d{\cal E}\;\delta(\hat{H}(z)-{\cal E})&=1
\end{align}
 inside   (\ref{eq:22}) to obtain
\begin{align}
  \label{eq:178}
  P^{\rm rest}_{\beta}({{\bf M}},\boldsymbol{\Pi})
  &=\int d{\cal E}\; P_\beta({\cal E})\;  P^{\rm rest}_{\cal {\cal E}}({{\bf M}},\boldsymbol{\Pi})
\end{align}
where the normalized probability is 
\begin{align}
  \label{eq:179}
  P_\beta({\cal E})&= \frac{e^{-\beta {\cal E}}\Omega^{\rm MT}({\cal E}) }{Z(\beta)}=  \frac{e^{S_B^{\rm MT}({\cal E})/k_B-\beta {\cal E}}}{Z(\beta)}  
\end{align}
This is the probability a system in the canonical ensemble has a value
of the energy given by ${\cal E}$, i.e.
\begin{align}
  \label{eq:50}
    P_\beta({\cal E})&=\int dz \rho^{\rm res}_\beta(z)\delta(\hat{H}(z)-{\cal E})
\end{align}
This probability is expected to be  highly peaked because both ${\cal E}$ and
$S^{\rm  MT}({\cal E})$ are  extensive quantities  expected to  scale as  the
number  $N$  of particles  in  the  body.  Setting the  derivative  of
(\ref{eq:179}) to  zero, the  maximum of  $P_\beta({\cal E})$ occurs  at ${\cal E}^*$
which is the solution of
\begin{align}
  \label{eq:180}
\beta&= \frac{1}{k_B} \frac{\partial S_B^{\rm MT}}{\partial {\cal E}}({\cal E}^*)= \frac{1}{k_BT^{\rm MT}({\cal E}^*)}
\end{align}
In the thermodynamic limit, $P_\beta({\cal E})\simeq\delta({\cal E}-{\cal E}^*)$ and
\begin{align}
  \label{eq:181}
  P^{\rm rest}_{\cal {\cal E}}({{\bf M}},\boldsymbol{\Pi})
  &\simeq P^{\rm rest}_{\beta({\cal E})}({{\bf M}},\boldsymbol{\Pi})
\end{align}
where $\beta({\cal E})$ is given by (\ref{eq:180}).
Now return to  (\ref{eq:22}), where we may separate position and momentum integrals as
\begin{align}
      \label{eq:161b}
      P^{\rm rest}_{\beta}({{\bf M}},\boldsymbol{\Pi})
      &=\frac{1}{Z(\beta)}\int dr     e^{-\beta\hat{\Phi}(z)}\delta\left(\hat{{\bf M}}(z)-{{\bf M}}\right)
        \delta\left(\hat{\bf R}(z) \right)
    \int dp e^{-\beta\sum_i\frac{{\bf p}^2_i}{2m_i}}    \delta\left(\hat{\bf P}(z) \right)
    \delta\left(\hat{\bf S}(z) \right)
\delta(\hat{\boldsymbol{\Pi}}(z)-\boldsymbol{\Pi}) 
\end{align}
Using (\ref{G0}), the momentum integral is $G^0({\bf P}=0,{\bf S}=0,\boldsymbol{\Pi},\beta)$ whose explicit expression in (\ref{eq:244}) gives
\begin{align}
  \label{eq:64b}
          P^{\rm rest}_{\beta}({{\bf M}},\boldsymbol{\Pi})
  &=\frac{1}{Z(\beta)}\int dr     e^{-\beta\hat{\Phi}(r)}\delta\left(\hat{{\bf M}}(r)-{{\bf M}}\right)
        \delta\left(\hat{\bf R}(r) \right)
    \nonumber\\
  &\times\left(\frac{\beta}{2\pi}\right)^{9/2}\prod_i^{N}\left(\frac{2\pi m_i}{\beta}\right)^{3/2}\frac{1}{{\rm M}^{3/2}
    \left(\det\hat{\bf I}(r)\right)^{1/2}
    \left(\det\hat{\bf T}(r)\right)^{1/2}}
    \exp\left\{-\frac{\beta}{2}\boldsymbol{\Pi}^T\esc\hat{\bf T}^{-1}(r)\esc\boldsymbol{\Pi}
    \right\}
\end{align}
The                                                        determinant
$\det\hat{\bf  I}=\hat{I}_1\hat{I}_2\hat{I}_3$  and,  as
shown     in     (\ref{TG}),     the     principal     mass     matrix
$\hat{\bf T}(z)=\hat{\mathbb{G}}(z)$  are functions of  the microstate
only through the  central moments.  The presence of  the Dirac delta
function  on  $\hat{{\bf  M}}$  allows one  to  substitute  the  phase
functions   $\hat{\bf   M}$   with    the   numerical   values   ${\bf
  M}$. Therefore, we have
\begin{align}
  \label{eq:70}
            P^{\rm rest}_{\beta}({{\bf M}},\boldsymbol{\Pi})
  &=\frac{1}{Z(\beta)}\left(\frac{\beta}{2\pi}\right)^{9/2}\prod_i^{N}\left(\frac{2\pi m_i}{\beta}\right)^{3/2}
    \frac{1}{{\rm M}^{3/2}    \left(I_1 I_2 I_3\right)^{1/2}}
    \frac{1}{  \left(M_1 M_2 M_3\right)^{1/2}}
\exp\left\{-\beta
    \sum_\alpha\frac{\Pi_\alpha^2}{2{ M}_\alpha}\right\}
    \nonumber\\
  &\times\int dr     e^{-\beta\hat{\Phi}(r)}\delta\left(\hat{{\bf M}}(r)-{{\bf M}}\right)
        \delta\left(\hat{\bf R}(r) \right)
\end{align}
Of  course, we  could do  the momentum  integrals in $Z(\beta)$ and some cancellations  of factors would
occur, but this is not necessary. Note that 
\begin{align}
  \label{eq:92}
  P^{\rm rest}_{\beta}({{\bf M}},\boldsymbol{\Pi})&=
  P^{\rm rest}_{\beta}({{\bf M}})  G^{\rm rest}_{\beta}({{\bf M}},\boldsymbol{\Pi})
\end{align}
where  the  normalized  Gaussian   probability  of  dilation  momentum
(conditional on ${{\bf M}}$) is
\begin{align}
  \label{eq:414}
    G^{\rm rest}_{\beta}({{\bf M}},\boldsymbol{\Pi})
  &=\frac{\beta^{3/2}}{(2\pi)^{3/2}(M_1 M_2 M_3)^{1/2}}
    \exp\left\{-\beta
    \sum_\alpha\frac{\Pi_\alpha^2}{2{M}_\alpha}\right\}
\end{align}

The physical meaning of $ P^{\rm rest}_{\beta}({{\bf M}})$ is simply the
marginal  equilibrium probability  of  finding  the principal  moments
${{\bf M}}$.  In fact,  because this probability  is normalized,  we may
integrate (\ref{eq:92}) with respect to $\boldsymbol{\Pi}$ and obtain
\begin{align}
  \label{eq:88}
  P^{\rm rest}_{\beta}({{\bf M}})&=\int d\boldsymbol{\Pi}
              P^{\rm rest}_{\beta}({{\bf M}},\boldsymbol{\Pi})
=\int dz \rho^{\rm rest}_{\beta}(z)
\delta\left(\hat{{\bf M}}(z)-{{\bf M}}\right)                                   
\end{align}
This probability could be measured,  in principle, from MD simulations
but  we  will consider  here  a  simple  approximation in  which  this
probability  has  a Gaussian  form.  For  example,  if we  assume  the
temperature  of the  body is  low  and particles  are executing  small
oscillations  around a  minimum, we  may expand  the potential  energy
around such a  minimum resulting in a Gaussian of  positions.  In this
way, the  resulting integral  is a  ``sectioned Gaussian''  similar to
(\ref{eq:134}).   The situation  is  somewhat  more delicate,  because
rotational and translational invariance  of the potential energy leads
to a  non-invertible matrix for  which (\ref{eq:134}) does  not exist.
Nevertheless, the calculation  can be performed and the  end result is
still  that  the  probability  $P^{\rm rest}_\beta({{\bf  M}})$  is  a
Gaussian of the form
\begin{align}
  \label{eq:110}
  P^{\rm rest}_\beta({\bf M})
  &=
    \frac{\beta^{3/2}}{\det(2\pi {\boldsymbol{\Sigma}})^{1/2}}
    \exp\left\{-\frac{\beta}{2}({{\bf M}}-{\bf M}^{\rm rest})^T\esc{\boldsymbol{\Sigma}}^{-1}\esc({{\bf M}}-{\bf M}^{\rm rest})\right\}
\end{align}
where  ${\bf M}^{\rm  rest}$ is  the rest  equilibrium average  of the
central moments  and $\boldsymbol{\Sigma}$ is  given by  the covariance  of the
central moments
\begin{align}
  \label{Cov}
\boldsymbol{\Sigma}&=\frac{1}{k_BT^{\rm MT}}\llangle  ({{\bf M}}-{\bf M}^{\rm rest})({{\bf M}}-{\bf M}^{\rm rest})^T\rrangle^{\rm eq}
\end{align}
\setcounter{my_equation_counter}{\value{equation}}
\section{Calculation of thermodynamic forces}
\setcounter{equation}{\value{my_equation_counter}}
\label{App:ThermForces}
In this section we show that
\begin{align}
  \label{App:7}
  {T}  \frac{\partial S_B}{\partial{\boldsymbol{\Lambda}}}
  &=-{\bf B}^{-T}\esc(\boldsymbol{\Omega}\times{\bf S}) +k_BT
    \frac{\partial}{\partial \boldsymbol{\Lambda}}\ln P^{\rm Haar}(\boldsymbol{\Lambda})
\\
  \label{App:380}
  T    \frac{\partial S_B}{\partial M_\alpha}
  &=\frac{1}{2}\nu_\alpha^2
    +2\left(    \boldsymbol{\Omega}_p^T\esc{\boldsymbol{\Omega}_p}-( {\Omega_p^{\alpha}})^2 \right)
  +    k_BT\frac{\partial}{\partial {
    M}_\alpha}\ln P^{\rm rest}_{\cal E}({{\bf M}},{\bf 0})
\end{align}
 The dependence of
the entropy (\ref{W:eq:282}) on $\boldsymbol{\Lambda}$ and ${\bf M}$ appears
explicitly through the probability and implicitly through the thermal energy
(\ref{Thermal-App}).  Using the chain rule then gives 
\begin{align}
  \label{eq:187}
  \frac{\partial S_B}{\partial{\boldsymbol{\Lambda}}}
  &= \frac{1}{T}
    \frac{\partial {\cal E}}{\partial{\boldsymbol{\Lambda}}} +
    k_B\frac{\partial}{\partial{\boldsymbol{\Lambda}}}\ln P^{\rm Haar}(\boldsymbol{\Lambda})
{\overset{(\ref{eq:150})}{=}} -\frac{1}{T}\frac{\partial K^{\rm rot}}{\partial\boldsymbol{\Lambda}}
+    k_B\frac{\partial}{\partial{\boldsymbol{\Lambda}}}\ln P^{\rm Haar}(\boldsymbol{\Lambda})
    \nonumber\\
  &{\overset{(\ref{eq:192})}{=}}-\frac{1}{T}{\bf B}^{-T}\esc(\boldsymbol{\Omega}\times{\bf S})
  + k_B\frac{\partial}{\partial{\boldsymbol{\Lambda}}}\ln P^{\rm Haar}(\boldsymbol{\Lambda})
    \\
  \label{eq:187d}  \frac{\partial S_B}{\partial{{\bf M}}}
  &= \frac{1}{T}
    \frac{\partial {\cal E}}{\partial{{\bf M}}} +
    k_B\frac{\partial}{\partial{{\bf M}}}
    \ln P^{\rm rest}_{\cal E} ({\boldsymbol{\Lambda}},{\bf M},{\bf 0})
\overset{(\ref{Thermal-App}) (\ref{eq:119})}{  =}-\frac{1}{T}  \frac{\partial }{\partial{{\bf M}}}
    \left(K^{\rm rot}+K^{\rm dil}\right) +
    k_B\frac{\partial}{\partial{{\bf M}}}
    \ln P^{\rm rest}_{\cal E} ({\bf M},{\bf 0})
\end{align}
where  ${\bf B}^{-T}$  denotes the  transpose  of the  inverse of  the
matrix ${\bf B}$. The derivative of  the logarithm of the Haar measure
is
\begin{align}
  \label{eq:365}
\frac{\partial}{\partial{\boldsymbol{\Lambda}}}\ln P^{\rm Haar}(\boldsymbol{\Lambda})
  &=-2f(\Lambda)\boldsymbol{\Lambda}
\end{align}
where $f(\Lambda)$
\begin{align}
  \label{eq:262}
  f(\Lambda) = \frac{1}{\Lambda^2}
\left ( 1 - \frac{\Lambda}{2}\cot\frac{\Lambda}{2}\right )
\end{align}

The derivatives of the kinetic energies with respect central moments are
\begin{align}
  \label{eq:375}
  \frac{\partial K^{\rm rot}}{\partial{{\bf M}}}  
  &=\frac{1}{2}
    \left(e^{-[\boldsymbol{\Lambda}]_\times}\esc{\bf S}\right)^T\esc
    \frac{\partial {\mathbb{I}}^{-1}}{\partial{{\bf M}}}\esc
    \left(e^{-[\boldsymbol{\Lambda}]_\times}\esc{\bf S}\right)
    \nonumber\\
    \frac{\partial K^{\rm dil}}{\partial {M}_\alpha}  &=-
\frac{1}{2}{\nu}_\alpha^2
\end{align}
Let us find a more explicit form of the derivative of the rotational kinetic energy by
looking at its components
\begin{align}
  \label{eq:376}
    \frac{\partial K^{\rm rot}}{\partial {M}_\alpha}  
  &=-\frac{1}{2}\left[e^{-[\boldsymbol{\Lambda}]_\times}\esc{\bf S}\right]^\mu
    \left[\mathbb{I}^{-1}\right]_{\mu\mu'}
    \frac{\partial \left[\mathbb{I}\right]_{\mu'\nu'}}{\partial {M}_\alpha}
    \left[\mathbb{I}^{-1}\right]_{\nu'\nu}
    \left[e^{-[\boldsymbol{\Lambda}]_\times}\esc{\bf S}\right]^\nu
\end{align}
By using that the intertia tensor can be expressed  in terms of
the gyration tensor $\hat{\mathbb{I}}= 4\left({\rm Tr}[\hat{\mathbb{G}}]\mathbb{1}-\hat{\mathbb{G}}\right)$,
we may evaluate the tensor
\begin{align}
  \label{eq:408}
\sum_{\mu'\nu'}      \left[\mathbb{I}^{-1}\right]_{\mu\mu'}
    \frac{\partial \left[\mathbb{I}\right]_{\mu'\nu'}}{\partial {M}_\alpha}
    \left[\mathbb{I}^{-1}\right]_{\nu'\nu}
  &=
\sum_{\mu'\nu'}      \left[\mathbb{I}^{-1}\right]_{\mu\mu'}
{\black  4}\left(\delta_{\mu'\nu'}   -\delta_{\mu'\nu'\alpha}\right)
    \left[\mathbb{I}^{-1}\right]_{\nu'\nu}
    \nonumber\\
&    ={\black  4}
\sum_{\mu'}      \left[\mathbb{I}^{-1}\right]_{\mu\mu'}
    \left[\mathbb{I}^{-1}\right]_{\mu'\nu}
-     \left[\mathbb{I}^{-1}\right]_{\mu\underline{\alpha}}
    \left[\mathbb{I}^{-1}\right]_{\underline{\alpha}\nu}
                  ={\black  4}\left(\frac{1}{I_{\underline{\mu}}^2}
\delta_{\underline{\mu}\nu}-\frac{1}
{I_{\underline{\alpha}}^2}
                  \delta_{\underline{\alpha}\mu}\delta_{\underline{\alpha}\nu}\right)
\end{align}
where the underlines indicate indices are not summed over.  Using
(\ref{eq:408}) in (\ref{eq:376}) gives
\begin{align}
  \label{eq:409}
      \frac{\partial K^{\rm rot}}{\partial {M}_\alpha} 
  &=-\sum_{\mu\nu}
    \frac{1}{2}\left[e^{-[\boldsymbol{\Lambda}]_\times}\esc{\bf S}\right]^\mu
    {\black  4}\left(
\frac{1}{I_{\mu}^2}\delta_{{\mu}\nu}-
\frac{1}{I_{\underline{\alpha}}^2}
                  \delta_{{\mu}\underline{\alpha}}\delta_{{\nu}\underline{\alpha}}
    \right)
    \left[e^{-[\boldsymbol{\Lambda}]_\times}\esc{\bf S}\right]^\nu
    \nonumber\\
  &=-{\black  2}\sum_\mu\left[e^{-[\boldsymbol{\Lambda}]_\times}\esc{\bf S}\right]^\mu
    \frac{1}{I_{\mu}^2}
    \left[e^{-[\boldsymbol{\Lambda}]_\times}\esc{\bf S}\right]^\mu
    +\frac{2}{I_{\underline{\alpha}}^2}
    \left[e^{-[\boldsymbol{\Lambda}]_\times}\esc{\bf S}\right]_{\underline{\alpha}}
    \left[e^{-[\boldsymbol{\Lambda}]_\times}\esc{\bf S}\right]_{\underline{\alpha}}
    \nonumber\\
   &=2\left( ( \boldsymbol{\Omega}_p^{\alpha})^2 - {\boldsymbol{\Omega}_p}^T\esc{\boldsymbol{\Omega}_p}\right)
\end{align}
where 
\begin{align}
  \label{eq:372}
  {\boldsymbol{\Omega}_p}\equiv
  e^{-[\boldsymbol{\Lambda}]_\times}\esc\boldsymbol{\Omega}
 \end{align}

Note that $\boldsymbol{\Omega}^T\esc\boldsymbol{\Omega}=\boldsymbol{\Omega}_p^T \esc\boldsymbol{\Omega}_p$.
Collecting terms together then gives
\begin{align}
  \label{eq:358}
T    \frac{\partial S_B}{\partial {M}_\alpha}
  &=2\left(    \boldsymbol{\Omega}_p^T\esc \boldsymbol{\Omega}_p
    -( \boldsymbol{\Omega}_p^\alpha )^2 \right)
+\frac{1}{2}{\nu}_\alpha^2
+    k_BT\frac{\partial\ln P^{\rm rest}_{\cal E}}{\partial{{M}_\alpha}}({{\bf M}},{\bf 0})
\end{align}

\setcounter{my_equation_counter}{\value{equation}}
\section{Calculation of the reversible terms }
\setcounter{equation}{\value{my_equation_counter}}
\label{App:ReviLLambda}

Consider first  the reversible  term corresponding to  the orientation
equation.   The   transform  of   conditional  expectations   to  rest
conditional    expectations    (\ref{eq:43}),    (\ref{eq:35}),    and
(\ref{W:BodyFrameInv})   gives   the    conditional   expectation   of
$i{\cal L}\hat{\boldsymbol{\Lambda}}$ as
\begin{align}
    \llangle  i{\cal L}\hat{\boldsymbol{\Lambda}}\rrangle^a
  &=
  \llangle {\cal G}
  \sum_{i}\frac{\partial \hat{\boldsymbol{\Lambda}}}{\partial {\bf r}_{i}}\esc{\bf v}_{i}\rrangle^{\boldsymbol{\Lambda}{\bf M}{\cal E}}_{\rm rest}
=\llangle
    \sum_{i}\frac{\partial \hat{\boldsymbol{\Lambda}}}{\partial {\bf r}'_{i}}\esc
        \left[{\bf v}'_{i}+{\bf V} - [{\bf r}'_{i}]_\times\esc\hat{\boldsymbol{\Omega}}
    +\frac{1}{m_i}\frac{\partial \hat{{\bf M}}}{\partial {\bf r}'_{i}}\esc\hat{\boldsymbol{\nu}}
    \right]
\rrangle^{\boldsymbol{\Lambda}{\bf M}{\cal E}}_{\rm rest}
\end{align}
The first  term, which is  linear in  ${\bf v}'_i$ gives  a conditional
average  where the  integrand  is  odd under  the  change of  variable
${\bf v}'_i\to  -{\bf v}'_i$, so  vanishes by symmetry. The  second term
with ${\bf V}$ vanishes because  of the translation invariant property
(\ref{trans}). The  fourth term  involving the dilation  velocity also
vanishes once (\ref{eq:H})  and (\ref{eq:449}) are used.  In the third
term with the spin velocity we use (\ref{eq:255}) to get
\begin{align}
  \label{eq:264}
  \llangle  i{\cal L}\hat{\boldsymbol{\Lambda}}\rrangle^a
  &=
    \llangle{\bf B}(\hat{\boldsymbol{\Lambda}}(z'))\esc\hat{\boldsymbol{\Omega}(z')}
    \rrangle^{\boldsymbol{\Lambda}{\bf M}{\cal E}}_{\rm rest}
=
  {\bf B}({\boldsymbol{\Lambda}})\esc{\boldsymbol{\Omega}}
\end{align}
where we have used $\langle f(\hat{A})\rangle^{a}=f(a)$.

Next, let us  consider the reversible term corresponding to the dilational
momentum.  The dilational force is $\hat{\boldsymbol{\cal K}}=i{\cal L}\hat{\boldsymbol{\Pi}}=i{\cal L}^2{\bf M}$, this is
\begin{align}
 \label{eq:91}
\hat{\boldsymbol{\cal K}}
  &=\sum_i\left ( \frac{\partial\hat{\boldsymbol{\Pi}}}{\partial{\bf r}_i}\esc {\bf v}_i
+\frac{1}{m_i}\frac{\partial\hat{\boldsymbol{\Pi}}}{\partial{\bf v}_i}\esc \hat{\bf F}_i
\right )
  =\sum_{i j}\frac{\partial^2 \hat{{\bf M}}}{\partial {\bf r}_{i}\partial {\bf r}_{j}}:{\bf v}_{i}{\bf v}_{j}
   +\sum_{i}\frac{1}{m_i}\frac{\partial \hat{{\bf M}}}{\partial {\bf r}_{i}}\esc\hat{\bf F}_{i }
\end{align}
in which $\hat{\bf F}_i$ is the force on particle $i$.  Applying $\cal G$ to
produce the change of variables (\ref{W:BodyFrameInv}) gives
\begin{align}
  \label{eq:94}
  {\cal G}\hat{\boldsymbol{\cal K}}
  =&-\sum_{i j}\left[
({\bf r}'_{i}\times\hat{\boldsymbol{\Omega}})\esc
\frac{\partial^2 \hat{{{\bf M}}}}{\partial {\bf r}'_{i}\partial {\bf r}'_{j}}
\esc\frac{1}{m_j}\frac{\partial \hat{{\bf M}}}{\partial {\bf r}'_{j}}
\esc\hat{\boldsymbol{\nu}}
+\frac{1}{m_i}\frac{\partial \hat{{\bf M}}}{\partial {\bf r}'_{i}}
\esc\hat{\boldsymbol{\nu}}\esc
\frac{\partial^2 \hat{{{\bf M}}}}{\partial {\bf r}'_{i}\partial {\bf r}'_{j}}
\esc ({\bf r}'_{j}\times\hat{\boldsymbol{\Omega}})\right]\nonumber\\
& +\sum_{i j}\left[({\bf r}'_{i}\times\hat{\boldsymbol{\Omega}})\esc
\frac{\partial^2 \hat{{{\bf M}}}}{\partial {\bf r}'_{i}\partial {\bf r}'_{j}}
\esc ({\bf r}'_{j}\times\hat{\boldsymbol{\Omega}})
+\hat{\boldsymbol{\nu}}\esc\frac{1}{m_i}\frac{\partial \hat{{\bf M}}}{\partial {\bf r}'_{i}}
\esc
\frac{\partial^2 \hat{{{\bf M}}}}{\partial {\bf r}'_{i}\partial {\bf r}'_{j}}
\esc\frac{1}{m_j}\frac{\partial \hat{{\bf M}}}{\partial {\bf r}'_{j}}
\esc\hat{\boldsymbol{\nu}}\right ]
\end{align}
where we  have already eliminated all  terms with $\bf V$  that vanish
due to  (\ref{trans2}) and all terms linear in the
  velocity  ${\bf v}_i'$, for which  the  corresponding conditional  average
  vanish by symmetry.
The first term is
\begin{align}
{\hat{\boldsymbol{\Omega}}}^\beta\sum_{i j}[{\bf r}'_i]_\times^{\gamma\beta}
\frac{\partial^2 \hat{M}_\alpha}
{\partial {{\bf r}'_i}^\gamma\partial {{\bf r}'_j}^{\gamma'}}
\frac{1}{m_j}\frac{\partial \hat{M}_{\alpha'}}
{\partial {{\bf r}'_j}^{\gamma'}} \hat{\nu}_{\alpha'}
        \overset{(\ref{eq:267})}{=}
{\hat{\boldsymbol{\Omega}}}^\beta              \sum_{ j}  \epsilon_{\gamma'\gamma\beta}  \frac{\partial\hat{M}_\alpha}{\partial {\bf r}^{\gamma'}_j}
      \frac{1}{m_j}\frac{\partial \hat{M}_{\alpha'}}
{\partial {{\bf r}'_j}^{\gamma}} \hat{\nu}_{\alpha'}
\end{align} while the second term is identical to this one, but with a minus sign. Therefore, the first two terms of (\ref{eq:94}) cancel each other. 
The remaining terms are
\begin{align}
\sum_{i j}[{\bf r}'_i]_\times^{\gamma\beta}{\hat{\boldsymbol{\Omega}}}^\beta
\frac{\partial^2 \hat{M}_\alpha}
{\partial {{\bf r}'_i}^\gamma\partial {{\bf r}'_j}^{\gamma'}}
[{\bf r}'_j]_\times^{\gamma'\beta'}{\hat{\boldsymbol{\Omega}}}^{\beta'}
&\overset{(\ref{eq:267})}{=}\sum_i
{\hat{\boldsymbol{\Omega}}}^\beta {\hat{\boldsymbol{\Omega}}}^{\beta'}
[{\bf r}'_i]_\times^{\gamma\beta}\epsilon_{\gamma\gamma'\beta'}
\frac{\partial\hat{M}_\alpha}{\partial {{\bf r}'_i}^{\gamma'}}
= \sum_i {\hat{\boldsymbol{\Omega}}}^\beta {\hat{\boldsymbol{\Omega}}}^{\beta'}
[{\bf r}'_i]_\times^{\alpha'}\epsilon_{\gamma\alpha'\beta}\epsilon_{\gamma\gamma'\beta'}
\frac{\partial\hat{M}_\alpha}{\partial {{\bf r}'_i}^{\gamma'}}\nonumber\\
&\overset{(\ref{eq:604})}{=} \sum_i {\hat{\boldsymbol{\Omega}}}^\beta {\hat{\boldsymbol{\Omega}}}^{\beta'}
{{\bf r}'_i}^{\alpha'}
\left [ \delta_{\alpha'\gamma'}\delta_{\beta\beta'}-
\delta_{\alpha'\beta'}\delta_{\beta\gamma'}\right ]
\frac{\partial\hat{M}_\alpha}{\partial {{\bf r}'_i}^{\gamma'}}\nonumber\\
&=\sum_i\left [
{\hat{\boldsymbol{\Omega}}}^\beta {\hat{\boldsymbol{\Omega}}}^\beta
{{\bf r}'_i}^{\gamma'}
\frac{\partial\hat{M}_\alpha}{\partial {{\bf r}'_i}^{\gamma'}}
-
{\hat{\boldsymbol{\Omega}}}^\beta {\hat{\boldsymbol{\Omega}}}^{\beta'}
{{\bf r}'_i}^{\beta'}
\frac{\partial\hat{M}_\alpha}{\partial {{\bf r}'_i}^\beta}
\right] \nonumber\\
&\overset{(\ref{eq:256})}{=}2
{\hat{\boldsymbol{\Omega}}}^\beta {\hat{\boldsymbol{\Omega}}}^\beta
\left[e^{-[\hat{\boldsymbol{\Lambda}}]_\times}\right]_{\underline{\alpha}\beta'}
\hat{\bf G}_{\beta'\gamma'}
\left[e^{-[\hat{\boldsymbol{\Lambda}}]_\times}\right]_{\underline{\alpha}\gamma'}
-2 {\hat{\boldsymbol{\Omega}}}^\beta {\hat{\boldsymbol{\Omega}}}^{\beta'}
\left[e^{-[\hat{\boldsymbol{\Lambda}}]_\times}\right]_{\underline{\alpha}\beta}
\left[e^{-[\hat{\boldsymbol{\Lambda}}]_\times}\right]_{\underline{\alpha}\gamma'}
\hat{\bf G}_{\gamma'\beta'}\nonumber\\
&=2 \hat{\boldsymbol{\Omega}}^{{T}}\esc \hat{\boldsymbol{\Omega}}
\hat{M}_\alpha - 2
\left[e^{-[\hat{\boldsymbol{\Lambda}}]_\times}\right]_{\underline{\alpha}\beta}
{\hat{\boldsymbol{\Omega}}}^\beta
\left[e^{-[\hat{\boldsymbol{\Lambda}}]_\times}\right]_{\underline{\alpha}\beta'}
{\hat{\boldsymbol{\Omega}}}^{\beta'}
\hat{M}_{\underline{\alpha}}\nonumber\\
&\overset{(\ref{eq:372})}{=}
2\left (\hat{\boldsymbol{\Omega}}_p^{{T}}\esc \hat{\boldsymbol{\Omega}}_p-
{\hat{\boldsymbol{\Omega}}_p}^{\underline{\alpha}}
{\hat{\boldsymbol{\Omega}}_p}^{\underline{\alpha}}\right)
\hat{M}_{\underline{\alpha}}(z')
\end{align}
and
\begin{align}
 \hat{\nu}_{\beta'}\sum_{i j} \frac{1}{m_i}\frac{\partial \hat{M}_{\beta'}}
{\partial {{\bf r}'_i}^\gamma}
\frac{\partial^2 \hat{M}_\alpha}
{\partial {{\bf r}'_i}^\gamma\partial {{\bf r}'_j}^{\gamma'}}
\frac{1}{m_j}\frac{\partial \hat{M}_{\alpha'}}
{\partial {{\bf r}'_j}^{\gamma'}} \hat{\nu}_{\alpha'}
&\overset{(\ref{eq:42b})}{=} \frac{1}{2}\sum_i \frac{1}{m_i}\frac{\partial \hat{M}_{\beta'}}
{\partial {{\bf r}'_i}^\gamma} \hat{\nu}_{\beta'}
\frac{\partial \hat{\bf T}^{\alpha\alpha'}}{\partial {{\bf r}'_i}^\gamma}
\hat{\nu}_{\alpha'}
\overset{(\ref{TG})}{=}\frac{1}{2} \hat{\nu}_{\beta'}
\hat{\nu}_{\underline{\alpha}}
\sum_i \frac{1}{m_i}\frac{\partial \hat{M}_{\beta'}}
{\partial {{\bf r}'_i}^\gamma}
\frac{\partial \hat{M}_{\underline{\alpha}}}{\partial {{\bf r}'_i}^\gamma}
\nonumber\\
&\overset{(\ref{eq:42b})}{=}\frac{1}{2} \hat{\nu}_{\beta'}
\hat{\nu}_{\underline{\alpha}}
\hat{\bf T}^{\beta'\underline{\alpha}}(z')
\overset{(\ref{TG})}{=}\frac{1}{2} \hat{\nu}_{\underline{\alpha}}
\hat{\nu}_{\underline{\alpha}}
\hat{M}_{\underline{\alpha}}(z')
\overset{(\ref{eq:85})}{=}\frac{1}{2}\hat{{\Pi}}_{\underline{\alpha}}
\hat{\nu}_{\underline{\alpha}}
\end{align}
Collecting all these results together then gives
\begin{align}
 \label{eq:194}
  {\cal G}\hat{\boldsymbol{\cal K}}
&=\hat{\boldsymbol{\cal K}}(z') -2\sum_{i j}{\bf v}'_{i}\esc
\frac{\partial^2 \hat{{{\bf M}}}}{\partial {\bf r}'_{i}\partial {\bf r}'_{j}}
\esc({\bf r}'_{j}\times\hat{\boldsymbol{\Omega}}) +
\frac{1}{2}\hat{{\Pi}}_{\underline{\alpha}}(z')
\hat{\nu}_{\underline{\alpha}}
+2\left (\hat{\boldsymbol{\Omega}}_p^{{T}}\esc \hat{\boldsymbol{\Omega}}_p-
{\hat{\boldsymbol{\Omega}}_p}^{\underline{\alpha}}
{\hat{\boldsymbol{\Omega}}_p}^{\underline{\alpha}}\right)
\hat{M}_{\underline{\alpha}}(z')
+\frac{1}{2} \hat{\nu}_{\underline{\alpha}}
\hat{\nu}_{\underline{\alpha}}
\hat{M}_{\underline{\alpha}}(z')
\end{align}
which when used in (\ref{eq:43}) gives the conditional expectation of the
dilational force to be
\begin{align}
  \label{eq:107}
  \llangle  \hat{\cal K}_\alpha\rrangle^a
  &= \llangle {\cal G}\hat{\cal K}_\alpha\rrangle^{\boldsymbol{\Lambda}{\bf M}{\cal E}}_{\rm rest}
  = \llangle \hat{\cal K}_\alpha\rrangle^{\boldsymbol{\Lambda}{\bf M}{\cal E}}_{\rm rest}  +
   2\left(\boldsymbol{\Omega}_p^{{T}}\esc\boldsymbol{\Omega}_p
  -\boldsymbol{\Omega}_p^{\underline{\alpha}}
    \boldsymbol{\Omega}_p^{\underline{\alpha}}
    \right){M}_{\underline{\alpha}}
    +  \frac{1}{2} \nu^2_{\underline{\alpha}}{M}_{\underline{\alpha}}
\end{align}
where the term linear in ${\bf v}'_i$ vanishes due to symmetry and the other
terms depend on the microstate only through the CG variables.
The    rest   conditional    average   of    the   dilational    force
$\llangle\hat{\boldsymbol{\cal K}}\rrangle^{\boldsymbol{\Lambda}{\bf M}{\cal E}}_{\rm  rest}$
involves  the  pair  forces between
particles, and an explicit expression  is difficult to obtain by brute
force, except  for very simple  linear force laws, i.e.   the harmonic
solid.

\setcounter{my_equation_counter}{\value{equation}}
\section{The reversibility condition}
\setcounter{equation}{\value{my_equation_counter}}
\label{App:RevCond}
 We collect  the reversible drift
in the form
\begin{align}
  \label{eq:62}
V(a)&\to  \left(
      \begin{array}{c}
        \llangle i{\cal L}\hat{\bf R}\rrangle^a
        \\
        \\
        \llangle i{\cal L}\hat{\boldsymbol{\Lambda}}\rrangle^a
        \\
        \\
        \llangle i{\cal L}\hat{{\bf M}}\rrangle^a
        \\
        \\
        \llangle i{\cal L}\hat{\bf P}\rrangle^a
        \\
        \\
        \llangle i{\cal L}\hat{\bf S}\rrangle^a
        \\
        \\
        \llangle i{\cal L}\hat{\boldsymbol{\Pi}}\rrangle^a
        \\
        \\
        \llangle i{\cal L}\hat{H}\rrangle^a
      \end{array}
  \right)
  =\left(
      \begin{array}{c}
       {\bf V}
        \\
        \\
{\bf B}\esc{\boldsymbol{\Omega}}
        \\
        \\
        \boldsymbol{\Pi}
        \\
        \\
        0
        \\
        \\
        0
        \\
        \\
        {\boldsymbol{\cal K}}
        \\
        \\
        0
      \end{array}
  \right)
\end{align}
The reversibility  condition
\begin{align}
  V^T(a)\esc\frac{\partial S_B}{\partial a}(a)+  k_B\frac{\partial \esc V}{\partial a}(a)=0
\end{align}
in this case takes the explicit form
\begin{align}
  \label{eq:394}
&  \llangle i{\cal L}\hat{\bf R}\rrangle^a\esc
    \frac{\partial S_B}{\partial{\bf R}}
    +   \llangle i{\cal L}\hat{\boldsymbol{\Lambda}}\rrangle^a\esc
    \frac{\partial S_B}{\partial{\boldsymbol{\Lambda}}}
    +        \llangle i{\cal L}\hat{{\bf M}}\rrangle^a\esc
    \frac{\partial S_B}{\partial{{\bf M}}}
    +   \llangle i{\cal L}\hat{\bf P}\rrangle^a\esc
     \frac{\partial S_B}{\partial{\bf P}}
     +        \llangle i{\cal L}\hat{\bf S}\rrangle^a\esc
     \frac{\partial S_B}{\partial{\bf S}}
     +        \llangle i{\cal L}\hat{\boldsymbol{\Pi}}\rrangle^a\esc
     \frac{\partial S_B}{\partial{\boldsymbol{\Pi}}}
+     \llangle i{\cal L}\hat{H}\rrangle^a\esc
     \frac{\partial S_B}{\partial E}
                                             \nonumber\\
  +&k_B        \frac{\partial }{\partial{\bf R}}\esc
     \llangle i{\cal L}\hat{\bf R}\rrangle^a
     +k_B        \frac{\partial }{\partial{\boldsymbol{\Lambda}}}\esc
     \llangle i{\cal L}\hat{\boldsymbol{\Lambda}}\rrangle^a
     +k_B        \frac{\partial }{\partial{{\bf M}}}\esc
     \llangle i{\cal L}\hat{{\bf M}}\rrangle^a
  +k_B        \frac{\partial }{\partial{\bf P}}\esc
     \llangle i{\cal L}\hat{\bf P}\rrangle^a
     +k_B        \frac{\partial }{\partial{\bf S}}\esc
     \llangle i{\cal L}\hat{\bf S}\rrangle^a
     +k_B     
     \frac{\partial }{\partial{\boldsymbol{\Pi}}}\esc
     \llangle i{\cal L}\hat{\boldsymbol{\Pi}}\rrangle^a
     \nonumber\\
  +& k_B             \frac{\partial }{\partial E}
     \llangle i{\cal L}\hat{H}\rrangle^a=0
\end{align}
which using (\ref{eq:163}) and (\ref{eq:62}) simplifies to
\begin{align}
  \label{eq:2}
  0  &=
       \frac{\partial S_B}{\partial{\boldsymbol{\Lambda}}}\esc{\bf B}\esc{\boldsymbol{\Omega}}
       +  \boldsymbol{\Pi} \esc         \frac{\partial S_B}{\partial{{\bf M}}}
       -      {\boldsymbol{\cal K}}\esc\frac{\boldsymbol{\nu}}{T}
       + k_B\frac{\partial}{\partial\boldsymbol{\Lambda}}
       \esc\left(
       {\bf B}({\boldsymbol{\Lambda}})\esc{\boldsymbol{\Omega}}\right)
  +   k_B\frac{\partial}{\partial\boldsymbol{\Pi}}\esc      {\boldsymbol{\cal K}}
\end{align}
By  using (\ref{W:eq:282}),  (\ref{W:eq:266}), and  (\ref{eq:25}), the
reversibility condition further simplifies to
\begin{align}
  \label{eq:236}
    0  &=
         \boldsymbol{\Pi} \esc         \frac{\partial S_B}{\partial{{\bf M}}}
       -      \frac{\boldsymbol{\nu}}{T}\esc{\boldsymbol{\cal K}}
              +   k_B\frac{\partial}{\partial\boldsymbol{\Pi}}\esc      {\boldsymbol{\cal K}}
\end{align}
For the last term  we  use   (\ref{eq:107})
\begin{align}
  \label{eq:112}
\frac{\partial}{\partial{\Pi}_\alpha}    \llangle  \hat{\cal K}_\alpha\rrangle^a
  &=\frac{\partial{\cal E}}{\partial{\Pi}_\alpha}\frac{\partial}{\partial{\cal E}}\llangle \hat{\cal K}_\alpha\rrangle^{\boldsymbol{\Lambda}{\bf M}{\cal E}}_{\rm rest}
    +   \sum_\alpha  {\nu}_{\alpha}
\overset{(\ref{Thermal-App})}{=}
    -{\nu}_{\alpha}\frac{\partial}{\partial{\cal E}}\llangle \hat{\cal K}_\alpha\rrangle^{\boldsymbol{\Lambda}{\bf M}{\cal E}}_{\rm rest}
    + \sum_\alpha  {\nu}_{\alpha}
\end{align}
which can be written in a compact form as
\begin{align}
  \label{eq:114}
  \frac{\partial}{\partial\boldsymbol{\Pi}}\esc    \llangle  \hat{\boldsymbol{\cal K}}\rrangle^a
  &=    \boldsymbol{\nu}\esc\left(-\frac{\partial}{\partial{\cal E}}\llangle \hat{\boldsymbol{\cal K}}\rrangle^{\boldsymbol{\Lambda}{\bf M}{\cal E}}_{\rm rest}+{\bf c}\right)
\end{align}
where ${\bf c}=(1,1,1)$. Using this in (\ref{eq:236}) along with (\ref{eq:85}) gives
\begin{align}
  \label{eq:75}
        0  &=
         \boldsymbol{\nu} \esc \left(T{\bf T}\esc        \frac{\partial S_B}{\partial{{\bf M}}}
       -      {\boldsymbol{\cal K}}
+ k_BT\left(-\frac{\partial}{\partial{\cal E}}\llangle \hat{\boldsymbol{\cal K}}\rrangle^{\boldsymbol{\Lambda}{\bf M}{\cal E}}_{\rm rest}+{\bf c}\right)\right)
\end{align}
This identity should be true for arbitrary values of $\boldsymbol{\nu}$ and we conclude
\begin{align}
  \label{eq:120}
{\boldsymbol{\cal K}}  &=
-k_BT\frac{\partial}{\partial{\cal E}}\llangle \hat{\boldsymbol{\cal K}}\rrangle^{\boldsymbol{\Lambda}{\bf M}{\cal E}}_{\rm rest}+k_BT{\bf c}+  {\bf T}\esc    T    \frac{\partial S_B}{\partial{{\bf M}}}
\end{align}
By using the form of the  mean dilational force (\ref{eq:107}) and the
derivative   of  the   entropy  (\ref{eq:358}),   a  quite   fortunate
cancellation of ``convective'' terms occur and (\ref{eq:120}) becomes
\begin{align}
  \label{eq:129}
  \frac{\partial}{\partial{\cal E}}\llangle \hat{\boldsymbol{\cal K}}\rrangle^{\boldsymbol{\Lambda}{\bf M}{\cal E}}_{\rm rest}
    &=-\frac{1}{k_BT} \llangle \hat{\boldsymbol{\cal K}}\rrangle^{\boldsymbol{\Lambda}{\bf M}{\cal E}}_{\rm rest}  
   +\left( {\bf T}\esc\frac{\partial}{\partial{{\bf M}}}\ln P^{\rm rest}_{\cal E}({{\bf M}},{\bf 0})+{\bf c}\right)
\end{align}
This  can  be understood  as  an  ordinary differential  equation  for
$\llangle  \hat{\boldsymbol{\cal K}}\rrangle^{\boldsymbol{\Lambda}{\bf
    M}{\cal E}}_{\rm rest} $ with the structure
\begin{align}
  \label{eq:221}
  x'(t)=-f(t)x(t)+g(t)
\end{align}
and with general solution
\begin{align}
  \label{eq:243}
  x(t)=e^{-\int_{t_0}^tf(t')dt'}\left(x(0)+\int_{t_0}^tdt''e^{\int_0^{t''}f(t')dt'}g(t'')\right)
\end{align}
Therefore, the solution of (\ref{eq:129}) is
\begin{align}
  \label{eq:292}
  \llangle \hat{\boldsymbol{\cal K}}\rrangle^{\boldsymbol{\Lambda}{\bf M}{\cal E}}_{\rm rest}
  &=e^{-\int_0^{\cal E}d{\cal E}'\frac{1}{k_BT({\cal E}')}}
    \left(\llangle \hat{\boldsymbol{\cal K}}\rrangle^{\boldsymbol{\Lambda}{\bf M}0}_{\rm rest}
    +    \int_0^{\cal E}d{\cal E}''e^{\int_0^{{\cal E}''}d{\cal E}'\frac{1}{k_BT({\cal E}')}}
\left( {\bf T}\esc\frac{\partial}{\partial{{\bf M}}}\ln P^{\rm rest}_{{\cal E}''}({{\bf M}},{\bf 0})+{\bf c}\right)    \right)
\end{align}
The definition of temperature as the derivative of the entropy with respect to the energy  implies
\begin{align}
  \label{eq:307}
  \int_0^{\cal E}d{\cal E}'\frac{1}{k_BT({\cal E}')}=\frac{S_B({\cal E})-S_B(0)}{k_B}
\end{align}
where we leave implicit the rest of the variable dependencies. Also, the value of the entropy and of any conditional expectation at the zero energy should vanish, so that
\begin{align}
  \label{eq:326}
  S_B(0)&=0
          \nonumber\\
    \llangle \hat{\boldsymbol{\cal K}}\rrangle^{\boldsymbol{\Lambda}{\bf M}0}_{\rm rest}&=0
\end{align}
Therefore, (\ref{eq:292}) takes the form
\begin{align}
  \label{eq:330}
    \llangle \hat{\boldsymbol{\cal K}}\rrangle^{\boldsymbol{\Lambda}{\bf M}{\cal E}}_{\rm rest}
  &=    \int_0^{\cal E}d{\cal E}''e^{\frac{S_B({\cal E}'')-S_B({\cal E})}{k_B}}
\left( {\bf T}\esc\frac{\partial}{\partial{{\bf M}}}\ln P^{\rm rest}_{{\cal E}''}({{\bf M}},{\bf 0})+{\bf c}\right)
\end{align}
The prefactor containing the entropies is, by using the exact result (\ref{W:eq:287})
\begin{align}
  \label{eq:341}
 e^{\frac{S_B({\cal E}'')-S_B({\cal E})}{k_B}}
  & =  e^{ \frac{S^{\rm MT}({\cal E}'')- S^{\rm MT}({\cal E})}{k_B}}
\frac{ P^{\rm rest}_{{\cal E}''}({{\bf M}},{\bf 0})    }{ P^{\rm rest}_{{\cal E}}({{\bf M}},{\bf 0})    }
\end{align}
so that
\begin{align}
  \label{eq:342}
      \llangle \hat{\boldsymbol{\cal K}}\rrangle^{\boldsymbol{\Lambda}{\bf M}{\cal E}}_{\rm rest}
  &=    \int_0^{\cal E}d{\cal E}'' e^{ \frac{S^{\rm MT}({\cal E}'')- S^{\rm MT}({\cal E})}{k_B}}
\frac{ P^{\rm rest}_{{\cal E}''}({{\bf M}},{\bf 0})    }{ P^{\rm rest}_{{\cal E}}({{\bf M}},{\bf 0})    }
\left( {\bf T}\esc\frac{\partial}{\partial{{\bf M}}}\ln P^{\rm rest}_{{\cal E}''}({{\bf M}},{\bf 0})+{\bf c}\right)
\end{align}
No approximations have  been taken in order to arrive  at this result,
which is rigorous and exact. Observe that the mean dilational force at
rest is independent of the orientation.  It is a simple calculation to
show    (\ref{eq:342})  satisfies  the   reversibility  condition
(\ref{eq:129}).

Of  course,  (\ref{eq:342})  is  still  formal,  because  it  requires
knowledge  of  the functional  forms  of  $S^{\rm MT}({\cal  E})$  and
$P^{\rm  rest}_{{\cal  E}}({\bf  M},0)$.   Once they  are  known,  the
additional non-trivial integral over the  energy needs to be performed
explicitly. This is not easy to  do in general, and some approximation
is required  in order to have  a more explicit result.   We assume the
macroscopic entropy  $S^{\rm MT}({\cal E})$ is  an increasing function
of the thermal energy in such  a way that the temperature $T^{\rm MT}$
is always positive.  We also assume the entropy scales  as the size of
the body and, therefore, for a relatively large body (composed of more
than,       say,        ten       particles)        the       function
$e^{\frac{S^{\rm  MT}({\cal  E}'')-S^{\rm   MT}({\cal  E})}{k_B}}$  is
highly  peaked  around ${\cal  E}''={\cal  E}$.   For example,  for  a
harmonic  solid for  which  the macroscopic  thermodynamic entropy  is
$  S^{\rm MT}(E)=3Nk_B\ln E+{\rm ctn}$ this factor is
\begin{align}
  \label{eq:344}
  e^{\frac{S^{\rm MT}({\cal E}'')-S^{\rm MT}({\cal E})}{k_B}}&=\left(\frac{{\cal E}''}{{\cal E}}\right)^{3N}
\end{align}
As we  increase $N$ for  ${\cal E}''<{\cal  E}$, the region  for which
this function  is significantly different from  zero gets concentrated
around ${\cal E}''\simeq {\cal E}$.  Therefore, we may approximate the
integral in (\ref{eq:342}) as
\begin{align}
  \label{eq:331}
      \llangle \hat{\boldsymbol{\cal K}}\rrangle^{\boldsymbol{\Lambda}{\bf M}{\cal E}}_{\rm rest}
  &\simeq\left(        \int_0^{\cal E}d{\cal E}''
e^{\frac{S^{\rm MT}({\cal E}'')-S^{\rm MT}({\cal E})}{k_B}}\right)
\left( {\bf T}\esc\frac{\partial}{\partial{{\bf M}}}\ln P^{\rm rest}_{\cal E}({{\bf M}},{\bf 0})+{\bf c}\right)
\end{align}
For the harmonic solid this integral is
\begin{align}
  \label{eq:346}
    \int_0^{\cal E}d{\cal E}''e^{\frac{S^{\rm MT}({\cal E}'')-S^{\rm MT}({\cal E})}{k_B}}
  &=\int_0^{\cal E}d{\cal E}''\left(\frac{{\cal E}''}{{\cal E}}\right)^{3N}=
{\cal E}    \int_0^1dx x^{3N}=\frac{{\cal E}  }{3N-1}\simeq k_BT^{\rm MT}
\end{align}
For more general models of the entropy we may expand the entropy to first order and write the prefactor as
\begin{align}
  \label{eq:347}
  \int_0^{\cal E}d{\cal E}''e^{\frac{S^{\rm MT}({\cal E}'')-S^{\rm MT}({\cal E})}{k_B}}\simeq
      \int_0^{\cal E}d{\cal E}''e^{-\beta({\cal E}-{\cal E}'')}=\frac{1}{\beta}\left(1-e^{-\beta{\cal E}}\right)\simeq k_BT^{\rm MT}
\end{align}
where we have taken ${\cal E}/k_BT^{\rm MT}\gg1$ for a macroscopic
body.   Therefore,  approximating  the   integral  of  the exponential  of
entropies with the temperature seems to be a good approximation. In this way,
(\ref{eq:331}) becomes
\begin{align}
  \label{eq:38}
      \llangle \hat{\boldsymbol{\cal K}}\rrangle^{\boldsymbol{\Lambda}{\bf M}{\cal E}}_{\rm rest}
  &\simeq k_BT^{\rm MT}\left( {\bf T}\esc\frac{\partial}{\partial{{\bf M}}}\ln P^{\rm rest}_{\cal E}({{\bf M}},{\bf 0})+{\bf c}\right)
\end{align}
This is a remarkable result  relating the rest conditional expectation
of  the dilational  force with  the derivative  of the  probability of
central  moments.  Using  (\ref{TG})  for the  matrix  ${\bf T}$,  the
$\alpha$  component of  the  mean dilational  force in  (\ref{eq:107})
becomes
\begin{align}
  \label{eq:46}
{\cal K}_\alpha
  &={M}_{\underline{\alpha}}\left( \frac{1}{2}{\nu}_{\underline{\alpha}}^2
    +
2\left(  \boldsymbol{\Omega}_p^T\esc\boldsymbol{\Omega}_p-({\boldsymbol{\Omega}_p^{\underline{\alpha}}})^2\right)\right)
    +k_BT^{\rm MT}\left({M}_{\underline{\alpha}}     \frac{\partial}{\partial{M}_{\underline{\alpha}}}\ln
    P^{\rm rest}_{\cal E}({{\bf M}},{\bf 0})+1\right)
\end{align}
where underlined repeated indices  do not follow Einstein's convention
and are  not summed over.  In  particular, for the Gaussian  model for
$P^{\rm  rest}_{\cal  E}({{\bf  M}},{\bf  0})$  we  obtain  the  fully
explicit result
\begin{align}
  \label{eq:121b}
  {\cal K}_\alpha
  &={M}_{\underline{\alpha}}\left(\frac{1}{2}{\nu}_{\underline{\alpha}}^2
+ 2\left(  {\boldsymbol{\Omega}_p^T\esc\boldsymbol{\Omega}_p-(\boldsymbol{\Omega}_p^{\underline{\alpha}})^2}\right)
+\frac{k_BT^{\rm MT}}{2{M}_{\underline{\alpha}}}\right)
    -{M}_{\underline{\alpha}}[{\boldsymbol{\Sigma}}^{-1}]^{\underline{\alpha}\beta}({M}_\beta-{M}_\beta^{\rm rest})
\end{align}

\newpage\color{black}
\setcounter{my_equation_counter}{\value{equation}}
\section{Calculation of the dissipative matrix}
\setcounter{equation}{\value{my_equation_counter}}
\label{App:Gamma}

\subsection{The element $\boldsymbol{\Gamma}_{_\Lambda}$ of the dissipative matrix}
We compute first the element $\boldsymbol{\Gamma}_{_\Lambda}$ of the 
dissipative  matrix  defined through the Green-Kubo formula
\begin{align}
  \label{eq:196}
    {\boldsymbol{\Gamma}}_{_\Lambda}(a)
  &=  \frac{1}{k_BT}  \int_0^\infty dt
\llangle {\cal Q}i{\cal L}\hat{\boldsymbol{\Lambda}}(0)
{\cal Q}i{\cal L}\hat{\boldsymbol{\Lambda}}^T(t)\rrangle^{a}
\end{align}
where $ {\cal Q}i{\cal L}\hat{\boldsymbol{\Lambda}}(t)$ stands for
\begin{align}
  \label{eq:329}
\exp\{i{\cal Q}{\cal L}t\}{\cal Q}i{\cal L}\hat{\boldsymbol{\Lambda}}(z)\simeq \exp\{i{\cal L}t\}{\cal Q}i{\cal L}\hat{\boldsymbol{\Lambda}}(z)
    ={\cal Q}i{\cal L}\hat{\boldsymbol{\Lambda}}(z_t)
\end{align}
where we have taken the usual approximation of replacing the projected dynamics with the real dynamics,
and $z_t$ is  the  solution of  Hamilton's  equations with  initial
condition $z$.  Therefore, the conditional correlation  in (\ref{eq:196}) is an average
over initial conditions that correspond  to the specific values $a$ of
the  CG variables  $\hat{A}(z)$. Performing the  change of  variables
(\ref{W:BodyFrameInv})  
\begin{align}
\label{eq:308}
i{\cal L}\hat{\boldsymbol{\Lambda}}= \sum_{i}
\frac{\partial \hat{\boldsymbol{\Lambda}}}{\partial{\bf r}_{i}}\esc{\bf v}_{i}
=\sum_{i}\left[\frac{\partial \hat{\boldsymbol{\Lambda}}}{\partial {\bf r}_{i}'}
\esc{\bf v}_{i}'
+ \frac{\partial \hat{\boldsymbol{\Lambda}}}{\partial {\bf r}_{i}'}\esc{\bf V} 
- \frac{\partial \hat{\boldsymbol{\Lambda}}}{\partial {\bf r}_{i}'}
  \esc ({\bf r}_{i}'\times\hat{\boldsymbol{\Omega}})
  +{ \frac{\partial \hat{\boldsymbol{\Lambda}}}{\partial {\bf r}_{i}'}}\frac{\partial \hat{{\bf M}}}{\partial {\bf r}'_{i}}\esc\hat{\boldsymbol{\nu}}
  \right]
&=\sum_{i} \frac{\partial \hat{\boldsymbol{\Lambda}}}{\partial {\bf r}_{i}'}
\esc{\bf v}_{i}' +
{\bf B}(\hat{\boldsymbol{\Lambda}})\esc\hat{\boldsymbol{\Omega}}
  \nonumber\\
  &=\sum_{i} \frac{\partial \hat{\boldsymbol{\Lambda}}}{\partial {\bf r}_{i}'}
\esc{\bf v}_{i}' +
\llangle i{\cal L}{\hat{\boldsymbol{\Lambda}}}\rrangle^{\hat{A}}
\end{align}
where (\ref{trans}) was used to eliminate the term with ${\bf V}$, (\ref{eq:H}), (\ref{eq:449}) to
eliminate the term with $\hat{\boldsymbol{\nu}}$, and (\ref{eq:255}) and
(\ref{eq:264})  to rewrite the last term.
Rearranging this equation gives the projected current for the orientation
\begin{align}
\label{eq:309}
{\cal Q}i{\cal L}\hat{\boldsymbol{\Lambda}} =
i{\cal L}{\hat{\boldsymbol{\Lambda}}}-
\llangle i{\cal L}{\hat{\boldsymbol{\Lambda}}}\rrangle^{\hat{A}}  =
\sum_{i}\frac{\partial \hat{{\boldsymbol{\Lambda}}}}{\partial {\bf r}_{i}'}
\esc{\bf v}_{i}'\overset{(\ref{eq:35})}{=}{\cal G}i{\cal L}{\hat{\boldsymbol{\Lambda}}}
\end{align}
Using this relation in (\ref{eq:43}) and noting ${\cal G}^2=1$ then gives
\begin{align}
  \label{eq:7-S}
  \llangle {\cal Q}i{\cal L}{\hat{\boldsymbol{\Lambda}}}(0){\cal Q}i{\cal L}
  {\hat{\boldsymbol{\Lambda}}}^T(t) \rrangle^a
  &=  \llangle i{\cal L}{\hat{\boldsymbol{\Lambda}}}(0)i{\cal L}
    {\hat{\boldsymbol{\Lambda}}}^T(t) \rrangle^{\boldsymbol{\Lambda}{{\bf M}}{\cal E}}_{\rm rest}    
\end{align}
In words,  this equation  states that  the conditional  correlation of
projected  currents  ${\cal Q}i{\cal  L}{\hat{\boldsymbol{\Lambda}}}$,
conditional                                                         on
$a={\bf       R},\boldsymbol{\Lambda},{\bf      M},{\bf       P},{\bf
  S},\boldsymbol{\Pi},E$, is  the same  as the rest  frame conditional
correlation          of          the          time          derivative
$i{\cal   L}\hat{\boldsymbol{\Lambda}}$,  which   is  conditional   on
$a'=\left\{{\bf    R}=0,\boldsymbol{\Lambda},{\bf    M},    {\bf    P}=0,{\bf
  S}=0,\boldsymbol{\Pi}={\bf 0},{\cal  E}\right\}$.  Using (\ref{eq:283}) to write
$i{\cal                              L}\hat{\boldsymbol{\Lambda}}=\dot{\hat{\boldsymbol{\Lambda}}}={\bf
  B}(\hat{\boldsymbol{\Lambda}})   \hat{\boldsymbol{\omega}}$, the 
dissipative matrix   $\boldsymbol{\Gamma}_{_\Lambda}(a)$ in (\ref{eq:196}) becomes
\begin{align}
  \label{eq:193}
      \boldsymbol{\Gamma}_{_\Lambda}(a)
  &=
    {\bf B}({\boldsymbol{\Lambda}})\esc
    \boldsymbol{\cal D}({\boldsymbol{\Lambda}},{{\bf M}}, {\cal E})
    \esc{\bf B}^T({\boldsymbol{\Lambda}}) 
\end{align}
  where we have introduced the angular diffusion tensor
\begin{align}
  \label{eq:430}
    \boldsymbol{\cal D}(\boldsymbol{\Lambda},{{\bf M}}, {\cal E})
  &    \equiv \frac{1}{k_BT}\int_0^\infty dt
    \llangle\hat{\boldsymbol{\omega}}(0)\hat{\boldsymbol{\omega}}^{T}(t)
    \rrangle^{\boldsymbol{\Lambda}{{\bf M}}{\cal E}}_{\rm rest} 
\end{align}
Because the angular velocity $\hat{\boldsymbol{\omega}}$ transforms as
a                  vector,                 the                  dyadic
$\hat{\boldsymbol{\omega}}(0)\hat{\boldsymbol{\omega}}^{T}(t)$
transforms as  a tensor.   This is  true even  for the  time dependent
angular    velocity   $\hat{\boldsymbol{\omega}}(t)$,    because   the
Hamiltonian   is    invariant   under   rotations.    As    shown   by
(\ref{W:eq:469})  and (\ref{W:eq:470})  in Sec.  \ref{App:RotInv},
this tensor takes the form
\begin{align}
  \label{eq:355}
      \boldsymbol{\cal D}(\boldsymbol{\Lambda},{\bf M},{\cal E})
  &=e^{[{{\boldsymbol{\Lambda}}}]_\times}\esc
    \boldsymbol{\cal D}_0({{\bf M}},{\cal E})\esc e^{-[{\boldsymbol{\Lambda}}]_\times}
\end{align}
where the ``rest'' angular diffusion tensor is
\begin{align}
      \label{W:eq:356}
  \boldsymbol{\cal D}_0({{\bf M}},{\cal E})
  &\equiv \frac{1}{k_BT}\int_0^\infty dt
    \llangle\hat{\boldsymbol{\omega}}_{0}(0)\hat{\boldsymbol{\omega}}_{0}^{T}(t)
    \rrangle^{\boldsymbol{\Lambda}{{\bf M}}{\cal E}}_{\rm rest}
\end{align}
where $\hat{\boldsymbol{\omega}}_0=e^{-[\hat{\boldsymbol{\Lambda}}]_\times}\hat{\boldsymbol{\omega}}$.  Inserting these expressions in (\ref{eq:193}) then gives
\begin{align}
  \label{eq:130}
      \boldsymbol{\Gamma}_{_\Lambda}(a)
  &=
    {\bf B}({\boldsymbol{\Lambda}})\esc
e^{[{{\boldsymbol{\Lambda}}}]_\times}\esc
    \boldsymbol{\cal D}_0({{\bf M}},{\cal E})\esc e^{-[{\boldsymbol{\Lambda}}]_\times}
    \esc{\bf B}^T({\boldsymbol{\Lambda}}) 
\end{align}
The interest in expression  (\ref{eq:130}) lies with the  fact that
all  the dependence  on the  orientation is  now explicit  because, as
explained   in   Sec.  \ref{App:RotInv}  the   rest   conditional
expectation of a tensor in the  principal frame does not depend on the
orientation.  However, (\ref{eq:130}) is  still formal  because the  actual
functional form of $\boldsymbol{\cal D}_0({{\bf  M}}, {\cal E})$ is  not known.
In general the  conditional correlation  in (\ref{W:eq:356})  cannot be
  computed analytically nor through simulations. For this reason, we need
  to approximate or model this quantity. 
  One simple possibility is to approximate the conditional expectations with
    ordinary equilibrium expectations
    \begin{align}
  \label{}
    \boldsymbol{\cal D}_0({{\bf M}},{\cal E})
  \simeq          \boldsymbol{\cal D}_0^{\rm eq}({\cal E})
\end{align}
where
  \begin{align}
  \label{}
    \boldsymbol{\cal D}_0^{\rm eq}({\cal E})
  &\equiv \frac{1}{k_BT}\int_0^\infty dt
    \llangle\hat{\boldsymbol{\omega}}_{0}(0)\hat{\boldsymbol{\omega}}_{0}^{T}(t)
    \rrangle^{{\cal E}}_{\rm rest}
\end{align}

\subsection{The element $\boldsymbol{\Gamma}_{_\Pi}$ of the dissipative matrix}
Let    us   consider    now   the    calculation   of    the   element
$\boldsymbol{\Gamma}_{_\Pi}$ of  the dissipative matrix, as  defined by
the Green-Kubo  expression
\begin{align}
      {\boldsymbol{\Gamma}}_{_\Pi}(a)
  &=  \frac{1}{k_BT}  \int_0^\infty dt
\llangle \left({\cal Q}i{\cal L}\hat{\boldsymbol{\Pi}}\right)
\left(\exp\{i{\cal Q}{\cal L}t\}{\cal Q}i{\cal L}\hat{\boldsymbol{\Pi}}\right)^T\rrangle^{a}
\end{align}
 According to  the definition
of the dilation force we have
\begin{align}
  \label{eq:106}
    \llangle {\cal Q}i{\cal L}{\hat{\boldsymbol{\Pi}}}(0){\cal Q}i{\cal L}
  {\hat{\boldsymbol{\Pi}}}^T(t) \rrangle^a
  &=    \llangle \left(\hat{\boldsymbol{\cal K}}(0)-\llangle \hat{\boldsymbol{\cal K}}\rrangle^a\right)
 \left(\hat{\boldsymbol{\cal K}}^T(t)-\llangle \hat{\boldsymbol{\cal K}}^T\rrangle^a\right)\rrangle^a
\end{align}
which when transformed into the rest conditional
expectations using (\ref{eq:43}) gives
\begin{align}
  \label{eq:332}
    \llangle \left(\hat{\boldsymbol{\cal K}}(0)-\llangle \hat{\boldsymbol{\cal K}}\rrangle^a\right)
  \left(\hat{\boldsymbol{\cal K}}^T(t)-\llangle \hat{\boldsymbol{\cal K}}^T\rrangle^a\right)\rrangle^a
  &=  \llangle \left({\cal G}\hat{\boldsymbol{\cal K}}(0)-\llangle {\cal G}\hat{\boldsymbol{\cal K}}\rrangle^{\boldsymbol{\Lambda}{\bf M}{\cal E}}\right)
 \left( {\cal G}\hat{\boldsymbol{\cal K}}^T(t)-\llangle  {\cal G}\hat{\boldsymbol{\cal K}}^T\rrangle^{\boldsymbol{\Lambda}{\bf M}{\cal E}}\right)\rrangle^{\boldsymbol{\Lambda}{\bf M}{\cal E}}_{\rm rest}
\end{align}
Using (\ref{eq:194}) and (\ref{eq:107}) shows    that      the      conditional      expectations
$\llangle\cdots\rrangle^a$  can be  substituted  with the  conditional
expectations
$\llangle\cdots\rrangle^{\boldsymbol{\Lambda}{{\bf  M}}{\cal  E}}_{\rm
  rest}$, so that analogously to (\ref{eq:7-S}),
\begin{align}
  \label{eq:182}
  \llangle {\cal Q}i{\cal L}{\hat{\boldsymbol{\Pi}}}(0){\cal Q}i{\cal L}
  {\hat{\boldsymbol{\Pi}}}^T(t) \rrangle^a
  &=  \llangle
    \left(\hat{\boldsymbol{\cal K}}(0)-\llangle \hat{\boldsymbol{\cal K}}\rrangle^{\boldsymbol{\Lambda}{{\bf M}}{\cal E}}_{\rm rest}\right)
    \left(\hat{\boldsymbol{\cal K}}(t)-\llangle \hat{\boldsymbol{\cal K}}\rrangle^{\boldsymbol{\Lambda}{{\bf M}}{\cal E}}_{\rm rest}\right)^T\rrangle^{\boldsymbol{\Lambda}{{\bf M}}{\cal E}}_{\rm rest}    
\end{align}
The corresponding element in the dissipative matrix is
\begin{align}
  \label{eq:72}
  \boldsymbol{\Gamma}_{_\Pi}(\boldsymbol{\Lambda},{{\bf M}},{\cal E})=\frac{1}{k_BT}\int_0^\infty dt  \llangle
  \left(\hat{\boldsymbol{\cal K}}(0)-\llangle \hat{\boldsymbol{\cal K}}\rrangle^{\boldsymbol{\Lambda}{{\bf M}}{\cal E}}_{\rm rest}\right)
  \left(\hat{\boldsymbol{\cal K}}(t)-\llangle \hat{\boldsymbol{\cal K}}\rrangle^{\boldsymbol{\Lambda}{{\bf M}}{\cal E}}_{\rm rest}\right)^T\rrangle^{\boldsymbol{\Lambda}{{\bf M}}{\cal E}}_{\rm rest}    
\end{align}
and   is   a    function   of   the   CG    variables   only   through
$\boldsymbol{\Lambda},{\bf M}$ and the  thermal energy ${\cal E}$. The
element (\ref{eq:72}) cannot be directly computed using equilibrium MD
simulations as it involves a conditional expectation.  We will
approximate the conditional expectation with an ordinary equilibrium expectation
\begin{align}
  \label{eq:429}
\boldsymbol{\Gamma}_{_\Pi}(\boldsymbol{\Lambda},{{\bf M}},{\cal E})
  \simeq        \boldsymbol{\Gamma}_{_\Pi}^{\rm eq}({\cal E})
\end{align}
where
  \begin{align}
  \label{eq:428}
    \boldsymbol{\Gamma}_{_\Pi}^{\rm eq}&\equiv\frac{1}{k_BT}\int_0^\infty dt  \llangle
  \left(\hat{\boldsymbol{\cal K}}(0)-\llangle \hat{\boldsymbol{\cal K}}\rrangle^{{\cal E}}\right)
  \left(\hat{\boldsymbol{\cal K}}(t)-\llangle \hat{\boldsymbol{\cal K}}\rrangle^{{\cal E}}\right)^T\rrangle^{{\cal E}}
\end{align}

\section{The stochastic drift}
\setcounter{equation}{\value{my_equation_counter}}
  \label{App:StochDrift}
For the  present level  of
  description,  the  stochastic  drift term $    V_\mu^{\rm sto}(a)=k_B\frac{\partial M_{\mu\nu}}{\partial  a_\nu}$
reduces to two       components:
  $\boldsymbol{V}^{\rm sto}_{_\Lambda}$,  affecting  the orientation
  dynamics,  and $\boldsymbol{V}^{\rm  sto}_\Pi$, affecting the
  dilational  momentum dynamics
\begin{align}
  \label{eq:269-2}
\boldsymbol{V}^{{\rm sto }}_{_\Lambda}
  &=k_B\frac{\partial}{\partial\boldsymbol{\Lambda}}\esc
    T\boldsymbol{\Gamma}_{_\Lambda}
    \nonumber\\
\boldsymbol{V}^{\rm sto}_{_\Pi}
  &=k_B\frac{\partial}{\partial\boldsymbol{\Pi}}\esc T\boldsymbol{\Gamma}_{_\Pi}
\end{align}
The second component $  \boldsymbol{V}^{\rm sto}_{_\Pi}$  is easy to compute
under the equilibrium approximation (\ref{eq:429}) giving
\begin{align}
  \label{eq:253}
  \boldsymbol{V}^{\rm sto}_{_\Pi}
    \overset{(\ref{eq:429})}{=}
    k_B\frac{\partial}{\partial\boldsymbol{\Pi}}\esc T \boldsymbol{\Gamma}_{_\Pi}^{\rm eq}
    =
    \frac{k_B}{C^{\rm MT}}\left(\frac{\partial}{\partial T}T \boldsymbol{\Gamma}_{_\Pi}^{\rm eq}\right)
    \esc \frac{\partial {\cal E}}{\partial \boldsymbol{\Pi}}
    \overset{(\ref{Thermal-App})}{=}
    -\frac{k_B}{C^{\rm MT}}\frac{\partial T \boldsymbol{\Gamma}_{_\Pi}^{\rm eq}}{\partial T}\esc
    \boldsymbol{\nu}
\end{align}
where $C^{\rm MT}$ is the heat capacity of the body.
We now compute  $\boldsymbol{V}^{{\rm sto }}_{_\Lambda}$ using (\ref{eq:130})
\begin{align}
        \boldsymbol{\Gamma}_{_\Lambda}(a)
  &=
    {\bf B}\esc
e^{[{{\boldsymbol{\Lambda}}}]_\times}\esc
    \boldsymbol{\cal D}_0\esc e^{-[{\boldsymbol{\Lambda}}]_\times}
    \esc{\bf B}^T
\overset{(\ref{BeL})}{=}
    {\bf B}^T\esc
   \boldsymbol{\cal D}_0\esc{\bf B}
    \label{eq:103b}
\end{align}
and observing the angular diffusion tensor
$  \boldsymbol{\cal  D}_0=\boldsymbol{\cal D}_0({{\bf  M}},{\cal  E})$
depends upon the  orientation  through the  thermal energy.  Therefore,
$ \boldsymbol{\Gamma}_{_\Lambda}(a)$ has an explicit dependence on the
orientation through ${\bf  B}({\boldsymbol{\Lambda}})$, and an implicit
dependence through the thermal energy in the angular diffusion tensor.
We will separate these
dependencies by writing the derivative in (\ref{eq:269-2}) as two terms with
the first carrying the implicit dependence (for which the chain rule will be
used) and the second carrying the explicit dependence, that is
\begin{align}
  \label{eq:279}
  \boldsymbol{V}^{\rm sto}_{_\Lambda}
=
    \frac{k_B}{C^{\rm MT}}\frac{\partial T\boldsymbol{\Gamma}_{_\Lambda}}{\partial T}\esc 
    \frac{\partial {\cal E}}{\partial\boldsymbol{\Lambda}}
        +    k_BT\frac{\partial\esc \boldsymbol{\Gamma}_{_\Lambda}} {\partial\boldsymbol{\Lambda}}
\overset{(\ref{Thermal-App})}{=}
  = -\frac{k_B}{C^{\rm MT}}\frac{\partial T\boldsymbol{\Gamma}_{_\Lambda}}{\partial T}\esc
    {\bf B}^{-T}\esc(\boldsymbol{\Omega}\times{\bf S})+ k_BT\frac{\partial\esc \boldsymbol{\Gamma}_{_\Lambda}} {\partial\boldsymbol{\Lambda}}
\end{align}
where the first and second terms carry the implicit and explicit
dependencies, respectively. Evaluating the explicit derivative dependence in (\ref{eq:279}) gives
\begin{align}
  \label{eq:48}
\frac{\partial\boldsymbol{\Gamma}^{\alpha\beta}_{_\Lambda}}{\partial{\Lambda}_\beta}
  &\overset{(\ref{eq:103b})}{=}
\frac{\partial}{\partial{\Lambda}_\beta}
\left(    {B}_{\alpha'\alpha}
   \boldsymbol{\cal D}^{\alpha'\beta'}_0{B}_{\beta'\beta}\right)
\overset{\mbox{\small chain rule}}{=}
  \left(\frac{\partial{B}_{\alpha'\alpha}}{\partial{\Lambda}_\beta}\right)
    \boldsymbol{\cal D}^{\alpha'\beta'}_0{B}_{\beta'\beta}
    +   {B}_{\alpha'\alpha}
    \boldsymbol{\cal D}^{\alpha'\beta'}_0
    \left(  \frac{\partial{B}_{\beta'\beta}}{\partial{\Lambda}_\beta}\right)
\end{align}
The second term can be computed as follows
\begin{align}
  \label{eq:363}
   {B}_{\alpha'\alpha}
    \boldsymbol{\cal D}^{\alpha'\beta'}_0
  \left(  \frac{\partial{B}_{\beta'\beta}}{\partial{\Lambda}_\beta}\right)
  &    \overset{(\ref{eq:606})}{=} {B}_{\alpha'\alpha}
     \boldsymbol{\cal D}^{\alpha'\beta'}_02f(\Lambda) {\Lambda}_{\beta'}
     \overset{(\ref{eq:224})}{=}
      {B}_{\alpha'\alpha}
     \boldsymbol{\cal D}^{\alpha'\beta'}_0{B}_{\beta'\beta}2f(\Lambda) {\Lambda}_{\beta}
     \overset{(\ref{eq:103b})}{=}
      \boldsymbol{\Gamma}_{_\Lambda}^{\alpha\beta}2f(\Lambda) {\Lambda}_{\beta}
                                                                                             \nonumber\\
&  \overset{(\ref{eq:365})}{=}
  -   \boldsymbol{\Gamma}_{_\Lambda}^{\alpha\beta}\frac{\partial}{\partial{{\Lambda}_\beta}}\ln P^{\rm Haar}(\boldsymbol{\Lambda})
\end{align}
The first term on the right hand side of (\ref{eq:48}) is computed in Sec.~\ref{App:ThermDrift} giving (\ref{Fin-Fth}).

\newpage
\setcounter{my_equation_counter}{\value{equation}}
\section{The thermal drift}
\setcounter{equation}{\value{my_equation_counter}}
\label{App:ThermDrift}
In this section, we compute the thermal drift
\begin{align}
  \label{Tdrift}
     {\bf F}^{\rm th}
    &\equiv  \frac{\partial\esc   \boldsymbol{\Gamma}_{_\Lambda}}{\partial\boldsymbol{\Lambda}}
+  \boldsymbol{\Gamma}_{_\Lambda}\esc   \frac{\partial}{\partial \boldsymbol{\Lambda}}\ln P^{\rm Haar}(\boldsymbol{\Lambda})
\end{align}
using (\ref{eq:363}) in (\ref{eq:48}) to give
\begin{align}
  \label{eq:366}
  {\bf F}^{\rm th}_\alpha
  & =\frac{\partial\boldsymbol{\Gamma}^{\alpha\beta}_{_\Lambda}}{\partial{\Lambda}_\beta}
                           +  \boldsymbol{\Gamma}_{_\Lambda}^{\alpha\beta}\frac{\partial}{\partial{{\Lambda}_\beta}}
                           \ln P^{\rm Haar}(\boldsymbol{\Lambda})
                           =    \left(\frac{\partial{B}_{\alpha'\alpha}}{\partial{\Lambda}_\beta}\right)
                           \boldsymbol{\cal D}^{\alpha'\beta'}_0{B}_{\beta'\beta}
\equiv\boldsymbol{\cal D}^{\alpha'\beta'}_0{\Xi}_{\beta'\alpha'\alpha}
\end{align}
which coincides with   the  first  term   of
(\ref{eq:48}), and where we have introduced the third order tensor
\begin{align}
  \label{eq:3}
  {\Xi}_{\beta'\alpha'\alpha}&\equiv
                                          {B}_{\beta'\beta}\frac{\partial{B}_{\alpha'\alpha}}{\partial{\Lambda}_\beta}
\end{align}
To compute ${\Xi}_{\beta'\alpha'\alpha}$, we insert (\ref{eq:155}) and (\ref{eq:156}) in (\ref{eq:3}) giving
\begin{align}
  \label{eq:26}
  {\Xi}_{\beta'\alpha'\alpha}
  &=\left[
    \delta_{\beta'\beta}-\frac{\Lambda}{2}[{\bf n}]_\times^{\beta'\beta}
    +f(\Lambda)\Lambda^2\left[[{\bf n}]_\times\esc[{\bf n}]_\times\right]_{\beta'\beta}
    \right]\nonumber\\
    &\quad\times\left[
    -\frac{1}{2}\epsilon_{\alpha'\beta\alpha} +
    f'(\Lambda)\Lambda^2
    \left [[{{\bf n}}]_\times\esc [{{\bf n}}]_\times
    \right ]^{\alpha'\alpha}{\bf n}_\beta
    + f(\Lambda) \Lambda \left(\delta_{\alpha'\beta}{\bf n}_{\alpha}
    +    \delta_{\alpha\beta}{\bf n}_{\alpha'} -2\delta_{\alpha'\alpha}{\bf n}_\beta \right)
    \right]
    \nonumber\\
  &= -\frac{1}{2}\epsilon_{\alpha'\beta'\alpha} +
    f'(\Lambda)\Lambda^2
    \left [[{{\bf n}}]_\times\esc [{{\bf n}}]_\times
    \right ]^{\alpha'\alpha}{\bf n}_{\beta'}
    + f(\Lambda)\Lambda \left(\delta_{\alpha'\beta'}{\bf n}_{\alpha}
    +    \delta_{\alpha\beta'}{\bf n}_{\alpha'} -2\delta_{\alpha'\alpha}{\bf n}_{\beta'} \right)
    \nonumber\\
  &\quad +\frac{\Lambda}{4}[{\bf n}]_\times^{\beta'\beta}\epsilon_{\alpha'\beta\alpha} 
    -\frac{f(\Lambda)\Lambda^2}{2} \left([{\bf n}]_\times^{\beta'\alpha'}{\bf n}_{\alpha}
    +  [{\bf n}]_\times^{\beta'\alpha}{\bf n}_{\alpha'} \right)
    \nonumber\\
  &\quad-\frac{f(\Lambda)\Lambda^2}{2}\left[[{\bf n}]_\times\esc[{\bf n}]_\times\right]_{\beta'\beta}
    \epsilon_{\alpha'\beta\alpha} 
    + f^2(\Lambda)\Lambda^3 \left(\left[[{\bf n}]_\times\esc[{\bf n}]_\times\right]_{\beta'\alpha'}
    {\bf n}_{\alpha}
    +\left[[{\bf n}]_\times\esc[{\bf n}]_\times\right]_{\beta'\alpha}
    {\bf n}_{\alpha'} \right)
\end{align}
Use the properties of the Levi-Civita symbol to get
\begin{align}
  \label{eq:56}
  [{\bf n}]_\times^{\beta'\beta}\epsilon_{\alpha'\beta\alpha}
  &  =  {\bf n}_\gamma\epsilon_{\beta'\gamma \beta}\epsilon_{\alpha'\beta\alpha}
    = - {\bf n}_\gamma\left(\delta_{\beta'\alpha'}\delta_{\gamma\alpha}
      -\delta_{\gamma\alpha'}\delta_{\beta'\alpha}\right)
    =\delta_{\beta'\alpha}{\bf n}_{\alpha'}
    - \delta_{\beta'\alpha'} {\bf n}_\alpha
    \nonumber\\
  \left[[{\bf n}]_\times\esc[{\bf n}]_\times\right]_{\beta'\beta}\epsilon_{\alpha'\beta\alpha}
  &=  [{\bf n}]_\times^{\beta'\sigma}[{\bf n}]_\times^{\sigma \beta}\epsilon_{\alpha'\beta\alpha}
    = [{\bf n}]_\times^{\beta'\sigma}\left(
    \delta_{\sigma\alpha}{\bf n}_{\alpha'}
    -\delta_{\sigma\alpha'}{\bf n}_\alpha\right)
    = 
    [{\bf n}]_\times^{\beta'\alpha}{\bf n}_{\alpha'}
    - [{\bf n}]_\times^{\beta'\alpha'}{\bf n}_\alpha
\end{align}
and insert in (\ref{eq:26}) to give
\begin{align}
  \label{eq:77}
  {\Xi}_{\beta'\alpha'\alpha}
  &= -\frac{1}{2}\epsilon_{\alpha'\beta'\alpha} +
    f'(\Lambda)\Lambda^2
    \left [[{{\bf n}}]_\times\esc [{{\bf n}}]_\times
    \right ]^{\alpha'\alpha}{\bf n}_{\beta'}
    + f(\Lambda) \Lambda\left(\delta_{\alpha'\beta'}{\bf n}_{\alpha}
    +    \delta_{\alpha\beta'}{\bf n}_{\alpha'} -2\delta_{\alpha'\alpha}{\bf n}_{\beta'} \right)
    \nonumber\\
  &\quad +\frac{\Lambda}{4}\left(\delta_{\beta'\alpha}{\bf n}_{\alpha'}
    - \delta_{\beta'\alpha'}{\bf n}_\alpha\right)
    -f(\Lambda)\Lambda^2 [{\bf n}]_\times^{\beta'\alpha}{\bf n}_{\alpha'}
    + f^2(\Lambda) \Lambda^3\left(\left[[{\bf n}]_\times\esc[{\bf n}]_\times\right]_{\beta'\alpha'}
    {\bf n}_{\alpha}
    +\left[[{\bf n}]_\times\esc[{\bf n}]_\times\right]_{\beta'\alpha}
    {\bf n}_{\alpha'} \right)
\end{align}

Observe that  there are bits  in this tensor that  are antisymmetric
under the interchange of indices $\alpha', \beta'$.  The contraction
of      these      bits      with     the      symmetric      tensor
$\boldsymbol{\cal D}_0^{\alpha'\beta'}$ will vanish.  Therefore, the
thermal drift (\ref{eq:366}) becomes
\begin{align}
  \label{eq:93}
  {\bf F}^{\rm th}_\alpha
  &=f'(\Lambda)\Lambda^2\boldsymbol{\cal D}^{\alpha'\beta'}_0    
    \left [[{{\bf n}}]_\times\esc [{{\bf n}}]_\times \right]_{\alpha'\alpha}{\bf n}_{\beta'}
    + f(\Lambda)\Lambda \left(
    \boldsymbol{\cal D}^{\beta'\beta'}_0 {\bf n}_{\alpha} +
    \boldsymbol{\cal D}^{\alpha'\alpha}_0 {\bf n}_{\alpha'} -
   2\boldsymbol{\cal D}^{\alpha\beta'}_0 {\bf n}_{\beta'} \right)
    \nonumber\\
  &\quad
    +\frac{\Lambda}{4}\left(
      \boldsymbol{\cal D}^{\alpha'\alpha}_0 {\bf n}_{\alpha'} -
      \boldsymbol{\cal D}^{\beta'\beta'}_0  {\bf n}_\alpha \right)
    -    f(\Lambda)\Lambda^2 [{\bf n}]_\times^{\beta'\alpha}\boldsymbol{\cal D}^{\alpha'\beta'}_0{\bf n}_{\alpha'}
    \nonumber\\
  &\quad +f^2(\Lambda)\Lambda^3 \left (
    {\rm Tr}\left[ \boldsymbol{\cal D}_0\esc[{\bf n}]_\times\esc[{\bf n}]_\times\right]
    {\bf n}_{\alpha}
    +\boldsymbol{\cal D}^{\alpha'\beta'}_0 \left[[{\bf n}]_\times\esc[{\bf n}]_\times\right]_{\beta'\alpha}
    {\bf n}_{\alpha'} \right )
\end{align}
Collecting and simplifying terms, using (\ref{eq:603}) and writing in vector
notation then gives
\begin{align}
  \label{eq:241}
  {\bf F}^{\rm th}
  &=\left(\frac{\Lambda}{4}-f(\Lambda)\Lambda\right)
\left(\boldsymbol{\cal D}_0\esc {\bf n} -{\rm Tr}[\boldsymbol{\cal D}_0]{\bf n}\right)
    + f(\Lambda)\Lambda^2[{{\bf n}}]_\times\esc    \boldsymbol{\cal D}_0 \esc{\bf n}
    +\left(f'(\Lambda)\Lambda^2+f^2(\Lambda)\Lambda^3\right)
    [{{\bf n}}]_\times\esc [{{\bf n}}]_\times\esc
    \boldsymbol{\cal D}_0 \esc{\bf n}
\nonumber\\
&\quad +f^2(\Lambda)\Lambda^3 {\rm Tr}\left[  \boldsymbol{\cal D}_0\esc[{\bf n}]_\times\esc[{\bf n}]_\times\right]
    {{\bf n}}
\end{align}
We seek a more compact form for the thermal drift.  We use (\ref{eq:600}) in the form
\begin{align}
  \label{eq:235}
  [{\bf n}]_\times\esc[{\bf n}]_\times={\bf n}{\bf n}^T-\mathbb{1}
\end{align}
to write the last two terms of (\ref{eq:241}) as
\begin{align}
\left(f'(\Lambda)\Lambda^2+f^2(\Lambda)\Lambda^3\right)
({\bf n}{\bf n}^T-\mathbb{1})\esc\boldsymbol{\cal D}_0\esc{{\bf n}}& +
f^2(\Lambda) \Lambda^3{\rm Tr}\left[
    ({\bf n}{\bf n}^T-\mathbb{1})\esc\boldsymbol{\cal D}_0\right] {{\bf n}}  
\nonumber\\
&= \left(f'(\Lambda)\Lambda^2+f^2(\Lambda)\Lambda^3\right)
    \left(({\bf n}^T\esc\boldsymbol{\cal D}_0\esc{\bf n}){{\bf n}}
    -\boldsymbol{\cal D}_0    \esc{{\bf n}}\right)
    +f^2(\Lambda)\Lambda^3 \left(({\bf n}^T\esc\boldsymbol{\cal D}_0\esc{\bf n})-{\rm Tr}[\boldsymbol{\cal D}_0]
    \right){{\bf n}}  
\end{align}
which when inserted in (\ref{eq:241}) gives
\begin{align}
  \label{eq:209}
  {\bf F}^{\rm th}
  &=
    F_1{\rm Tr}[\boldsymbol{\cal D}_0]{{\bf n}}
    +    F_2 \boldsymbol{\cal D}_0\esc{{\bf n}}
    +F_3({\bf n}^T\esc\boldsymbol{\cal D}_0\esc{\bf n}){{\bf n}}
    + f(\Lambda)\Lambda^2[{{\bf n}}]_\times\esc\boldsymbol{\cal D}_0\esc{\bf n}
\end{align}
where
  \begin{align}
    \label{eq:247}
F_1&\equiv        - \frac{\Lambda}{4}+f(\Lambda)\Lambda    -f^2(\Lambda)\Lambda^3
     \nonumber\\
F_2&\equiv    \frac{\Lambda}{4}-f(\Lambda)\Lambda-f'(\Lambda)\Lambda^2-f^2(\Lambda)\Lambda^3 
     \nonumber\\
    F_3&\equiv     f'(\Lambda)\Lambda^2+2f^2(\Lambda)\Lambda^3=-F_1-F_2
\end{align}
Using (\ref{eq:262}) for $f(\Lambda)$, and with the help of Mathematica, we find
\begin{align}
  \label{eq:258}
  F_1&=\frac{1}{2}\frac{\sin\Lambda -\Lambda}{1-\cos\Lambda}
              \nonumber\\
F_2&=\cot\left(\frac{\Lambda}{2}\right)\left(1-\frac{\Lambda}{2}\cot\left(\frac{\Lambda}{2}\right)\right)
\end{align}
Using the last identity (\ref{eq:247}) we may write (\ref{eq:209}) as
\begin{align}
    \label{Fin-Fth}
   {\bf F}^{\rm th}
    &=
      F_1\left[{\rm Tr}[\boldsymbol{\cal D}_0]{{\bf n}}-({\bf n}^T\esc\boldsymbol{\cal D}_0\esc{\bf n}){{\bf n}}\right]
            +F_2 \left[\boldsymbol{\cal D}_0\esc{{\bf n}}-({\bf n}^T\esc\boldsymbol{\cal D}_0\esc{\bf n}){{\bf n}}\right]
      +      f(\Lambda)\Lambda^2[{{\bf n}}]_\times\esc    \boldsymbol{\cal D}_0\esc{{\bf n}}
      \nonumber\\
    &=F_1{\rm Tr}\left[\left[\mathbb{1}-{\bf n}{\bf n}^T\right]\esc\boldsymbol{\cal D}_0\right]{\bf n}
+      F_2 \left[\mathbb{1}-{\bf n}{\bf n}^T\right]\esc\boldsymbol{\cal D}_0\esc{{\bf n}}
      +      f(\Lambda)\Lambda^2[{{\bf n}}]_\times\esc    \boldsymbol{\cal D}_0\esc{{\bf n}}
\end{align}

\setcounter{my_equation_counter}{\value{equation}}
  \section{The cross  product}
\setcounter{equation}{\value{my_equation_counter}}
    \label{App:cross}

    We summarize  a
  number of properties  of the cross product matrix  and cross product
  of two vectors, as they are used in many occasions in the paper. For
  arbitrary vectors  $\bf a$, $\bf  b$, and $\bf c$,  scalar $\kappa$,
  and rotation matrix $\boldsymbol{\cal Q}$ one has
\begin{align}
\label{vecprod}
  [{\bf a}]_\times\esc{\bf b}
  &={\bf a}\times{\bf b}\\
  {\bf a}^T\esc [{\bf b}]_\times
  &=({\bf a}\times{\bf b})^T\\
\label{eq:603}
  [{\bf a}]_\times^T
  &=-[{\bf a}]_\times= [-{\bf a}]_\times\\
  [{\bf a}+{\bf b}]_\times
  &=[{\bf a}]_\times+[{\bf b}]_\times\\
  [\kappa{\bf a}]_\times
  &=\kappa[{\bf a}]_\times\\
\label{eq:600}
  [{\bf a}]_\times\esc[{\bf b}]_\times
  &={\bf b} {\bf a}^T - ( {\bf a}\esc {\bf b})\mathbb{1}\\
  [{\bf a}]_\times\esc[{\bf b}]_\times\esc[{\bf c}]_\times
  &={\bf b} {\bf a}^T\esc [{\bf c}]_\times -( {\bf a}\esc {\bf b})[{\bf c}]_\times\\
  &={\bf b} ({\bf a}\times{\bf c})^T-( {\bf a}\esc {\bf b})[{\bf c}]_\times\\
  \label{eq:319}
  [{\bf a}]_\times\esc[{\bf a}]_\times\esc[{\bf a}]_\times
  &=-a^2[{\bf a}]_\times\\
  \label{eq:601}
[\boldsymbol{\cal Q}\esc{\bf a}]_\times
  &=\boldsymbol{\cal Q}\esc[{\bf a}]_\times\esc \boldsymbol{\cal Q}^T\\
  \label{Qaxb}
  \boldsymbol{\cal Q}\esc({\bf a}\times{\bf b})
  &=
      (\boldsymbol{\cal Q}\esc{\bf a}\times\boldsymbol{\cal Q}\esc{\bf b})
\\\label{eq:602}
  \frac{\partial}{\partial {\bf a}^\gamma}[{\bf a}]_\times^{\alpha\beta}
  &=\epsilon_{\alpha\gamma\beta}
\end{align}

\color{black}

\setcounter{my_equation_counter}{\value{equation}}
\section{Results concerning angular velocity, orientation, and gyration tensor}
\setcounter{equation}{\value{my_equation_counter}}
\label{App:DerRot}
In this section, we  derive a  number of  results involving the
derivative   of   the   rotation   matrix  with   respect   to   the
orientation. This allows us to  relate the angular velocity with the
time  derivative  of the  orientation,  and  the derivative  of  the
inertia tensor  with  respect to  the  orientation.  These
results are used in the main text.

\subsection{The derivative of the rotation matrix with respect to the orientation}

We  need an  explicit expression  for the  derivative of  the rotation
matrix  with respect  to  the  orientation.  While  this  can be  done
directly from Rodrigues  formula, we take here  the following somewhat
shorter and  more elegant route.  Using  a formula given by  Wilcox in
\cite{Wilcox1967}, namely
\begin{align}
  \frac{\partial e^{{\bf A}(\lambda)}}{\partial \lambda} &=
\int_0^1 dx e^{(1-x){\bf A}(\lambda)}\esc
  \frac{\partial{\bf A}(\lambda)}{\partial \lambda} \esc
e^{x{\bf A}(\lambda)}
\label{WilcoxLambda}
\end{align}
the derivative of the rotation matrix with respect to the orientation is
\begin{align}
  \label{eq:60}
  \frac{\partial   \left[e^{[\boldsymbol{\Lambda}]_\times}\right]_{\alpha\beta}}{\partial{\Lambda}_\gamma}
  &=\int_0^1 dx \left[e^{(1-x)[\boldsymbol{\Lambda}]_\times}\right]_{\alpha\alpha'}
  \frac{\partial[\boldsymbol{\Lambda}]^{\alpha'\beta'}_\times}{\partial{\Lambda}_\gamma} 
\left[e^{x[\boldsymbol{\Lambda}]_\times}\right]_{\beta'\beta}
    \nonumber\\
  &=\int_0^1 dx \left[e^{(1-x)[\boldsymbol{\Lambda}]_\times}\right]_{\alpha\alpha'}
\epsilon_{\alpha'\gamma\beta'}
    \left[e^{x[\boldsymbol{\Lambda}]_\times}\right]_{\beta'\beta}
    \nonumber\\
  &=\int_0^1 dx \left[e^{(1-x)[\boldsymbol{\Lambda}]_\times}\right]_{\alpha\alpha'}
\epsilon_{\alpha'\gamma\beta'}
    \left[e^{(x-1)[\boldsymbol{\Lambda}]_\times}\right]_{\beta'\delta}
    \left[e^{[\boldsymbol{\Lambda}]_\times}\right]_{\delta\beta}
    \nonumber\\
  &=\int_0^1 dx \left[e^{x[\boldsymbol{\Lambda}]_\times}\right]_{\alpha\alpha'}
\epsilon_{\alpha'\gamma\beta'}
    \left[e^{-x[\boldsymbol{\Lambda}]_\times}\right]_{\beta'\delta}
    \left[e^{[\boldsymbol{\Lambda}]_\times}\right]_{\delta\beta}
\end{align}
where a simple change of variables $1-x\to x$ has been performed in the last equation. Because
the Levi-Civita symbol is rotationally invariant, that is
\begin{align}
  \label{eq:96}
  \left[e^{-x[{\boldsymbol{\Lambda}}]_\times}\right]_{\alpha'\alpha}
  \left[e^{-x[{\boldsymbol{\Lambda}}]_\times}\right]_{\beta'\delta}
  \left[e^{-x[{\boldsymbol{\Lambda}}]_\times}\right]_{\gamma\gamma'}
  \epsilon_{\alpha'\gamma\beta'} &=      \epsilon_{\alpha\gamma'\delta}
\end{align}
we have after some rearrangement
\begin{align}
  \label{eq:98}
      \left[e^{x[{\boldsymbol{\Lambda}}]_\times}\right]_{\alpha\alpha'}
    \left[e^{-x[{\boldsymbol{\Lambda}}]_\times}\right]_{\beta'\delta}
                \epsilon_{\alpha'\gamma\beta'} &=      \epsilon_{\alpha\gamma'\delta} \left[e^{-x[{\boldsymbol{\Lambda}}]_\times}\right]_{\gamma'\gamma}
\end{align}
Substituting (\ref{eq:98}) in (\ref{eq:60}) we obtain the final expression of the
derivative of the rotation matrix with respect to the orientation
\begin{align}
  \label{eq:99}
  \frac{\partial}{\partial{\Lambda}_{\gamma}}
\left[e^{[{\boldsymbol{\Lambda}}]_\times}\right]_{\alpha\beta}
  &=\epsilon_{\alpha\gamma'\beta'}
    {B}_{\gamma'\gamma}^{-1}({\boldsymbol{\Lambda}}) 
    \left[e^{[{\boldsymbol{\Lambda}}]_\times}\right]_{\beta'\beta}
\end{align}
where we have introduced the matrix
\begin{align}
  \label{eq:281}
  {\bf B}^{-1}({\boldsymbol{\Lambda}})  &\equiv\int_0^1dx   e^{x[{\boldsymbol{\Lambda}}]_\times}
\end{align}
This matrix can be explicitly computed by using Rodrigues' formula as
\begin{align}
  \label{eq:138}
  {\bf B}^{-1}({\boldsymbol{\Lambda}})=
  &\int_0^1dx  \left(\mathbb{1}
    +\sin(x{\Lambda}) [{\bf n}]_\times
    +\left(1-\cos(x{\Lambda})\right) [{\bf n}]_\times\esc [{\bf n}]_\times\right)
=\mathbb{1}+a({\Lambda}) [{\bf n}]_\times+b({\Lambda})[{\bf n}]_\times\esc [{\bf n}]_\times 
\end{align}
where 
\begin{align}
  \label{eq:30}
      \nonumber\\
  a({\Lambda})&=\frac{1-\cos{\Lambda}}{{\Lambda}}
     \nonumber\\
  b({\Lambda})&=1-\frac{\sin{\Lambda}}{{\Lambda}}
\end{align}
Using the property (\ref{eq:319}) it easy to check that the inverse
matrix has the same structure as the original matrix
\begin{align}
\label{eq:139}
{\bf B}({\boldsymbol{\Lambda}}) &= \mathbb{1}+p [{\bf n}]_\times+
q[{\bf n}]_\times\esc[{\bf n}]_\times
\end{align}
where
\begin{align}
  \label{eq:140}
p&=\frac{-a}{(1-b)^2+a^2}=-\frac{\Lambda}{2}
     \nonumber\\
  q&=\frac{a^2+b^2-b}{(1-b)^2+a^2}
     =1-\frac{\Lambda}{2}\frac{\sin\Lambda}{(1-\cos\Lambda)}
     =1-\frac{\Lambda}{2}\cot\frac{\Lambda}{2}
\end{align}
{An immediate consequence of the form (\ref{eq:139}) is
  \begin{align}
    \label{eq:224}
    {\bf B}(\boldsymbol{\Lambda})\esc\boldsymbol{\Lambda}=\boldsymbol{\Lambda}
  \end{align}}
The following identities follow from (\ref{eq:99}) and
all express  how the rotation matrix changes with the orientation
\begin{align}
  \label{eq:183}
\frac{\partial}{\partial{\Lambda}_{\gamma}}
\left[e^{[{\boldsymbol{\Lambda}}]_\times}\right]_{\alpha\beta}
{B}_{\gamma\gamma'}({\boldsymbol{\Lambda}})
&=\epsilon_{\alpha\gamma'\beta'} \left[e^{[{\boldsymbol{\Lambda}}]_\times}\right]_{\beta'\beta}\nonumber\\
\frac{\partial}{\partial{\Lambda}_{\gamma}}
\left[e^{-[{\boldsymbol{\Lambda}}]_\times}\right]_{\alpha\beta}
{B}_{\gamma\gamma'}({\boldsymbol{\Lambda}})
&=\epsilon_{\beta\gamma'\beta'} \left[e^{-[{\boldsymbol{\Lambda}}]_\times}\right]_{\alpha\beta'}\nonumber\\
\left[e^{[{\boldsymbol{\Lambda}}]_\times}\right]_{{\beta'\alpha}}
\frac{\partial}{\partial{\Lambda}_{\gamma}}
\left[e^{-[{\boldsymbol{\Lambda}}]_\times}\right]_{{\alpha}\beta}
&=\epsilon_{\beta\gamma'{\beta'}}{B}^{-1}_{\gamma'\gamma}({\boldsymbol{\Lambda}})\nonumber\\
\left(\frac{\partial}{\partial{\Lambda}_{\gamma}}
\left[e^{[{\boldsymbol{\Lambda}}]_\times}\right]_{\alpha\beta}\right)
\left[e^{-[{\boldsymbol{\Lambda}}]_\times}\right]_{\beta\alpha'}
&=\epsilon_{\alpha\gamma'\alpha'} {B}^{-1}_{\gamma'\gamma}({\boldsymbol{\Lambda}})
\end{align}
From these expressions, evaluated at $\boldsymbol{\Lambda}=0$, we obtain the following useful identities.
\begin{align}
\label{eq:457}
  \left . \frac{\partial}{\partial{\Lambda}_{\gamma}}
\left[e^{[{\boldsymbol{\Lambda}}]_\times}\right]_{\alpha\beta}
\right |_{{\boldsymbol{\Lambda}}=0}
&= \epsilon_{\alpha\gamma\beta}\nonumber\\
\left . \frac{\partial}{\partial{\Lambda}_{\gamma}}
\left[e^{-[{\boldsymbol{\Lambda}}]_\times}\right]_{\alpha\beta}
\right |_{{\boldsymbol{\Lambda}}=0}
&= \epsilon_{\beta\gamma\alpha}
\end{align}

\subsection{Properties of ${\bf B}$}

We now  consider some properties  of
${\bf  B}(\boldsymbol{\Lambda})$.  From  (\ref{eq:138}) and
(\ref{eq:139}), we have the symmetry properties
\begin{align}
  \label{eq:216}
  {\bf B}(-{\boldsymbol{\Lambda}})
  &=  {\bf B}^T({\boldsymbol{\Lambda}})
    \nonumber\\
  {\bf B}^{-1}(-{\boldsymbol{\Lambda}})
  &=  {\bf B}^{-T}({\boldsymbol{\Lambda}})
\end{align}
and the interesting property
\begin{align}
  \label{eq:489}
  {\bf B}\esc e^{[{\boldsymbol{\Lambda}}]_\times}={\bf B}^T
\end{align}
whose  proof    is  as  follows.  First,  to  alleviate
notation, write  Rodrigues' formula in the form
\begin{align}
  \label{eq:163}
e^{[{\boldsymbol{\Lambda}}]_\times}&=\mathbb{1}+s[{\bf n}]_\times+v[{\bf n}]_\times\esc[{\bf n}]_\times
\end{align}
where
\begin{align}
  \label{eq:164}
  s&=\sin\Lambda
     \nonumber\\
  v&=1-\cos\Lambda
\end{align}
then starting with the left hand side of (\ref{eq:489}) and using
(\ref{eq:319}) gives
\begin{align}
  \label{BeL}
  {\bf B}\esc e^{[{\boldsymbol{\Lambda}}]_\times}&=
(\mathbb{1}+p[{\bf n}]_\times+q[{\bf n}]_\times\esc[{\bf n}]_\times)\esc
(\mathbb{1}+s[{\bf n}]_\times+v[{\bf n}]_\times\esc[{\bf n}]_\times)\cr
&=\mathbb{1}+(s+p-pv-qs)[{\bf n}]_\times+(v+ps+q-qv)[{\bf n}]_\times\esc[{\bf n}]_\times\cr
&=\mathbb{1}-p[{\bf n}]_\times+q[{\bf n}]_\times\esc[{\bf n}]_\times\cr
&={\bf B}^T
\end{align}
where using (\ref{eq:140}) and (\ref{eq:164}) gives
\begin{align}
  \label{eq:73}
  qv-ps
  &=v
    \nonumber\\
  qs+pv&=s+2p
\end{align}
We will need the derivative of ${\bf B}$ with respect to 
${\boldsymbol{\Lambda}}$, so begin by expressing (\ref{eq:139}) as
\begin{align}
\label{eq:155}
{\bf B}({\boldsymbol{\Lambda}})\overset{(\ref{eq:262})}{=}
\mathbb{1} - \frac{1}{2}[{\boldsymbol{\Lambda}}]_\times +
f(\Lambda) [{\boldsymbol{\Lambda}}]_\times\esc
[{\boldsymbol{\Lambda}}]_\times
\end{align}
Taking the derivative of (\ref{eq:155}) and using (\ref{eq:602}) along with
component expressions in Einstein summation convention where appropriate gives
\begin{align}
\label{eq:156}
\frac{\partial{B}_{\alpha\beta}({\boldsymbol{\Lambda}})}
{\partial{\Lambda}_\gamma}
&=-\frac{1}{2}\epsilon_{\alpha\gamma\beta} +
\frac{df }{d\Lambda}\frac{{\Lambda}_\gamma}{\Lambda}
[{\boldsymbol{\Lambda}}]^{\alpha\alpha'}_\times
[{\boldsymbol{\Lambda}}]^{\alpha'\beta}_\times + f(\Lambda) \left (
\epsilon_{\alpha\gamma\alpha'} [{\boldsymbol{\Lambda}}]^{\alpha'\beta}_\times +
[{\boldsymbol{\Lambda}}]^{\alpha\alpha'}_\times \epsilon_{\alpha'\gamma\beta}\right )\cr
&=-\frac{1}{2}\epsilon_{\alpha\gamma\beta} +
\frac{1}{\Lambda}\frac{df }{d\Lambda}
[{\boldsymbol{\Lambda}}]^{\alpha\alpha'}_\times
[{\boldsymbol{\Lambda}}]^{\alpha'\beta}_\times {\Lambda}_\gamma
+ f(\Lambda) \left (
\epsilon_{\alpha\gamma\alpha'} \epsilon_{\alpha'\beta'\beta}{\Lambda}_{\beta'} +
\epsilon_{\alpha\beta'\alpha'} \epsilon_{\alpha'\gamma\beta}{\Lambda}_{\beta'}\right )\cr
&=-\frac{1}{2}\epsilon_{\alpha\gamma\beta} +
\frac{1}{\Lambda}\frac{df }{d\Lambda}
[{\boldsymbol{\Lambda}}]^{\alpha\alpha'}_\times
[{\boldsymbol{\Lambda}}]^{\alpha'\beta}_\times {\Lambda}_\gamma
+ f(\Lambda)  {\Lambda}_{\beta'} \left (
\delta_{\alpha\beta'}\delta_{\gamma\beta} - \delta_{\alpha\beta}\delta_{\beta'\gamma} +
\delta_{\alpha\gamma}\delta_{\beta'\beta} - \delta_{\alpha\beta}\delta_{\gamma\beta'}
\right )\cr
&=-\frac{1}{2}\epsilon_{\alpha\gamma\beta} +
\frac{1}{\Lambda}\frac{df }{d\Lambda}
\left [[{\boldsymbol{\Lambda}}]_\times\esc [{\boldsymbol{\Lambda}}]_\times
\right ]^{\alpha\beta}{\Lambda}_\gamma + f(\Lambda)  \left (
\delta_{\alpha\gamma}{\Lambda}_\beta +
\delta_{\beta\gamma}{\Lambda}_\alpha -
2\delta_{\alpha\beta}{\Lambda}_\gamma \right )
\end{align}
where the following property of the Levi-Civita symbol has been used as needed
\begin{align}
\label{eq:604}
\epsilon_{ijk}\epsilon_{\ell mn}=
\delta_{i\ell}[\delta_{jm}\delta_{kn}-\delta_{jn}\delta_{km}] -
\delta_{im}[\delta_{j\ell}\delta_{kn}-\delta_{jn}\delta_{k\ell}] +
\delta_{in}[\delta_{j\ell}\delta_{km}-\delta_{jm}\delta_{k\ell}]
\end{align}
Contracting over the indices $\alpha$ and $\gamma$ (or equivalently over $\beta$
and $\gamma$) then gives
\begin{align}
\label{eq:606}
\frac{\partial}{\partial{\boldsymbol{\Lambda}}}\esc
{\bf B}({\boldsymbol{\Lambda}})=
\frac{\partial}{\partial{\boldsymbol{\Lambda}}}\esc
{\bf B}^T({\boldsymbol{\Lambda}}) = 2f(\Lambda)  {\boldsymbol{\Lambda}}=
\left( \frac{2}{\Lambda} - \cot\frac{\Lambda}{2}\right ){\bf n}
\end{align}
Further multiplying this result by ${\bf B}^{-1}$ and using (\ref{eq:138}) gives
\begin{align}
\label{eq:173}
\left ( \frac{\partial}{\partial{\boldsymbol{\Lambda}}}\esc
{\bf B}({\boldsymbol{\Lambda}})\right ) \esc
{\bf B}^{-1}({\boldsymbol{\Lambda}}) = \left( \frac{2}{\Lambda} -
\cot\frac{\Lambda}{2}\right ){\bf n} =
-\frac{\partial}{\partial {\boldsymbol{\Lambda}}}
\ln\left(\frac{1-\cos\Lambda}{\Lambda^2}\right)
\end{align}
which in component form is
  \begin{align}
    \label{eq:723}
      \frac{\partial {B}_{\alpha\beta}}{\partial{\Lambda}_\alpha}{B}^{-1}_{\beta\gamma}
  & =   \left(\frac{2}{\Lambda}  -\cot\frac{\Lambda}{2}\right){\bf n}_\gamma=-\frac{\partial }{\partial {\Lambda}_\gamma}\ln\left(\frac{1-\cos\Lambda}{\Lambda^2}\right)
  \end{align}

\subsection{The  relationship between the angular  velocity and the time
  derivative of the orientation}
  \label{App:velang}
  From the definition of the angular velocity matrix in terms of the rotation matrix we have
  (we omit hats  denoting phase functions
  for the sake of clarity)
\begin{align}
  \label{eq:501}
  [{\boldsymbol{\omega}}]_\times
  &=\left(\frac{d}{dt}e^{[{\boldsymbol{\Lambda}}]_\times}\right)\esc e^{-[{\boldsymbol{\Lambda}}]_\times}
    =-e^{[{\boldsymbol{\Lambda}}]_\times}\esc \left(\frac{d}{dt}e^{-[{\boldsymbol{\Lambda}}]_\times}\right)
\end{align}
and applying (\ref{WilcoxLambda})
\begin{align}
  \label{eq:22b}
    \frac{d}{dt}e^{[{\boldsymbol{\Lambda}}]_\times}=\int_0^1dx e^{(1-x)[{\boldsymbol{\Lambda}}]_\times}\esc
\left[  \frac{d{\boldsymbol{\Lambda}}}{dt}\right]_\times\esc e^{x[{\boldsymbol{\Lambda}}]_\times}
\end{align}
Using this relation in (\ref{eq:501}) along with a change
of variable $x\rightarrow 1-x$ gives
\begin{align}
  \label{eq:285}
  [{\boldsymbol{\omega}}]_\times
  &=\int_0^1dx   e^{x[{\boldsymbol{\Lambda}}]_\times}\esc
    [\dot{{\boldsymbol{\Lambda}}}]_\times \esc e^{-x[{\boldsymbol{\Lambda}}]_\times}
\overset{(\ref{eq:601})}{=}\int_0^1dx  
    \left[ e^{x[{\boldsymbol{\Lambda}}]_\times}\esc
\dot{{\boldsymbol{\Lambda}}}\right]_\times
    =
    \left[ \int_0^1dx  e^{x[{\boldsymbol{\Lambda}}]_\times}\esc
\dot{{\boldsymbol{\Lambda}}}\right]_\times 
    \overset{(\ref{eq:281})}{=}
    \left[{\bf B}^{-1}(\boldsymbol{\Lambda})\esc
\dot{{\boldsymbol{\Lambda}}}\right]_\times 
    \end{align}
Therefore, we obtain a
neat expression  relating the  angular velocity  \textit{vector} with  the time
derivative of the orientation
\begin{align}
  \label{eq:136}
  {\boldsymbol{\omega}}&=  {\bf B}^{-1}({\boldsymbol{\Lambda}})\esc
\dot{{\boldsymbol{\Lambda}}}\\
  \label{eq:283}
\dot{{\boldsymbol{\Lambda}}}&=  {\bf B}({\boldsymbol{\Lambda}})\esc    {\boldsymbol{\omega}}
\end{align}
Another expression  used in the main text is
\begin{align}
  \label{eq:28}
  e^{-[\boldsymbol{\Lambda}]_\times}\esc\frac{d}{dt}\left(
  e^{[\boldsymbol{\Lambda}]_\times}\right)
  &\overset{(\ref{eq:501})}{=}
    e^{-[\boldsymbol{\Lambda}]_\times}\esc
    [\boldsymbol{\omega}]_\times      e^{[\boldsymbol{\Lambda}]_\times}
\overset{(\ref{eq:601})}{=}
    \left[e^{-[\boldsymbol{\Lambda}]_\times}\esc\boldsymbol{\omega}\right]_\times
    =\left[\boldsymbol{\omega}_{0}\right]_\times
\end{align}

For the sake of completeness, we recover known expressions that relate
angle and  axis variables with  the angular velocity of  the principal
axis. Noting that
$2{{\bf n}} \esc {\dot{{{\bf n}}}}=d{{\bf n}}^2/dt=0$, and by using
\begin{align}
  \label{eq:132}
  [{\bf n}]_\times  \cdot \dot{{\boldsymbol{\Lambda}}}
  &={\Lambda}({{\bf n}}\times\dot{{\bf n}})
    \nonumber\\
  [{\bf n}]_\times\esc[{\bf n}]_\times
  \cdot \dot{{\boldsymbol{\Lambda}}}
  &={\bf n}\times({\bf n}\times\dot{{\boldsymbol{\Lambda}}})
   ={\bf n}({\bf n}\esc\dot{{\boldsymbol{\Lambda}}})                                          -\dot{{\boldsymbol{\Lambda}}}
   ={\bf n}({\bf n}\esc(\dot{\Lambda}{\bf n}+\Lambda\dot{{\bf n}})-\dot{\Lambda}{\bf n}-\Lambda\dot{{\bf n}}
   ={\bf n}\dot{\Lambda}-\dot{\Lambda}{\bf n}-\Lambda\dot{{\bf n}}=-\Lambda\dot{{\bf n}}
\end{align}
in (\ref{eq:136}) with (\ref{eq:138}) we have
\begin{align}
  \label{eq:149}
    {\boldsymbol{\omega}}
    &= {\dot{{\Lambda}}}{\bf n}
      + \sin{\Lambda} {\dot{{\bf n}}}+(1-\cos{\Lambda})({\bf n}\times{\dot{{\bf n}}})
\end{align}
which coincides with Eq. (3.52) of Ref. \cite{Olguin2019}.  On  the other hand,
from (\ref{eq:139}) and (\ref{eq:140}) in (\ref{eq:283}) we obtain
\begin{align}
        \label{eq:64}
  \dot{{\boldsymbol{\Lambda}}}
  &={\boldsymbol{\omega}}
    +\frac{\Lambda}{2} ({\boldsymbol{\omega}}\times {\bf n})
    +\left(1-\frac{\Lambda}{2}\cot\frac{\Lambda}{2}\right)
    {\bf n}\times ({\bf n}\times {\boldsymbol{\omega}})
\end{align}
Since $\dot{\Lambda}={\bf n}\esc\dot{{\boldsymbol{\Lambda}}}$ and from
(\ref{eq:132})
$\dot{{\bf                n}}=-[{\bf               n}]_\times\esc[{\bf
  n}]_\times\cdot\dot{{\boldsymbol{\Lambda}}}/\Lambda$,  (\ref{eq:64})
can be used with (\ref{eq:319}) to show that
\begin{align}
  \label{eq:226}
  \dot{\Lambda}&={\boldsymbol{\omega}}\esc{\bf n}
                \nonumber\\
  \dot{{\bf n}}&=\frac{1}{2}{\boldsymbol{\omega}}\times{\bf n}-\frac{1}{2}\cot\frac{\Lambda}{2}{\bf n}\times({\bf n}\times{\boldsymbol{\omega}})
\end{align}
This coincides with Eq.~(4.11.a) of Ref. \cite{Olguin2019}.

\subsection{Derivative of the inertia tensor  with respect to the orientation}
\label{App:DerI}
The derivative of  the inertia tensor with respect  to the orientation
is a third order tensor given by
\begin{align}
  \label{eq:169}
  \frac{\partial}{\partial{\Lambda}_\gamma}{\bf I}^{-1}_{\alpha\beta}
  &= \frac{\partial}{\partial{\Lambda}_\gamma}
    \left([e^{[\boldsymbol{\Lambda}]_\times}]_{\alpha\alpha'}
    \mathbb{I}^{-1}_{\alpha'\beta'}[e^{-[\boldsymbol{\Lambda}]_\times}]_{\beta'\beta}\right)
   \nonumber\\
&=\left(\frac{\partial[e^{[\boldsymbol{\Lambda}]_\times}]_{\alpha\alpha'}}{\partial{\Lambda}_\gamma}\right)
\mathbb{I}^{-1}_{\alpha'\beta'}[e^{-[\boldsymbol{\Lambda}]_\times}]_{\beta'\beta}
    +[e^{[\boldsymbol{\Lambda}]_\times}]_{\alpha\alpha'}
                 \mathbb{I}^{-1}_{\alpha'\beta'}
\left(\frac{\partial[e^{-[\boldsymbol{\Lambda}]_\times}]_{\beta'\beta}}{\partial{\Lambda}_\gamma}\right)
\nonumber\\
 &=\frac{\partial[e^{[\boldsymbol{\Lambda}]_\times}]_{\alpha\alpha'}}{\partial{\Lambda}_\gamma}
\left[    e^{-[\boldsymbol{\Lambda}]_\times}\right]_{\alpha'\beta'}{\bf I}^{-1}_{\beta'\beta}
    +{\bf I}^{-1}_{\alpha\alpha'}
\left[e^{[\boldsymbol{\Lambda}]_\times}\right]_{\alpha'\beta'}\frac{\partial[e^{-[\boldsymbol{\Lambda}]_\times}]_{\beta'\beta}}{\partial{\Lambda}_\gamma}
\end{align}
We may now use the last two identities in (\ref{eq:183})
to write  (\ref{eq:169}) in the form
\begin{align}
  \label{eq:108}
  \frac{\partial}{\partial{\Lambda}_\gamma}{\bf I}^{-1}_{\alpha\beta}
  &=\epsilon_{\alpha\gamma'\beta'}{B}^{-1}_{\gamma'\gamma}{\bf I}^{-1}_{\beta'\beta}
+{\bf I}^{-1}_{\alpha\alpha'}\epsilon_{\beta\gamma'\alpha'}{B}^{-1}_{\gamma'\gamma}
\end{align}
Note that both the left and right hand sides are symmetric under the permutation of
$\alpha,\beta$.
Now, (\ref{eq:108}) can be used to get the derivative of the rotational kinetic energy as
\begin{align}
  \label{eq:170}
  \frac{\partial K^{\rm rot}}{\partial {\Lambda}_\gamma}
  &=  \frac{\partial}{\partial{\Lambda}_\gamma}
    \left(\frac{1}{2}  {\bf S}^\alpha{\bf I}^{-1}_{\alpha\beta}  {\bf S}^\beta\right)
    =
    \left(\frac{1}{2}  {\bf S}^\alpha\left[\epsilon_{\alpha\gamma'\beta'}{B}^{-1}_{\gamma'\gamma}{\bf I}^{-1}_{\beta'\beta}
+{\bf I}^{-1}_{\alpha\alpha'}\epsilon_{\beta\gamma'\alpha'}{B}^{-1}_{\gamma'\gamma}\right] 
 {\bf S}^\beta\right)
    \nonumber\\
    &=
\frac{1}{2}  \left[{\bf S}^\alpha\epsilon_{\alpha\gamma'\beta'}{B}^{-1}_{\gamma'\gamma}\boldsymbol{\Omega}^{\beta'}
+\boldsymbol{\Omega}^{\alpha'}\epsilon_{\beta\gamma'\alpha'}{B}^{-1}_{\gamma'\gamma}{\bf S}^\beta\right] 
    \nonumber\\
  &=\left(\boldsymbol{\Omega}\times{\bf S}\right)^{\gamma'}{B}^{-1}_{\gamma'\gamma}
\end{align}
or in vector form
\begin{align}
  \label{eq:192}
    \frac{\partial K^{\rm rot}}{\partial \boldsymbol{\Lambda}} =
    {\bf B}^{-T}( \boldsymbol{\Lambda})\esc
 (\boldsymbol{\Omega}\times{\bf S}) 
\end{align}
Multiplying  (\ref{eq:108}) by  $B_{\gamma\alpha}$ and  summing
over  $\alpha$ and  $\gamma$, and  noting  the full  contraction of  a
product of symmetric and antisymmetric tensors vanishes, gives
\begin{align}
  \label{eq:261}
{B}_{\gamma\alpha}
\frac{\partial{\bf I}^{-1}_{\alpha\beta}}
{\partial{\Lambda}_\gamma}=0
\end{align}
which implies
\begin{align}
  \label{eq:25}
  {B}_{\gamma\alpha}\frac{\partial\boldsymbol{\Omega}^\alpha}{\partial{\Lambda}_\gamma}=0
\end{align}
a result that will be used later.

\newpage
\setcounter{my_equation_counter}{\value{equation}}
\section{Effect of symmetries on orientation and central moments}
\setcounter{equation}{\value{my_equation_counter}}

\label{App:Sym}
\subsection{Translations}
The  gyration tensor  is invariant under a  translation, that
is,     if     ${\bf     r}_{i}'={\bf    r}_{i}+{\bf     a}$,     then
$\hat{\bf  G}(\{{\bf  r}'_{i}\})=\hat{\bf G}(\{{\bf  r}_{i}\})$.  This
implies the orientation is also  invariant under a global translation,
so
\begin{align}
  \hat{\boldsymbol{\Lambda}} (\{{\bf r}_{i }+{\bf a}\})&=
  \hat{\boldsymbol{\Lambda}} (\{{\bf r}_{i }\})
\label{la}
\end{align}
where ${\bf a}$  is an arbitrary translation. As a  consequence, if we
take the derivative of both sides  of (\ref{la}) with respect to ${\bf
  a}$ and then set ${\bf a}=0$, we have
\begin{align}
  \sum_{i }\frac{ \partial \hat{\boldsymbol{\Lambda}} }{\partial {\bf r}_{i }}&=0
\label{trans}
\end{align}
We have a similar expression for translational invariance for the central moments
\begin{align}
  \sum_{i }\frac{ \partial \hat{{\bf M}} }{\partial {\bf r}_{i }}
  &=0                                                                                               
\label{trans2}
\end{align}
\subsection{Rotations}
\label{App:Orientation}
  The orientation $\hat{\boldsymbol{\Lambda}}$ does not transform as a
  vector under rotations.  To see  this consider an arbitrary rotation
  matrix  $\boldsymbol{\cal  Q}$  from the  inertial  reference  frame
  $\cal  S$  to  a  frame  ${\cal  S}^\prime$  so  that  the  particle
  coordinates change accordingly
\begin{align}
{\bf r}_i^\prime&=\boldsymbol{\cal Q}\esc{\bf r}_i\cr
{\bf r}_i&=\boldsymbol{\cal Q}^T\esc{\bf r}_i^\prime
\end{align}
Let $\hat{\boldsymbol{\cal R}}$ be  the rotation matrix
from $\cal S$ to the frame  ${\cal S}^0$ which diagonalizes the moment
of  inertia tensor.   Let  $\hat{\boldsymbol{\cal  R}}^\prime$ be  the
rotation  matrix  from  ${\cal  S}^\prime$ to  ${\cal  S}^0$  so  that
$\hat{\boldsymbol{\cal
    R}}^\prime=e^{-[\hat{\boldsymbol{\Lambda}}(\{{\bf
    r}_i^\prime\})]_\times}$.  This  expression follows from  the fact
that $\hat{\boldsymbol{\Lambda}}$  provides the  orientation necessary
to  transform to  ${\cal S}^0$  regardless the  current frame.   Hence
$\hat{\boldsymbol{\Lambda}}(\{{\bf r}_i^\prime\})$  is the orientation
of  ${\cal   S}^0$  relative   to  ${\cal  S}^\prime$.    Because  the
transformation  ${\cal   S}\rightarrow{\cal  S}^0$  is  the   same  as
${\cal S}\rightarrow{\cal S}^\prime\rightarrow{\cal  S}^0$, it follows
that
\begin{align}
  \hat{\boldsymbol{\cal R}}&=\hat{\boldsymbol{\cal R}}^\prime\esc
                               \boldsymbol{\cal Q}\cr
\label{eq:239}
e^{-[\hat{\boldsymbol{\Lambda}}(\{{\bf r}_i\})]_\times}&=
e^{-[\hat{\boldsymbol{\Lambda}}(\{\boldsymbol{\cal Q}\esc{\bf r}_i\})]_\times}\esc \boldsymbol{\cal Q}
\end{align}
Using the series expansion of the exponential and
properties (\ref{eq:601}) we may easily arrive at
\begin{align}
  \label{eq:24b}
  e^{-[\hat{\boldsymbol{\Lambda}}(\{{\bf r}_i\})]_\times}
  &=
    \boldsymbol{\cal Q}\esc
    \boldsymbol{\cal Q}^Te^{-[\hat{\boldsymbol{\Lambda}}(\{\boldsymbol{\cal Q}\esc{\bf r}_i\})]_\times}\esc \boldsymbol{\cal Q}
    \nonumber\\
    &=
    \boldsymbol{\cal Q}\esc
    e^{-\boldsymbol{\cal Q}^T\esc\  [\hat{\boldsymbol{\Lambda}}(\{\boldsymbol{\cal Q}\esc{\bf r}_i\})]_\times\ \esc\ \boldsymbol{\cal Q}}
      \nonumber\\
  &=
    \boldsymbol{\cal Q}\esc
    e^{-[\boldsymbol{\cal Q}^T\esc\ \hat{\boldsymbol{\Lambda}}(\{\boldsymbol{\cal Q}\esc{\bf r}_i\})]_\times}
\end{align}
Should the  matrix $ \boldsymbol{\cal Q}$  multiplying the exponential
on the right hand side of  (\ref{eq:24b}) not be present, this equation
would                            imply                          
$\hat{\boldsymbol{\Lambda}}(\{\boldsymbol{\cal              Q}\esc{\bf
  r}_i\})=\boldsymbol{\cal     Q}\esc\hat{\boldsymbol{\Lambda}}(\{{\bf
  r}_i\})$, which is the way a  vector would transform. Because of the
presence of  $ \boldsymbol{\cal Q}$ multiplying  the exponential, this
is not true, and the orientation does not transform as a vector.

Equation (\ref{eq:239}) allows us  to obtain an important relationship
for  the  orientation.  Take  the  derivative  of (\ref{eq:239})  with
respect  to  the   orientation  ${\bf  A}$  of   the  rotation  matrix
$\boldsymbol{\cal  Q}=e^{[{\bf  A}]_\times}$.    The left  hand  side  does not  depend  on
${\bf A}$ so
\begin{align}
  \label{eq:242}
\frac{\partial}{\partial {\bf A}} \left(e^{-[\hat{\boldsymbol{\Lambda}}(Q r)]_\times}\esc\boldsymbol{\cal Q}\right)&=  0
\end{align}
which can be arranged in the form
\begin{align}
  \label{eq:459}
\left(\frac{\partial}{\partial {\bf A}} \boldsymbol{\cal Q}\right)\esc\ \boldsymbol{\cal Q}^T +
e^{[\hat{\boldsymbol{\Lambda}}(Q r)]_\times}\esc\ 
\frac{\partial}{\partial {\bf A}} \left(e^{-[\hat{\boldsymbol{\Lambda}}(Q r)]_\times}\right)
&=  0
\end{align}
Using (\ref{eq:457}) for the
  derivative of  $\boldsymbol{\cal Q}$ evaluated at ${\bf A}=0$ gives
\begin{align}
  \label{eq:711}
\left.  \frac{\partial \left[e^{-[{\bf A}]_\times}\right]_{\alpha\beta}}{\partial {\bf A}^\gamma}\right|_{{\bf A}=0
  }
  &=  \epsilon_{\beta\gamma\alpha}
\end{align}
Now use (\ref{WilcoxLambda}) for the derivative of  the second
  term in (\ref{eq:459}) and write in component form as
\begin{align}
  \label{eq:276}
\left[e^{[{\boldsymbol{\Lambda}}(Q r)]_\times}\right]_{\alpha\delta}
\frac{\partial}{\partial {\bf A}^\gamma}
\left[e^{-\left[ \hat{\boldsymbol{\Lambda}}(Q r)\right]_\times}\right ]_{\delta\beta} =
  \int_0^1dx \left[e^{x[\hat{\boldsymbol{\Lambda}}(Q r)]_\times}\right]_{\alpha\alpha'}
  \left(  \frac{\partial}{\partial {\bf A}^\gamma}\left[ \hat{\boldsymbol{\Lambda}}(Q r)\right]^{\alpha'\beta'}_\times\right) \left[e^{-x[\hat{\boldsymbol{\Lambda}}(Q r)]_\times}\right]_{\beta'\beta}
\end{align}
The chain rule gives the following expression
\begin{align}
  \label{eq:273}
  \frac{\partial}{\partial{\bf A}^\gamma}  \left[\hat{\boldsymbol{\Lambda}}(Q r)\right]_\times^{\alpha'\beta'}
  &=\epsilon_{\alpha'\gamma'\beta'}\sum_i  \frac{\partial \hat{\Lambda}_{\gamma'}}{\partial{\bf r}_i^{\alpha''}}(Q r) \frac{\partial}{\partial{\bf A}^\gamma}\left (\boldsymbol{\cal Q}^{\alpha''\beta''}\right ){\bf r}_i^{\beta''}
\end{align}
Evaluated at ${\bf A}=0$ and using (\ref{eq:457}) this becomes
\begin{align}
  \label{eq:274}
  \left.    \frac{\partial}{\partial{\bf A}^\gamma}  \left[\hat{\boldsymbol{\Lambda}}(Q r)\right]_\times^{\alpha'\beta'}\right|_{{\bf A}=0}
  &=\epsilon_{\alpha'\gamma'\beta'}\sum_i  \frac{\partial \hat{\Lambda}_{\gamma'}}{\partial{\bf r}_i^{\alpha''}}(r)
    \epsilon_{\alpha''\gamma\beta''}{\bf r}_i^{\beta''}
    =-\epsilon_{\alpha'\gamma'\beta'}
    \hat{O}_{\gamma'\gamma}(r)
\end{align}
where we have introduced
\begin{align}
  \label{eq:407}
    \hat{O}_{\gamma'\gamma}(r)&\equiv\sum_i^N      \frac{\partial \hat{\Lambda}_{\gamma'}}{\partial {\bf r}^{\alpha''}_i} [{\bf r}_i]_\times^{\alpha''\gamma}    
  \end{align}
Putting all these terms in (\ref{eq:459}) and evaluating at ${\bf A}=0$ gives
  \begin{align}
    \label{eq:277}
\int_0^1dx \left[e^{x[\hat{\boldsymbol{\Lambda}}(r)]_\times}\right]_{\alpha\alpha'}\left[
    -\epsilon_{\alpha'\gamma'\beta'}
    \hat{O}_{\gamma'\gamma}\right] \left[e^{-x[\hat{\boldsymbol{\Lambda}}(r)]_\times}\right]_{\beta'\beta}
    &=  -\epsilon_{\beta\gamma\alpha}
  \end{align}
Matching the integrand to that of (\ref{eq:60}) followed by the use of
 (\ref{eq:183}) then gives
  \begin{align}
\hat{O}_{\gamma'\gamma}
\left ( \frac{\partial}{\partial{\Lambda}_{\gamma'}}
\left[e^{[{\boldsymbol{\Lambda}}]_\times}\right]_{\alpha\beta'}\right )
\left[e^{-[\hat{\boldsymbol{\Lambda}}]_\times}\right]_{\beta'\beta}
    &= \epsilon_{\beta\gamma\alpha}\nonumber\\
\hat{O}_{\gamma'\gamma}\epsilon_{\alpha\alpha'\beta}
{B}_{\alpha'\gamma'}^{-1}({\boldsymbol{\Lambda}})
    &= \epsilon_{\beta\gamma\alpha}\nonumber\\
-\epsilon_{\beta\alpha'\alpha}
{B}_{\alpha'\gamma'}^{-1}({\boldsymbol{\Lambda}})
\hat{O}_{\gamma'\gamma}
&= \epsilon_{\beta\gamma\alpha}
  \end{align}
  that implies
  \begin{align}
    \label{eq:160b}
    {B}_{\beta'\gamma'}^{-1}(\hat{\boldsymbol{\Lambda}})\hat{O}_{\gamma'\gamma}=-\delta_{\beta'\gamma}
  \end{align}
In  conclusion, the  transformation  property of  the orientation
(\ref{eq:239}) is reflected into the following mathematical identity
\begin{align}
 \label{eq:255}
  \hat{\bf O}
  &= \sum_i^N      \frac{\partial \hat{\boldsymbol{\Lambda}}}{\partial {\bf r}_i}\esc [{\bf r}_i]_\times
    =-{\bf B}\left(     \hat{\boldsymbol{\Lambda}}\right)
\end{align}
This   equation  shows   that   the  matrix   valued  phase   function
$\hat{\bf  O}$  depends on  the  configuration  $r$ only  through  the
orientation.

  Let us
now consider  the transformation  properties of the  central moments
and of the  dilation momentum.  Because the  eigenvalues are invariant
under rotations, we know that
\begin{align}
  \label{eq:47}
\hat{{\bf M}}(Qr)=\hat{{\bf M}}(r)
\end{align}
By   following  an   argument  identical   to  the   one  leading   to
(\ref{eq:255})  we  may  take  the derivative  of  (\ref{eq:47})  with
respect to ${\bf  A}$ and evaluate the final result  at ${\bf A}=0$ to
get the identity
\begin{align}
0=\left.\frac{\partial}{\partial {\bf A}}  \hat{{\bf M}} \left({\cal Q}r\right)\right|_{{\bf A}=0}=
  \left.\sum_{i }  \frac{\partial \hat{{\bf M}} }{\partial {\bf r}_i}\esc\frac{\partial e^{[{\bf A}]_\times}}{\partial {\bf A}}\right|_{{\bf A}=0}\esc {\bf r}_{i }
  \overset{(\ref{eq:711})}{=}
\sum_{i }  [{\bf r}_{i }]_\times\esc\frac{\partial \hat{{\bf M}} }{\partial {\bf r}_i}
\label{dsym}
\end{align}
On the other hand, if we take the derivative of (\ref{eq:47}) and
use the chain rule we get
\begin{align}
  \label{eq:44}
\frac{\partial}{\partial {\bf r}^\alpha_i}  \hat{{\bf M}}(r)= 
  \frac{\partial}{\partial {\bf r}^\alpha_i}  \hat{{\bf M}}(Qr)
  &=\sum_j \frac{\partial \hat{{\bf M}}}{\partial {\bf r}_j^{\beta}}(Qr)
    \boldsymbol{\cal Q}^{\beta \alpha'}\frac{\partial {\bf r}_j^{\alpha'}}{\partial {\bf r}_i^{\alpha}}
    =\frac{\partial \hat{{\bf M}}}{\partial {\bf r}_i^{\beta}}(Qr)
    \boldsymbol{\cal Q}^{\beta \alpha}
\end{align}
 This implies, recalling that $\boldsymbol{\cal Q}^{\alpha\alpha'}\boldsymbol{\cal Q}^{\beta\alpha'}=\delta_{\alpha\beta}$, 
\begin{align}
  \label{eq:49}
  \frac{\partial \hat{{\bf M}}}{\partial {\bf r}_i^{\alpha}}(Qr)
=  \boldsymbol{\cal Q}^{\alpha\beta }\frac{\partial\hat{{\bf M}}}{\partial {\bf r}^\beta_i}  (r)  
\end{align}
As a consequence,  the dilation momentum satisfies
\begin{align}
  \label{eq:51}
  \hat{\boldsymbol{\Pi}}(Qz)
  &=\sum_i\frac{\partial \hat{{\bf M}}}{\partial {\bf r}_i^\alpha}(Qr)
    \boldsymbol{\cal Q}^{\alpha\alpha'}{\bf v}_i^{\alpha'}  =
    \sum_i \boldsymbol{\cal Q}^{\alpha\beta }\frac{\partial\hat{{\bf M}}}{\partial {\bf r}^\beta_i}  (r)
\boldsymbol{\cal Q}^{\alpha\alpha'}{\bf v}_i^{\alpha'}
    =    \sum_i \frac{\partial\hat{{\bf M}}}{\partial {\bf r}^\beta_i}  (r)
{\bf v}_i^{\beta}=\hat{\boldsymbol{\Pi}}(z)
\end{align}
The dilation momentum  is invariant under rotation  of the coordinates
as a consequence  of the central moments also  being invariant under
rotation.   As   a  further   development,  take  the   derivative  of
(\ref{eq:49})  with  respect  to  ${\bf  A}^\gamma$  and  evaluate  at
${\bf A}=0$. The left hand side gives
\begin{align}
  \label{eq:263}
\left .  \frac{\partial}{\partial{\bf A}^\gamma}    \frac{\partial \hat{{\bf M}}}{\partial {\bf r}_i^{\alpha}}(Qr)\right|_{{\bf A}=0}
  &=
    \left . \sum_j   \frac{\partial^2 \hat{{\bf M}}}
    {\partial {\bf r}_i^{\alpha}\partial {\bf r}_j^{\beta}}(Qr)
    \frac{\partial {\boldsymbol{\cal Q}}^{\beta\beta'}{\bf r}_j^{\beta'}}{\partial{\bf A}^\gamma}\right|_{{\bf A}=0}
\overset{(\ref{eq:711})}{=}
    \sum_j   \frac{\partial^2 \hat{{\bf M}}}
    {\partial {\bf r}_i^{\alpha}\partial {\bf r}_j^{\beta}}(r)
    \epsilon_{\beta\gamma\beta'}{\bf r}_j^{\beta'}
    =-\sum_j   \frac{\partial^2 \hat{{\bf M}}}
    {\partial {\bf r}_i^{\alpha}\partial {\bf r}_j^{\beta}}(r)[{\bf r}_j]_\times^{\beta\gamma}
\end{align}
The right hand side gives, using (\ref{eq:711})
\begin{align}
    \label{eq:266}
\left.    \frac{\partial}{\partial{\bf A}^\gamma}\boldsymbol{\cal Q}^{\alpha\beta }
  \right|_{{\bf A}=0}\frac{\partial\hat{{\bf M}}}{\partial {\bf r}^\beta_i}  (r)  
&=\epsilon_{\alpha\gamma\beta}  \frac{\partial\hat{{\bf M}}}{\partial {\bf r}^\beta_i}  (r)  
\end{align}
Equating both sides we obtain the identity
\begin{align}
  \label{eq:267}
  \sum_j   \frac{\partial^2 \hat{{\bf M}}}
  {\partial {\bf r}_i^{\alpha}\partial {\bf r}_j^{\beta}}(r)[{\bf r}_j]_\times^{\beta\gamma}
  &=\epsilon_{\alpha\beta\gamma}  \frac{\partial\hat{{\bf M}}}{\partial {\bf r}^\beta_i}(r)  
\end{align}
This identity reflects rotational invariance at the level of second derivatives of the central moments. 

\newpage

\setcounter{my_equation_counter}{\value{equation}}
\section{Microscopic expressions}
\setcounter{equation}{\value{my_equation_counter}}
\label{App:Mic}
In this section  we consider the explicit form for  a number of phase
functions in terms  of the positions and velocities  of the particles.
These    expressions    are    required     in    the    main    text.
We will need the transformation of coordinates from the lab to the principal axis system
\begin{align}
  \label{rvmic}
      {\bf r}_{0i}&=\hat{\boldsymbol{\cal R}}\esc({\bf r}_i-\hat{\bf R})
                  \nonumber\\
    {\bf v}_{0i}&\equiv\frac{d{\bf r}_{0i}}{dt}=\hat{\boldsymbol{\cal R}}\esc\left( {\bf v}_i-\hat{\bf V}-[\hat{\boldsymbol{\omega}}]_\times\esc({\bf r}_i-\hat{\bf R})\right)
                  \nonumber\\
    {\bf v}_{pi}&=\hat{\boldsymbol{\cal R}}({\bf v}_i-{\bf V})
                  \nonumber\\
    [\hat{\boldsymbol{\omega}}]_\times
                &\equiv-{\hat{\boldsymbol{\cal R}}}^{T}\esc\frac{d{\hat{\boldsymbol{\cal R}}}}{dt}
                  =\frac{d{\hat{\boldsymbol{\cal R}}}^{T}}{dt}\esc{\hat{\boldsymbol{\cal R}}}
                  \nonumber\\
    [\hat{\boldsymbol{\omega}}_0]_\times
                &\equiv{\hat{\boldsymbol{\cal R}}}\esc\frac{d{\hat{\boldsymbol{\cal R}}}^{T}}{dt}
                  =-\frac{d{\hat{\boldsymbol{\cal R}}}}{dt}\esc{\hat{\boldsymbol{\cal R}}}^T
                  =\hat{\boldsymbol{\cal R}}\esc      [\hat{\boldsymbol{\omega}}]_\times\esc \hat{\boldsymbol{\cal R}}^T
  \end{align}

\subsection{Dilational  and  angular  velocities  in  microscopic terms}

The      dilational      $\hat{\boldsymbol{\nu}}$     and      angular
$\hat{\boldsymbol{\omega}}$    velocities    are    closely    related
concepts. In order  to find explicit expressions for both  in terms of
the positions and momenta of  the particles, we consider the following
sequence of identities
  \begin{align}
    \label{eq:6}
    i{\cal L}\hat{\mathbb{G}}
    &=   i{\cal L}\left(\hat{\boldsymbol{\cal R}}\esc\hat{\bf G} \esc \hat{\boldsymbol{\cal R}}^T\right)
\overset{\mbox{\tiny chain rule}}{=} 
\left(i{\cal L}\hat{\boldsymbol{\cal R}}\right)\esc\hat{\bf G} \esc \hat{\boldsymbol{\cal R}}^T
      +\hat{\boldsymbol{\cal R}}\esc
      \left(i{\cal L}\hat{\bf G} \right)\esc \hat{\boldsymbol{\cal R}}^T
      +\hat{\boldsymbol{\cal R}}\esc
      \hat{\bf G}\esc  \left(i{\cal L}\hat{\boldsymbol{\cal R}}^T \right)
      \nonumber\\
&=\left(i{\cal L}\hat{\boldsymbol{\cal R}}\right)\esc \hat{\boldsymbol{\cal R}}^T\esc \hat{\boldsymbol{\cal R}}\esc\hat{\bf G} \esc \hat{\boldsymbol{\cal R}}^T
      +\hat{\boldsymbol{\cal R}}\esc
      \left(i{\cal L}\hat{\bf G} \right)\esc \hat{\boldsymbol{\cal R}}^T
      +\hat{\boldsymbol{\cal R}}\esc
      \hat{\bf G}\esc  \hat{\boldsymbol{\cal R}}^T\esc \hat{\boldsymbol{\cal R}} \left(i{\cal L}\hat{\boldsymbol{\cal R}}^T \right)
      \nonumber\\
&= \left(i{\cal L}\hat{\boldsymbol{\cal R}}\right)\esc\hat{\boldsymbol{\cal R}}^T\esc \hat{\mathbb{G}}
      +\hat{\boldsymbol{\cal R}}\esc
      \left(i{\cal L}\hat{\bf G} \right)\esc \hat{\boldsymbol{\cal R}}^T
      + \hat{\mathbb{G}}\esc \hat{\boldsymbol{\cal R}}\esc \left(i{\cal L}\hat{\boldsymbol{\cal R}}^T \right)
      \nonumber\\
&\overset{(\ref{rvmic})}{=}-[\hat{\boldsymbol{\omega}}_0]_\times \esc \hat{\mathbb{G}}
      +\hat{\boldsymbol{\cal R}}\esc\left(i{\cal L}\hat{\bf G} \right)\esc \hat{\boldsymbol{\cal R}}^T
+\hat{\mathbb{G}}\esc[\hat{\boldsymbol{\omega}}_0]_\times
  \end{align}
  Reordering terms gives
  \begin{align}
    \label{eq:215}
    i{\cal L}\hat{\mathbb{G}}+[\hat{\boldsymbol{\omega}}_0]_\times \esc \hat{\mathbb{G}}-\hat{\mathbb{G}}\esc[\hat{\boldsymbol{\omega}}_0]_\times
    &=\hat{\boldsymbol{\cal R}}\esc\left(i{\cal L}\hat{\bf G} \right)\esc \hat{\boldsymbol{\cal R}}^T
  \end{align}
which in matrix form becomes
  \begin{align}
    \label{eq:219}
    \begin{pmatrix}
      \hat{\Pi}_x
      & (\hat{M}_1-\hat{M}_2)\hat{\boldsymbol{\omega}}_{0}^z
      & (\hat{M}_3-\hat{M}_1)\hat{\boldsymbol{\omega}}_{0}^y \\
(\hat{M}_1-\hat{M}_2)\hat{\boldsymbol{\omega}}_{0}^z 
      &       \hat{\Pi}_y
      & (\hat{M}_2-\hat{M}_3)\hat{\boldsymbol{\omega}}_{0}^x \\
(\hat{M}_3-\hat{M}_1)\hat{\boldsymbol{\omega}}_{0}^y 
      &      (\hat{M}_2-\hat{M}_3)\hat{\boldsymbol{\omega}}_{0}^x 
      &  \hat{\Pi}_z
      \end{pmatrix}      
    &=\hat{\boldsymbol{\cal R}}\esc\left(i{\cal L}\hat{\bf G} \right)\esc \hat{\boldsymbol{\cal R}}^T
  \end{align}
  As the  gyration tensor depends  on particle's positions,  the right
  hand side is explicit in particle positions and velocities giving
  \begin{align}
     \hat{\boldsymbol{\cal R}}\esc
    \left(i{\cal L}\hat{\bf G} \right)\esc \hat{\boldsymbol{\cal R}}^T
    &=
      \hat{\boldsymbol{\cal R}}\esc
      \frac{1}{4}\sum_im_i(({\bf r}_i-\hat{\bf R})({\bf v}_i-\hat{\bf V})^T+({\bf v}_i-\hat{\bf V})({\bf r}_i-\hat{\bf R})^T)\esc \hat{\boldsymbol{\cal R}}^T
      \nonumber\\
    &=
      \frac{1}{4}\sum_im_i({\bf r}_{0i}{\bf v}^T_{pi}+{\bf v}_{pi}{\bf r}^T_{0i})
  \end{align}
Observe that  the diagonal
  elements   in  (\ref{eq:219})   contain   the  dilational   momentum
 and  the  off-diagonal  terms  contain  the  angular
  velocity  in   the  principal   axis.   Therefore,   the  dilational
  velocities are given from the diagonal components
  \begin{align}
\label{eq:260}     \hat{  \nu}_x
    &\equiv\frac{\hat{{\Pi}}_x}{\hat{M}_1}= \frac{1}{2\hat{M}_1} \sum_i m_i {\bf r}_{0i}^{x}{\bf v}_{pi}^{x}
      \nonumber\\
    \hat{  \nu}_y
    &\equiv\frac{\hat{{\Pi}}_y}{\hat{M}_2}= \frac{1}{2\hat{M}_2} \sum_i m_i {\bf r}_{0i}^{y}{\bf v}_{pi}^{y}
      \nonumber\\
    \hat{  \nu}_z
    &\equiv\frac{\hat{{\Pi}}_z}{\hat{M}_3}=  \frac{1}{2\hat{M}_3} \sum_i m_i {\bf r}_{0i}^{z}{\bf v}_{pi}^{z}
  \end{align}
  From (\ref{eq:219}), the off diagonal terms give the angular velocity as
  \begin{align}
     \label{eq:275}
    \hat{\boldsymbol{\omega}}_0^x&=                                    \frac{1}{4\left(\hat{M}_2 - \hat{M}_3\right)}
                                   \sum_i m_i ({\bf v}_{pi}^y {\bf r}_{0i}^z + {\bf r}_{0i}^y {\bf v}_{pi}^z)
                                   =\frac{1}{\left(\hat{I}_3 - \hat{I}_2\right)}
                                   \sum_i m_i({\bf v}_{pi}^y {\bf r}_{0i}^z + {\bf r}_{0i}^y {\bf v}_{pi}^z)
                                   \nonumber\\
    \hat{\boldsymbol{\omega}}_0^y &=
                                    \frac{1}{4\left(\hat{M}_3 - \hat{M}_1\right)}
                                    \sum_i m_i({\bf v}_{pi}^x {\bf r}_{0i}^z + {\bf r}_{0i}^x {\bf v}_{pi}^z)
                                    =\frac{1}{\left(\hat{I}_1 - \hat{I}_3\right)}
                                    \sum_i m_i({\bf v}_{pi}^x {\bf r}_{0i}^z+{\bf r}_{0i}^x {\bf v}_{pi}^z)
\nonumber\\
\hat{\boldsymbol{\omega}}_0^z &=
\frac{1}{4\left(\hat{M}_1-\hat{M}_2\right)}
\sum_i m_i({\bf v}_{pi}^x {\bf r}_{0i}^y + {\bf r}_{0i}^x {\bf v}_{pi}^y)
=\frac{1}{\left(\hat{I}_2 - \hat{I}_1\right)}
\sum_i m_i({\bf v}_{pi}^x {\bf r}_{0i}^y + {\bf r}_{0i}^x {\bf v}_{pi}^y)
\end{align}
Equations  (\ref{eq:260})  and   (\ref{eq:275})  give  the  dilational
momentum  and angular  velocity in  terms  of the  coordinates of  the
particles  and  the principal  moments  and  orientation (through  the
rotation matrix).   As the latter  are CG variables their  presence is
easy to deal with inside conditional expectations.

\subsection{First derivatives of orientation  and central moments}
\label{App:DerOr}
{We seek to evaluate  $\frac{\partial\hat{\boldsymbol{\Lambda}}}{\partial {\bf r}_i}$ and $\frac{\partial\hat{\bf M}}{\partial {\bf r}_i}$.}
In  principle both  the  orientation and  principal  moments are  very
complicated functions of the coordinates  of the particles and a brute
force calculation  seems not  feasible.  However,  a couple  of tricks
allow  us to  find the  explicit dependence  for derivatives  of these
quantities.          From the diagonalization of the gyration tensor,
$    \hat{\bf G}=e^{[\hat{\boldsymbol{\Lambda}}]_\times}\esc
\hat{\mathbb{G}}\esc e^{-[\hat{\boldsymbol{\Lambda}}]_\times}$
      we        know
$\hat{\bf G}(z)={\bf G}(\hat{\boldsymbol{\Lambda}}(z),\hat{\bf M}(z))$
so that using the chain rule gives
\begin{align}
  \label{eq:251}
  \frac{\partial \hat{\bf G}_{\alpha\beta}}{\partial {\bf r}^\gamma_i}
  &=
    \frac{\partial {\bf G}_{\alpha\beta}}{\partial {\Lambda}_\mu}
    \frac{\partial \hat{\Lambda}_\mu}{\partial {\bf r}^\gamma_i}
    +\frac{\partial {\bf G}_{\alpha\beta}}{\partial {M}_\mu}
    \frac{\partial \hat{M}_\mu}{\partial {\bf r}^\gamma_i}
\end{align}
The derivatives of $\hat{\bf G}$ can be found explicitly, so this equation can be solved for 
$    \frac{\partial \hat{\Lambda}_\mu}{\partial {\bf r}^\gamma_i}
,    \frac{\partial \hat{M}_\mu}{\partial {\bf r}^\gamma_i}$.
The derivative in the first term is
\begin{align}
  \label{eq:169b}
    \frac{\partial{\bf G}_{\alpha\beta}}{\partial{\Lambda}_\mu}
  &= \frac{\partial}{\partial{\Lambda}_\mu}
    \left([e^{[\boldsymbol{\Lambda}]_\times}]_{\alpha\alpha'}
    \mathbb{G}^{\alpha'\beta'}[e^{-[\boldsymbol{\Lambda}]_\times}]_{\beta'\beta}\right)
    \nonumber\\
  &=\left(\frac{\partial[e^{[\boldsymbol{\Lambda}]_\times}]_{\alpha\alpha'}}{\partial{\Lambda}_\mu}\right)
    \mathbb{G}^{\alpha'\beta'}[e^{-[\boldsymbol{\Lambda}]_\times}]_{\beta'\beta}
    +[e^{[\boldsymbol{\Lambda}]_\times}]_{\alpha\alpha'}
\mathbb{G}^{\alpha'\beta'}
    \left(\frac{\partial[e^{-[\boldsymbol{\Lambda}]_\times}]_{\beta'\beta}}
    {\partial{\Lambda}_\mu}\right)
    \nonumber\\
    &=\frac{\partial[e^{[\boldsymbol{\Lambda}]_\times}]_{\alpha\alpha'}}{\partial{\Lambda}_\mu}
    \left[    e^{-[\boldsymbol{\Lambda}]_\times}\right]_{\alpha'\beta'}{\bf G}_{\beta'\beta}
    +{\bf G}_{\alpha\alpha'}
    \left[e^{[\boldsymbol{\Lambda}]_\times}\right]_{\alpha'\beta'}
    \frac{\partial[e^{-[\boldsymbol{\Lambda}]_\times}]_{\beta'\beta}}{\partial{\Lambda}_\mu}
\end{align}
We may now use the last  two identities in (\ref{eq:183}) to write (\ref{eq:169b}) in the form
\begin{align}
  \label{eq:108b}
  \frac{\partial{\bf G}_{\alpha\beta}}{\partial{\Lambda}_\mu}
  &=\left[\epsilon_{\alpha\gamma'\beta'}{\bf G}_{\beta'\beta}
    +{\bf G}_{\alpha\alpha'}\epsilon_{\beta\gamma'\alpha'}\right]{B}^{-1}_{\gamma'\mu}
\end{align}
The  derivative in the second term is
\begin{align}
  \label{eq:31}
  \frac{\partial {\bf G}_{\alpha\beta}}{\partial{M}_\mu}
  &=
    \left[e^{[\boldsymbol{\Lambda}]_\times}\right]_{\alpha\alpha'}
    \frac{\partial \mathbb{G}_{\alpha'\beta'}}{\partial{M}_\mu}
\left[e^{-[\boldsymbol{\Lambda}]_\times}\right]_{\beta'\beta}
=\left[e^{[\boldsymbol{\Lambda}]_\times}\right]_{\alpha\underline{\mu}}
\left[e^{-[\boldsymbol{\Lambda}]_\times}\right]_{\underline{\mu}\beta}
\end{align}
A simple way to solve the  system (\ref{eq:251}) is by transforming to
the principal axis system by multiplying both sides from the left with
$e^{-[\hat{\boldsymbol{\Lambda}}]_\times}$  and  from the  right  with
$e^{[\hat{\boldsymbol{\Lambda}}]_\times}$ to give
\begin{align}
  \label{eq:333}
    \left[e^{-[\hat{\boldsymbol{\Lambda}}]_\times}\right]_{\alpha\alpha'}    \frac{\partial \hat{\bf G}_{\alpha'\beta'}}{\partial {\bf r}^\gamma_i}    \left[e^{[\hat{\boldsymbol{\Lambda}}]_\times}\right]_{\beta'\beta}
  &=
    \left[e^{-[\hat{\boldsymbol{\Lambda}}]_\times}\right]_{\alpha\alpha'}    \frac{\partial {\bf G}_{\alpha'\beta'}}{\partial {\Lambda}_\mu}    \left[e^{[\hat{\boldsymbol{\Lambda}}]_\times}\right]_{\beta}
    \frac{\partial \hat{\Lambda}_\mu}{\partial {\bf r}^\gamma_i}
    \nonumber\\
  &+    \left[e^{-[\hat{\boldsymbol{\Lambda}}]_\times}\right]_{\alpha\alpha'}
    \frac{\partial {\bf G}_{\alpha'\beta'}}{\partial {M}_\mu}    \left[e^{[\hat{\boldsymbol{\Lambda}}]_\times}\right]_{\beta'\beta}
    \frac{\partial \hat{M}_\mu}{\partial {\bf r}^\gamma_i}
\end{align}
Let us compute each term in this equation. Using (\ref{eq:108b}) the first term is
\begin{align}
  \label{eq:334}
& \left[e^{-[\hat{\boldsymbol{\Lambda}}]_\times}\right]_{\alpha\alpha'}    \frac{\partial {\bf G}_{\alpha'\beta'}}{\partial {\Lambda}_\mu}    \left[e^{[\hat{\boldsymbol{\Lambda}}]_\times}\right]_{\beta'\beta}
    \nonumber\\
&\overset{(\ref{eq:108b})}{=}
    \left[e^{-[\hat{\boldsymbol{\Lambda}}]_\times}\right]_{\alpha\alpha'}
    \left[\epsilon_{\alpha'\gamma'\beta''}{\bf G}_{\beta''\beta'}
    +    {\bf G}_{\alpha'\alpha''}\epsilon_{\beta'\gamma'\alpha''}\right]
    B^{-1}_{\gamma'\mu}
    \left[e^{[\hat{\boldsymbol{\Lambda}}]_\times}\right]_{\beta'\beta}
    \nonumber\\
  &=
     \left[e^{-[\hat{\boldsymbol{\Lambda}}]_\times}\right]_{\alpha\alpha'}
   \epsilon_{\alpha'\gamma'\beta''}
    {\bf G}_{\beta''\beta'}\left[e^{[\hat{\boldsymbol{\Lambda}}]_\times}\right]_{\beta'\beta}
    B^{-1}_{\gamma'\mu}
  + \left[e^{-[\hat{\boldsymbol{\Lambda}}]_\times}\right]_{\alpha\alpha'}
     {\bf G}_{\alpha'\alpha''}\epsilon_{\beta'\gamma'\alpha''}
     \left[e^{[\hat{\boldsymbol{\Lambda}}]_\times}\right]_{\beta'\beta}
    B^{-1}_{\gamma'\mu}
    \nonumber\\
&=
     \left[e^{-[\hat{\boldsymbol{\Lambda}}]_\times}\right]_{\alpha\alpha'}
     \epsilon_{\alpha'\gamma'\beta''}
     \left[e^{[\hat{\boldsymbol{\Lambda}}]_\times}\right]_{\beta''\beta'}
     \hat{\mathbb{G}}^{\beta'\beta}
     {B}^{-1}_{\gamma'\mu}
     +\hat{\mathbb{G}}^{\alpha\alpha'}     \left[e^{-[\hat{\boldsymbol{\Lambda}}]_\times}\right]_{\alpha'\alpha''}
     \epsilon_{\beta'\gamma'\alpha''}
     \left[e^{[\hat{\boldsymbol{\Lambda}}]_\times}\right]_{\beta'\beta}
     {B}^{-1}_{\gamma'\mu}
    \nonumber\\
&\overset{(\ref{eq:98})}{=}
      \epsilon_{\alpha\gamma''\beta''}
      \hat{\mathbb{G}}^{\beta''\beta}          \left[e^{-[\hat{\boldsymbol{\Lambda}}]_\times}\right]_{\gamma''\gamma'}
      {B}^{-1}_{\gamma'\mu}
      - \hat{\mathbb{G}}^{\alpha\alpha'}     
      \epsilon_{\alpha'\gamma''\beta}\left[e^{-[\hat{\boldsymbol{\Lambda}}]_\times}\right]_{\gamma''\gamma'}
      {B}^{-1}_{\gamma'\mu}
    \nonumber\\
&=
      \epsilon_{\underline{\alpha}\gamma''\underline{\beta}}
\left ( \hat{M}_{\underline{\beta}} - \hat{M}_{\underline{\alpha}}\right )
\left[e^{-[\hat{\boldsymbol{\Lambda}}]_\times}\right]_{\gamma''\gamma'}
      {B}^{-1}_{\gamma'\mu}
    \end{align}
Using (\ref{eq:31}) the second term is
\begin{align}
  \label{eq:339}
  \left[e^{-[\hat{\boldsymbol{\Lambda}}]_\times}\right]_{\alpha\alpha'}
  \frac{\partial {\bf G}_{\alpha'\beta'}}{\partial {M}_\mu}    \left[e^{[\hat{\boldsymbol{\Lambda}}]_\times}\right]_{\beta'\beta}
  &=
  \left[e^{-[\hat{\boldsymbol{\Lambda}}]_\times}\right]_{\alpha\alpha'}
\left[e^{[\boldsymbol{\Lambda}]_\times}\right]_{\alpha'\underline{\mu}}
\left[e^{-[\boldsymbol{\Lambda}]_\times}\right]_{\underline{\mu}\beta'}
  \left[e^{[\hat{\boldsymbol{\Lambda}}]_\times}\right]_{\beta'\beta}
=\delta_{\alpha\underline{\mu}}\delta_{\beta\underline{\mu}}=
    \delta_{\alpha\beta\mu}
\end{align}
where  the three indexed Kronecker delta  $\delta_{\alpha\beta\mu}$    takes    the    value    1    if
$\alpha=\beta=\mu$  and 0  otherwise. Putting  together (\ref{eq:333})
with (\ref{eq:334}) and (\ref{eq:339}) gives
\begin{align}
  \label{eq:126}
  \left[e^{-[\hat{\boldsymbol{\Lambda}}]_\times}\right]_{\alpha\alpha'}
  \frac{\partial    \hat{\bf G}_{\alpha'\beta'}}{\partial {\bf r}^\gamma_i}
  \left[e^{[\hat{\boldsymbol{\Lambda}}]_\times}\right]_{\beta'\beta}
  &=
      \epsilon_{\underline{\alpha}\gamma''\underline{\beta}}
\left ( \hat{M}_{\underline{\beta}} - \hat{M}_{\underline{\alpha}}\right )
    \left[e^{-[\hat{\boldsymbol{\Lambda}}]_\times}\right]_{\gamma''\gamma'}
      {B}^{-1}_{\gamma'\mu}  \frac{\partial \hat{\Lambda}_\mu}{\partial {\bf r}^\gamma_i}
    +\delta_{\alpha\beta\mu}
    \frac{\partial \hat{M}_\mu}{\partial {\bf r}^\gamma_i}
\end{align}
We now  use the  definition of the  gyration tensor in terms  of particle
positions to
express  the left  hand side  of (\ref{eq:126}),
\begin{align}
  \label{eq:286}
       \frac{\partial    \hat{\bf G}_{\alpha'\beta'}}{\partial {\bf r}^\gamma_i}
       &=    
         \sum_j m_j\frac{\partial    }{\partial {\bf r}^\gamma_i}({\bf r}_j^{\alpha'}-\hat{\bf R}^{\alpha'})({\bf r}_j^{\beta'}-\hat{\bf R}^{\beta'})
+            
         \sum_j m_j({\bf r}_j^{\alpha'}-\hat{\bf R}^{\alpha'})\frac{\partial    }{\partial {\bf r}^\gamma_i}({\bf r}_j^{\beta'}-\hat{\bf R}^{\beta'})
         \nonumber\\
            &=    
         \sum_j m_j\left(\delta_{ij}\delta_{\gamma\alpha'}-\frac{m_i}{M}\delta_{\gamma\alpha'}\right)({\bf r}_j^{\beta'}-\hat{\bf R}^{\beta'})
+            
              \sum_j m_j({\bf r}_j^{\alpha'}-\hat{\bf R}^{\alpha'})
              \left(\delta_{ij}\delta_{\gamma\beta'}-\frac{m_i}{M}\delta_{\gamma\beta'}\right)
         \nonumber\\
            &= m_i\delta_{\gamma\alpha'}({\bf r}_i^{\beta'}-\hat{\bf R}^{\beta'})+m_i({\bf r}_i^{\alpha'}-\hat{\bf R}^{\alpha'})\delta_{\gamma\beta'}
     \end{align}
Therefore
\begin{align}
  \label{eq:328}
    \left[e^{-[\hat{\boldsymbol{\Lambda}}]_\times}\right]_{{\alpha}\alpha'}
    \frac{\partial    \hat{\bf G}_{\alpha'\beta'}}{\partial {\bf r}^\gamma_i}
    &\left[e^{[\hat{\boldsymbol{\Lambda}}]_\times}\right]_{\beta'\beta}
      \nonumber\\
  &={\black  \frac{1}{4}}\left[e^{-[\hat{\boldsymbol{\Lambda}}]_\times}\right]_{{\alpha}\gamma}
    m_i     \left({\bf r}^{\beta'}_{i}-  \hat{\bf R}^{\beta'}\right)
    \left[e^{-[\hat{\boldsymbol{\Lambda}}]_\times}\right]_{\beta\beta'}
    +  \frac{1}{4}\left[e^{-[\hat{\boldsymbol{\Lambda}}]_\times}\right]_{{\alpha}\alpha'}
    m_i     \left({\bf r}^{\alpha'}_{i}-  \hat{\bf R}^{\alpha'}\right)
    \left[e^{-[\hat{\boldsymbol{\Lambda}}]_\times}\right]_{\beta\gamma}
    \nonumber\\
  &=\frac{1}{4}\left[e^{-[\hat{\boldsymbol{\Lambda}}]_\times}\right]_{{\alpha}\gamma}
    m_i    {\bf r}^{\beta}_{0i}
    +  \frac{1}{4}
    \left[e^{-[\hat{\boldsymbol{\Lambda}}]_\times}\right]_{\beta\gamma}    m_i    {\bf r}^{\alpha}_{0i}  
\end{align}
where the last line shows the expression in the principal axis frame.  Putting this in (\ref{eq:126}) gives
\begin{align}
  \label{eq:440}
      \epsilon_{\underline{\alpha}\gamma''\underline{\beta}}
\left ( \hat{M}_{\underline{\beta}} - \hat{M}_{\underline{\alpha}}\right )
    \left[e^{-[\hat{\boldsymbol{\Lambda}}]_\times}\right]_{\gamma''\gamma'}
      {B}^{-1}_{\gamma'\mu}  \frac{\partial \hat{\Lambda}_\mu}{\partial {\bf r}^\gamma_i}
    +
    \delta_{\alpha\beta\mu}
  \frac{\partial \hat{M}_\mu}{\partial {\bf r}^\gamma_i}
  &= \frac{1}{4}\left[e^{-[\hat{\boldsymbol{\Lambda}}]_\times}\right]_{{\alpha}\gamma}
    m_i    {\bf r}^{\beta}_{0i}
    +  \frac{1}{4}
  \left[e^{-[\hat{\boldsymbol{\Lambda}}]_\times}\right]_{\beta\gamma}    m_i    {\bf r}^{\alpha}_{0i}
\end{align}
This  is  the  original  system of  equations  (\ref{eq:251})  for  the
unknowns
$    \frac{\partial   \hat{\Lambda}_\mu}{\partial    {\bf
    r}^\gamma_i},               \frac{\partial               \hat{M}_{\alpha}}{\partial  {\bf  r}^\gamma_i}$ written in  the
principal axis system.  In this  reference frame, the system separates
nicely in  its diagonal  and off-diagonal elements,  and allows  us to
find the solution trivially.  When $\alpha=\beta$ we have
  \begin{align}
    \label{eq:425}
    \frac{\partial \hat{M}_\alpha}{\partial {\bf r}^\gamma_i}
    &={\black  \frac{1}{2}}
    \left[e^{-[\hat{\boldsymbol{\Lambda}}]_\times}\right]_{\underline{\alpha}\gamma}
    \left[e^{-[\hat{\boldsymbol{\Lambda}}]_\times}\right]_{\underline{\alpha}\beta'}
    m_i     \left({\bf r}^{\beta'}_{i}-  \hat{\bf R}^{\beta'}\right)
=\frac{1}{2}
\left[e^{-[\hat{\boldsymbol{\Lambda}}]_\times}\right]_{\underline{\alpha}\gamma}
m_i {\bf r}^{\underline{\alpha}}_{0i}
  \end{align}
This result can also be obtained from the diagonalization of the gyration tensor,  by noting $\mathbb{G}$ is
just $\bf G$ expressed in the principal axis system so that
\begin{align}
\label{eq:284}
\hat{\mathbb{G}}^{\alpha\beta}=\delta_{\alpha\beta\gamma}\hat{M}_\gamma=
\delta_{\underline{\alpha}\beta}\hat{M}_{\underline{\alpha}}
=\frac{1}{4}\sum_i m_i {\bf r}_{0i}^{\alpha}{\bf r}_{0i}^{\beta}
\end{align}
and $\frac{\partial \hat{M}_\alpha}{\partial {\bf r}^\gamma_i}=
\left[e^{-[\hat{\boldsymbol{\Lambda}}]_\times}\right]_{\alpha'\gamma}
\frac{\partial \hat{M}_\alpha}{\partial {\bf r}^{\alpha'}_{0i}}$.
When $\alpha\neq\beta$, the factor of
$\frac{\partial\hat{M}_\mu}{\partial{\bf  r}^\gamma_i}$
in (\ref{eq:440}) vanishes, yielding an equation involving only
$\frac{\partial\hat{\Lambda}_\mu}{\partial{\bf r}^\gamma_i}$ which
in principle can be used to isolate the latter. We do not pursue
this calculation, because the only place where it would be needed is
in the calculation of $\hat{\bf L}^{\mu\nu}$ in (\ref{eq:L}), but this
matrix is computed using a direct argument.

For future reference, we note that (\ref{eq:425}) implies
\begin{align}
  \label{eq:256}
  \sum_i     \frac{\partial \hat{M}_\alpha}{\partial {\bf r}^\mu_i}    ({\bf r}^\nu_i-\hat{\bf R}^\nu)
  &={\black  \frac{1}{2}}
\sum_i
    \left[e^{-[\hat{\boldsymbol{\Lambda}}]_\times}\right]_{\underline{\alpha}\mu}
    \left[e^{-[\hat{\boldsymbol{\Lambda}}]_\times}\right]_{\underline{\alpha}\beta'}
    m_i     \left({\bf r}^{\beta'}_{i}-  \hat{\bf R}^{\beta'}\right)    ({\bf r}^\nu_i-\hat{\bf R}^\nu)
=
    2\left[e^{-[\hat{\boldsymbol{\Lambda}}]_\times}\right]_{\underline{\alpha}\mu}
    \left[e^{-[\hat{\boldsymbol{\Lambda}}]_\times}\right]_{\underline{\alpha}\beta'}
    \hat{\bf G}_{\beta'\nu}
\end{align}

\subsection{The dilational force}
\label{App:Krp}
The dilation momentum can be determined from the 
derivative of (\ref{eq:284}) giving
\begin{align}
\label{eq:78}
\hat{{\Pi}}_\alpha & =\frac{d}{dt}\hat{M}_\alpha = \frac{1}{2}
\sum_i m_i {\bf r}_{0i}^{\underline{\alpha}}{\bf v}_{0i}^{\underline{\alpha}}
\end{align}
The dilation force is the time derivative of the
dilation momentum, so using (\ref{eq:78}) gives
\begin{align}
\label{eq:42}
\hat{\cal K}_\alpha& =\frac{d}{dt}\hat{  {\Pi}}_\alpha 
= \frac{1}{2} \sum_i m_i\left (
 {\bf v}_{0i}^{\underline{\alpha}}{\bf v}_{0i}^{\underline{\alpha}} +
{\bf r}_{0i}^{\underline{\alpha}}{\bf a}_{0i}^{\underline{\alpha}} \right )
\end{align}
where  ${\bf a}_{0i}$  is  the  acceleration of  particle  $i$ in  the
principal axis  frame, and is  calculated from the time  derivative of
${\bf  v}_{0i}$  in (\ref{rvmic})  as  (remembering  $\hat{\bf V}$  is
constant)
\begin{align}
\label{eq:42a}
{\bf a}_{0i}&=\frac{d}{dt}{\bf v}_{0i}\overset{(\ref{eq:601})}{=}
\frac{d}{dt}\left ( e^{-[\hat{\boldsymbol{\Lambda}}]_\times}\esc
\left( {\bf v}_i-\hat{\bf V}\right )- [\hat{\boldsymbol{\omega}}_0]_\times
\esc {\bf r}_{0i}\right )
\nonumber\\
&= \left ( \frac{d}{dt} e^{-[\hat{\boldsymbol{\Lambda}}]_\times}\right )\esc
\left( {\bf v}_i-\hat{\bf V}\right ) +
 e^{-[\hat{\boldsymbol{\Lambda}}]_\times}\esc\frac{d{\bf v}_i}{dt} -
\left [\frac{d\hat{\boldsymbol{\omega}}_0}{dt}\right ]_\times\esc{\bf r}_{0i}
- [\hat{\boldsymbol{\omega}}_0]_\times\esc{\bf v}_{0i}
\nonumber\\
&\overset{(\ref{rvmic})(\ref{eq:601})}{=}
-[\hat{\boldsymbol{\omega}}_0]_\times\esc{\bf v}_{pi} + \frac{1}{m_i}{\bf F}_{0i} -
\left [\frac{d\hat{\boldsymbol{\omega}}_0}{dt}\right ]_\times\esc{\bf r}_{0i}
- [\hat{\boldsymbol{\omega}}_0]_\times\esc\left ( {\bf v}_{pi} -
[\hat{\boldsymbol{\omega}}_0]_\times\esc{\bf r}_{0i}\right )
\nonumber\\
&=\frac{1}{m_i}{\bf F}_{0i} - 2 [\hat{\boldsymbol{\omega}}_0]_\times\esc{\bf v}_{pi} +
 [\hat{\boldsymbol{\omega}}_0]_\times\esc[\hat{\boldsymbol{\omega}}_0]_\times
\esc{\bf r}_{0i} -
\left [\frac{d\hat{\boldsymbol{\omega}}_0}{dt}\right ]_\times\esc{\bf r}_{0i}
\nonumber\\
&\overset{(\ref{rvmic})(\ref{eq:601})}{=}
\frac{1}{m_i}{\bf F}_{0i} - 2 [\hat{\boldsymbol{\omega}}_0]_\times\esc{\bf v}_{0i} -
 [\hat{\boldsymbol{\omega}}_0]_\times\esc[\hat{\boldsymbol{\omega}}_0]_\times
\esc{\bf r}_{0i} -
\left [\frac{d\hat{\boldsymbol{\omega}}_0}{dt}\right ]_\times\esc{\bf r}_{0i}
\end{align}
where from (\ref{rvmic}) we have
${\bf v}_0 = {\bf v}_p-[\hat{\boldsymbol{\omega}}_0]_\times\esc{\bf r}_0$ and
\begin{align}
\label{eq:42d}
{\bf F}_{0i} = e^{-[\hat{\boldsymbol{\Lambda}}]_\times}\esc{\bf F}_i=
e^{-[\hat{\boldsymbol{\Lambda}}]_\times}\esc\frac{d{\bf p}_i}{dt}
=-e^{-[\hat{\boldsymbol{\Lambda}}]_\times}\esc\frac{\partial \Phi(r)}
{\partial{\bf r}_i}
\end{align}
is the force on particle $i$,  due to the other particles, transformed
to the principal axis frame.  From
(\ref{eq:42}),   two   main   effects   drive  the   change   in   the
dilational moments.   The last term  shows forces in  the radial
direction can increase or decrease  the moments, and these forces (see
(\ref{eq:42a}))  arise  from  intermolecular  interactions  among  the
particles, Coriolis and centrifugal effects, and angular acceleration.
However, even  in the absence  of interactions and for  a non-rotating
frame, the first term of (\ref{eq:42}) shows moments will increase due
to the velocities  of particles.  This simply  represents the tendency
of a group of moving, non-interacting particles to disperse over time,
thereby leading to increasing  dilational moments.  Thus, a body
with a stable shape is only possible if ${\bf F}_{0i}$ is large enough
to compensate all the terms causing the dilational moments to increase.  Finally,
observe that
\begin{align}
\sum_i m_i {\bf r}^{\underline{\alpha}}_{0i}
\left[\frac{d\hat{\boldsymbol{\omega}}_0}{dt}\right ]_\times^{\underline{\alpha}\alpha'}
{\bf r}^{\alpha'}_{0i}\overset{(\ref{eq:284})}{=}
\left[\frac{d\hat{\boldsymbol{\omega}}_0}{dt}\right ]_\times^{\underline{\alpha}\alpha'}
\hat{\mathbb{G}}^{\underline{\alpha}\alpha'}=
\left[\frac{d\hat{\boldsymbol{\omega}}_0}{dt}\right ]_\times^{\underline{\alpha}\underline{\alpha}}
\hat{M}_{\underline{\alpha}}=0
\end{align}
since  the diagonal  elements of  the cross  product matrix  are zero.
This shows the angular acceleration  in (\ref{eq:42a}) does not affect
the value of $\hat{\cal K}_\alpha$.

\subsection{Some matrices involving first derivatives with respect particle's  positions}
\label{App:TH}
In  this section  we compute  explicitly the  functional form  of the
following matrices  that involve first derivatives  of orientation and
central moments with respect to particle positions

\begin{align}
  \label{eq:L}  \hat{\bf L}^{\mu\nu}
  &\equiv
    {B}^{-1}_{\mu\mu'} 
    {B}^{-1}_{\nu\nu'}  \sum_i\frac{1}{m_i}
    \frac{\partial \hat{\Lambda}_{\mu'}}{\partial {\bf r}^\gamma_i}
    \frac{\partial \hat{\Lambda}_{\nu'}}{\partial {\bf r}^\gamma_i}
\\\label{eq:H}  \hat{\bf H}^{\mu\nu}
  &\equiv\sum_i\frac{1}{m_i}
  \frac{\partial \hat{\Lambda}_\mu}{\partial {\bf r}_i^\gamma}
  \frac{\partial \hat{M}_\nu}{\partial {\bf r}_i^\gamma}
  \\
  \label{eq:T}
    \hat{\bf T}^{\mu\nu}
  &\equiv\sum_i\frac{1}{m_i}\frac{\partial \hat{M}_\mu}{\partial {\bf r}_i^\gamma}
    \frac{\partial \hat{M}_\nu}{\partial {\bf r}_i^\gamma}
\end{align}
To get the explicit form, consider (\ref{eq:440}) with the following two suitable renaming of indices
  \begin{align}
    \label{eq:393}
      \epsilon_{\underline{\alpha}\mu''\underline{\beta}}
\left ( \hat{M}_{\underline{\beta}} - \hat{M}_{\underline{\alpha}}\right )
    \left[e^{-[\hat{\boldsymbol{\Lambda}}]_\times}\right]_{\mu''\mu'}
      {B}^{-1}_{\mu'\mu}  \frac{\partial \hat{\Lambda}_\mu}{\partial {\bf r}^\gamma_i}
    +
\delta_{\alpha\beta\mu}
    \frac{\partial \hat{M}_\mu}{\partial {\bf r}^\gamma_i}
  &=   \frac{1}{4}\left[e^{-[\hat{\boldsymbol{\Lambda}}]_\times}\right]_{{\alpha}\gamma}
    m_i    {\bf r}^{\beta}_{0i}
    +  \frac{1}{4}
  \left[e^{-[\hat{\boldsymbol{\Lambda}}]_\times}\right]_{\beta\gamma}    m_i    {\bf r}^{\alpha}_{0i}
    \nonumber\\
      \epsilon_{\underline{\alpha'}\nu''\underline{\beta'}}
\left ( \hat{M}_{\underline{\beta'}} - \hat{M}_{\underline{\alpha'}}\right )
    \left[e^{-[\hat{\boldsymbol{\Lambda}}]_\times}\right]_{\nu''\nu'}
      {B}^{-1}_{\nu'\nu}  \frac{\partial \hat{\Lambda}_\nu}{\partial {\bf r}^\gamma_i}
    +
\delta_{\alpha'\beta'\nu}
    \frac{\partial \hat{M}_\nu}{\partial {\bf r}^\gamma_i}
  &=     \frac{1}{4}\left[e^{-[\hat{\boldsymbol{\Lambda}}]_\times}\right]_{\alpha'\gamma}
    m_i    {\bf r}^{\beta'}_{0i}
    +  \frac{1}{4}
  \left[e^{-[\hat{\boldsymbol{\Lambda}}]_\times}\right]_{\beta'\gamma}    m_i    {\bf r}^{\alpha'}_{0i}
  \end{align}
  Multiply together  each side  of each  equation, sum  over $\gamma$,
  divide by $m_i$, and sum over all particles $i$ to get
  \begin{align}
    \label{eq:395}
&\sum_i\frac{1}{m_i}
      \left[
      \epsilon_{\underline{\alpha}\mu''\underline{\beta}}
\left ( \hat{M}_{\underline{\beta}} - \hat{M}_{\underline{\alpha}}\right )
    \left[e^{-[\hat{\boldsymbol{\Lambda}}]_\times}\right]_{\mu''\mu'}
      {B}^{-1}_{\mu'\mu}  \frac{\partial \hat{\Lambda}_\mu}{\partial {\bf r}^\gamma_i}
    +
\delta_{\alpha\beta\mu}
    \frac{\partial \hat{M}_\mu}{\partial {\bf r}^\gamma_i}\right]
      \nonumber\\
  &\times\left[
      \epsilon_{\underline{\alpha'}\nu''\underline{\beta'}}
\left ( \hat{M}_{\underline{\beta'}} - \hat{M}_{\underline{\alpha'}}\right )
    \left[e^{-[\hat{\boldsymbol{\Lambda}}]_\times}\right]_{\nu''\nu'}
      {B}^{-1}_{\nu'\nu}  \frac{\partial \hat{\Lambda}_\nu}{\partial {\bf r}^\gamma_i}
    +
    \delta_{\alpha'\beta'\nu}
    \frac{\partial \hat{M}_\nu}{\partial {\bf r}^\gamma_i}\right]&           \nonumber\\
    &=\sum_i \frac{1}{m_i}   \left[   \frac{1}{4}\left[e^{-[\hat{\boldsymbol{\Lambda}}]_\times}\right]_{{\alpha}\gamma}
    m_i    {\bf r}^{\beta}_{0i}
    +  \frac{1}{4}
  \left[e^{-[\hat{\boldsymbol{\Lambda}}]_\times}\right]_{\beta\gamma}    m_i    {\bf r}^{\alpha}_{0i}
\right]\left[     \frac{1}{4}\left[e^{-[\hat{\boldsymbol{\Lambda}}]_\times}\right]_{\alpha'\gamma}
    m_i    {\bf r}^{\beta'}_{0i}
    +  \frac{1}{4}
  \left[e^{-[\hat{\boldsymbol{\Lambda}}]_\times}\right]_{\beta'\gamma}    m_i    {\bf r}^{\alpha'}_{0i}
\right]          
  \end{align}
Now use (\ref{eq:L}), (\ref{eq:H}), (\ref{eq:T}) and (\ref{eq:284}) to
simplify this as
\begin{align}
  \label{eq:398}
&\epsilon_{\underline{\alpha}\mu''\underline{\beta}}
\left ( \hat{M}_{\underline{\beta}} - \hat{M}_{\underline{\alpha}}
\right )
\epsilon_{\underline{\alpha'}\nu''\underline{\beta'}}
\left ( \hat{M}_{\underline{\beta'}} - \hat{M}_{\underline{\alpha'}}
\right )
    \left[e^{-[\hat{\boldsymbol{\Lambda}}]_\times}\right]_{\mu''\mu'}
    \left[e^{-[\hat{\boldsymbol{\Lambda}}]_\times}\right]_{\nu''\nu'}
    \hat{\bf L}^{\mu'\nu'}
    \nonumber\\
  &+ \epsilon_{\underline{\alpha}\mu''\underline{\beta}}
\left ( \hat{M}_{\underline{\beta}} - \hat{M}_{\underline{\alpha}}\right )
    \left[e^{-[\hat{\boldsymbol{\Lambda}}]_\times}\right]_{\mu''\mu'}
      {B}^{-1}_{\mu'\mu} \delta_{\alpha'\beta'\nu} \hat{\bf H}^{\mu\nu}
      \nonumber\\
  &+ \epsilon_{\underline{\alpha'}\nu''\underline{\beta'}}
\left ( \hat{M}_{\underline{\beta'}} - \hat{M}_{\underline{\alpha'}}\right )
    \left[e^{-[\hat{\boldsymbol{\Lambda}}]_\times}\right]_{\nu''\nu'}
 {B}^{-1}_{\nu'\nu} \delta_{\alpha\beta\mu} \hat{\bf H}^{\nu\mu}
+
    \delta_{\alpha\beta\mu}
    \delta_{\alpha'\beta'\nu}\hat{\bf T}^{\mu\nu}
                   \nonumber\\
  &=   \frac{1}{4}  \left(
\delta_{\alpha\alpha'}\delta_{\underline{\beta}\beta'}
\hat{M}_{\underline{\beta}}
+  2 \delta_{\alpha\beta'}\delta_{\alpha'\underline{\beta}}
\hat{M}_{\underline{\beta}}
+  \delta_{\beta\beta'}\delta_{\underline{\alpha}\alpha'} 
\hat{M}_{\underline{\alpha}}
  \right)
\end{align}
Quite fortunately, this apparently complicated system of equations  is,  in
fact, trivial, as can be seen by looking at particular components.  If
$\alpha=\beta$ and $\alpha'=\beta'$ then we have
\begin{align}
  \label{eq:399}
\delta_{\alpha\alpha\mu} \delta_{\alpha'\alpha'\nu} \hat{\bf T}^{\mu\nu}
  &=\hat{\bf T}^{\alpha\alpha'}=
\frac{1}{4}\left (
\delta_{\underline{\alpha}\alpha'}\hat{M}_{\underline{\alpha}}
+  2 \delta_{\underline{\alpha}\alpha'}\hat{M}_{\underline{\alpha}}
+  \delta_{\underline{\alpha}\alpha'}\hat{M}_{\underline{\alpha}}\right )
= \delta_{\underline{\alpha}\alpha'}\hat{M}_{\underline{\alpha}}
\end{align}
so that
\begin{align}
  \label{TG}
  \hat{\bf T}&=
               \left(
               \begin{array}{ccc}
                 \hat{M}_1
                 & 0
                 & 0 \\
                 0
                 &  \hat{M}_2
                 & 0  \\
                 0 & 
                   0 &  \hat{M}_3 \\
               \end{array}
  \right)=\hat{\mathbb{G}}
\end{align}
Setting $\alpha=\beta$ and $\alpha'\neq\beta'$ in (\ref{eq:398}) gives
\begin{align}
  \label{eq:227}
\epsilon_{\underline{\alpha'}\nu''\underline{\beta'}}
\left ( \hat{M}_{\underline{\beta'}} - \hat{M}_{\underline{\alpha'}}\right )
    \left[e^{-[\hat{\boldsymbol{\Lambda}}]_\times}\right]_{\nu''\nu'}
 {B}^{-1}_{\nu'\nu} \hat{\bf H}^{\nu\alpha}=0
\end{align}
which, provided the moments are different, implies  $\hat{\bf H}$
vanishes identically, that is
\begin{align}
  \label{eq:449}
    \hat{\bf H}^{\mu\nu}&=0
\end{align}

Finally, 
by suitably introducing
\begin{align}
  \label{eq:225}
  \hat{\mathbb{L}}^{\mu\nu}
  &\equiv
    \left[e^{-[\hat{\boldsymbol{\Lambda}}]_\times}\right]_{\mu\mu'}
    \left[e^{-[\hat{\boldsymbol{\Lambda}}]_\times}\right]_{\nu\nu'}
    \hat{\bf L}^{\mu'\nu'}
\end{align}
then (\ref{eq:398}) becomes for $\alpha\neq\beta$ and $\alpha'\neq\beta'$,  
\begin{align}
\label{eq:400}
\epsilon_{\underline{\alpha}\mu\underline{\beta}}
\left ( \hat{M}_{\underline{\beta}} - \hat{M}_{\underline{\alpha}}
\right )
\epsilon_{\underline{\alpha'}\nu\underline{\beta'}}
\left ( \hat{M}_{\underline{\beta'}} - \hat{M}_{\underline{\alpha'}}
\right )
    \hat{\mathbb{L}}^{\mu\nu}
  =   \frac{1}{4}  \left(
\delta_{\alpha\alpha'}\delta_{\underline{\beta}\beta'}
\hat{M}_{\underline{\beta}}
+  2 \delta_{\alpha\beta'}\delta_{\alpha'\underline{\beta}}
\hat{M}_{\underline{\beta}}
+  \delta_{\beta\beta'}\delta_{\underline{\alpha}\alpha'} 
\hat{M}_{\underline{\alpha}}
  \right)
\end{align}
Provided the moments are different, (\ref{eq:400}) can be solved by taking
different combinations of components and using (\ref{eq:604}) to simplify the
product of Levi-Civita symbols.  For example, setting
$\alpha\neq\alpha'\neq\beta\neq\beta'$ shows that both sides of (\ref{eq:400})
vanish, giving no useful information about $\hat{\mathbb{L}}$.  Setting
$\alpha=\alpha'$, $\beta\neq\beta'$, $\alpha\neq\beta$ and $\alpha'\neq\beta'$
reduces (\ref{eq:400}) to
\begin{align}
-\delta_{\mu\underline{\beta'}}\delta_{\nu\underline{\beta}}
\left ( \hat{M}_{\underline{\beta}} - \hat{M}_{\underline{\alpha}}
\right )
\left ( \hat{M}_{\underline{\beta'}} - \hat{M}_{\underline{\alpha'}}
\right )
    \hat{\mathbb{L}}^{\mu\nu} = 0
\end{align}
giving $\hat{\mathbb{L}}^{\beta'\beta}=0$ when $\beta\neq\beta'$. The
matrix $\hat{\mathbb{L}}$  is a diagonal matrix, hence the voided font used to denote it. Finally,
setting $\alpha=\alpha'$, $\beta=\beta'$, $\alpha\neq\beta$ and
$\alpha'\neq\beta'$ reduces (\ref{eq:400}) to
\begin{align}
\left(\delta_{\mu\nu}-
\delta_{\mu\underline{\beta}}\delta_{\nu\underline{\beta}}-
\delta_{\mu\underline{\alpha}}\delta_{\nu\underline{\alpha}}\right )
\left ( \hat{M}_{\underline{\beta}} - \hat{M}_{\underline{\alpha}}
\right )^2 \hat{\mathbb{L}}^{\mu\nu} = \frac{1}{4}\left (\hat{M}_\beta+
\hat{M}_\alpha\right )
\end{align}
giving the diagonal element of $\hat{\mathbb{L}}$ as
\begin{align}
\hat{\mathbb{L}}^{\mu\mu}
-\hat{\mathbb{L}}^{\underline{\alpha}\underline{\alpha}}
-\hat{\mathbb{L}}^{\underline{\beta}\underline{\beta}}
= \frac{\hat{M}_{\underline{\alpha}}+\hat{M}_{\underline{\beta}}}
{4\left ( \hat{M}_{\underline{\beta}} - \hat{M}_{\underline{\alpha}}
\right )^2}
\end{align}
where summation over $\mu$ is implied but not over $\alpha$ and $\beta$.
Finally, we have
\begin{align}
  \label{eq:229}
  \hat{\mathbb{L}}=
               \left(
               \begin{array}{ccc}
                \frac{\hat{M}_2+\hat{M}_3}{4(\hat{M}_2-\hat{M}_3)^2}
                 & 0
                 & 0 \\
                 0
                 &  \frac{\hat{M}_1+\hat{M}_3}{4(\hat{M}_1-\hat{M}_3)^2}
                 & 0  \\
                 0 & 
                   0 & \frac{\hat{M}_1+\hat{M}_2}{4(\hat{M}_1-\hat{M}_2)^2}  \\
               \end{array}
  \right)
\end{align}

\setcounter{my_equation_counter}{\value{equation}}
\section{Momentum integrals}
\setcounter{equation}{\value{my_equation_counter}}
\label{App:Submanifold}
\subsection{The general formula}
Here consider  the evaluation of the  following ``sectioned Gaussian''
integral
\begin{align}
  \label{eq:134}
  G(x,J)&=\int dy \delta(\Pi y-x) \exp\left\{-\frac{1}{2}y^T Ay+J^Ty\right\}
\end{align}
where              $y\in\mathbb{R}^n$,             $J\in\mathbb{R}^n$,
$A\in\mathbb{R}^{n\times    n}$   is    symmetric   and    invertible,
$x\in\mathbb{R}^m$,   and   $\Pi\in\mathbb{R}^{m\times    n}$   is   a
rectangular matrix.   The integral  is restricted  by the  Dirac delta
function to the hyperplane defined by $\Pi y=x$. By taking derivatives
with  respect   to  $J$   we  will  generate   moments  of   the  form
(\ref{eq:18}).  Note that  in this section, $\Pi$ is  a parameter and
$G$ a function that should not  be confused with the dilation momentum
and gyration tensor which appear with the same symbols elsewhere.

Begin by introducing the Fourier transform of the
Dirac delta function  in (\ref{eq:134})
\begin{align}
  \label{Jeq:116}
G(x,J)
  &=
    \int \frac{d\lambda}{(2\pi)^m} \exp\left\{-i\lambda^T x\right\}
    \int dy \exp\left\{-\frac{1}{2}y^T Ay +J^Ty\right\} 
      \exp\left\{i\lambda^T \Pi y\right\}
      \nonumber\\
&=\int \frac{d\lambda}{(2\pi)^m} \exp\left\{-i\lambda^T x\right\} \tilde{G}(y,J)
\end{align}
where 
\begin{align}
  \label{Jeq:115}
    \tilde{G}(\lambda,J)
  &=  \int dy \exp\left\{-\frac{1}{2}y^T Ay+\left(J+i\Pi^T \lambda\right)^Ty\right\}
\nonumber\\
&
=\int dy \exp\left\{ -\frac{1}{2}
\left ( y-A^{-1}(J+i\Pi^T \lambda)\right)^T A
\left ( y-A^{-1}(J+i\Pi^T \lambda)\right)
+\frac{1}{2} \left ( J+i\Pi^T \lambda\right)^T A^{-1}
\left ( J+i\Pi^T \lambda\right)
\right\}
\nonumber\\
&
=\frac{(2\pi)^{n/2}}{\det A^{1/2}}\exp\left\{
\frac{1}{2}\left(J+ i\Pi^T\lambda\right)^TA^{-1}\left(J+i\Pi^T\lambda\right)
\right\}
\end{align}
noting that $(J+ i\Pi^T\lambda)^Ty=y^T(J+ i\Pi^T\lambda)$ and where the
last line was obtained using
 the Gaussian integral
\begin{align}
\int dy' \exp\left\{-\frac{1}{2}y'^T Ay'\right\}
&=\frac{(2\pi)^{n/2}}{\det A^{1/2}}
\label{Gaussian}
\end{align}
Inserting (\ref{Jeq:115})  in (\ref{Jeq:116}) one gets
\begin{align}
  G(x,J)&=\frac{(2\pi)^{n/2}}{\det A^{1/2}}e^{\frac{1}{2}J^TA^{-1}J}
        \int \frac{d\lambda}{(2\pi)^m} \exp\left\{-i\lambda^T (x-\Pi A^{-1}J) 
-\frac{1}{2}\lambda^T\Pi  A^{-1}\Pi^T\lambda\right\}
        \nonumber\\
&=\frac{(2\pi)^{n/2}}{\det A^{1/2}}e^{\frac{1}{2}J^TA^{-1}J}
\int \frac{d\lambda}{(2\pi)^m} \exp \left\{ -\frac{1}{2}
\left (\lambda + i\overline{A}(x-\Pi A^{-1}J)\right )^T \overline{A}^{-1} 
\left (\lambda + i\overline{A}(x-\Pi A^{-1}J)\right )\right\}
\nonumber\\
&\qquad\qquad\qquad\qquad\times\exp\left\{
-\frac{1}{2}(x-\Pi A^{-1}J)^T \overline{A} (x-\Pi A^{-1}J)
\right\}
        \nonumber\\
  &=\frac{(2\pi)^{n/2}}{\det A^{1/2}}
\frac{\det{\overline{A}}^{1/2}}{(2\pi)^{m/2}}
\exp\left\{-\frac{1}{2}(x-\Pi A^{-1}J)^T \overline{A} (x-\Pi A^{-1}J)+\frac{1}{2}J^TA^{-1}J\right\}
\label{apomJ}
\end{align}
where  $\overline{A}\in\mathbb{R}^{m\times m}$ is symmetric and assumed
invertible, and is given by
\begin{align}
  \overline{A}&\equiv \left[\Pi A^{-1}\Pi^T\right]^{-1}
\label{overA}
\end{align}
The calculation of the Gaussian integrals suggests
the change of variables
\begin{align}
  \label{eq:100}
  y'=y-A^{-1} \Pi^T [\Pi  A^{-1} \Pi^T]^{-1} x
\end{align}
that has the property
\begin{align}
  \label{eq:234}
  \Pi y'=\Pi y- x
\end{align}
and, therefore, 
\begin{align}
  \label{eq:240}
  \delta(\Pi y- x)=\delta(\Pi y')
\end{align}
This change of variables will turn to be instrumental in the calculation of the entropy.
\subsection{The momentum integrals in conditional expectations}
The results  of this  subsection are  used in  the calculation  of the
equilibrium  probability $P^{\rm  rest}_{\cal E}({\bf M},\boldsymbol{\Pi})$
in  (\ref{eq:64b})   and  the  equilibrium
approximation to the angular  diffusion tensor $\boldsymbol{\cal D}_0$.
In  this  subsection we  compute  the  following integrals over momenta
\begin{align}
  G^{(0)}({\bf P},{\bf S},\boldsymbol{\Pi},\beta)
  &=\int dp \Delta(p,{\bf P},{\bf S},\boldsymbol{\Pi},\beta)
    \label{G0}
    \\
{\bf  G}^{(1)}_i({\bf P},{\bf S},\boldsymbol{\Pi},\beta)
  &=\int dp \frac{{\bf p}_{i}}{m_{i}}  \Delta(p,{\bf P},{\bf S},\boldsymbol{\Pi},\beta)
\label{G1}  \\  
{\bf  G}^{(2)}_{ij}({\bf P},{\bf S},\boldsymbol{\Pi},\beta)
  &  =\int dp \frac{{\bf p}_{i}}{m_{i}}\frac{{\bf p}^T_{j}}{m_{j}}
    \Delta(p,{\bf P},{\bf S},\boldsymbol{\Pi},\beta)
\label{G2}\end{align}
where 
\begin{align}
  \label{eq:45}
  \Delta(p,{\bf P},{\bf S},\boldsymbol{\Pi},\beta)
  &\equiv
  \delta\left(\sum_{i}{\bf p}_{i}-{\bf P}\right)
  \delta\left(\sum_{i}[{\bf r}'_{i}]_\times\esc{\bf p}_{i}-{\bf S} \right)
  \delta\left(\sum_{i}\frac{\partial \hat{{\bf M}}}{\partial m_{i}{\bf r}_{i}}\esc{\bf p}_{i}-\boldsymbol{\Pi}\right)
e^{-\beta\sum_i^N\frac{{\bf p}_i^2}{2m_i}+\sum_i{\bf J}_i\esc\frac{{\bf p}_i}{m_i}}
\end{align}
and ${\bf  r}'_{i}={\bf r}_{i}-\hat{\bf R}$.
Observe that
\begin{align}
  \label{eq:18}
  {\bf  G}^{(1)}_i&=\frac{\partial G^{(0)}}{\partial {\bf J}_i}
                    \nonumber\\
    {\bf  G}^{(2)}_i&=\frac{\partial^2 G^{(0)}}{\partial {\bf J}_i\partial {\bf J}_j}
\end{align}
so $G^{(0)}$ acts as a generating function.

\subsection{Calculation of ${G}^0({\bf P},{\bf S},\boldsymbol{\Pi},\beta)$}
Let us translate now the general result (\ref{apomJ}) into momentum
variables by identifying from (\ref{eq:45}) the following
\begin{align}
  n&=3N
\nonumber\\
m&=9
\nonumber\\
x&\to {\bf P},{\bf S},\boldsymbol{\Pi}
\nonumber\\
y&\to {\bf p}_{1},\cdots {\bf p}_{N}
\nonumber\\
J&\to \frac{{\bf J}_{1}}{m_1},\cdots \frac{{\bf J}_{N}}{m_N}
\end{align}
The matrix  $A$ has the diagonal form
\begin{align}
A&\to \beta\left(\begin{array}{ccc}
\frac{1}{m_1}\mathbb{1}_3&\cdots&0\\
\vdots&\ddots&\vdots\\
0&\cdots&\frac{1}{m_N}\mathbb{1}_3
\end{array}\right)  
\end{align}
where $\mathbb{1}_3$ is the  $3\times3$ identity matrix. The inverse and determinant of $A$ are
\begin{align}
A^{-1}&\to \frac{1}{\beta}\left(\begin{array}{ccc}
m_1\mathbb{1}_3&\cdots&0\\
\vdots&\ddots&\vdots\\
0&\cdots&{m_N}\mathbb{1}_3
\end{array}\right)  
,\qquad \det A=\prod_{i}^{3N}\frac{\beta}{m_{i}}
\end{align}
The product
\begin{align}
  \label{eq:90}
  A^{-1}J \to \frac{1}{\beta}\left(\begin{array}{ccc}
m_1\mathbb{1}_3&\cdots&0\\
\vdots&\ddots&\vdots\\
0&\cdots&{m_N}\mathbb{1}_3
\end{array}\right)  \left(\begin{array}{c}
\frac{{\bf J}_1}{m_1}\\
\vdots\\
\frac{{\bf J}_N}{m_N}\\
\end{array}\right) =\frac{1}{\beta} \left(\begin{array}{c}
{\bf J}_1\\
\vdots\\
{\bf J}_N\\
\end{array}\right) 
\end{align}
so that
\begin{align}
  \label{eq:109b}
J^T    A^{-1}J \to \frac{1}{\beta}\sum_i\frac{1}{m_i}{\bf J}^2_i
\end{align}
The rectangular matrix $\Pi$ is best appreciated from its action,
\begin{align}
x=\Pi y &\to \left(
  \begin{array}{c}
{\bf P}\\
    {\bf S} \\
    \boldsymbol{\Pi}
  \end{array}
\right)=
\left(\begin{array}{cccc}
 \mathbb{1}_3  & \mathbb{1}_3& \cdots &\mathbb{1}_3\\
        {[}{\bf r}'_{1}{]}_\times&{[}{\bf r}'_{2}{]}_\times&\cdots &{[}{\bf r}'_{N}{]}_\times\\
        \frac{\partial \hat{{\bf M}}}{\partial m_{1}{\bf r}_{1}}
            &\frac{\partial \hat{{\bf M}}}{\partial m_{2}{\bf r}_{2}}&\cdots
                          &\frac{\partial \hat{{\bf M}}}{\partial m_{N}{\bf r}_{N}}
      \end{array}\right)
\left(
  \begin{array}{c}
{\bf p}_{1}\\
\vdots\\
{\bf p}_{N}
  \end{array}
\right)
\end{align}
Therefore, $\Pi$
depends only on particle positions so is constant  as  far  as  the momentum
variables are concerned. The product
\begin{align}
  \label{eq:50b}
  \Pi A^{-1}J &\to \frac{1}{\beta}
\left(\begin{array}{cccc}
 \mathbb{1}_3  & \mathbb{1}_3& \cdots &\mathbb{1}_3\\
        {[}{\bf r}'_{1}{]}_\times&{[}{\bf r}'_{2}{]}_\times&\cdots &{[}{\bf r}'_{N}{]}_\times\\
        \frac{\partial \hat{{\bf M}}}{\partial m_{1}{\bf r}_{1}}
            &\frac{\partial \hat{{\bf M}}}{\partial m_{2}{\bf r}_{2}}&\cdots
                          &\frac{\partial \hat{{\bf M}}}{\partial m_{N}{\bf r}_{N}}
      \end{array}\right)
\left(
  \begin{array}{c}
{\bf J}_{1}\\
\vdots\\
{\bf J}_{N}
  \end{array}
\right)
= \frac{1}{\beta}\left( \begin{array}{c}
\sum_i{\bf J}_i\\
\sum_i{[}{\bf r}'_{i}{]}_\times\esc{\bf J}_i \\
\sum_i\frac{\partial \hat{{\bf M}}}{\partial m_{i}{\bf r}_{i}}\esc{\bf J}_i
  \end{array}
\right)
\end{align}

The $9\times9$ matrix $ \Pi  A^{-1}\Pi^T$ in (\ref{overA}) is obtained
by block-wise matrix multiplication
\begin{align}
  \Pi A^{-1}\Pi^T&\to  \frac{1}{\beta}                 \left(\begin{array}{cccc}
                           \mathbb{1}_3& \mathbb{1}_3& \cdots &\mathbb{1}_3\\
                           {[}{\bf r}'_{1}{]}_\times&{[}{\bf r}'_{2}{]}_\times&\cdots &{[}{\bf r}'_{N}{]}_\times\\
                           \frac{\partial \hat{{\bf M}}}{\partial m_{1}{\bf r}_{1}}
                                      &\frac{\partial \hat{{\bf M}}}{\partial m_{2}{\bf r}_{2}}&\cdots
                          &\frac{\partial \hat{{\bf M}}}{\partial m_{N}{\bf r}_{N}}
\end{array}\right)\left(\begin{array}{ccc}
m_1\mathbb{1}_3&\cdots&0\\
\vdots&\ddots&\vdots\\
0&\cdots&{m_N}\mathbb{1}_3
\end{array}\right)  
\left(\begin{array}{ccc}
        \mathbb{1}_3  & {[}{\bf r}'_{1}{]}^T_\times&        \left(\frac{\partial \hat{{\bf M}}}{\partial m_{1}{\bf r}_{1}}\right)^T
        \\
\vdots & \vdots& \vdots\\
        \mathbb{1}_3&{[}{\bf r}'_{N}{]^T}_\times&
                \left( \frac{\partial \hat{{\bf M}}}{\partial m_{N}{\bf r}_{N}}\right)^T
\end{array}\right)
\nonumber\\
&=\frac{1}{\beta}\left(\begin{array}{cccc}
 \mathbb{1}_3  & \mathbb{1}_3& \cdots &\mathbb{1}_3\\
{[}{\bf r}'_{1}{]}_\times&{[}{\bf r}'_{2}{]}_\times&\cdots &{[}{\bf r}'_{N}{]}_\times
\\
                                                       \frac{\partial \hat{{\bf M}}}{\partial m_{1}{\bf r}_{1}}
            &\frac{\partial \hat{{\bf M}}'}{\partial m_{2}{\bf r}_{2}}&\cdots
                          &\frac{\partial \hat{{\bf M}}}{\partial m_{N}{\bf r}_{N}}\end{array}\right)
\left(\begin{array}{ccc}
m_1 \mathbb{1}_3  & m_1{[}{\bf r}'_{1}{]}_\times &        \frac{\partial \hat{{\bf M}}}{\partial{\bf r}_{1}}\\
\vdots & \vdots& \vdots\\
m_N\mathbb{1}_3&m_N{[}{\bf r}'_{N}{]}_\times &        \frac{\partial \hat{{\bf M}}}{\partial{\bf r}_{N}}
\end{array}\right)
=\frac{1}{\beta}\left(\begin{array}{ccc}
M \mathbb{1}_3  & 0& 0\\
          0&\hat{\bf I}& 0\\
         0& 0& \hat{\bf T}\end{array}\right)
               \label{Block}
\end{align}
where  (\ref{trans2}) and  (\ref{dsym})  have been  used,  as well  as
(\ref{eq:42b}).   In   fact,  the  definition  of   $\hat{\bf  T}$  in
(\ref{eq:42b})  has its  origin  in (\ref{Block}).  Using this  result
gives $\overline{A}$ and its determinant as
\begin{align}
  \label{eq:30b}
  \overline{A}=\left[  \Pi A^{-1}\Pi^T\right]^{-1}
  &\to
    \beta\left(\begin{array}{ccc}
M^{-1} \mathbb{1}_3  & 0& 0\\
          0&\hat{\bf I}^{-1}& 0\\
         0& 0& \hat{\bf T}^{-1}\end{array}\right)
,\qquad \det\overline{A}=
\frac{\beta^9}{M^{3}\left(\det\hat{\bf I}\right)\left(\det\hat{\bf T}\right)}
\end{align}
and the quadratic form becomes
\begin{align}
  \label{eq:30c}
&(x-\Pi A^{-1}J)^T\overline{A}(x-\Pi A^{-1}J)
   \nonumber\\
  \to &\beta\left(
       \left({\bf P}-\frac{1}{\beta}\sum_i{\bf J}_i\right)^T,
       \left({\bf S}- \frac{1}{\beta}\sum_i{[}{\bf r}'_{i}{]}_\times\esc{\bf J}_i\right)^T,
        \left(\boldsymbol{\Pi}
        -\frac{1}{\beta}   \sum_i\frac{\partial \hat{{\bf M}}}{\partial m_{i}{\bf r}_{i}} \esc{\bf J}_i\right)^T
        \right)
        \nonumber\\
  &\times\left(\begin{array}{ccc}
M^{-1} \mathbb{1}_3  & 0& 0\\
          0&\hat{\bf I}^{-1}& 0\\
        0& 0& \hat{\bf T}^{-1}\end{array}\right)
\left(\begin{array}{c}
{\bf P}-\frac{1}{\beta}\sum_i{\bf J}_i
        \\
{\bf S}- \frac{1}{\beta}\sum_i{[}{\bf r}'_{i}{]}_\times\esc{\bf J}_i
        \\
        \boldsymbol{\Pi}
        - \frac{1}{\beta} \sum_i \frac{\partial \hat{{\bf M}}}{\partial m_{i}{\bf r}_{i}}   \esc{\bf J}_i
      \end{array}\right)
  \nonumber\\
  &=   \beta\left({\bf P}-\frac{1}{\beta}\sum_i{\bf J}_i\right)^T
    M^{-1}    \left({\bf P}-\frac{1}{\beta}\sum_i{\bf J}_i\right)
    +    \beta   \left({\bf S}- \frac{1}{\beta}\sum_i{[}{\bf r}'_{i}{]}_\times\esc{\bf J}_i\right)^T
    \esc\hat{\bf I}^{-1} \esc
    \left({\bf S}- \frac{1}{\beta}\sum_i{[}{\bf r}'_{i}{]}_\times\esc{\bf J}_i\right)
    \nonumber\\
  &+   \beta   \left(\boldsymbol{\Pi}
    - \frac{1}{\beta}\sum_i \frac{\partial \hat{{\bf M}}}{\partial m_{i}{\bf r}_{i}} \esc{\bf J}_i\right)^T
    \esc\hat{\bf T}^{-1}\esc       \left(\boldsymbol{\Pi}
    -  \frac{1}{\beta}       \sum_i \frac{\partial \hat{{\bf M}}}{\partial m_{i}{\bf r}_{i}} \esc{\bf J}_i\right)
\end{align}
Therefore, translating (\ref{apomJ}) gives
\begin{align}
  \label{eq:105}
  G^{(0)}({\bf P},{\bf S},\boldsymbol{\Pi},\beta)
  &=\frac{( \frac{2\pi}{\beta})^{3(N-3)/2}\prod^N_i(m_i)^{3/2}}
    {\sqrt{M^{3}\det\hat{\bf I}\det\hat{\bf T}}}
    \exp\left\{\frac{1}{2\beta}\sum_i\frac{1}{m_i}{\bf J}^2_i\right\}
    \nonumber\\
  & \times  
    \exp\left\{-\frac{\beta}{2}
    \left({\bf P} -\frac{1}{\beta}\sum_i{\bf J}_i\right)^T
    M^{-1}
    \left({\bf P}-\frac{1}{\beta}\sum_i{\bf J}_i\right)\right\}
    \nonumber\\
  & \times  
    \exp\left\{-\frac{\beta}{2}
    \left({\bf S}-  \frac{1}{\beta}\sum_i{[}{\bf r}'_{i}{]}_\times\esc{\bf J}_i\right)^T
    \esc\hat{\bf I}^{-1}\esc
    \left({\bf S}- \frac{1}{\beta}\sum_i{[}{\bf r}'_{i}{]}_\times\esc{\bf J}_i\right)\right\}
    \nonumber\\
  & \times  
    \exp\left\{   -\frac{\beta}{2}
    \left(\boldsymbol{\Pi} - \frac{1}{\beta}\sum_i \frac{1}{m_i}\frac{\partial \hat{{\bf M}}}{\partial{\bf r}_{i}} \esc{\bf J}_i\right)^T
    \esc\hat{\bf T}^{-1}\esc
    \left(\boldsymbol{\Pi} - \frac{1}{\beta} \sum_i \frac{1}{m_i}\frac{\partial \hat{{\bf M}}}{\partial{\bf r}_{i}}\esc{\bf J}_i\right)
\right\}                                                 
\end{align}
As a final remark and for future reference,  the  change   of  variables (\ref{eq:100})
in the present case takes the form
\begin{align}
  \label{eq:52b}
  \left(\begin{array}{c}{\bf p}'_1\\\vdots\\{\bf p}'_N\end{array}\right)
  &=\left(\begin{array}{c}{\bf p}_1\\\vdots\\{\bf p}_N\end{array}\right)
    -\frac{1}{\beta}\left(\begin{array}{ccc}
m_1\mathbb{1}_3&\cdots&0\\
\vdots&\ddots&\vdots\\
0&\cdots&{m_N}\mathbb{1}_3
\end{array}\right)  
\left(\begin{array}{ccc}
        \mathbb{1}_3  & {[}{\bf r}'_{1}{]}_\times^T&        \left(\frac{\partial \hat{{\bf M}}}{\partial m_{1}{\bf r}_{1}}\right )^T
        \\
\vdots & \vdots& \vdots\\
        \mathbb{1}_3&{[}{\bf r}'_{N}{]}_\times^T&
                          \left(\frac{\partial \hat{{\bf M}}}{\partial m_{N}{\bf r}_{N}}\right )^T
\end{array}\right)  \beta\left(\begin{array}{ccc}
M^{-1} \mathbb{1}_3  & 0& 0\\
          0&\hat{\bf I}^{-1}& 0\\
         0& 0& \hat{\bf T}^{-1}\end{array}\right)
\left(  \begin{array}{c}
{\bf P}\\
    {\bf S} \\
    \boldsymbol{\Pi}
  \end{array}
\right)\end{align}
so that
\begin{align}
 {\bf p}_{i}'
  &={\bf p}_{i} - m_{i}{\bf V} + m_{i} ({\bf r}_{i}-{\bf R})\times\hat{\boldsymbol{\Omega}}
    -\frac{\partial \hat{{\bf M}}}{\partial {\bf r}_{i}}\esc\hat{\boldsymbol{\nu}}
\label{BodyFrame}
\end{align}
where the dot product in the last term is over the indices of $\hat{\bf M}$ and $\hat{\boldsymbol{\nu}}$, and the different velocities are
\begin{align}
  {\bf V}
  &=M^{-1}{\bf P}
    \nonumber\\
  \hat{\boldsymbol{\Omega}}
  &=\hat{\bf I}^{-1}\esc{\bf S}
    \nonumber\\
  \hat{\boldsymbol{\nu}}&=\hat{\bf T}^{-1}\esc\boldsymbol{\Pi}
    \label{velos}
\end{align}
This motivated the change of variables (\ref{W:BodyFrame}) introduced in Sec. \ref{App:Entropy}.

\subsection{Calculation of ${\bf G}^{(1)}({\bf P},{\bf S},\boldsymbol{\Pi},\beta)$, ${\bf G}^2({\bf P},{\bf S},\boldsymbol{\Pi},\beta)$}
In component form, (\ref{eq:105}) becomes (repeated index summation convention used)
  \begin{align}
    \label{eq:249}
      G^{(0)}({\bf P},{\bf S},\boldsymbol{\Pi},\beta)
  &=\frac{(\frac{2\pi}{\beta})^{3(N-3)/2}\prod^N_i(m_i)^{3/2}}
    {\sqrt{M^{3}\det\hat{\bf I}\det\hat{\bf T}}}
    \exp\left\{\frac{1}{2\beta }\sum_i\frac{1}{ m_i}{\bf J}^\mu_i{\bf J}^\mu_i\right\}
    \nonumber\\
  & \times  
    \exp\left\{-\frac{\beta}{2M}  
    \left({\bf P}^\mu-\frac{1}{\beta}\sum_i{\bf J}^\mu_i\right)
    \left({\bf P}^\mu-\frac{1}{\beta}\sum_j{\bf J}^\mu_j\right)\right\}
    \nonumber\\
  & \times  
    \exp\left\{-\frac{\beta}{2}
    \left({\bf S}^\mu- \frac{1}{\beta}\sum_i{[}{\bf r}'_{i}{]}_\times^{\mu\sigma}{\bf J}^\sigma_i\right)
    \left[\hat{\bf I}^{-1}\right]^{\mu\nu}
    \left({\bf S}^\nu- \frac{1}{\beta}\sum_j{[}{\bf r}'_{j}{]}^{\nu\sigma'}_\times{\bf J}^{\sigma'}_j\right)
    \right\}
    \nonumber\\
  & \times  
    \exp\left\{   -\frac{\beta}{2}
    \left({\Pi}_\mu -\frac{1}{\beta}\sum_i \frac{1}{m_i}\frac{\partial \hat{M}_\mu}{\partial{\bf r}_{i}^{\sigma}} {\bf J}^\sigma_i\right)
    \left[\hat{\bf T}^{-1}\right]^{\mu\nu}
    \left({\Pi}_\nu- \frac{1}{\beta}\sum_j \frac{1}{m_j}\frac{\partial \hat{M}_\nu}{\partial {\bf r}^{\sigma'}_{j}}{\bf J}^{\sigma'}_j\right)
\right\}                                                 
\end{align}
From (\ref{eq:18}) and (\ref{eq:249}) we have
\begin{align}
  \label{eq:242b}
{\bf  G}^{(1)\gamma}_i
  =\frac{\partial G^{(0)}}{\partial {\bf J}^\gamma_i}
  =     G^{(0)}&\left[
  \frac{1}{\beta m_i}{\bf J}_i^\gamma
   + M^{-1}    \left({\bf P}^\gamma-\frac{1}{\beta}\sum_{i'}{\bf J}^\gamma_{i'}\right)
- {[}{\bf r}'_{i}{]}^{\gamma\mu}_\times \left[\hat{\bf I}^{-1}\right]^{\mu\nu}\left({\bf S}^\nu
    -\frac{1}{\beta}\sum_{i'}{[}{\bf r}'_{i'}{]}^{\nu\sigma'}_\times{\bf J}^{\sigma'}_{i'}\right)
    \right.
    \nonumber\\
  &\left.+ \frac{1}{m_i}
    \frac{\partial \hat{M}_\mu}{\partial{\bf r}^\gamma_{i}}\left[\hat{\bf T}^{-1}\right]^{\mu\nu}
    \left({\Pi}_\nu
    - \frac{1}{\beta} \sum_{i'} \frac{1}{m_{i'}}\frac{\partial \hat{M}_\nu}{\partial {\bf r}^\sigma_{i'}}
    {\bf J}^\sigma_{i'}\right)
\right]
\end{align}
The second derivative term is
\begin{align}
  \label{eq:271}
  {\bf  G}^{(2)\gamma\gamma'}_{ij}
  &  =\frac{\partial^2G^{(0)}}{\partial {\bf J}^\gamma_i\partial {\bf J}_j^{\gamma'}}
    =\frac{\partial{\bf G}_i^{(1)\gamma}}{\partial {\bf J}_j^{\gamma'}}
      \nonumber\\
&=  \frac{\partial G^{(0)}}{\partial {\bf J}_j^{\gamma'}}   \left[
    \frac{1}{\beta m_i}{\bf J}_i^\gamma
    + M^{-1}    \left({\bf P}^\gamma-\frac{1}{\beta}\sum_{i'}{\bf J}^\gamma_{i'}\right)
    -  {[}{\bf r}'_{i}{]}^{\gamma\mu}_\times \left[\hat{\bf I}^{-1}\right]^{\mu\nu}\left({\bf S}^\nu
    -\frac{1}{\beta}\sum_{i'}{[}{\bf r}'_{i'}{]}^{\nu\sigma'}_\times{\bf J}^{\sigma'}_{i'}\right)
    \right.
    \nonumber\\
  &\left.+\frac{1}{m_i}
    \frac{\partial \hat{M}_\mu}{\partial{\bf r}^\gamma_{i}}\left[\hat{\bf T}^{-1}\right]^{\mu\nu}
    \left({\Pi}_\nu
    - \frac{1}{\beta} \sum_{i'} \frac{1}{m_{i'}}\frac{\partial \hat{M}_\nu}{\partial {\bf r}^\sigma_{i'}}
    {\bf J}^\sigma_{i'}\right)
    \right]
    \nonumber\\
  &+G^{(0)}\frac{\partial}{\partial {\bf J}_j^{\gamma'}}\left[
    \frac{1}{\beta m_i}{\bf J}_i^\gamma
    + M^{-1}    \left({\bf P}^\gamma-\frac{1}{\beta}\sum_{i'}{\bf J}^\gamma_{i'}\right)
    -          {[}{\bf r}'_{i}{]}^{\gamma\mu}_\times \left[\hat{\bf I}^{-1}\right]^{\mu\nu}\left({\bf S}^\nu
    -\frac{1}{\beta}\sum_{i'}{[}{\bf r}'_{i'}{]}^{\nu\sigma'}_\times{\bf J}^{\sigma'}_{i'}\right)
    \right.
    \nonumber\\
  &\left.+ \frac{1}{m_i}
    \frac{\partial \hat{M}_\mu}{\partial{\bf r}^\gamma_{i}}\left[\hat{\bf T}^{-1}\right]^{\mu\nu}
    \left({\Pi}_\nu
    - \frac{1}{\beta} \sum_{i'} \frac{1}{m_{i'}}\frac{\partial \hat{M}_\nu}{\partial {\bf r}^\sigma_{i'}}
    {\bf J}^\sigma_{i'}\right)
    \right]
\end{align}
that is\begin{align}
      \label{eq:282}
  {\bf  G}^{(2)\gamma\gamma'}_{ij}&= G^{(0)}  \left[
    \frac{1}{\beta m_i}{\bf J}_i^\gamma
    + M^{-1}    \left({\bf P}^\gamma-\frac{1}{\beta}\sum_{i'}{\bf J}^\gamma_{i'}\right)
    -{[}{\bf r}'_{i}{]}^{\gamma\mu}_\times \left[\hat{\bf I}^{-1}\right]^{\mu\nu}\left({\bf S}^\nu
    -\frac{1}{\beta}\sum_{i'}{[}{\bf r}'_{i'}{]}^{\nu\sigma'}_\times{\bf J}^{\sigma'}_{i'}\right)
    \right.
    \nonumber\\
  &\left.+\frac{1}{m_i}
    \frac{\partial \hat{M}_\mu}{\partial{\bf r}^\gamma_{i}}\left[\hat{\bf T}^{-1}\right]^{\mu\nu}
    \left({\Pi}_\nu
    - \frac{1}{\beta} \sum_{i'} \frac{1}{m_{i'}}\frac{\partial \hat{M}_\nu}{\partial {\bf r}^\sigma_{i'}}
    {\bf J}^\sigma_{i'}\right)
    \right]
    \nonumber\\
  &\times\left[
 \frac{1}{\beta m_j}{\bf J}_j^{\gamma'}
    + M^{-1}    \left({\bf P}^{\gamma'}-\frac{1}{\beta}\sum_{i'}{\bf J}^{\gamma'}_{i'}\right)
    -          {[}{\bf r}'_{j}{]}^{\gamma'\mu'}_\times \hat{\bf I}^{-1}_{\mu'\nu'}\left({\bf S}^{\nu'}
    -\frac{1}{\beta}\sum_{i'}{[}{\bf r}'_{i'}{]}^{\nu'\sigma''}_\times{\bf J}^{\sigma''}_{i'}\right)
    \right.
    \nonumber\\
  &\left.+ \frac{1}{m_j}
    \frac{\partial \hat{M}_{\mu'}}{\partial {\bf r}^{\gamma'}_{j}}\hat{\bf T}^{-1}_{\mu'\nu'}
    \left({\Pi}_{\nu'}
    - \frac{1}{\beta} \sum_{i'} \frac{1}{m_{i'}}\frac{\partial \hat{M}_{\nu'}}{\partial {\bf r}^{\sigma'}_{i'}}
    {\bf J}^{\sigma'}_{i'}\right)
    \right]
    \nonumber\\
  &+G^{(0)}\left[
    \frac{1}{\beta m_i}\delta_{ij}\delta_{\gamma\gamma'}
   + M^{-1}    \left(-\frac{1}{\beta}\sum_{i'}\frac{\partial}{\partial {\bf J}_j^{\gamma'}}{\bf J}^\gamma_{i'}\right)
    -{[}{\bf r}'_{i}{]}^{\gamma\mu}_\times \left[\hat{\bf I}^{-1}\right]^{\mu\nu}
  \left(-  \frac{1}{\beta}\sum_{i'}{[}{\bf r}'_{i'}{]}^{\nu\sigma'}_\times\frac{\partial}{\partial {\bf J}_j^{\gamma'}}{\bf J}^{\sigma'}_{i'}\right)
    \right.
    \nonumber\\
  &\left.+ \frac{1}{m_i}
    \frac{\partial \hat{M}_\mu}{\partial{\bf r}^\gamma_{i}}\left[\hat{\bf T}^{-1}\right]^{\mu\nu}
    \left(  - \frac{1}{\beta} \sum_{i'} \frac{1}{m_{i'}}\frac{\partial \hat{M}_\nu}{\partial {\bf r}^\sigma_{i'}}
   \frac{\partial}{\partial {\bf J}_j^{\gamma'}} {\bf J}^\sigma_{i'}\right)
    \right]
\end{align}
which simplifies to
\begin{align}
  \label{eq:343}
    {\bf  G}^{(2)\gamma\gamma'}_{ij}
  &=G^{(0)}  \left \{ \left[
    \frac{1}{\beta m_i}{\bf J}_i^\gamma
    + M^{-1}    \left({\bf P}^\gamma-\frac{1}{\beta}\sum_{i'}{\bf J}^\gamma_{i'}\right)
    -{[}{\bf r}'_{i}{]}^{\gamma\mu}_\times \hat{\bf I}^{-1}_{\mu\nu}\left({\bf S}^\nu
    -\frac{1}{\beta}\sum_{i'}{[}{\bf r}'_{i'}{]}^{\nu\sigma'}_\times{\bf J}^{\sigma'}_{i'}\right)
    \right.\right.
    \nonumber\\
  &\left.+ \frac{1}{m_i}
    \frac{\partial \hat{M}_\mu}{\partial{\bf r}^\gamma_{i}}\hat{\bf T}^{-1}_{\mu\nu}
    \left({\Pi}_\nu
    - \frac{1}{\beta} \sum_{i'} \frac{1}{m_{i'}}\frac{\partial \hat{M}_\nu}{\partial {\bf r}^\sigma_{i'}}
    {\bf J}^\sigma_{i'}\right)
    \right]
    \nonumber\\
  &\times\left[
    \frac{1}{\beta m_j}{\bf J}_j^{\gamma'}
    + M^{-1}    \left({\bf P}^{\gamma'}-\frac{1}{\beta}\sum_{i'}{\bf J}^{\gamma'}_{i'}\right)
    -          {[}{\bf r}'_{j}{]}^{\gamma'\mu'}_\times \hat{\bf I}^{-1}_{\mu'\nu'}\left({\bf S}^{\nu'}
    -\frac{1}{\beta}\sum_{i'}{[}{\bf r}'_{i'}{]}^{\nu'\sigma''}_\times{\bf J}^{\sigma''}_{i'}\right)
    \right.
    \nonumber\\
  &\left.+ \frac{1}{m_j}
    \frac{\partial \hat{M}_{\mu'}}{\partial {\bf r}^{\gamma'}_{j}}\hat{\bf T}^{-1}_{\mu'\nu'}
    \left({\Pi}_{\nu'}
    - \frac{1}{\beta} \sum_{i'} \frac{1}{m_{i'}}\frac{\partial \hat{M}_{\nu'}}{\partial {\bf r}^{\sigma'}_{i'}}
    {\bf J}^{\sigma'}_{i'}\right)
    \right]
    \nonumber\\
  &+\left .
    \frac{1}{\beta m_i}\delta_{ij}\delta_{\gamma\gamma'}
    -\frac{1}{\beta} M^{-1}    \delta_{\gamma\gamma'}
    +\frac{1}{\beta}{[}{\bf r}'_{i}{]}^{\gamma\mu}_\times \hat{\bf I}^{-1}_{\mu\nu} {[}{\bf r}'_{j}{]}^{\nu\gamma'}_\times
    -\frac{1}{\beta m_i m_j} 
    \frac{\partial \hat{M}_\mu}{\partial{\bf r}^\gamma_{i}}\hat{\bf T}^{-1}_{\mu\nu}
    \frac{\partial \hat{M}_\nu}{\partial {\bf r}^{\gamma'}_{j}}
    \right \}
\end{align}
Evaluated at ${\bf J}=0$ these quantities become
\begin{align}
  \label{eq:244}
  G^{(0)}({\bf P},{\bf S},\boldsymbol{\Pi},\beta)
  &=\frac{( \frac{2\pi}{\beta})^{3(N-3)/2}\prod^N_i(m_i)^{3/2}}
    {\sqrt{M^{3}\det\hat{\bf I}\det\hat{\bf T}}}
    \exp\left\{-\beta\frac{{\bf P}^2}{2M} -\frac{\beta}{2}        {\bf S}^T \esc\hat{\bf I}^{-1} \esc     {\bf S}  -\frac{\beta}{2}   \boldsymbol{\Pi}^T
    \esc\hat{\bf T}^{-1}\esc    \boldsymbol{\Pi}
    \right\}                                                 
    \nonumber\\
{\bf  G}^{(1)\gamma}_i({\bf P},{\bf S},\boldsymbol{\Pi},\beta)
&  =    G^{(0)}\left[
    M^{-1}    {\bf P}^\gamma
-          {[}{\bf r}'_{i}{]}^{\gamma\mu}_\times \hat{\bf I}^{-1}_{\mu\nu}{\bf S}^\nu
+    \frac{1}{m_i}\frac{\partial \hat{M}_\mu}{\partial{\bf r}^\gamma_{i}}\hat{\bf T}^{-1}_{\mu\nu}
  {\Pi}_\nu
\right]
    \nonumber\\
    {\bf  G}^{(2)\gamma\gamma'}_{ij}({\bf P},{\bf S},\boldsymbol{\Pi},\beta)
  &=G^{(0)}  \left\{ \left[
     M^{-1}    {\bf P}^\gamma
    -{[}{\bf r}'_{i}{]}^{\gamma\mu}_\times \hat{\bf I}^{-1}_{\mu\nu}{\bf S}^\nu
    +\frac{1}{m_i}\frac{\partial \hat{M}_\mu}{\partial{\bf r}^\gamma_{i}}\hat{\bf T}^{-1}_{\mu\nu}
   {\Pi}_\nu
       \right]\right .
    \nonumber\\
  &\times\left[
        M^{-1}    {\bf P}^{\gamma'}
    -          {[}{\bf r}'_{j}{]}^{\gamma'\mu'}_\times \hat{\bf I}^{-1}_{\mu'\nu'}{\bf S}^{\nu'}
    +   \frac{1}{m_j}\frac{\partial \hat{M}_{\mu'}}{\partial{\bf r}^{\gamma'}_{i}}\hat{\bf T}^{-1}_{\mu'\nu'}
    {\Pi}_{\nu'}
        \right]
    \nonumber\\
  &+\left .\frac{1}{\beta}\left[
    \frac{1}{m_i}\delta_{ij}\delta_{\gamma\gamma'}
   - M^{-1}    \delta_{\gamma\gamma'}
    +{[}{\bf r}'_{i}{]}^{\gamma\mu}_\times \hat{\bf I}^{-1}_{\mu\nu} {[}{\bf r}'_{j}{]}^{\nu\gamma'}_\times
    - \frac{1}{m_i m_j}\frac{\partial \hat{M}_\mu}{\partial{\bf r}^\gamma_{i}}\hat{\bf T}^{-1}_{\mu\nu}
    \frac{\partial \hat{M}_\nu}{\partial {\bf r}^{\gamma'}_{j}}
    \right]\right \}
\end{align}
Evaluated at rest, that is ${\bf P}={\bf S}=\boldsymbol{\Pi}=0$, these quantities become
\begin{align}
  \label{eq:220}
  G^{(0)}
  &=\frac{( \frac{2\pi}{\beta})^{3(N-3)/2}\prod^N_i(m_i)^{3/2}}
    {\sqrt{M^{3}\det\hat{\bf I}\det\hat{\bf T}}}
    \nonumber\\
\frac{1}{G^{(0)}}  \int dp {\bf v}^\mu_{i}
    \Delta(p,{\bf 0},{\bf 0},{\bf 0},\beta)
&=  0
                                              \nonumber\\
  \frac{1}{G^{(0)}}  \int dp {\bf v}^\mu_{i}{\bf v}_j^\nu
    \Delta(p,{\bf 0},{\bf 0},{\bf 0},\beta)
&= \frac{1}{\beta}\left[
    \frac{1}{m_i}\delta_{ij}\delta_{\mu\nu}
    - M^{-1}    \delta_{\mu\nu}
    +{[}{\bf r}'_{i}{]}^{\mu\mu'}_\times \hat{\bf I}^{-1}_{\mu'\nu'} {[}{\bf r}'_{j}{]}^{\nu'\nu}_\times
    - \frac{1}{m_i m_j}\frac{\partial \hat{M}_{\mu'}}{\partial{\bf r}^\mu_{i}}\hat{\bf T}^{-1}_{\mu'\nu'}
 \frac{\partial \hat{M}_{\nu'}}{\partial {\bf r}^{\nu}_{j}}
    \right]
\end{align}

\color{black}
\subsection{Sectioned Gaussian integrals for equilibrium averages}
\label{App:Gaussian}
The  results   of  this  subsection   allow  us  to   demonstrate  the
equipartition theorem in  Sec. \ref{App:Equipartition}. We compute
sectioned Gaussians similar  to the those in the  previous section but
with no restriction on the dilational momentum, that is
\begin{align}
  g^{(0)}({\bf P},{\bf S},\beta)
  &=\int dp \Delta(p,{\bf P},{\bf S},\beta)
    \label{G0b}
    \\
{\bf  g}^{(1)}_i({\bf P},{\bf S},\beta)
  &=\int dp \frac{{\bf p}_{i}}{m_{i}}  \Delta(p,{\bf P},{\bf S},\beta)
\label{G1b}  \\  
{\bf  g}^{(2)}_{ij}({\bf P},{\bf S},\beta)
  &  =\int dp \frac{{\bf p}_{i}}{m_{i}}\frac{{\bf p}^T_{j}}{m_{j}}
    \Delta(p,{\bf P},{\bf S},,\beta)
\label{G2b}\end{align}
where
\begin{align}
  \label{eq:45b}
  \Delta(p,{\bf P},{\bf S},\beta)
  &\equiv
  \delta\left(\sum_{i}{\bf p}_{i}-{\bf P}\right)
  \delta\left(\sum_{i}[{\bf r}'_{i}]_\times\esc{\bf p}_{i}-{\bf S} \right)
e^{-\beta\sum_i^N\frac{{\bf p}_i^2}{2m_i}+\sum_i{\bf J}_i\esc\frac{{\bf p}_i}{m_i}}
\end{align}
The results for ${\bf J}=0$ can be obtained by integrating (\ref{eq:244}) over
$\boldsymbol{\Pi}$, so that
\begin{align}
g^{(0)}({\bf P},{\bf S},\beta) &=
\int d\boldsymbol{\Pi}\; G^{(0)}({\bf P},{\bf S},\boldsymbol{\Pi},\beta)=
\frac{(\frac{2\pi}{\beta})^{3(N-2)/2}\prod^N_i(m_i)^{3/2}}
{\sqrt{M^{3}\det\hat{\bf I}}}
\exp\left\{-\beta\frac{{\bf P}^2}{2M} -
\frac{\beta}{2}{\bf S}^T \esc\hat{\bf I}^{-1} \esc{\bf S}\right\}
\nonumber\\
{\bf g}^{(1)\gamma}_i({\bf P},{\bf S},\beta) &=
\int d\boldsymbol{\Pi}\;{\bf G}^{(1)\gamma}_i({\bf P},{\bf S},\boldsymbol{\Pi},\beta)
= g^{(0)}\left[ M^{-1}{\bf P}^\gamma -
[{\bf r}'_i]^{\gamma\mu}_\times\hat{\bf I}^{-1}_{\mu\nu}{\bf S}^\nu
\right]
\nonumber\\
{\bf  g}^{(2)\gamma\gamma'}_{ij}({\bf P},{\bf S},\beta) &=
\int d\boldsymbol{\Pi}\;{\bf G}^{(2)\gamma\gamma'}_{ij}({\bf P},{\bf S},\boldsymbol{\Pi},\beta)
=g^{(0)}  \left\{ \left[ M^{-1}{\bf P}^\gamma -
[{\bf r}'_i]^{\gamma\mu}_\times\hat{\bf I}^{-1}_{\mu\nu}{\bf S}^\nu
       \right]\right .
\nonumber\\
&\quad\times\left[ M^{-1}{\bf P}^{\gamma'} -
[{\bf r}'_j]^{\gamma'\mu'}_\times\hat{\bf I}^{-1}_{\mu'\nu'}{\bf S}^{\nu'}
        \right]
    \nonumber\\
  &\quad+\left .\frac{1}{\beta}\left[
\frac{1}{m_i}\delta_{ij}\delta_{\gamma\gamma'} - M^{-1}\delta_{\gamma\gamma'}
+[{\bf r}'_i]^{\gamma\mu}_\times\hat{\bf I}^{-1}_{\mu\nu}[{\bf r}'_j]^{\nu\gamma'}_\times
    \right]\right \}
\nonumber\\
&- g^{(0)} \frac{1}{\beta m_i m_j}\frac{\partial \hat{M}_\mu}{\partial {\bf r}^\gamma_i}
\hat{\bf T}^{-1}_{\mu\nu}
\frac{\partial \hat{M}_\nu}{\partial{\bf r}^{\gamma'}_j}
+\frac{1}{m_i m_j}\frac{\partial\hat{M}_\mu}{\partial{\bf r}^\gamma_{i}}
\hat{\bf T}^{-1}_{\mu\nu}
\frac{\partial \hat{M}_{\mu'}}{\partial{\bf r}^{\gamma'}_i}
\hat{\bf T}^{-1}_{\mu'\nu'}
\int d\boldsymbol{\Pi}\;\boldsymbol{\Pi}_\nu\;\boldsymbol{\Pi}_{\nu'}
G^{(0)}({\bf P},{\bf S},\boldsymbol{\Pi},\beta)
\nonumber\\
&=g^{(0)}  \left\{ \left[ M^{-1}{\bf P}^\gamma -
[{\bf r}'_i]^{\gamma\mu}_\times\hat{\bf I}^{-1}_{\mu\nu}{\bf S}^\nu
\right] \left[ M^{-1}{\bf P}^{\gamma'} -
[{\bf r}'_j]^{\gamma'\mu'}_\times\hat{\bf I}^{-1}_{\mu'\nu'}{\bf S}^{\nu'}
\right]\right .\nonumber\\
&\quad+\left .\frac{1}{\beta}\left[
\frac{1}{m_i}\delta_{ij}\delta_{\gamma\gamma'} - M^{-1}\delta_{\gamma\gamma'}
+[{\bf r}'_i]^{\gamma\mu}_\times\hat{\bf I}^{-1}_{\mu\nu}[{\bf r}'_j]^{\nu\gamma'}_\times
    \right]\right \}
\end{align}
where (\ref{Gaussian}) has been used to obtain the first result, terms with a
single factor of $\boldsymbol{\Pi}$ integrate to zero because the integrands
are odd functions, and in the last result, the final two terms cancel after
integration by parts is used in the integral to yield
$g^{(0)}\beta^{-1}\hat{\bf T}^{\nu'\nu}$.

Evaluated at rest, that is ${\bf P}={\bf S}=0$, we obtain
\begin{align}
  \label{eq:315}
    g^{(0)}
  &=\frac{( \frac{2\pi}{\beta})^{3(N-2)/2}\prod^N_i(m_i)^{3/2}}
    {\sqrt{M^{3}\det\hat{\bf I}}}
    \nonumber\\
  \frac{1}{g^{(0)}}  \int dp {\bf v}^\mu_{i}
  \Delta(p,{\bf 0},{\bf 0},\beta)
  &=  0
    \nonumber\\
  \frac{1}{g^{(0)}}  \int dp {\bf v}^\mu_{i}{\bf v}_j^\nu
  \Delta(p,{\bf 0},{\bf 0},\beta)
  &= \frac{1}{\beta}\left[
    \frac{1}{m_i}\delta_{ij}\delta_{\mu\nu}
    - M^{-1}    \delta_{\mu\nu}
    +{[}{\bf r}'_{i}{]}^{\mu\mu'}_\times \hat{\bf I}^{-1}_{\mu'\nu'} {[}{\bf r}'_{j}{]}^{\nu'\nu}_\times
    \right]
\end{align}
\subsection{The equipartition theorem}
\label{App:Equipartition}

The equipartition theorem gives an expression for the rest equilibrium
average  of the  translational  kinetic energy.  The rest  equilibrium
average is

   \begin{align}
  \label{eq:27}
  \llangle K^{\rm trans}\rrangle^{\cal E} =
  \llangle\sum_i\frac{m_i}{2}{\bf v}_i^2\rrangle^{\cal E}_{\rm rest}
  &=\int dz \rho_{\cal E}^{\rm rest}(z)\sum_i\frac{m_i}{2}{\bf v}_i^2
\end{align}
where the microcanonical ensemble is given in (\ref{W:eq:289}). Under equivalence of ensembles we
may use the canonical ensemble (\ref{eq:16}) instead, and then
\begin{align}
  \label{eq:71}
  \llangle K^{\rm trans}\rrangle^{\cal E} =
  \llangle\sum_i\frac{m_i}{2}{\bf v}_i^2\rrangle^{\cal E}_{\rm rest}
  &=\frac{1}{Z(\beta)}\int dz
    \delta\left(\hat{\bf R}(z) \right)
    \delta\left(\hat{\bf P}(z) \right)
    \delta\left(\hat{\bf S}(z) \right)
    e^{-\beta\hat{H}(z)}\sum_i\frac{m_i}{2}{\bf v}_i^2
\end{align}
This integral is of the form (\ref{eq:315}). 
Set $i=j$ and $\mu=\nu$, multiply by $m_i/2$, and sum over $\mu$ and all particles in (\ref{eq:315}) to obtain
\begin{align}
  \label{eq:424}
\frac{1}{g^{(0)}}  \int dp \sum_i\frac{m_i}{2}{\bf v}^\mu_{i}{\bf v}_i^\mu
      \Delta(p,{\bf 0},{\bf 0},\beta)
    &=   \frac{1}{2\beta}
      \left[
      \sum_im_i    \left(\frac{1}{m_i}    -     M^{-1}\right)\delta_{\mu\mu}
      +\sum_im_i     {[}{\bf r}'_{i}{]}_\times^{\mu\mu'}\left[\hat{\bf I}^{-1} \right]^{\mu'\nu'} {[}{\bf r}'_{i}{]}_\times^{\nu'\mu}
\right]
      \nonumber\\
    &=  \frac{1}{2\beta}
      \left(3N-3
      -\left[\hat{\bf I}\right]^{\mu'\nu'}\left[\hat{\bf I}^{-1} \right]^{\mu'\nu'}
\right)
=\frac{3(N-2)}{2\beta}
  \end{align}
where (\ref{eq:603}) have been used to simplify the
expression.
  Therefore, we obtain
\begin{align}
    \label{eq:168}
  \llangle K^{\rm trans}\rrangle^{\cal E}    =
  \llangle \sum_i\frac{m_i}{2}{\bf v}_i^2\rrangle^{\cal E}_{\rm rest}    =\frac{3(N-2)}{2\beta}
       \end{align}  where $\beta=\beta({\cal E})$ is the function given in (\ref{eq:180}). This
  theorem allows us to compute the temperature function $T^{\rm MT}({\cal E})$ in MD simulations.

  \subsection{The conditional covariance of angular velocity}
  \label{Sec:CondCovAng}
  The conditional covariance of the angular velocity $\hat{\boldsymbol{\omega}}$ is 
\begin{align}
  \label{eq:389}
    \llangle\hat{\boldsymbol{\omega}}^\alpha\hat{\boldsymbol{\omega}}^{\beta}
    \rrangle^{\boldsymbol{\Lambda}{{\bf M}}{\cal E}}_{\rm rest}
    \overset{(\ref{eq:136})}{=}
{B}^{-1}_{\alpha\alpha'}{B}^{-1}_{\beta\beta'}
    \llangle
    i{\cal L}\hat{\Lambda}_{\alpha'} i{\cal L}\hat{\Lambda}_{\beta'}
    \rrangle^{\boldsymbol{\Lambda}{{\bf M}}{\cal E}}_{\rm rest}
   ={B}^{-1}_{\alpha\alpha'}{B}^{-1}_{\beta\beta'}
    \llangle
    \sum_{i}\frac{\partial\hat{\Lambda}_{\alpha'}}{\partial{\bf r}_i^{\mu}}{\bf v}_i^\mu
    \sum_{j}\frac{\partial\hat{\Lambda}_{\beta'}}{\partial{\bf r}_j^{\nu}}{\bf v}_j^\nu
    \rrangle^{\boldsymbol{\Lambda}{{\bf M}}{\cal E}}_{\rm rest}
\end{align}
Using the equivalence of ensembles, the momentum part of the integrals
is given by (\ref{eq:220}), so that
\begin{align}
  \label{eq:391}
  &\llangle
    \sum_{i}\frac{\partial\hat{\Lambda}_{\alpha'}}{\partial{\bf r}_i^{\mu}}{\bf v}_i^\mu
    \sum_{j}\frac{\partial\hat{\Lambda}_{\beta'}}{\partial{\bf r}_j^{\nu}}{\bf v}_j^\nu
    \rrangle^{\boldsymbol{\Lambda}{{\bf M}}{\cal E}}_{\rm rest}
    \nonumber\\
  &= k_BT\llangle
    \sum_{ij}\frac{\partial\hat{\Lambda}_{\alpha'}}{\partial{\bf r}_i^{\mu}}
    \frac{\partial\hat{\Lambda}_{\beta'}}{\partial{\bf r}_j^{\nu}}
    \left[
    \frac{1}{m_i}\delta_{ij}\delta_{\mu\nu}
    - M^{-1}    \delta_{\mu\nu}
    +{[}{\bf r}'_{i}{]}^{\mu\mu'}_\times \hat{\bf I}^{-1}_{\mu'\nu'} {[}{\bf r}'_{j}{]}^{\nu'\nu}_\times
    - \frac{\partial \hat{M}_{\mu'}}{\partial m_{i}{\bf r}^\mu_{i}}\hat{\bf T}^{-1}_{\mu'\nu'}
    \frac{\partial \hat{M}_{\nu'}}{\partial m_{i'}{\bf r}^{\nu}_{j}}
    \right]      
    \rrangle^{\boldsymbol{\Lambda}{{\bf M}}{\cal E}}_{\rm rest}
\end{align}
This gives four terms that we discuss separately. The first term is
\begin{align}
  \label{eq:345}
\llangle  \sum_{ij}\frac{\partial\hat{\Lambda}_{\alpha'}}{\partial{\bf r}_i^{\mu}}
  \frac{\partial\hat{\Lambda}_{\beta'}}{\partial{\bf r}_j^{\nu}}
  \frac{1}{m_i}\delta_{ij}\delta_{\mu\nu}
  \rrangle^{\boldsymbol{\Lambda}{{\bf M}}{\cal E}}_{\rm rest}
  &=\llangle
    \sum_{i}     \frac{1}{m_i}
    \frac{\partial\hat{\Lambda}_{\alpha'}}{\partial{\bf r}_i^{\mu}}
    \frac{\partial\hat{\Lambda}_{\beta'}}{\partial{\bf r}_i^{\mu}}
    \rrangle^{\boldsymbol{\Lambda}{{\bf M}}{\cal E}}_{\rm rest}
\end{align}
The second term vanishes due to the translation invariance property (\ref{trans}).
The third term becomes, by virtue of (\ref{eq:255}) and the fact  the inertia tensor is a function of the CG variables,
\begin{align}
  \label{eq:402}
  & \llangle
    \sum_{i}\frac{\partial\hat{\Lambda}_{\alpha'}}{\partial{\bf r}_i^{\mu}}      {[}{\bf r}'_{i}{]}^{\mu\mu'}_\times 
    \sum_j\frac{\partial\hat{\Lambda}_{\beta'}}{\partial{\bf r}_j^{\nu}}{[}{\bf r}'_{j}{]}^{\nu'\nu}_\times
    \hat{\bf I}^{-1}_{\mu'\nu'} \rrangle^{\boldsymbol{\Lambda}{{\bf M}}{\cal E}}_{\rm rest}
    =-{B}_{\alpha'\mu'}{B}_{\beta'\nu'}{\bf I}^{-1}_{\mu'\nu'}
\end{align}
Finally,  we recognize  in the  fourth term  the matrix  $\hat{\bf H}$
defined  in  (\ref{eq:H})  that  vanishes  identically,  according  to
(\ref{eq:449}).
Collecting these results gives
\begin{align}
  \label{eq:392}
    \llangle\hat{\boldsymbol{\omega}}^\alpha\hat{\boldsymbol{\omega}}^{\beta}
    \rrangle^{\boldsymbol{\Lambda}{{\bf M}}{\cal E}}_{\rm rest}
  &=k_BT\left({B}^{-1}_{\alpha\alpha'}{B}^{-1}_{\beta\beta'}
    \llangle
    \sum_{i}     \frac{1}{m_i}
    \frac{\partial\hat{\Lambda}_{\alpha'}}{\partial{\bf r}_i^{\mu}}
    \frac{\partial\hat{\Lambda}_{\beta'}}{\partial{\bf r}_i^{\mu}}      \rrangle^{\boldsymbol{\Lambda}{{\bf M}}{\cal E}}_{\rm rest}  
    -{\bf I}^{-1}_{\alpha\beta}\right)
\end{align}
We recognize the matrix ${\bf L}$ defined in (\ref{eq:L}) in the first
term, and  note from (\ref{eq:225})  and (\ref{eq:229}) that it  is an
explicit function of CG variables, so that
\begin{align}
      \label{eq:415}
    \llangle\hat{\boldsymbol{\omega}}\hat{\boldsymbol{\omega}}^T
    \rrangle^{\boldsymbol{\Lambda}{{\bf M}}{\cal E}}_{\rm rest}
  &{=}k_BT\left({\bf L}-{\bf I}^{-1}\right)
\end{align}
The conditional covariance of the angular velocity in the principal axis frame is
\begin{align}
  \label{eq:00-S}
      \llangle\hat{\boldsymbol{\omega}}_0^\alpha\hat{\boldsymbol{\omega}}_0^{\beta}
    \rrangle^{\boldsymbol{\Lambda}{{\bf M}}{\cal E}}_{\rm rest}
  &=\left[e^{-[\boldsymbol{\Lambda}]_\times}\right]_{\alpha\alpha'}
    \left[e^{-[\boldsymbol{\Lambda}]_\times}\right]_{\beta\beta'}
    \llangle\hat{\boldsymbol{\omega}}^{\alpha'}\hat{\boldsymbol{\omega}}^{\beta'}
    \rrangle^{\boldsymbol{\Lambda}{{\bf M}}{\cal E}}_{\rm rest}
=kT(\mathbb{L}-\mathbb{I}^{-1})
\end{align}
where the last expression was obtained by inserting (\ref{eq:415}) into
(\ref{eq:00-S}), and using  (\ref{eq:225}).
